\documentstyle[11pt,mydefs,epsfig]{article}
\begin{document}

%
%
%
%
%
%
\message{FEYNMAN:  For generating Feynman Diagrams in LaTex}
\message{Mark 1.0 Last Altered by MJSL 2/89}
\setlength{\unitlength}{0.01pt}
\gdef\Feynmanlength{\setlength{\unitlength}{0.01pt}}  
\gdef\unlock{\catcode`\@=11}
\gdef\lock{\catcode`\@=12}
\global\newcount\LINETYPE
\global\newcount\LINEDIRECTION
\global\newcount\LINECONFIGURATION
\newcommand{\LTYPE}{\LINETYPE}
\newcommand{\LDIR}{\LINEDIRECTION}
\newcommand{\LCONFIG}{\LINECONFIGURATION}
\global\LINETYPE=1  \global\LINEDIRECTION=0  \global\LINECONFIGURATION=0
\global\newcount\fermion    \fermion=1
\global\newcount\scalar     \scalar=2
\global\newcount\photon     \photon=3
\global\newcount\gluon      \gluon=4
\global\newcount\SPECIAL    \SPECIAL=5
\gdef\N{0}  \gdef\NE{1}  \gdef\E{2}   \gdef\SE{3}
\gdef\S{4}  \gdef\SW{5}  \gdef\W{6}   \gdef\NW{7}
\global\newcount\REG            \global\REG=0
\global\newcount\FLIPPED        \global\FLIPPED=1
\global\newcount\CURLY          \global\CURLY=2
\global\newcount\FLIPPEDCURLY   \global\FLIPPEDCURLY=3
\global\newcount\FLAT           \global\FLAT=4
\global\newcount\FLIPPEDFLAT    \global\FLIPPEDFLAT=5
\global\newcount\CENTRAL        \global\CENTRAL=6
\global\newcount\FLIPPEDCENTRAL \global\FLIPPEDCENTRAL=7
\gdef\LONGPHOTON{6}             \gdef\FLIPPEDLONG{7}
\global\newcount\SQUASHEDGLUON  \global\SQUASHEDGLUON=8
\gdef\SQUASHED{\SQUASHEDGLUON}
%
\newcount\adjx \adjx=0
\newcount\adjy \adjy=0
\global\newdimen\BIGPHOTONS     \BIGPHOTONS=0pt  
\gdef\bigphotons{\global\BIGPHOTONS=12pt}
\global\newdimen\THICKPHOTONS     \THICKPHOTONS=0pt  
\global\newdimen\THICKPHOTONSWITCH    \THICKPHOTONSWITCH=0pt
\gdef\THICKPHOTONTEST{
\THICKPHOTONSWITCH=0pt
\ifdim\THICKPHOTONS=0pt \relax
  \else \ifnum\LTYPE=3
           \ifnum\LDIR=2 \THICKPHOTONSWITCH=1pt \fi 
           \ifnum\LDIR=6 \THICKPHOTONSWITCH=1pt \fi 
        \fi
\fi
}  
\gdef\THICKLINES{\thicklines  \THICKPHOTONS=1pt}
\gdef\THINLINES{\thinlines  \THICKPHOTONS=0pt}
\global\newcount\phantomswitch   \global\phantomswitch=0
\global\newcount\stemlength   \global\stemlength=275   
\global\newcount\absstemlength        
\global\newcount\stemlengthx          
\global\newcount\stemlengthy          
\newdimen\FRONTSTEM  \FRONTSTEM=0pt   
\newdimen\BACKSTEM   \BACKSTEM=0pt    
\newdimen\EITHERSTEM \EITHERSTEM=0pt  
\gdef\frontstemmed{\FRONTSTEM=1pt}            
\gdef\backstemmed{\BACKSTEM=1pt}              
\gdef\stemmed{\FRONTSTEM=1pt  \BACKSTEM=1pt}    
\global\newcount\arrowlength                
\global\newdimen\ATTIP   \global\ATTIP=0pt  
\global\newdimen\ATBASE  \global\ATBASE=1pt 
\global\newcount\unitboxnumber  
\global\newcount\unitboxnumberpo  
\global\newcount\particlelengthx  
\gdef\plengthx{\particlelengthx}
\global\newcount\particlelengthy  
\gdef\plengthy{\particlelengthy}
\global\newcount\boxlengthx  
\global\newcount\boxlengthy  
\global\newcount\particleadjustx  
\global\newcount\particleadjusty  
\global\newcount\particlelength   
\global\newcount\particlefrontx
\gdef\pfrontx{\particlefrontx}
\global\newcount\PFRONTx
\global\newcount\particlefronty
\gdef\pfronty{\particlefronty}
\global\newcount\PFRONTy
\global\newcount\particlebackx
\gdef\pbackx{\particlebackx}
\global\newcount\particlebacky
\gdef\pbacky{\particlebacky}
\global\newcount\particlemidx
\gdef\pmidx{\particlemidx}
\global\newcount\particlemidy
\gdef\pmidy{\particlemidy}
\global\newcount\seglength  \global\newcount\gaplength
\global\gaplength=850  
\global\seglength=1416  
\global\newcount\Xone    \global\newcount\Yone    
\global\newcount\Xtwo    \global\newcount\Ytwo    
\global\newcount\Xthree  \global\newcount\Ythree  
\global\newcount\Xfour   \global\newcount\Yfour   
\global\newcount\Xfive   \global\newcount\Yfive   
\global\newcount\Xsix    \global\newcount\Ysix    
\global\newcount\Xseven  \global\newcount\Yseven  
\global\newcount\Xeight  \global\newcount\Yeight  
%
%
\newsavebox{\lastline}  
\global\newcount\numlineparts   
\global\newcount\upperlineadjx  \upperlineadjx=0  
\global\newcount\upperlineadjy  \upperlineadjy=0  
\global\newcount\lowerlineadjx  \lowerlineadjx=0  
\global\newcount\lowerlineadjy  \lowerlineadjy=0  
\global\newcount\thirdlineadjx  \thirdlineadjx=0  
\global\newcount\thirdlineadjy  \thirdlineadjy=0  
\global\newcount\fourthlineadjx \fourthlineadjx=0  
\global\newcount\fourthlineadjy \fourthlineadjy=0  
\global\newcount\unitboxwidth   \unitboxwidth=1000
\global\newcount\unitboxheight  \unitboxheight=0  
\global\newcount\numupperunits  \numupperunits=8  
\global\newcount\numlowerunits  \numlowerunits=8  
\global\newcount\numthirdunits  \numthirdunits=8  
\global\newcount\numfourthunits \numfourthunits=8  
\global\newcount\fermioncount   \global\fermioncount=0
\global\newcount\scalarcount    \global\scalarcount=0
\global\newcount\photoncount    \global\photoncount=0
\global\newcount\gluoncount     \global\gluoncount=0
\global\newcount\SPECIALcount   \global\SPECIALcount=0
\global\newcount\vertexcount    \global\vertexcount=-1
%
\global\newcount\XDIR
\global\newcount\YDIR
\gdef\SETDIR{  
\ifcase\LDIR
     \global\XDIR=0  \global\YDIR=1   
\or  \global\XDIR=1  \global\YDIR=1   
\or  \global\XDIR=1  \global\YDIR=0   
\or  \global\XDIR=1  \global\YDIR=-1  
\or  \global\XDIR=0  \global\YDIR=-1  
\or  \global\XDIR=-1 \global\YDIR=-1  
\or  \global\XDIR=-1 \global\YDIR=0   
\or  \global\XDIR=-1 \global\YDIR=1   
\else\DIRECTERROR
\fi}  
\gdef\moduloeight#1{
\ifnum#1>7 \global\advance #1 by -8
\relax
\moduloeight#1
\relax
\else \relax
\fi}
\gdef\multroothalf#1{\global\multiply #1 by 7071 \global\divide #1 by 10000}
\gdef\negate#1{\global\multiply #1 by -1}
\gdef\double#1{\global\multiply #1 by 2}
\gdef\slanttest(#1,#2){
\ifodd\LDIR
\multiply #1 by 7071  \divide #1 by 10000
\multiply #2 by 7071  \divide #2 by 10000
\fi
}
\gdef\gslanttest(#1,#2){
\ifodd\LDIR
\multroothalf#1
\multroothalf#2
\fi
}
%
%
\gdef\setplength{ 
\global\particlelengthx=\unitboxwidth
\global\particlelengthy=\unitboxheight
\global\multiply \particlelengthx by \unitboxnumber
\global\multiply \particlelengthy by \unitboxnumber
\global\advance \particlelengthx by \particleadjustx
\global\advance \particlelengthy by \particleadjusty
}
\gdef\boxlengthdefault{  
\global\boxlengthx=\plengthx
\global\boxlengthy=\plengthy
\ifnum\plengthx<0 \global\multiply\boxlengthx by -1 \fi
\ifnum\plengthy<0 \global\multiply\boxlengthy by -1 \fi
}
\gdef\rearcoords{  
\global\particlebacky=\particlefronty
\global\particlebackx=\particlefrontx
\global\advance \particlebackx by \particlelengthx
\global\advance \particlebacky by \particlelengthy
}
\gdef\midcoords{  
\global\particlemidy=\particlefronty
\global\particlemidx=\particlefrontx
\global\stemlengthx=\particlelengthx  
\global\stemlengthy=\particlelengthy
\global\divide\stemlengthx by 2
\global\divide\stemlengthy by 2
\global\advance \particlemidx by \stemlengthx
\global\advance \particlemidy by \stemlengthy
}
\gdef\setparticle{\setplength\rearcoords\midcoords\boxlengthdefault}  
%
\gdef\setcoords(#1,#2,#3)(#4,#5,#6)[#7,#8]{
\global\upperlineadjx=#1
\global\lowerlineadjx=#2
\global\thirdlineadjx=#3
\global\upperlineadjy=#4
\global\lowerlineadjy=#5
\global\thirdlineadjy=#6
\global\unitboxwidth=#7
\global\unitboxheight=#8
}
%
%
%
\gdef\drawoldpic#1(#2,#3){  
\global\particlefrontx=#2
\global\particlefronty=#3
\rearcoords
\midcoords
\put(#2,#3){\usebox{#1}}
}
\gdef\drawsavedline`#1' as #2[#3#4](#5,#6)[#7]{
\global\LINETYPE=#2
\global\LINEDIRECTION=#3
\global\LINECONFIGURATION=#4
\global\particlefrontx=#5
\global\particlefronty=#6
\global\unitboxnumber=#7
\selectcase
\rearcoords
\midcoords
\ifnum\phantomswitch=0 \drawas{#1}\fi
}

\gdef\startphantom{\phantomswitch=1} 
\gdef\stopphantom{\phantomswitch=0}  

\gdef\drawas#1{
\global\savebox{#1}(\boxlengthx,\boxlengthy){
\setlength{\unitlength}{0.01pt}
\begin{picture}(\boxlengthx,\boxlengthy)
\multiput(\upperlineadjx,\upperlineadjy)(\unitboxwidth,\unitboxheight)
{\numupperunits}{\upperunitbox}
\ifnum\numlineparts > 1  
\multiput(\lowerlineadjx,\lowerlineadjy)(\unitboxwidth,\unitboxheight)
{\numlowerunits}{\lowerunitbox}
\fi
\ifnum\numlineparts > 2  
\multiput(\thirdlineadjx,\thirdlineadjy)(\unitboxwidth,\unitboxheight)
{\numthirdunits}{\thirdunitbox}
\fi
\ifnum\numlineparts > 3  
\multiput(\fourthlineadjx,\fourthlineadjy)(\unitboxwidth,\unitboxheight)
{\numfourthunits}{\lowerunitbox}
\fi
\end{picture} }
\global\PFRONTx=\pfrontx  \global\PFRONTy=\pfronty   
\SETFRONTSTEM
\THICKPHOTONTEST
\ifdim\THICKPHOTONSWITCH=1pt\global\advance\PFRONTy by 20  \fi
\put(\PFRONTx,\PFRONTy) {\usebox{#1}}   
\ifdim\THICKPHOTONSWITCH=1pt
\global\advance\PFRONTy by -40
\put(\PFRONTx,\PFRONTy) {\usebox{#1}}   
\global\advance \PFRONTy by 20  
\fi  
\SETBACKSTEM
\seglength=1416   \gaplength=850   
}
%
%

\gdef\drawandsaveline`#1' as #2[#3#4](#5,#6)[#7]{
\global\newsavebox{#1}
\drawsavedline`#1' as #2[#3#4](#5,#6)[#7]
}

\gdef\drawline#1[#2#3](#4,#5)[#6]{   
\drawsavedline`\lastline' as #1[#2#3](#4,#5)[#6]}

\gdef\saveas#1{  
\global\newsavebox#1
\drawas#1}
%
%
%
\gdef\TYPEERROR{\message{*** ERROR IN PARTICLE TYPE SELECTION ***}
\message{+++ Try with line type \fermion,\scalar,\photon,\gluon
(see manual) +++}\SETERR}
\gdef\DIRECTERROR{\SETERR\message{*** ERROR IN PARTICLE DIRECTION SELECTION
***}
\message{+++ Try again with direction N, NE, E, SE  etc. or see manual +++}}
\gdef\UNIMPERROR{\message{*** ERROR IN PARTICLE OPTIONS SELECTION ***}
\message{
+++ The requested options combination has not yet been implemented +++}\SETERR}
\gdef\SETERR{\gdef\upperunitbox{{\tiny Error}}  
\gdef\lowerunitbox{\relax}
\gdef\thirdunitbox{\relax}
}
\gdef\neglengthcheck{\ifnum\unitboxnumber < 1
\message{   *** ERROR:  PARTICLE OF NEGATIVE OR ZERO LENGTH REQUESTED. ***   }
\message{   ***         TAKING ABSOLUTE VALUE. ***   }\negate\unitboxnumber
\fi}
\gdef\selectcase{
\neglengthcheck   
\SETDIR
\ifcase\LINETYPE
\TYPEERROR  
\or \selectfermion  
\or \selectscalar   
\or \selectphoton   
\or \selectgluon    
\or \selectspecial  
\else \TYPEERROR \fi  }
\gdef\selectfermion{
\ifnum\fermioncount=0 \input FERMIONSETUP \fi
\global\advance\fermioncount by 1  
\ALLfermion
}
\gdef\selectscalar{
\ifnum\scalarcount=0 \input SCALARSETUP \fi
\global\advance\scalarcount by 1  
\ALLscalar
}
\gdef\selectphoton{   
\ifnum\photoncount=0 
\newcount\numwiggles    \newcount\numwigglespo
\global\newcount\photonlengthx
\global\newcount\photonlengthy
\global\newcount\photonfrontx  
\global\newcount\photonfronty  
\global\newcount\photonbackx
\global\newcount\photonbacky
\newcount\halfwigglelength
\global\font\Twelverom=cmr12
\global\font\Tenrom=cmr10
\gdef\Lbr{{\Twelverom(}}   \gdef\Rbr{{\Twelverom)}}
\gdef\SLbr{{\Tenrom(}}     \gdef\SRbr{{\Tenrom)}}
\gdef\Smile{{\large$\smile$}}  
\gdef\Frown{{\large$\frown$}}  
\ifdim\BIGPHOTONS>0pt  \gdef\Smile{$\smile$} \gdef\Frown{$\frown$} \fi
%
\gdef\selectphoton{   
\global\advance\photoncount by 1  
\global\photonfrontx=\particlefrontx   
\global\photonfronty=\particlefronty   
\ifnum\unitboxnumber > 50
\message{   *** WARNING *** Photon with
\the\unitboxnumber\space half-wiggles requested ***   }
\ifnum\unitboxnumber > 150
\message{   *** Reducing photon length to 10 half-wiggles (max 150) ***   }
\ifnum\unitboxnumber > 1000
\message{   *** Probable Cause:  Photon selected instead of Fermion ***   }
\fi \global\unitboxnumber=10 \fi \fi  
\numwiggles=\unitboxnumber
\divide\numwiggles by 2
\global\unitboxnumberpo=\numwiggles 
\global\multiply \unitboxnumberpo by -1
\numwigglespo=\unitboxnumber
\advance\numwigglespo by \unitboxnumberpo 
\global\numlineparts = 2  
\global\numupperunits=\numwigglespo  
\global\numlowerunits=\numwiggles  
\particleadjustx=0  
\particleadjusty=0  
\ifcase\LINEDIRECTION
     \Nphoton    
\or  \NEphoton   
\or  \Ephoton    
\or  \SEphoton   
\or  \Sphoton    
\or  \SWphoton   
\or  \Wphoton    
\or  \NWphoton   
\else\DIRECTERROR \fi
\setplength
\global\divide\plengthx by 2  \global\divide\plengthy by 2
\rearcoords  \boxlengthdefault   \midcoords
\global\photonbackx=\pbackx  
\global\photonbacky=\pbacky  
\global\photonlengthx=\plengthx  
\global\photonlengthy=\plengthy  
}
\gdef\SETUNITBOX(#1)[#2][#3]{ 
\gdef\upperunitbox{\oval(#1,#1)[#2]}
\gdef\lowerunitbox{\oval(#1,#1)[#3]}
}
\gdef\Nphoton{  
\ifcase\LINECONFIGURATION  
\setcoords(-490,-250,0)(260,1250,0)[0,2000]
\gdef\upperunitbox{\SLbr}   \gdef\lowerunitbox{\SRbr}
\particleadjusty=10
\or 
\setcoords(-271,-501,0)(250,1250,0)[0,2000]
\gdef\upperunitbox{\SRbr}   \gdef\lowerunitbox{\SLbr}
\or 
\particleadjusty=0
\setcoords(-501,-351,0)(300,1400,0)[0,2200]
\gdef\upperunitbox{\Lbr}   \gdef\lowerunitbox{\Rbr}
\or 
\setcoords(-353,-499,0)(300,1400,0)[0,2200]
\gdef\upperunitbox{\Rbr}   \gdef\lowerunitbox{\Lbr}
\or 
\setcoords(-481,-371,0)(280,1300,0)[0,2000]
\gdef\upperunitbox{\Lbr}   \gdef\lowerunitbox{\Rbr}
\particleadjusty=150
\ifnum\numwiggles=\number\numwigglespo \particleadjustx=-50 \fi
\or 
\setcoords(-321,-391,0)(280,1300,0)[0,2000]
\gdef\upperunitbox{\Rbr}   \gdef\lowerunitbox{\Lbr}
\particleadjusty=150
\ifnum\numwiggles=\number\numwigglespo \particleadjustx=80 \fi
\or 
\setcoords(-490,-260,0)(300,1500,0)[0,2400]
\gdef\upperunitbox{\Lbr}   \gdef\lowerunitbox{\Rbr}
\or 
\setcoords(-301,-531,0)(300,1500,0)[0,2400]
\gdef\upperunitbox{\Rbr}   \gdef\lowerunitbox{\Lbr}
\else \UNIMPERROR
\fi
}
\gdef\NEphoton{    
\ifcase\LINECONFIGURATION  
\setcoords(425,425,0)(1250,0,0)[1250,1250]       \SETUNITBOX(1250)[br][tl]
\ifnum\numwigglespo > \number \numwiggles \particleadjustx=15 \fi
\or 
\setcoords(1050,-200,0)(625,625,0)[1250,1250]    \SETUNITBOX(1250)[tl][br]
\ifnum\numwigglespo > \number \numwiggles \particleadjustx=25 \fi
\or 
\setcoords(500,500,0)(1400,0,0)[1400,1400]       \SETUNITBOX(1400)[br][tl]
\or 
\setcoords(1200,-200,0)(700,700,0)[1400,1400]    \SETUNITBOX(1400)[tl][br]
\or 
\setcoords(400,400,0)(1200,0,0)[1200,1200]       \SETUNITBOX(1200)[br][tl]
\or 
\setcoords(1000,-200,0)(600,600,0)[1200,1200]    \SETUNITBOX(1200)[tl][br]
\else \UNIMPERROR
\fi
\numupperunits=\numwiggles   \numlowerunits=\numwigglespo
}
\gdef\Ephoton{    
\ifcase\LINECONFIGURATION  
\setcoords(-285,715,0)(-150,-400,0)[2005,0]
\gdef\upperunitbox{\Frown}   \gdef\lowerunitbox{\Smile}
\or  
\setcoords(-285,715,0)(-420,-170,0)[2005,0]
\gdef\upperunitbox{\Smile}   \gdef\lowerunitbox{\Frown}
\else \UNIMPERROR
\fi
\particleadjustx=-15 
}
\gdef\SEphoton{   
\ifcase\LINECONFIGURATION  
\setcoords(-200,1050,0)(-625,-625,0)[1250,-1250] \SETUNITBOX(1250)[tr][bl]
\ifnum\numwigglespo > \number \numwiggles \particleadjustx=25 \fi
\or 
\setcoords(425,425,0)(0,-1250,0)[1250,-1250]     \SETUNITBOX(1250)[bl][tr]
\ifnum\numwigglespo > \number \numwiggles \particleadjustx=15 \fi
\or 
\setcoords(-200,1200,0)(-700,-700,0)[1400,-1400] \SETUNITBOX(1400)[tr][bl]
\or 
\setcoords(500,500,0)(0,-1400,0)[1400,-1400]     \SETUNITBOX(1400)[bl][tr]
\or 
\setcoords(-200,1000,0)(-600,-600,0)[1200,-1200] \SETUNITBOX(1200)[tr][bl]
\particleadjustx=-20
\or 
\setcoords(420,420,0)(0,-1200,0)[1200,-1200]     \SETUNITBOX(1200)[bl][tr]
\particleadjustx=40
\else \UNIMPERROR
\fi
}
\gdef\Sphoton{  
\ifcase\LINECONFIGURATION  
\setcoords(-252,-490,0)(-740,-1740,0)[0,-2000]
\gdef\upperunitbox{\SRbr}   \gdef\lowerunitbox{\SLbr}
\or 
\setcoords(-490,-260,0)(-740,-1740,0)[0,-2002]
\gdef\upperunitbox{\SLbr}   \gdef\lowerunitbox{\SRbr}
\or 
\setcoords(-299,-449,0)(-870,-1970,0)[0,-2200]
\gdef\upperunitbox{\Rbr}    \gdef\lowerunitbox{\Lbr}
\particleadjusty=-95
\or 
\setcoords(-517,-371,0)(-900,-2000,0)[0,-2200]
\gdef\upperunitbox{\Lbr}    \gdef\lowerunitbox{\Rbr}
\particleadjusty=-165
\or 
\setcoords(-299,-409,0)(-885,-1905,0)[0,-2000]
\gdef\upperunitbox{\Rbr}   \gdef\lowerunitbox{\Lbr}
\particleadjustx=50     \particleadjusty=-380
\ifodd\unitboxnumber\relax\else\particleadjustx=250 \particleadjusty=-400 \fi
\or 
\setcoords(-519,-449,0)(-900,-1920,0)[0,-2000]
\gdef\upperunitbox{\Lbr}   \gdef\lowerunitbox{\Rbr}
\particleadjusty=-370
\ifodd\unitboxnumber\relax\else\particleadjustx=-240 \particleadjusty=-400 \fi
\or 
\gdef\upperunitbox{\Rbr}   \gdef\lowerunitbox{\Lbr}
\setcoords(-325,-555,0)(-900,-2100,0)[0,-2400]
\particleadjusty=-40
\or 
\setcoords(-505,-275,0)(-900,-2100,0)[0,-2400]
\gdef\upperunitbox{\Lbr}   \gdef\lowerunitbox{\Rbr}
\particleadjusty=-30  
\else \UNIMPERROR
\fi
}
\gdef\SWphoton{  
\ifcase\LINECONFIGURATION  
\setcoords(-825,-825,0)(0,-1250,0)[-1250,-1250]     \SETUNITBOX(1250)[br][tl]
\or 
\setcoords(-175,-1425,0)(-625,-625,0)[-1250,-1250]  \SETUNITBOX(1250)[tl][br]
\or 
\setcoords(-900,-900,0)(0,-1410,0)[-1400,-1400]     \SETUNITBOX(1400)[br][tl]
\or 
\setcoords(-200,-1600,0)(-700,-700,0)[-1400,-1400]  \SETUNITBOX(1400)[tl][br]
\or 
\setcoords(-800,-800,0)(0,-1200,0)[-1200,-1200]     \SETUNITBOX(1200)[br][tl]
\or 
\setcoords(-200,-1400,0)(-600,-600,0)[-1200,-1200]  \SETUNITBOX(1200)[tl][br]
\else \UNIMPERROR
\fi
}
\gdef\Wphoton{
\ifcase\LINECONFIGURATION 
\setcoords(-2245,-1245,0)(-150,-400,0)[-2005,0]
\gdef\upperunitbox{\Frown}   \gdef\lowerunitbox{\Smile}
\or 
\setcoords(-2245,-1245,0)(-400,-150,0)[-2005,0]
\gdef\upperunitbox{\Smile}   \gdef\lowerunitbox{\Frown}
\else \UNIMPERROR
\fi
\particleadjustx=57 
\ifnum\numwigglespo=\number\numwiggles \particleadjustx=0  \fi
\numlowerunits=\numwigglespo   \numupperunits=\numwiggles
}
\gdef\NWphoton{  
\ifcase\LINECONFIGURATION  
\setcoords(-200,-1425,0)(625,625,0)[-1250,1250]   \SETUNITBOX(1250)[bl][tr]
\or 
\setcoords(-825,-825,0)(0,1250,0)[-1250,1250]     \SETUNITBOX(1250)[tr][bl]
\ifnum\numwigglespo > \number \numwiggles \particleadjusty=-15 \fi
\or 
\setcoords(-200,-1600,0)(700,700,0)[-1400,1400]   \SETUNITBOX(1400)[bl][tr]
\or 
\setcoords(-900,-900,0)(0,1400,0)[-1400,1400]     \SETUNITBOX(1400)[tr][bl]
\or 
\setcoords(-200,-1400,0)(600,600,0)[-1200,1200]   \SETUNITBOX(1200)[bl][tr]
\or 
\setcoords(-800,-800,0)(0,1200,0)[-1200,1200]     \SETUNITBOX(1200)[tr][bl]
\else \UNIMPERROR
\fi
}
  \fi
\selectphoton
}
\gdef\selectgluon{   
\ifnum\gluoncount=0 
\global\newcount\gluonlength
\global\newcount\gluonlengthx
\global\newcount\gluonlengthy
\global\newcount\gluonfrontx  
\global\newcount\gluonfronty  
\global\newcount\gluonbackx
\global\newcount\gluonbacky
%
\gdef\setunitbox(#1)[#2][#3](#4)[#5]{
\gdef\upperunitbox{\oval(#1,#1)[#2]}
\gdef\lowerunitbox{\oval(401,401)[#3]}
\gdef\thirdunitbox{\oval(#4,#4)[#5]}
}
\gdef\selectgluon{  
\global\advance\gluoncount by 1  
\global\gluonfrontx=\particlefrontx   
\global\gluonfronty=\particlefronty   
\global\particleadjustx=0     \global\particleadjusty=0
\ifnum\unitboxnumber > 40
\message{   *** WARNING *** Gluon with
\the\unitboxnumber\space loops requested ***   }
\ifnum\unitboxnumber > 85
\message{   *** Reducing gluon length to 6 loops (max 85) ***   }
\ifnum\unitboxnumber > 1000
\message{   *** Probable Cause:  Gluon selected instead of Fermion ***   }
\fi \global\unitboxnumber=6 \fi \fi  
\global\unitboxnumberpo=\unitboxnumber  
\global\advance\unitboxnumberpo by 1 
\global\numlineparts = 3
\global\numupperunits=\unitboxnumber
\global\numlowerunits=\unitboxnumber
\global\numthirdunits=\unitboxnumber
\ifcase\LINEDIRECTION
\Ngluon    
\or  \NEgluon  
\or  \Egluon   
\or  \SEgluon
\or  \Sgluon
\or  \SWgluon
\or  \Wgluon
\or  \NWgluon
\else\DIRECTERROR \fi
\setparticle
\global\gluonlengthx=\particlelengthx  \global\gluonlengthy=\particlelengthy
\global\gluonbackx=\particlebackx      \global\gluonbacky=\particlebacky
}
\gdef\Ngluon{   
\ifcase\LINECONFIGURATION   
\setcoords(600,540,600)(20,620,1220)[0,1050]
\setunitbox(1600)[tl][r](1600)[bl]
\particleadjusty=195
\or 
\setcoords(-990,-930,-990)(12,615,1215)[0,1050]
\setunitbox(1600)[tr][l](1600)[br]
\particleadjusty=195
\or 
\setcoords(440,390,440)(-10,415,840)[0,850]
\setunitbox(1250)[tl][r](1250)[bl]
\particleadjustx=0
\particleadjusty=-10
\or 
\setcoords(-820,-770,-820)(-25,400,825)[0,850]  
\particleadjusty=-10  
\setunitbox(1250)[tr][l](1250)[br]
\or \UNIMPERROR  
\or \UNIMPERROR  
\or 
\numupperunits=\unitboxnumberpo
\numlowerunits=\unitboxnumber
\numthirdunits=\unitboxnumberpo
\setcoords(-200,-200,-200)(616,1041,616)[0,850]
\setunitbox(1250)[tl][r](1250)[bl]
\particleadjusty=1238
\particleadjusty=1233
\or 
\numupperunits=\unitboxnumberpo
\numlowerunits=\unitboxnumber
\numthirdunits=\unitboxnumberpo
\setcoords(-200,-200,-200)(620,1045,620)[0,850]
\setunitbox(1250)[tr][l](1250)[br]
\particleadjusty=1245
\else \UNIMPERROR 
\fi
}
\gdef\NEgluon{
\numupperunits=\unitboxnumberpo
\numlowerunits=\unitboxnumber
\numthirdunits=\unitboxnumber
\ifcase\LINECONFIGURATION
\setcoords(900,900,900)(0,900,900)[900,900]
\setunitbox(2200)[tl][tr](401)[b]
\particleadjustx=1100     \particleadjusty=1100
\or 
\setcoords(-180,720,720)(1090,1091,1091)[900,900]
\setunitbox(2200)[br][tr](401)[l]
\particleadjustx=1110     \particleadjusty=1050
\else \UNIMPERROR 
\fi
}
\gdef\Egluon{     
\ifcase\LINECONFIGURATION
\setcoords(-210,390,990)(-800,-745,-800)[1050,0]  
\setunitbox(1600)[tr][b](1600)[tl]
\particleadjustx=130  
\or 
\setcoords(-210,390,990)(800,745,800)[1050,0]  
\setunitbox(1600)[br][t](1600)[bl]
\particleadjustx=130
\or 
\setcoords(-200,225,650)(-625,-575,-625)[850,0]
\setunitbox(1250)[tr][b](1250)[tl]
\or 
\setcoords(-200,225,650)(625,575,625)[850,0]
\setunitbox(1250)[br][t](1250)[bl]
\or 
\setcoords(-200,430,1060)(-830,-780,-830)[1260,0]
\setunitbox(1660)[tr][b](1660)[tl]
\or 
\setcoords(-200,430,1060)(830,780,830)[1260,0]
\setunitbox(1660)[br][t](1660)[bl]
\or 
\numupperunits=\unitboxnumberpo
\numlowerunits=\unitboxnumber
\numthirdunits=\unitboxnumberpo
\setcoords(440,865,440)(0,50,0)[850,0]
\setunitbox(1250)[tr][b](1250)[tl]
\particleadjustx=1260
\or 
\numupperunits=\unitboxnumberpo
\numlowerunits=\unitboxnumber
\numthirdunits=\unitboxnumberpo
\setcoords(430,855,430)(0,-50,0)[850,0]
\setunitbox(1250)[br][t](1250)[bl]
\particleadjustx=1250
\or 
\setcoords(-160,440,1040)(-600,-550,-600)[1200,0]
\gdef\upperunitbox{\oval(1600,1200)[tr]}
\gdef\thirdunitbox{\oval(1600,1200)[tl]}
\gdef\lowerunitbox{\oval(401,401)[b]}
\else \UNIMPERROR
\fi
}
\gdef\SEgluon{
\numupperunits=\unitboxnumberpo
\numlowerunits=\unitboxnumber
\numthirdunits=\unitboxnumber
\ifcase\LINECONFIGURATION
\setcoords(-200,700,700)(-1100,-1100,-1100)[900,-900]
\setunitbox(2200)[tr][br](401)[l]
\particleadjustx=1100     \particleadjusty=-1100
\or 
\setcoords(890,890,890)(0,-900,-900)[900,-900]
\setunitbox(2200)[bl][br](401)[t]
\particleadjustx=1050     \particleadjusty=-1100
\else \UNIMPERROR 
\fi
}
\gdef\Sgluon{   
\ifcase\LINECONFIGURATION  
\setcoords(-1000,-940,-1000)(0,-595,-1195)[0,-1050]
\setunitbox(1600)[br][l](1600)[tr]
\particleadjusty=-150
\or 
\setcoords(605,545,605)(-20,-615,-1215)[0,-1050]
\setunitbox(1600)[bl][r](1600)[tl]
\particleadjusty=-150
\or 
\setcoords(-820,-770,-820)(0,-425,-850)[0,-850]
\setunitbox(1250)[br][l](1250)[tr]
\or 
\setcoords(440,390,440)(0,-425,-850)[0,-850]
\setunitbox(1250)[bl][r](1250)[tl]
\or \UNIMPERROR 
\or \UNIMPERROR
\or 
\numupperunits=\unitboxnumberpo
\numlowerunits=\unitboxnumber
\numthirdunits=\unitboxnumberpo
\setcoords(-180,-180,-180)(-635,-1060,-635)[0,-850]
\setunitbox(1250)[br][l](1250)[tr]
\particleadjusty=-1290
\or 
\numupperunits=\unitboxnumberpo
\numlowerunits=\unitboxnumber
\numthirdunits=\unitboxnumberpo
\setcoords(-180,-180,-180)(-635,-1060,-635)[0,-850]
\setunitbox(1250)[bl][r](1250)[tl]
\particleadjusty=-1290
\else \UNIMPERROR 
\fi
}
\gdef\SWgluon{
\numupperunits=\unitboxnumberpo
\numlowerunits=\unitboxnumber
\numthirdunits=\unitboxnumber
\ifcase\LINECONFIGURATION
\setcoords(-1300,-1300,-1300)(0,-900,-900)[-900,-900]
\setunitbox(2200)[br][bl](401)[t]
\particleadjustx=-1100     \particleadjusty=-1100
\or 
\setcoords(-215,-1115,-1115)(-1107,-1107,-1107)[-900,-900]
\setunitbox(2200)[tl][bl](401)[r]
\particleadjustx=-1120     \particleadjusty=-1120
\else \UNIMPERROR 
\fi
}
\gdef\Wgluon{   
\ifcase\LINECONFIGURATION
\setcoords(-190,-790,-1390)(800,745,800)[-1050,0]
\setunitbox(1600)[bl][t](1600)[br]
\particleadjustx=-150  
\or 
\setcoords(-190,-790,-1390)(-800,-745,-800)[-1050,0]
\setunitbox(1600)[tl][b](1600)[tr]
\particleadjustx=-150  
\or 
\setcoords(-200,-625,-1050)(625,575,625)[-850,0]
\setunitbox(1250)[bl][t](1250)[br]
\or 
\setcoords(-200,-625,-1050)(-625,-575,-625)[-850,0]
\setunitbox(1250)[tl][b](1250)[tr]
\or 
\setcoords(-230,-860,-1490)(830,780,830)[-1260,0]
\setunitbox(1660)[bl][t](1660)[br]
\or 
\setcoords(-230,-860,-1490)(-830,-780,-830)[-1260,0]
\setunitbox(1660)[tl][b](1660)[tr]
\or 
\numupperunits=\unitboxnumberpo
\numlowerunits=\unitboxnumber
\numthirdunits=\unitboxnumberpo
\setcoords(-825,-1250,-825)(0,-50,0)[-850,0]
\setunitbox(1250)[bl][t](1250)[br]
\particleadjustx=-1250
\or  
\numupperunits=\unitboxnumberpo
\numlowerunits=\unitboxnumber
\numthirdunits=\unitboxnumberpo
\setcoords(-825,-1250,-825)(0,50,0)[-850,0]
\setunitbox(1250)[tl][b](1250)[tr]
\particleadjustx=-1250
\else \UNIMPERROR 
\fi
}
\gdef\NWgluon{
\numupperunits=\unitboxnumberpo
\numlowerunits=\unitboxnumber
\numthirdunits=\unitboxnumber
\ifcase\LINECONFIGURATION
\setcoords(-200,-1100,-1100)(1100,1100,1100)[-900,900]
\setunitbox(2200)[bl][tl](401)[r]
\particleadjustx=-1110   \particleadjusty=1100
\or  
\setcoords(-1309,-1309,-1309)(-15,885,885)[-900,900]
\setunitbox(2200)[tr][tl](401)[b]
\particleadjustx=-1120   \particleadjusty=1065
\else \UNIMPERROR 
\fi
}
%
%
%
\gdef\gluonlink{    
\input GLUONLINKS   
\gluonlink}  
\gdef\gluoncap{    
\input GLUONLINKS   
\gluoncap}  
  \fi
\selectgluon
}
\gdef\selectspecial{\UNIMPERROR}
%
%
\gdef\checkvertex{ 
\ifnum\vertexcount=-1   \input VERTEX  \fi}
\gdef\drawvertex#1[#2#3](#4,#5)[#6]{\checkvertex\drawvertex#1[#2#3](#4,#5)[#6]}
\gdef\vertexcap#1{\checkvertex\vertexcap#1}
\gdef\vertexcaps{\checkvertex\vertexcaps}
\gdef\vertexlink#1{\checkvertex\vertexlink#1}
\gdef\vertexlinks{\checkvertex\vertexlinks}
\gdef\stemvertex#1{\checkvertex\stemvertex#1}
\gdef\stemvertices{\checkvertex\stemvertices}
\gdef\flipvertex{\checkvertex\flipvertex}
%
%
\global\arrowlength=349  
\gdef\drawarrow[#1#2](#3,#4){
\global\LDIR=#1
\SETDIR
\global\boxlengthx=#3  
\global\boxlengthy=#4  
\ifdim#2=1pt  
\adjx=\arrowlength      \adjy=\arrowlength
\multiply\adjx by \XDIR \multiply\adjy by \YDIR  
\slanttest(\adjx,\adjy)
\global\advance\boxlengthx by \adjx    \global\advance\boxlengthy by \adjy
\fi
\ifnum\phantomswitch=0\put(\boxlengthx,\boxlengthy){\vector(\XDIR,\YDIR){0}}\fi
}  
%
%
\gdef\SETFRONTSTEM{
\EITHERSTEM=\FRONTSTEM   \advance\EITHERSTEM by \BACKSTEM
\ifdim\EITHERSTEM>0pt
\global\stemlengthx=\stemlength   \global\stemlengthy=\stemlength
\global\absstemlength=\stemlength
\SETDIR
\gslanttest(\stemlengthx,\stemlengthy)
\gslanttest(\absstemlength,\REG)  
\ifnum\XDIR=0 \stemlengthx=0 \fi
\ifnum\YDIR=0 \stemlengthy=0 \fi
\global\multiply\stemlengthx by \XDIR
\global\multiply\stemlengthy by \YDIR
\ifdim\FRONTSTEM=1pt
\ifnum\phantomswitch=0
          \put(\pfrontx,\pfronty){\line(\XDIR,\YDIR){\absstemlength}}\fi
\global\advance\plengthx by \stemlengthx
\global\advance\plengthy by \stemlengthy
\global\advance\PFRONTx by \stemlengthx
\global\advance\PFRONTy by \stemlengthy
\global\advance\pmidx by \stemlengthx
\global\advance\pmidy by \stemlengthy
\global\advance\pbackx by \stemlengthx
\global\advance\pbacky by \stemlengthy
\ifnum\LTYPE=3
\global\photonfrontx=\PFRONTx  \global\photonfronty=\PFRONTy
\global\photonbackx=\pbackx    \global\photonbacky=\pbacky
\fi  
\ifnum\LTYPE=4
\global\gluonfrontx=\PFRONTx  \global\gluonfronty=\PFRONTy
\global\gluonbackx=\pbackx    \global\gluonbacky=\pbacky
\fi  
\fi  
\fi  
}    
\gdef\SETBACKSTEM{
\ifdim\BACKSTEM=1pt
\ifnum\phantomswitch=0
       \put(\pbackx,\pbacky){\line(\XDIR,\YDIR){\absstemlength}}\fi
\global\advance\plengthx by \stemlengthx
\global\advance\plengthy by \stemlengthy
\global\advance\pbackx by \stemlengthx
\global\advance\pbacky by \stemlengthy
\fi  
\global\stemlength=275  \FRONTSTEM=0pt  \BACKSTEM=0pt 
}    
\gdef\drawloop#1[#2#3](#4,#5){  
\input LOOPS  
\drawloop#1[#2#3](#4,#5)}
\Feynmanlength  



\begin{titlepage}
\centerline{}
\vskip2.cm

\hskip 10.5cm {\sl LAPTH-761/99}
\vskip .4cm
\begin{center}
{\Large\bf Perturbative Quantum Field Theory
in the}\\ 
{\Large\bf String-Inspired Formalism}
\vspace{1.0cm}

\centerline{\large Christian Schubert}
\vspace{.4cm}
{\it
Laboratoire d'Annecy-le-Vieux
de Physique Th{\'e}orique LAPTH\\
Chemin de Bellevue,
BP 110\\
F-74941 Annecy-le-Vieux CEDEX,
France\\
} 
\vskip.5cm
{\it
Instituto de Fisica y Matem\'aticas
\footnote{After 1. 10. 2000.}
\\
Universidad Michoacana de San Nicol\'as de Hidalgo\\ 
Apdo. Postal 2-82\\
C.P. 58040, Morelia, Michoac\'an, M\'exico\\
schubert@itzel.ifm.umich.mx\\
}
\vskip.5cm
{\it
California Institute for Physics and Astrophysics\\
366 Cambridge Ave., Palo Alto, California, US}
\vskip .9cm
To appear in {\it Physics Reports}
\vspace{30pt}

{\large \bf Abstract}
\end{center}

\begin{quotation}
\noindent
We review the status and present range of
applications of
the ``string -- inspired'' approach to
perturbative quantum field theory. 
This formalism offers the possibility
of computing effective actions and S-matrix
elements in a way which is similar in spirit
to string perturbation theory, and bypasses much of
the apparatus of standard second-quantized field
theory. Its development was initiated by Bern and
Kosower, originally with the aim of simplifying
the calculation of scattering amplitudes in
quantum chromodynamics and quantum gravity. 
We give a short account of the original derivation of the
Bern-Kosower rules from string theory. Strassler's
alternative approach in terms of first-quantized 
particle path integrals is then used to generalize the formalism
to more general field theories, and, in the abelian
case, also to higher loop orders. A considerable number of
sample calculations are presented in detail, with an emphasis
on quantum electrodynamics. 
\end{quotation}

\end{titlepage}


\tableofcontents
\section{Introduction: Strings vs. Particles, First vs. 
Second Quantization}
\renewcommand{\theequation}{1.\arabic{equation}}
\setcounter{equation}{0}

One of the main motivations for the study of string
theory is the 
fact that it reduces to quantum field theory
in the limit where the tension along the string becomes
infinite \cite{scherk,nevsch,yoneya,schsch}.
In this limit all massive modes of the
string get suppressed, and one remains with the
massless modes.
Those can be identified with ordinary massless particles
such as gauge bosons, gravitons, or massless spin-$1\over 2$
fermions. 
Moreover, the interactions taking place between those massless modes
turn out to be familiar from 
quantum field theory.
In particular, it came as a pleasant surprise 
that a consistent theory of fundamental strings must
by necessity include quantum gravity ~\cite{gsw}.

Regardless of whether string theory is realized in nature
or not,
those mathematical facts already
lend us a new perspective on quantum field theory,
which we now find embedded
in a theory which is not only vastly more complex,
but also structurally different.
In particular, a major difference between string theory and
particle theory is that, in string theory, the full perturbative S-matrix
is calculable in first quantization using the Polyakov path integral,
which describes the propagation of a single string in a given
background. 
An adequate second quantized
field theory for strings 
was built after considerable efforts 
~\cite{wittencubic,zwiebach}, allowing one to
define off-shell amplitudes with the correct
factorization properties, and even to 
compute nonperturbative effects in string theory
\cite{kossam,senzwi,moetay}. 
Nevertheless, so far the second quantized approach has not
led to improvements over the first quantized formalism
as far as the calculation of perturbative string scattering
amplitudes is concerned.

In ordinary quantum field theory, of course, perturbative
calculations are usually performed using second quantization, and
Feynman diagrams. First quantized alternatives have been
developed already at the very inception of relativistic
quantum field theory ~\cite{feynman:pr80}, 
and will, in fact, be the main subject of the present review.
However it appears that, until recently,
they were hardly ever seriously considered as an efficient 
tool for 
standard perturbative calculations.
This discrepancy between string and particle theory
becomes something of a puzzle as soon as one considers
the latter as a limiting special case of the former.
And we would like to convince
the reader in the following that this apparent paradox
owes more to historical development than
to mathematical fact.

If string theory reduces to field theory in
the infinite string tension limit, then clearly
it should be possible to obtain S-matrix elements
in certain  field theory models
by analyzing the corresponding
string scattering amplitudes in this limit.
It goes without saying that 
the calculation of string amplitudes is generally much
more difficult than the calculation of the corresponding field theory
amplitudes, so that the practical value of such a procedure 
may appear far 
from obvious. Nevertheless, it turns out to be sufficiently
motivated by the different organization of both types of
amplitudes, and by the different methods available for
their computation. 

As early as 1972 Gervais and Neveu observed that string theory
generates Feynman rules for Yang-Mills theory 
in a special gauge that
has certain calculational advantages \cite{gerneu}.
At the loop level, the first explicit calculation along these lines was
performed in 1982 by Green, Schwarz and Brink, who obtained the
one -- loop 4 -- gluon amplitude in $N = 4$ Super Yang-Mills theory from
the low energy limit of superstring theory ~\cite{grscbr}.
However, serious interest in this subject commenced
only in 1988, when it was shown by several authors 
that the one-loop
$\beta$ -- function for Yang-Mills theory can be extracted
from the genus one partition function of an open string coupled
to a background gauge field 
~\cite{minahan,kaplunovsky,mettse,berkos:prd38}.
This calculation gives also some insight into the well-known
fact that this $\beta$ -- function coefficient 
vanishes for $D=26$, the critical dimension of the bosonic
string, when calculated in dimensional regularization
~\cite{nepomechie}.

A systematic investigation of the infinite string tension limit
was undertaken in the following years by Bern and Kosower 
\cite{berkos:prl166,berkos:npb362,berkos:npb379}.
This was done again with a view on application to
non-abelian gauge theory, however now to the
computation of complete on-shell scattering amplitudes. 
Again the idea was to calculate, say, gauge boson scattering
amplitudes in an appropriate string model containing
$SU(N_c)$ gauge theory, up to the point where one has obtained
an explicit parameter integral representation for the
amplitude considered.
At this stage one performs the infinite string tension limit,
which should eliminate all contributions due to
propagating massive modes, and lead to a parameter integral
representation for the corresponding field theory
amplitude.

In the present work, 
we will concentrate on a different and more elementary approach to
this formalism, which does not rely on the 
calculation of string amplitudes any more, and uses
string theory only as a guiding principle.
Only a sketchy account will therefore be given of
string perturbation theory, and the reader is referred to
the literature for the details ~\cite{gsw,dhopho}.

The basic tool for the calculation of string scattering amplitudes
is the Polyakov path integral. In the simplest case, the closed bosonic
string propagating in flat spacetime, this integral is of the
form

\begin{figure}[ht]
\begin{center}
\begin{picture}(0,0)%
\epsfig{file=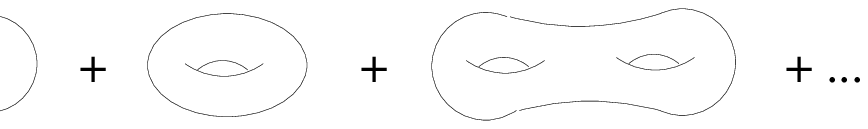}%
\end{picture}%
\setlength{\unitlength}{0.00087500in}%
\begingroup\makeatletter\ifx\SetFigFont\undefined
\def\x#1#2#3#4#5#6#7\relax{\def\x{#1#2#3#4#5#6}}%
\expandafter\x\fmtname xxxxxx\relax \def\y{splain}%
\ifx\x\y   
\gdef\SetFigFont#1#2#3{%
  \ifnum #1<17\tiny\else \ifnum #1<20\small\else
  \ifnum #1<24\normalsize\else \ifnum #1<29\large\else
  \ifnum #1<34\Large\else \ifnum #1<41\LARGE\else
     \huge\fi\fi\fi\fi\fi\fi
  \csname #3\endcsname}%
\else
\gdef\SetFigFont#1#2#3{\begingroup
  \count@#1\relax \ifnum 25<\count@\count@25\fi
  \def\x{\endgroup\@setsize\SetFigFont{#2pt}}%
  \expandafter\x
    \csname \romannumeral\the\count@ pt\expandafter\endcsname
    \csname @\romannumeral\the\count@ pt\endcsname
  \csname #3\endcsname}%
\fi
\fi\endgroup
\begin{picture}(3634,1536)(201,-850)
\end{picture}
\caption{\label{stringloop} 
The loop expansion in string perturbation theory.}
\end{center}
\end{figure}

\bear
&\Bigl\langle&
V_1\cdots V_N
\Bigr\rangle
\sim
\sum_{\rm top}
\int{\cal D}h
\int{\cal D}x(\sigma,\tau)
V_1\cdots V_N
\,
\e^{-S[x,h]}
\non\\
\label{ppi}
\ear
\no
This path integral corresponds to first quantization in the
sense that it describes a single string propagating in
a given background.
The 
parameters $\sigma,\tau$
parametrize the world sheet surface swept out
by the string in its motion,
and the integral 
$\int{\cal D}x(\sigma,\tau)$
has to be performed
over the space of all embeddings of the
string world sheet with a fixed topology into
spacetime. The integral $\int{\cal D}h$ is over the
space of all world sheet metrics, and the sum over
topologies
$\sum_{\rm top}$ corresponds to the loop expansion in
field theory (fig. \ref{stringloop}).  

\noindent
\no
If the closed string is assumed to be oriented,
there is only one topology at any fixed order of loops.

In the case that the background is simply Minkowski spacetime
the world sheet action is given by

\be
S[x,h]=-{1\over 4\pi \alpha'}
\int d\sigma d\tau
\sqrt{h}
h^{\alpha\beta}{\eta}_{\mu\nu}
\partial_{\alpha}x^{\mu}
\partial_{\beta}x^{\nu}
\label{S}
\ee\no
where ${1\over 2\pi\alpha'}$ is the
string tension.
Note that in Polyakov's formulation 
the action is quadratic in the coordinate field $x$.

\begin{figure}[ht]
\begin{center}
\begin{picture}(10000,0)
\epsfig{file=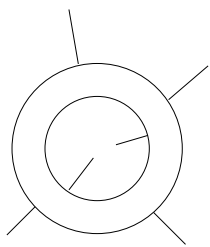}%
\end{picture}%
\setlength{\unitlength}{0.00087500in}%
\begingroup\makeatletter\ifx\SetFigFont\undefined
\def\x#1#2#3#4#5#6#7\relax{\def\x{#1#2#3#4#5#6}}%
\expandafter\x\fmtname xxxxxx\relax \def\y{splain}%
\ifx\x\y   
\gdef\SetFigFont#1#2#3{%
  \ifnum #1<17\tiny\else \ifnum #1<20\small\else
  \ifnum #1<24\normalsize\else \ifnum #1<29\large\else
  \ifnum #1<34\Large\else \ifnum #1<41\LARGE\else
     \huge\fi\fi\fi\fi\fi\fi
  \csname #3\endcsname}%
\else
\gdef\SetFigFont#1#2#3{\begingroup
  \count@#1\relax \ifnum 25<\count@\count@25\fi
  \def\x{\endgroup\@setsize\SetFigFont{#2pt}}%
  \expandafter\x
    \csname \romannumeral\the\count@ pt\expandafter\endcsname
    \csname @\romannumeral\the\count@ pt\endcsname
  \csname #3\endcsname}%
\fi
\fi\endgroup
\begin{picture}(2534,1536)(201,-850)
\end{picture}
\caption{\label{anninsert}
Vertex operators inserted on the boundary of the annulus. 
}
\end{center}
\end{figure}

The vertex operators 
$V_1,\ldots,V_N$ represent the scattering 
string states.
In the case of the open string, which is the more relevant one
for our purpose, the world sheet has a boundary, and the vertex
operators are inserted on this boundary.
For instance, for the open oriented string at the
one-loop level the world sheet is just an annulus,
and a vertex operator may be integrated along either one of
the two boundary components (fig. \ref{anninsert}).

\no
The vertex operators most relevant for us are
of the form

\bear
V^{\phi}[k]&=&\int d\tau\,{\e}^{ik\cdot x(\tau)}
\label{scalarvertop}\\
V^A[k,\varepsilon,a] &=&\int d\tau \,T^{a}\varepsilon\cdot\dot x(\tau)
\,{\e}^{ik\cdot x(\tau)}
\label{gluonvertop}
\ear\no
They represent a scalar and a gauge boson particle with definite
momentum $k$ and polarization vector $\varepsilon$. 
$T^a$ is a generator
of the gauge group in some representation. 
The integration variable
$\tau$ parametrizes the boundary in question.
Since the action is Gaussian, 
$\int{\cal D}x$ can be performed by Wick contractions,

\bear
\Bigl\langle
x^{\mu}(\tau_1)x^{\nu}(\tau_2)
\Bigr\rangle
=G(\tau_1,\tau_2)\,\eta^{\mu\nu}
\; 
\label{wickstring}
\ear\no
$G$ denotes the Green's function for the Laplacian
on the annulus, restricted to its boundary,
and
$\eta^{\mu\nu}$ the Lorentz metric.

In $D=26$,
the critical dimension of the
bosonic string, the Polyakov path integral
is conformally invariant. 
The remaining path integral over the 
infinite dimensional space of all world sheet
metrics $h$ can then be reduced to the space of conformal equivalence 
classes, which is finite dimensional. The actual integration
domain, moduli space, is somewhat smaller, since a further
discrete symmetry group has to be taken into account.
At this stage, then, the amplitude is in a form suitable for
performing the infinite string tension limit.
It turns out that, in this limit, only certain corners of the 
whole moduli
space contribute. The amplitude thus splinters into a number of pieces,
which individually are parameter integrals of the same type
encountered in field theory Feynman diagram calculations.

\begin{figure}[ht]
\begin{center}
\begin{picture}(-20000,0)%
\epsfig{file=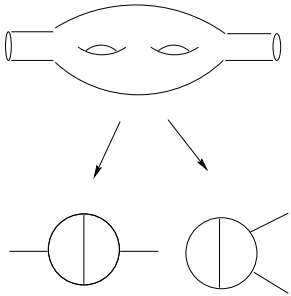}%
\end{picture}%
\setlength{\unitlength}{0.00087500in}%
\begingroup\makeatletter\ifx\SetFigFont\undefined
\def\x#1#2#3#4#5#6#7\relax{\def\x{#1#2#3#4#5#6}}%
\expandafter\x\fmtname xxxxxx\relax \def\y{splain}%
\ifx\x\y   
\gdef\SetFigFont#1#2#3{%
  \ifnum #1<17\tiny\else \ifnum #1<20\small\else
  \ifnum #1<24\normalsize\else \ifnum #1<29\large\else
  \ifnum #1<34\Large\else \ifnum #1<41\LARGE\else
     \huge\fi\fi\fi\fi\fi\fi
  \csname #3\endcsname}%
\else
\gdef\SetFigFont#1#2#3{\begingroup
  \count@#1\relax \ifnum 25<\count@\count@25\fi
  \def\x{\endgroup\@setsize\SetFigFont{#2pt}}%
  \expandafter\x
    \csname \romannumeral\the\count@ pt\expandafter\endcsname
    \csname @\romannumeral\the\count@ pt\endcsname
  \csname #3\endcsname}%
\fi
\fi\endgroup
\begin{picture}(7634,1536)(201,-850)
\end{picture}
\caption{\label{stringdiagram} 
Infinite string tension limit of a string diagram.}
\end{center}
\end{figure}

For illustration, consider the following two-point
``Feynman diagram'' for the closed string (fig. \ref{stringdiagram}).

\vspace{1cm}

\vspace{-15pt}
\noindent

In the $\alpha'\to 0$ limit this Riemann surface gets squeezed
to a Feynman graph, although not to a single one; two
Feynman diagrams of different topologies emerge.
This proliferation becomes, of course, much worse at higher
orders. Moreover, in gauge theory or gravity it
is further enhanced by the existence of quartic and higher order
vertices, which lead to many more possible topologies.
The generating string theories thus have a much smaller number
of ``Feynman diagrams'' then the limiting field theories,
which is another major motivation for the use of string-techniques
in field theory.

The uses of the Polyakov path integral
are not restricted to  
the calculation of scattering amplitudes.
As pointed out by Fradkin and Tseytlin ~\cite{fratse},
it is equally useful for the calculation of string effective
actions.
For instance, an open oriented 
string propagating in the background
of a Yang-Mills field $A$ would generate
an effective action for this background field given by the
following modification of the Polyakov path integral,

\bear
\Gamma[A]&\sim&
\sum_{\rm top}
\int{\cal D}h
\int{\cal D}x(\sigma,\tau)
\,
\e^{-S_0 -S_I}
\; \non\\
S_0 &=&
-{1\over 4\pi \alpha'}
\int_M
d\sigma d\tau
\sqrt{h}
h^{\alpha\beta}{\eta}_{\mu\nu}
\partial_{\alpha}x^{\mu}
\partial_{\beta}x^{\nu}
\; \non\\
S_I &=&
\int_{\partial M}
d\tau
\,
ie\,\dot x^{\mu}
A_{\mu}(x(\tau))
\label{fratsepi}
\ear
\no
The sum now extends over all oriented bounded manifolds.
The free term $S_0$ is the same as above,
eq.(\ref{S}), and the
interaction term $S_I$
has to be integrated along all components
of the boundary. For simplicity, we have
written the interaction term for the abelian case;
in the non-abelian case, a colour trace and path ordering
would have to be included. This will be discussed later
on in the field theory context.

Metsaev and Tseytlin calculated
the one-loop path integral exactly
for the constant field strength case ~\cite{mettse},
and verified that the $\alpha'\to 0$ limit coincides
with the corresponding effective action in
Yang-Mills theory. In particular, the correct $\beta$
- function coefficient can be read off from the
$F_{\mu\nu}^2$ -- term.
This procedure is not completely
rigorous, though, since the open 
bosonic string theory
cannot be consistently truncated
from 26 down to four dimensions.
In their analysis of the $N$ - gluon amplitude \cite{berkos:npb379},
Bern and Kosower therefore used, instead of the
open string, a certain heterotic string model
containing $SU(N_c)$ Yang-Mills theory in the
infinite string tension limit. 
This allows for a consistent reduction to
four dimensions, at the price of a more
complicated representation of this amplitude.
By an explicit analysis of the infinite string tension
limit, they succeeded in deriving a novel type
of parameter integral representation for the
on-shell
$N$ - gluon amplitude in Yang-Mills theory,
at the tree- and one-loop level.
Moreover, they established a set of rules which allows one
to construct this parameter integral, for
any number of gluons and choice of helicities,
without referring to string theory any more.

While those rules are very different from the 
corresponding field theoretic
Feynman rules, the precise connection and
equivalence between both sets of rules
were soon established ~\cite{berdun}.
Moreover, once an understanding of the rules
had been reached, it emerged that the
consistency requirements motivating the
choice of the heterotic string were not really
relevant in their derivation. An alternative
derivation using the naive truncated open string
was given ~\cite{bern:plb296}, and even yielded a
somewhat simpler set of rules.

This set of rules will be discussed in detail in
chapter 2. For the moment, let us just mention
some advantages of the ``Bern-Kosower Rules''
as compared to the Feynman rules:

\begin{enumerate}
\item
Superior organization of gauge invariance.

\item

Absence of loop momentum, which reduces the number of
kinematic invariants from the beginning, and allows for a
particularly efficient use of the spinor helicity method.

\item
The method combines nicely with spacetime
supersymmetry.

\item
Calculations of scattering amplitudes 
with the same external states but
particles of
different spin circulating in the loop
are more closely related than usual.

\end{enumerate}
\no
The last two points are, of course, closely related.
The efficiency of these rules 
has been demonstrated by the first
complete calculation of the one -- loop five -- gluon
amplitude ~\cite{bediko5glu}. A similar
set of rules for graviton scattering was derived
from  closed string theory in ~\cite{bedush}.
Those have been used for the first 
calculation of the complete 
one -- loop four -- graviton amplitude
in quantum gravity ~\cite{dunnor}.

Since the
Bern-Kosower rules do not refer to string theory any more,
the question naturally arises
whether it should not be possible
to re-derive them completely inside field theory.
Obviously, such a re-derivation should be attempted starting from
a first-quantized formulation of ordinary field theory,
rather than from standard quantum field theory.
As we mentioned in the beginning, such formulations have been
known for decades, albeit only for a very limited number of models.
Already in 1950,
in the appendix A of his famous paper
``Mathematical Formulation of the Quantum Theory of Electromagnetic
Interaction'' ~\cite{feynman:pr80}, Feynman presented such a
formalism for the case of scalar quantum electrodynamics,
``for its own interest as an alternative to the formulation of
second quantization''. What he states here is that
the amplitude for a charged scalar particle to move,
under the influence of the external potential $A_{\mu}$,
from point $x_{\mu}$ to $x_{\mu}'$ in Minkowski space
is given by

\begin{figure}[ht]
\vbox to 4.5cm{\vfill\hbox to 15.8cm{\hfill
\epsffile{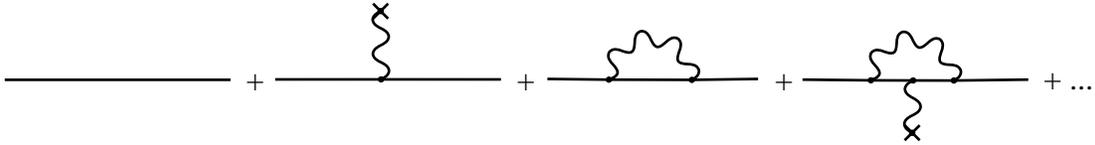}
\hfill}\vfill}\vskip-.4cm
\caption[dum]{Sum of Feynman diagrams represented by a single
path integral.
\hphantom{xxxxxxxxxxxxxxx}}
\label{totalelectronprop}
\end{figure}

\bear
\int_0^{\infty}
ds
\int
_{x(0)=x}^{x(s)=x'}
{\cal D}x(\tau)
&&\!\!\!\!\!\!\!\!\!
\exp
\Bigl(
-{1\over 2}
im^2s\Bigr)
\exp
\biggl[
-{i\over 2}
\int_0^sd\tau
{({dx_{\mu}\over d\tau})}^2
-i\int_0^s
d\tau
{dx_{\mu}\over d\tau}
A_{\mu}(x(\tau))\nonumber\\
&&
-{i\over 2}
e^2
\int_0^sd\tau
\int_0^sd\tau'
{dx_{\mu}\over d\tau}
{dx_{\nu}\over d\tau'}
\delta_{+}^{\mu\nu}(x(\tau)-x(\tau'))
\biggr]
\label{feynform}
\ear
That is,
for a fixed value of the variable $s$ (which can be identified with
Schwinger proper time) one can construct the amplitude as a certain
quantum mechanical path integral. This path integral has to be performed
on the set of all open
trajectories running from $x$ to $x'$ in the fixed
proper time $s$. The action 
consists of the familiar kinetic term, and two interaction terms.
Of those the first represents the interaction with the external field,
to all orders in the field, while the second one describes an
arbitrary number of virtual photons emitted and re-absorbed along the
trajectory of the particle ($\delta_{+}$ denotes the photon
propagator). In second quantized field theory, this amplitude would thus
correspond to the infinite sequence of Feynman diagrams
shown in fig. \ref{totalelectronprop}.

\par
\par

\noindent

As Feynman proceeds to show,
this representation extends in an obvious way to the case
of an arbitrary fixed number of scalar particles, moving in
an external potential and exchanging internal photons, and thus 
to the complete S-matrix for scalar quantum
electrodynamics. Every scalar line or loop is then 
separately described by a path integral such as the one above.
The path integrals are coupled by an arbitrary number of
photon insertions.
The derivation of this type of path
integral will be discussed in detail in chapter 3. 

In the present work, we are mainly concerned with path integrals
for closed loops. Let us therefore rewrite Feynman's formula
for the case of a single closed loop
in the
external field, with no internal photon corrections. What we
have at hand then is simply a representation of the
one-loop effective action for the
Maxwell field
\footnote{
The proper time
parameter
$s$ has been rescaled and Wick rotated,
$s\rightarrow -i2T$. The spacetime metric will also
be taken as Euclidean, except when stated otherwise 
(upper and lower indices will be used purely for
typographical convenience).
Moreover, we anticipate dimensional
regularization and thus usually continue to $D$ Euclidean
dimensions.}
,

\begin{equation}
\Gamma\lbrack A\rbrack = \int_0^{\infty}
{dT\over T} \, {\rm e}^{-m^2T}
\int {\cal D}x\, 
{\rm exp} \left[ 
- \int_0^T \!\!\! d\tau \left( {1\over 4}{\dot x}^2 
+ ieA_{\mu}\dot x^{\mu} 
\right) \right]
\label{scalarpi}
\end{equation}
\no
The path integral runs now over the space of closed trajectories
with period $T$, $x^{\mu}(T)=x^{\mu}(0)$. 

The comparison of the Fradkin-Tseytlin
path integral
(\ref{fratsepi}) with the
path integral
(\ref{scalarpi}) shows that
the former is clearly 
a string theoretic
generalization of the latter.
Conversely, at least at the one-loop level it is not difficult to
show that the Feynman path integral is
precisely the
infinite string tension limit of the
Fradkin-Tseytlin path integral. 
Naively, one can think of the annulus
in fig. \ref{anninsert} being squeezed to its boundary. 

The 
path integral representation 
eq.(\ref{scalarpi}) generalizes in various
ways to spinor quantum electrodynamics. 
In the
fermion loop case, one has a basic choice
to make in the treatment of the spin degrees of freedom.
Those can
be incorporated  
either by explicit $\gamma$ - matrices
~\cite{feynman:pr84,bardur}, or by
Grassmann variables 
~\cite{fradkin,casalbuoninc,casalbuoniplb,bermar},
which carry the same algebraic properties.
The first, ``bosonized'', version may be preferable for certain
purposes such as the evaluation of path integrals by numerical
or saddle point approximation.
Nevertheless,
we will generally use the second alternative, since it
offers the possibility to use worldline supersymmetry 
in a computationally
meaningful way.

Supersymmetric worldline Lagrangians were constructed 
soon after the
advent of supersymmetry.
This 
led to an intensive study of field
theories in $1+0$ dimensions, and
to the discovery that the worldline
Lagrangian appropriate for the description
of a Dirac particle is precisely the
one for $N=1$ supergravity ~\cite{bdzdh,brdiho}.
As a consequence of that work, the generalization of our
path integral
eq.(\ref{scalarpi}) for the one-loop effective action
to spinor QED 
can be written as a super path integral 
\cite{fradkin,borcas,baboca,fragit}

 \begin{eqnarray}
\Gamma\lbrack A\rbrack &  = &- \half {\displaystyle\int_0^{\infty}}
{dT\over T}
e^{-m^2T}
{\displaystyle\int}_P 
{\cal D} x
{\displaystyle\int}_A 
{\cal D}\psi\nonumber\\
& \phantom{=}
&\times
{\rm exp}\biggl [- \int_0^T d\tau
\Bigl ({1\over 4}{\dot x}^2 + {1\over
2}\psi\cdot\dot\psi
+ ieA_{\mu}\dot x^{\mu} - ie
\psi^{\mu}F_{\mu\nu}\psi^{\nu}
\Bigr )\biggr ]\nonumber\\
\label{spinorpi}
\end{eqnarray}
\noindent
In addition to the integral over
the periodic
functions $x^{\mu}(\tau )$,
we have now a second path integral
over the functions $\psi^{\mu}(\tau )$,
which are Grassmann valued and
antiperiodic
(the periodicity properties are expressed
by the subscripts $P,A$ on the path integral).
The global minus sign accounts for the
Fermi statistics of the spinor loop. 

Comparing this formula with eq.(\ref{scalarpi}),
it becomes immediately clear that
one can think of this double path integral as
breaking up the Dirac spinor into an 
``orbital part'' and a ``spin part''.
The former is represented by the same
coordinate path integral as the
scalar particle, the latter by the
additional Grassmann path integral.

In writing this path integral, we have
already gauge-fixed the local 
one-dimensional
supergravity. This leaves over a
global supersymmetry,
``worldline
supersymmetry'',

\bear
&&\delta x^\mu=-2\eta\psi^\mu\nonumber\\
&&\delta\psi^\mu=\eta\dot x^\mu
\label{wlsusy}
\ear
with a constant Grassmann parameter $\eta$.
As we will see later on, the existence of
this symmetry has far-reaching
calculational consequences.

A vast amount of literature is available on 
this type of relativistic particle
Lagrangians, and the corresponding
path integrals.
However, only
a minor part of it is concerned
with 
attempts to use them
as a tool for actual calculations
in quantum field theory.
Much of particularly the early literature
emphasizes the one-dimensional 
over
the spacetime point of view, or is concerned with
the formal properties of such worldline field
theories. In particular, one-dimensional 
field theories are often used 
for a comparative study of the various known
quantization procedures 
(see, e.g., ~\cite{holten:npb457}).

Of those applications which have come to this author's 
notice,
let us mention the work of Halpern et al. 
~\cite{halsie,hajese}, who 
proposed to use first
quantized path integrals for a construction of the
strong-coupling expansion in non-abelian gauge
theory.
More recently, various attempts have been
made to apply worldline path integrals 
to nonperturbative calculations, 
using various approximation
schemes for the path integral.
See, e.g., 
\cite{nietjo} for scalar field theory, ~\cite{stkakt}
for heavy-meson theory,
~\cite{kksw,antonov} for 
QCD, and ~\cite{rossch}
for meson-nucleon theory 
applications.
Some
applications to QED
can be found in 
~\cite{friedbook,afalma,baboca,frgish,simtjo,gitshv,gitzla},
to statistical physics in ~\cite{dotsenko}.

Probably best-known is, however, the
application to the calculation
of anomalies and index densities
~\cite{cecgir,alvarezgaume,alvwit,friwin,bastianelli,basvan,bpsv:npb446,bpsv:npb459}.
Here a number of special cases of
the
Atiyah-Singer index theorem 
could be reproduced in an 
elementary way by rewriting supertraces
of heat kernels for the
corresponding operators in terms
of supersymmetric 
particle path integrals
\cite{alvarezgaume,alvwit,friwin}.
Nevertheless, despite this remarkable
success it seems that, until recently, the first
quantized formalism was never seriously considered
as a competitor to the usual Feynman diagrammatic
approach with regard to everyday life calculations
of scattering amplitudes or effective actions.

The
principle of how 
one might reproduce ordinary perturbation
theory in the first quantized formalism,
simply by mimicking string perturbation
theory, was
already sketched in 
chapter 9 of Polyakov's book ~\cite{polbook}.
However it was only after the work of Bern and
Kosower, when it had become clear that 
techniques from first-quantized string
perturbation theory {\it do} have the potential
to improve on the efficiency of field
theory calculations, that such an
approach was seriously investigated by
Strassler ~\cite{strassler1,strasslerthesis}.

\begin{figure}[ht]
\begin{center}
\begin{picture}(-20000,0)%
\epsfig{file=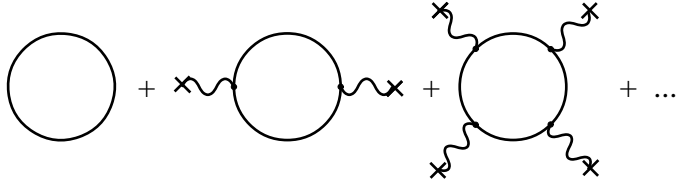}%
\end{picture}%
\setlength{\unitlength}{0.00087500in}%
\begingroup\makeatletter\ifx\SetFigFont\undefined
\def\x#1#2#3#4#5#6#7\relax{\def\x{#1#2#3#4#5#6}}%
\expandafter\x\fmtname xxxxxx\relax \def\y{splain}%
\ifx\x\y   
\gdef\SetFigFont#1#2#3{%
  \ifnum #1<17\tiny\else \ifnum #1<20\small\else
  \ifnum #1<24\normalsize\else \ifnum #1<29\large\else
  \ifnum #1<34\Large\else \ifnum #1<41\LARGE\else
     \huge\fi\fi\fi\fi\fi\fi
  \csname #3\endcsname}%
\else
\gdef\SetFigFont#1#2#3{\begingroup
  \count@#1\relax \ifnum 25<\count@\count@25\fi
  \def\x{\endgroup\@setsize\SetFigFont{#2pt}}%
  \expandafter\x
    \csname \romannumeral\the\count@ pt\expandafter\endcsname
    \csname @\romannumeral\the\count@ pt\endcsname
  \csname #3\endcsname}%
\fi
\fi\endgroup
\begin{picture}(7634,1536)(20,-850)
\end{picture}
\caption{\label{scalarloopexpand} 
Expanding the path integral in powers of the
background field.}
\end{center}
\end{figure}

Let us demonstrate the method for the
example of the scalar loop, eq.(\ref{scalarpi}).
The basic idea is simple: We will evaluate
this path integral 
in precisely the same way as one calculates the
Polyakov path integral in string theory,
i.e.
in a one-dimensional perturbation
theory. If we expand the
``interaction exponential'',

\bear
{\rm exp}\Bigl[
-\int_0^Td\tau\, ieA_{\mu}\dot x^{\mu}
\Bigr]
&=&\sum_{N=0}^{\infty}
{{(-ie)}^N\over N!}
\prod_{i=1}^N
\int_0^Td\tau_i
\biggl[
\dot x^{\mu}(\tau_i)
A_{\mu}(x(\tau_i))
\biggr]
\non\\
\label{expandint}
\ear\no
the individual terms correspond to Feynman diagrams
describing a fixed number of
interactions of the scalar loop with
the external field (fig. \ref{scalarloopexpand}).

\vspace{-5pt}

\no
By standard field theory,
the corresponding $N$ -- photon
correlator is then obtained by
specializing to a background
consisting of 
a sum of plane waves with definite
polarizations,

\be
A_{\mu}(x)=
\sum_{i=1}^N
\varepsilon_{i\mu}
\e^{ik_i\cdot x}
\label{planewavebackground}
\ee\no
and picking out the term containing every
$\varepsilon_i $ once (this also removes
the ${1\over N!}$ in eq.(\ref{expandint})).
We find thus 
exactly the same 
photon
vertex operator used in string perturbation
theory, eq. (\ref{gluonvertop}), inserted on a circle
instead on the boundary of the annulus.

At this stage the path integral has become Gaussian,
which reduces its evaluation to the task
of Wick contracting the expression

\be
\biggl\langle
\dot x_1^{\mu_1}\e^{ik_1\cdot x_1}
\cdots
\dot x_N^{\mu_N}\e^{ik_N\cdot x_N}
\biggr\rangle
\label{scalqedwick}
\ee
\no
The Green's function to be used is now
simply the one for the second-derivative
operator, acting on
periodic functions. To derive 
this Green's function, first observe that 
$\int{\cal D}x(\tau)$ contains the constant functions,
which we must get rid of to obtain a well-defined
inverse. Let us therefore restrict our integral
over the space of all loops by fixing the average
or ``center of mass''
position $x_0^{\mu}$ of the loop,

\begin{equation}
x_0^{\mu}\equiv {1\over T}\int_0^T d\tau\, x^{\mu}(\tau)
\label{defx0}
\end{equation}
\no
For effective action calculations this reduces the
effective action to the effective Lagrangian.
In scattering amplitude calculations,
the integral over $x_0$ just gives
momentum conservation.
The reduced path integral $\int{\cal D}y(\tau)$
over $y(\tau)\equiv x(\tau) - x_0$
has an invertible kinetic operator. The
inverse is easily seen to be,
up to an irrelevant constant,

\vspace{-8pt}
\be
2\bigl\langle\tau_1\mid
{\Bigl({d\over d\tau}\Bigr)}^{-2}
\mid\tau_2\bigr\rangle
=
G_B(\tau_1,\tau_2)
\label{calcG}
\ee
\vspace{-8pt}

\no
with the ``bosonic'' worldline Green's function

\be
G_B(\tau_1,\tau_2)=\mid \tau_1-\tau_2\mid 
-{{(\tau_1-\tau_2)}^2\over T} 
\label{defG}
\ee
\no
(a ``fermionic'' 
worldline
Green's function $G_F$ will be introduced later on).
For the performance of the Wick contractions, it is
convenient to formally exponentiate all the $\dot x_i$'s, 
writing

\be
\varepsilon_i\cdot
\dot x_i\,\e^{ik_i\cdot x_i}
=
\e^{\varepsilon_i\cdot\dot x_i
+ik_i\cdot x_i}
\mid_{{\rm lin}(\varepsilon_i)}
\label{formexp}
\ee
\no
This allows one to rewrite the product of $N$ photon vertex
operators as an exponential. Then one needs only
to ``complete the square'' to arrive at the following
closed expression for the one-loop
$N$ - photon amplitude in scalar QED,

\begin{eqnarray}
\Gamma_{\rm scal}[k_1,\varepsilon_1;\ldots;k_N,\varepsilon_N]
&=&
{(-ie)}^N
{(2\pi )}^D\delta (\sum k_i)
{\dps\int_{0}^{\infty}}{dT\over T}
{(4\pi T)}^{-{D\over 2}}
e^{-m^2T}
\prod_{i=1}^N \int_0^T 
d\tau_i
\nonumber\\
&&
\!\!\!\!\!\!\!
\times
\exp\biggl\lbrace\sum_{i,j=1}^N 
\Bigl\lbrack  \half G_{Bij} k_i\cdot k_j
-i\dot G_{Bij}\varepsilon_i\cdot k_j
+\half\ddot G_{Bij}\varepsilon_i\cdot\varepsilon_j
\Bigr\rbrack\biggr\rbrace
\mid_{\rm multi-linear}
\nonumber\\
\label{scalarqedmaster}
\end{eqnarray}
\no
Here it is understood that only the terms linear
in all the $\varepsilon_1,\ldots,\varepsilon_N$
have to be taken. 
Besides the Green's function $G_B$ also its first and
second derivatives appear,

\begin{eqnarray}
\dot G_B(\tau_1,\tau_2) &=& {\rm sign}(\tau_1 - \tau_2)
- 2 {{(\tau_1 - \tau_2)}\over T}\nonumber\\
\ddot G_B(\tau_1,\tau_2)
&=& 2 {\delta}(\tau_1 - \tau_2)
- {2\over T}\quad 
\label{GdGdd}
\end{eqnarray}
\noindent
Dots generally denote a
derivative acting on the first variable,
$\dot G_B(\tau_1,\tau_2) \equiv {\partial\over
{\partial\tau_1}}G_B(\tau_1,\tau_2)$, 
and we abbreviate
$G_{Bij}\equiv G_B(\tau_i,\tau_j)$ etc.
The factor ${[4\pi T]}^{-{D\over 2}}$
represents the free Gaussian path integral
determinant factor.

The expression (\ref{scalarqedmaster})
which we have arrived at
in this quite elementary way
is identical with the ``Bern-Kosower
Master Formula'' 
for the special case considered
\cite{berkos:npb379,berntasi}.
We will discuss various generalizations
and applications of this formula later on.
For now, the important point to note is
that we have at hand here a single
unifying generating functional for the
one-loop photon S-matrix -- something for which
no known analogue exists in standard field
theory.

How does this master integrand relate to
the integrals appearing in an
ordinary Feynman parameter calculation
of this amplitude? 
Note that in (\ref{scalarqedmaster}) every
photon leg is integrated around the loop
independently. 
As we will see in detail later on,
once one restricts the integration domain
to a fixed ordering 
$\tau_{i_1}>\tau_{i_2}>\cdots>\tau_{i_N}$, 
it is not difficult to identify the
integrand with the 
corresponding 
Feynman parameter integral.
In particular, there is an exact 
correspondence between the $\delta$
-- function appearing in the second
derivative of $G_B$, 
and the
seagull--vertex of scalar quantum
electrodynamics.
However the complete integral 
does not represent any particular Feynman diagram, 
with a fixed ordering of the external legs,
but
{\sl the sum of them} (fig. \ref{sumofdiagrams}):

\begin{figure}[ht]
\begin{center}
\begin{picture}(0,0)%
\epsfig{file=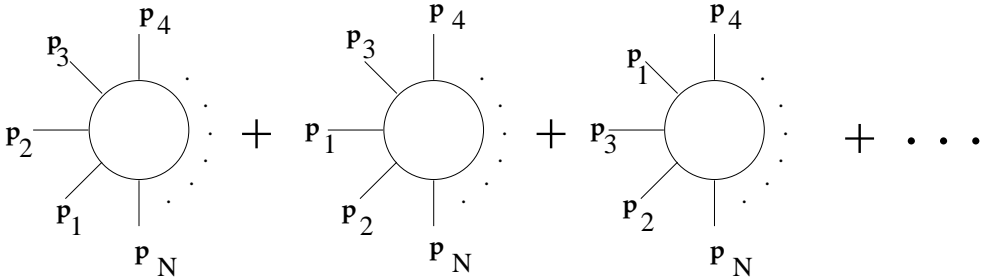}%
\end{picture}%
\setlength{\unitlength}{0.00087500in}%
\begingroup\makeatletter\ifx\SetFigFont\undefined
\def\x#1#2#3#4#5#6#7\relax{\def\x{#1#2#3#4#5#6}}%
\expandafter\x\fmtname xxxxxx\relax \def\y{splain}%
\ifx\x\y   
\gdef\SetFigFont#1#2#3{%
  \ifnum #1<17\tiny\else \ifnum #1<20\small\else
  \ifnum #1<24\normalsize\else \ifnum #1<29\large\else
  \ifnum #1<34\Large\else \ifnum #1<41\LARGE\else
     \huge\fi\fi\fi\fi\fi\fi
  \csname #3\endcsname}%
\else
\gdef\SetFigFont#1#2#3{\begingroup
  \count@#1\relax \ifnum 25<\count@\count@25\fi
  \def\x{\endgroup\@setsize\SetFigFont{#2pt}}%
  \expandafter\x
    \csname \romannumeral\the\count@ pt\expandafter\endcsname
    \csname @\romannumeral\the\count@ pt\endcsname
  \csname #3\endcsname}%
\fi
\fi\endgroup
\begin{picture}(3634,1736)(201,-850)
\end{picture}
\caption{\label{sumofdiagrams} 
Sum of one-loop diagrams with permuted legs.}
\end{center}
\end{figure}

\noindent
This fact
may not seem particularly relevant at the one-loop
level. However it is important to note that
the path integral representation
eq.(\ref{scalarpi}) and 
the resulting integral
representation
eq.(\ref{scalarqedmaster})
are valid off-shell
\footnote{
 This fact was not
obvious in the original string-theoretic
derivation of (\ref{scalarqedmaster}),
since before the infinite string tension limit
the requirement of
conformal invariance forces the 
external states to be on-shell.}.
We can therefore use this formula to
sew together a pair of legs,
say, legs number $1$ and
$N$, and obtain 
a parameter integral representing
the complete two-loop
$(N-2)$ -- photon amplitude
(fig. \ref{complete2loopphotamp}):

\begin{figure}[ht]
\begin{center}
\begin{picture}(0,0)%
\epsfig{file=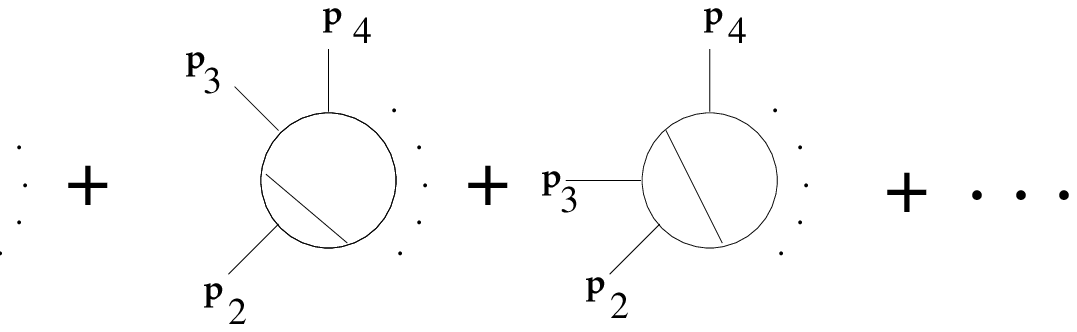}%
\end{picture}%
\setlength{\unitlength}{0.00087500in}%
\begingroup\makeatletter\ifx\SetFigFont\undefined
\def\x#1#2#3#4#5#6#7\relax{\def\x{#1#2#3#4#5#6}}%
\expandafter\x\fmtname xxxxxx\relax \def\y{splain}%
\ifx\x\y   
\gdef\SetFigFont#1#2#3{%
  \ifnum #1<17\tiny\else \ifnum #1<20\small\else
  \ifnum #1<24\normalsize\else \ifnum #1<29\large\else
  \ifnum #1<34\Large\else \ifnum #1<41\LARGE\else
     \huge\fi\fi\fi\fi\fi\fi
  \csname #3\endcsname}%
\else
\gdef\SetFigFont#1#2#3{\begingroup
  \count@#1\relax \ifnum 25<\count@\count@25\fi
  \def\x{\endgroup\@setsize\SetFigFont{#2pt}}%
  \expandafter\x
    \csname \romannumeral\the\count@ pt\expandafter\endcsname
    \csname @\romannumeral\the\count@ pt\endcsname
  \csname #3\endcsname}%
\fi
\fi\endgroup
\begin{picture}(3634,1536)(201,-850)
\end{picture}
\caption{\label{complete2loopphotamp} 
Sum of two -- loop diagrams with different topologies.}
\end{center}
\end{figure}

\begin{figure}[ht]
\begin{center}
{}~\epsfig{file=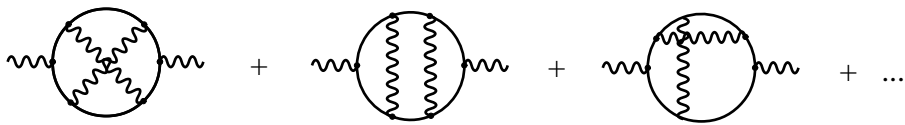}
\end{center}
\caption[]
{
\label{3loopbetadiag}
Sum of diagrams contributing to the 3-loop QED $\beta$ -- function.
}
\end{figure}

\noindent
\no
This is interesting, as we have at hand 
a single
integral formula for a sum containing 
many diagrams of
different topologies. We may think of it as a remnant
of the ``less fragmented'' nature of string perturbation
theory mentioned before (fig. \ref{stringdiagram}). 
Moreover, it calls certain well-known cancellations to mind
which happen in gauge theory due to the fact that the
Feynman diagram calculation splits a gauge invariant amplitude
into gauge non-invariant pieces. For instance, to obtain
the 3-loop $\beta$ -- function coefficient for
quenched (single spinor - loop)
QED, one needs to calculate the sum of diagrams
shown in fig. \ref{3loopbetadiag}.

\no
Performing this calculation in, say, dimensional regularization, one finds
that

\begin{enumerate}
\item
All poles of order
higher than ${1\over\varepsilon}$ cancel.

\item
Individual diagrams give contributions to 
the $\beta$ -- function proportional to $\zeta (3)$ 
which cancel in the sum, leaving a 
rational coefficient.
\end{enumerate}
\no
The first property is known
to be a consequence of gauge invariance,
and to hold true to all orders of perturbation theory
~\cite{jowiba}.
The second one, i.e. the absence of irrational numbers
in the quenched QED $\beta$ -- function, has been
explicitly verified to four-loop order
in spinor QED ~\cite{rosner,gkls:plb256},
and to three-loop order in scalar QED ~\cite{brdekr}. 
Recently arguments from knot theory have been
given which link both properties ~\cite{brdekr},
indicating that this property
should hold to all orders, too.
As every practician in quantum field theory knows,
similar extensive cancellations abound in 
calculations in gauge theory.

It seems therefore very natural to apply the Bern-Kosower
formalism to this type of calculation. 
However, in its original version the
Bern-Kosower formalism was confined to tree--level
and one--loop amplitudes. 
The extension of this formalism beyond one-loop
is obviously desirable, and
has already been attempted along quite different lines:
 
In the original
approach of Bern and Kosower,
going beyond one loop would imply finding
the particle theory limits 
of higher genus
string amplitudes, a formidable task 
considering the complicated
structure of moduli space for genus
higher than one.
While
a suitable representation of
the $N$ - gluon amplitude at arbitrary genus
was already given in \cite{roland},
and substantial progress was 
achieved in the analysis of the
infinite string tension limit
~\cite{dlmmr:plb388,dlmmr:npb469,magrus,limape,mape,frmaru,korsch:string}, 
so far this line of work has not yet led to
the formulation of
multiloop Bern-Kosower type rules.

Another, and
in some sense opposite route has been taken by 
Lam ~\cite{lamqed}, who
sets out with the usual Feynman parameter 
integral representation of 
multiloop diagrams, and uses the electric circuit
analog ~\cite{bjorken,lamleb}
to transform those into the Koba-Nielsen type 
representation which one would expect
from a string-type calculation. 
 Yet another approach has been followed by 
McKeon ~\cite{mckeon:ap224}, who 
proposes to perform multiloop
calculations by writing
a separate worldline path integral for 
every internal propagator of a 
diagram.
A Hamiltonian approach was considered
in ~\cite{siopsis}.

In principle
one could, of course, also construct
Bern-Kosower type multiloop formulae
using the explicit sewing procedure indicated above.
However, we will describe another multiloop formalism
here, proposed by M.G. Schmidt and the author
~\cite{ss2,ss3},
which is based on a more efficient way of inserting
propagators into one-loop amplitudes. 
This approach is a direct generalization of Strassler's
one-loop formalism, and preserves its main properties.
Its distinguishing features are the following:

\begin{enumerate}
\item
We will generalize eq.(\ref{defG}) for the 
one-loop Green's function to the construction of
Green's functions defined on multiloop graphs.

\item
The superfield formalism for the fermion loop will 
carry over to the multiloop level.

\item
All field theory vertices will be represented by
worldline quantities.

\end{enumerate}

Though not explicitly referring to string theory
any more, 
the resulting formalism may
still be called ``string-inspired'' in the sense that it
has a natural interpretation in terms of
a one - dimensional field
theory defined on graphs. 
As one would expect from a 
generalization of the Bern-Kosower method, it 
allows one to derive well-organized 
parameter integral representations for dimensionally
regularized off-shell amplitudes, 
without the need for computing
momentum integrals or Dirac traces.    

At the multiloop level, this
formalism has been worked out comprehensively for scalar field theories
~\cite{ss2,sato1,rolsat:npb480,rolsat:npb515,satsch1,sato4} 
as well as for scalar and spinor
QED ~\cite{ss3,dashsu,rescsc,zako}. 
In those models, in principle it applies to
the calculation of arbitrary
off-shell amplitudes involving only spin $0$ and spin $1$
scattering states, or of the corresponding effective
Lagrangians.
More recently along these lines preliminary results
have been obtained also for Yang-Mills Theory at the
two-loop level ~\cite{sato2,satsch2}.

Our applications center around 
the photon S-matrix in quantum electrodynamics as our
main paradigm. Here the formalism has been developed
to a point where it shows some distinct advantages
over the more standard
methods, in particular for problems involving 
constant external fields.
At the one-loop level, we present also a number of calculations
involving Yukawa and axial couplings, as well as non-abelian
gauge fields. 

The material is organized as follows.
In chapter 2 we state
the Bern-Kosower rules 
for the case of gluon scattering,
and shortly sketch their derivation from
the open bosonic string.

In chapter 3 we give derivations for the 
most basic worldline Lagrangians used
in this work, describing the
coupling of particles with spin $0$, $\half$,
and $1$ to external gauge fields.
The starting point in this derivation
is always the proper-time representation of the
one-loop effective action in terms
of the one-loop functional determinant.
We discuss in particular detail the path integral
representation of spin -- 1 particles
~\cite{strassler1,rescsc}, since here the application
of the string -- inspired technique requires a
non-standard approach.

The principle of how to calculate such path integrals
within the ``string-inspired formalism'' is then
explained in chapter 4. The advantages of the
technique compared to the standard Feynman
diagram technique are then demonstrated 
using the example of the QED and QCD vacuum
polarizations, and QCD gluon --  gluon scattering.
We investigate the systematics of
the Bern-Kosower partial integration procedure
for the general $N$ - photon / $N$ gluon amplitudes, and
determine the structure of the resulting
integrand. We also clarify the relation of the
worldline parameter integrals to the ones
arising in standard Feynman parameter calculations
of the same amplitudes.

Chapter 5 is devoted to the treatment of QED amplitudes
in a constant electromagnetic background field. This case
is given special attention since it provides a particularly
natural application of the string-inspired technique
\cite{ss1,gussho1,gussho2,shaisultanov,adlsch,rescsc}.

In chapter 6 we consider more general field theories.
While worldline path integral representations have
been known and investigated for decades for 
the case of spin -- 0 and spin -- $\half$ particles minimally
coupled to gauge and gravitational
backgrounds, worldline Lagrangians describing
the coupling of a Dirac fermion
to a general background consisting of a scalar,
pseudoscalar, vector, axialvector and antisymmetric tensor
field were obtained only recently 
~\cite{mnss1,dhogag1,mnss2,dhogag2,mcksch}.
In the present review we restrict ourselves to two
special cases which admit particularly elegant
formulations, namely the scalar -- pseudoscalar
and the vector -- axialvector amplitudes.

Chapter 7 deals with the application of the formalism
to the calculation of effective actions in 
the higher derivative or heat-kernel expansion
\cite{ss1,fss1,fhss4}. 
While our discussion concentrates on the gauge theory
case, we also shortly discuss some mathematical subtleties
which arise in the generalization of the formalism to
curved backgrounds, and which were clarified only
very recently.

In chapter 8 we generalize the worldline formalism
to the calculation of multiloop amplitudes in 
scalar self-interacting field theories.
This generalization is based on the concept of
Green's functions defined on graphs, and follows
the string analogy closely. 
We derive explicit expressions for those Green's functions
for a large class of graphs, the so-called
``Hamiltonian graphs'', and verify for some
simple two-loop examples that their application
reproduces the same amplitudes as the corresponding
Feynman diagram calculations.

This formalism is then generalized to quantum electrodynamics
in chapter 9, where in principle it applies to the
calculation of the whole photon S-matrix.
At the two-loop level, we present a 
detailed recalculation 
\cite{rescsc,frss}
of the 
Euler-Heisenberg Lagrangians 
in scalar and spinor QED, including the
$\beta$ -- function coefficients. An interesting
cancellation which occurs in the spinor QED
case is explained by an analysis of the
renormalization procedure.

In the conclusions we give a short overview over the
present range of applicability of the worldline
technique, and point out some possible future directions.

There are several technical appendices. 
Appendix \ref{conv} contains a summary
of the conventions used in the present work
\footnote{Those differ in some
points from previous work by this author.}, 
including the rules for continuation 
from Euclidean to Minkowski space. 
In appendix \ref{greendet} we give detailed calculations
of the various worldline propagators which are used
in the main text. 
Appendix \ref{q2to6} contains a more detailed discussion of
the partial integration procedure introduced in section
\ref{NphotonNgluon}, and a summary of the resulting worldline integrands 
up to the six-point case.
The results are used in appendix \ref{proof} for a simple
proof of the basic fermionic replacement rule (\ref{fermion}).
In appendix \ref{boxint} we explain a technique for the
calculation of four-point massless on-shell tensor 
parameter integrals,
following \cite{bedikopent}. Finally, appendix \ref{formulas}
contains a collection of formulas which we have found
useful, or at least amusing.

\section{The Bern -- Kosower Formalism }
\renewcommand{\theequation}{2.\arabic{equation}}
\setcounter{equation}{0}

This chapter is devoted to 
a statement of the Bern-Kosower rules,
and to
a short account of
their derivation from string theory.
We follow not the original derivation from the
heterotic string
\cite{berkos:prl166,berkos:npb362,berkos:npb379}
but the simpler one using
the open bosonic string, as given
in ~\cite{bern:plb296,berntasi}.

\subsection{The Infinite String Tension Limit}

For the calculation of the one-loop $N$ - gluon amplitude
for the open bosonic string, one 
inserts $N$ copies of the gluon vertex operator
eq.(\ref{gluonvertop})
on the boundary of the annulus, fig. \ref{anninsert}. Then
one has to compute the Polyakov path integral
over the space of all embeddings of
the annulus into spacetime, and the integral over
moduli space. In the case of the annulus there is
only one modular parameter ~\cite{gsw},

\be
\tau=-\half \ln (q)
\label{defmodpar}
\ee
\no
where $q$ can be interpreted as the square of the
ratio of the two radii defining the annulus.
Since the two-dimensional
worldsheet theory is free, the path integral
can be computed
by a repeated application of Wick's theorem, using the
Green's function $G_B^{\rm ann}$ for the Laplacian on the annulus.
If we assume all of the vertex operators to be
on the same boundary, one has explicitly

\be
\langle
y^{\mu}(\tau_1)y^{\nu}(\tau_2)
\rangle
=g^{\mu\nu}
G^{\rm ann}_B(\tau_1,\tau_2)
=-g^{\mu\nu}
\biggl[
{\rm ln}
|2 \sinh(\tau_{12})|
-{\tau_{12}^2\over \tau}
-4q\sinh^2(\tau_{12})
\biggr]
+O(q^2)
\label{defGann}
\ee\no
where $\tau$ denotes the length of the boundary,
$\tau_i$ the location of the 
i-th vertex operator
along the boundary, $\tau_N =0$,
and $\tau_{ij}\equiv\tau_i-\tau_j$.
One obtains a parameter integral 
(compare eq.(\ref{scalarqedmaster}))

\begin{eqnarray}
\Gamma[k_1,\varepsilon_1;\ldots;k_N,\varepsilon_N] &\sim&
{(\alpha')}^{({N\over 2}-2)}
{\dps\int_{0}^{\infty}}d\tau
{\tau}^{-{D\over 2}}
Z(\tau)
\prod_{i=1}^{N-1} \int_0^{\tau} 
d\tau_i\,
\theta(\tau_i-\tau_{i+1})
\nonumber\\
&&
\!\!\!\!\!\!\!\hspace{-50pt}
\times
\exp\biggl\lbrace\sum_{i<j=1}^N 
\bigl\lbrack \alpha' G_{Bij}^{\rm ann} k_i\cdot k_j
+\half
\sqrt{\alpha'}\dot G_{Bij}^{\rm ann}
(k_i\cdot\varepsilon_j -k_j\cdot\varepsilon_i)
-\fourth
\ddot G_{Bij}^{\rm ann}\varepsilon_i\cdot\varepsilon_j
\bigr\rbrack\biggr\rbrace
\mid_{\rm multi-linear}
\nonumber\\
\label{bkmaster}
\end{eqnarray}
\no
Here we have omitted the color trace and some global
factors. 
$Z(\tau)$ is essentially 
the string vacuum partition function,
and
given by

\be
Z=q^{-1}\prod_{n=1}^{\infty}
{(1-q^n)}^{-2}
= q^{-1}+2+ O(q)
\label{defZ}
\ee\no
The analysis of the infinite string tension limit 
$\alpha'\to 0$ 
can be
simplified by first removing all second derivatives
$\ddot G_{Bij}^{\rm ann}$ by suitable partial
integrations in the variables $\tau_i$. This is always
possible ~\cite{berkos:npb362}, and will be 
discussed later on in the field theory
context. The integrand then becomes homogeneous in $\alpha'$.
The possible boundary terms 
appearing in the integration by parts 
can be made to vanish by a suitable
analytic continuation in the external momenta.

In the $\alpha'\to 0$ limit, first one has to extract massless poles
in the S-matrix, which can appear for regions where
$\tau_i\to\tau_j$. Those are of the form

\be
\int d\tau_i{1\over {\tau_{ij}}^{1+\alpha'k_i\cdot k_j}}
\quad
{\stackrel{\sy{\alpha'\to 0}}{\longrightarrow}}
\quad
-{1\over\alpha'k_i\cdot k_j}
\label{extractpoles}
\ee\no
and yield the so-called ``tree'' or ``pinch'' contributions.
The limit itself is to be taken on the sum of the unpinched
expression together with all pinch contributions. It is consumed
by taking
$\tau,|\tau_{ij}|\to\infty$, which corresponds
to the ratio of radii approaching 1, and thus to the
annulus being squeezed to a field theory loop.
Analyzing $G^{\rm ann}_B$ and $\dot G^{\rm ann}_B$ in this
limit, one finds that they can be replaced by

\bear
{\rm exp}\Bigl[G^{\rm ann}_B(\tau_{12})\Bigr]
 &\rightarrow & 
{\rm const.}\times {\rm exp}
\Bigl(
{\tau_{12}^2\over\tau}
-|\tau_{12}|
\Bigr) \label{Gannlim}\\
\dot G^{\rm ann}_B(\tau_{12})&\rightarrow &
-{\rm sign} (\tau_{12})
+2{\tau_{12}\over\tau}
+2\,{\rm sign}(\tau_{12})
\Bigl(
q\,\e^{2|\tau_{12}|}
-\e^{-2|\tau_{12}|}
\Bigr)
\label{Gdotannlim}
\ear\no
Since this limit corresponds to 
$q\to 0$, and the expansion of $Z$
as a power series in $q$ starts with a $q^{-1}$, 
there
are again two types of contributions. 

The first type arises by picking the next-to-leading
constant term in $Z$. Then only the leading order terms
from the integrand can survive the limit,
so that $\dot G^{\rm ann}_B$ is further truncated to
the first two terms of
eq.(\ref{Gdotannlim}).

The second type is obtained by combining the
leading order term from $Z$ with a next-to-leading
term in the integrand. Then 
subleading terms in the $\dot G^{\rm ann}_{Bij}$ 
can potentially contribute, however it turns out that
a too strong suppression of the integrand
for $q\to 0$ can be avoided only 
for those terms in the integrand which contain
a closed cycle of $\dot G^{\rm ann}_{Bij}$'s,
i.e. a factor

\be
\dot G^{\rm ann}_{Bi_1i_2}
\dot G^{\rm ann}_{Bi_2i_3}
\cdots
\dot G^{\rm ann}_{Bi_ni_1}
\label{defcycle}
\ee\no
Even then, the cycle can only survive the limit
if the indices follow the ordering of the
external legs. 
Of course it is also possible to combine
the leading order terms from both
$Z$ and the integrand. This produces terms
which diverge in the limit. Those are discarded,
since they can be identified as being due 
to an unphysical tachyonic scalar
circulating in the loop.

\no
In this way one arrives at the Bern-Kosower 
rules for the gluon loop as given
in \cite{berkos:npb379,bern:plb296}.
In the following we will give those
rules in a slightly different version,
which is more in line with the worldline
path integral approach with respect to the treatment
of the color algebra.
The original string-based approach naturally yields
the gluon amplitudes in color-decomposed
form, i.e. with the group theory factors expressed
in terms of the fundamental instead of the adjoint
representation \cite{bergie,manpar}. 
If in the above derivation the gluon vertex operators
are taken to be in the fundamental representation,
as it was done in \cite{berkos:npb379,bern:plb296},
then the resulting field theory amplitude
represents the so-called ``leading color
partial amplitude'', where leading refers to
the large $N_c$ limit of $SU(N_c)$ gauge theory.
(There is an extra overall factor of $N_c$ in this approach, coming 
from a second index in the fundamental representation, which is untouched
by all the gluon vertex operators, leading to tr(1)$\,=\,N_c$.)
The missing subleading amplitudes would be obtained
by the inclusion of string theory amplitudes
with vertex operator insertions on both boundaries
of the annulus, although it turns out that they
can also be constructed directly as sums of 
permutations of the leading color amplitude 
\cite{bddk,dedima}.

The form of the spin 1 rules which we give here 
instead does not use color decomposition;
in eqs.(\ref{partialamplitude}),(\ref{totalamplitude})
below, as well as in the remainder of this review,
a color matrix $T^a$ always refers to the adjoint
representation in the gluon loop case.
The equivalence of these rules to the original
version of \cite{berkos:npb379,bern:plb296}
follows from recent work on the color
decomposition \cite{dedima}.

The analogous rules for the spin $\half$
loop can be derived by repeating the same analysis
for the open superstring.
Remarkably, this leads to a 
rule which allows one to infer all
contributions from worldsheet fermions to the final
integrand from the purely bosonic terms.
This ``replacement rule'' will play a prominent
role in many of our applications. Its validity
can be shown to be a consequence of worldsheet
supersymmetry ~\cite{berntasi}.
For the spin 0 loop, one finds only the
first type of contributions above.

\subsection{The Bern-Kosower Rules for Gluon Scattering}

For the statement of the rules, we rewrite them
in the conventions used throughout
the remainder of this work. In particular, we
work in the Euclidean.

The Bern-Kosower rules give a prescription for
the construction of 
integral representations for one-loop
photon or gluon scattering amplitudes
in non-abelian gauge theory
~\cite{berkos:npb379}.
We write them down here for the case of
a non-abelian gauge theory with massless
scalars and fermions.
To obtain the one-loop on-shell 
amplitude for 
the scattering of N gluons
\footnote{By ``gluon'' we denote any non-abelian
gauge boson.},
with momenta $k_i$ and polarization vectors
$\varepsilon_i$,
the following steps have to be taken:

\vskip.4cm

{\bf step 1}

\noindent
Consider the following kinematic expression:

\begin{equation}
K = \int \prod_{i=1}^N du_i \prod_{i<j}
{\rm exp} \Bigl[G_{Bij}k_i\cdot k_j 
+i\dot G_{Bij} (k_i\cdot\varepsilon_j
-k_j\cdot\varepsilon_i )
+ \ddot G_{Bij}
\varepsilon_i\cdot \varepsilon_j
\Bigr]\,\,\,
\bigg \vert_{{\rm multi-linear}}
\label{K}
\end{equation}

\noindent
where ``multi-linear'' means that only terms linear
in each of the $\varepsilon_1,\ldots,\varepsilon_N$
are to be kept.

\vskip15pt
{\bf step 2}

\noindent
An on-shell
gluon has only two physical polarizations, denoted by
``+'' and ``--''. 
Consider one helicity amplitude at a time, and 
denote, for example,
by A(+,+,--,...,--) the amplitude for the process
where the first two gluons have the same helicity, and
all remaining ones the opposite one. Each helicity amplitude is
separately gauge invariant, i. e. insensitive to a redefinition of
any of the polarization vectors by a transformation

\begin{equation}
\varepsilon_i^{\pm\mu}
\rightarrow
\varepsilon_i^{\pm\mu}
+ \lambda k_i^{\mu}
\label{trafo}
\end{equation}

\noindent
This freedom can be used to 
choose -- for given external momenta 
$\lbrace k_1,\ldots, k_N\rbrace$ --
a set of polarization vectors 
which makes a maximal number of the
invariants $k_i\cdot\varepsilon_j$ and
$\varepsilon_i\cdot\varepsilon_j$
vanish. A systematic way of finding such
a set of polarization vectors for a
given choice of helicities is provided
by the {\sl Spinor Helicity Method}
(see, e.g., \cite{berntasi,gaswubook}). At the end, all
surviving invariants are rewritten as functions
of the external  momenta alone.

\vskip18pt
{\bf step 3}
\vskip1pt
\noindent
Expand out the kinematic expression, and perform integrations
by parts, till all double derivatives of $G_B$ are
removed (ignore boundary terms). 
We have now an expression

\begin{equation}
\int \prod_{i=1}^N du_i 
K_{\rm red}
\prod_{i<j}
{\rm exp} \bigl[G_{Bij}k_i\cdot k_j
\bigr]\,\,\,
\quad 
\label{Kred}
\end{equation}

\noindent
where $K_{\rm red}$, the ``reduced kinematic factor'', 
is a sum of terms that are products of $\dot G_{Bij}$'s,
and of dot products $k_i\cdot k_j, k_i\cdot\varepsilon_j,
\varepsilon_i\cdot\varepsilon_j$.

\vskip18pt
{\bf step 4}
\vskip1pt
\noindent
Draw all possible labelled $\phi^3$ 1-loop diagrams 
$D_i$ with N external legs,

\vskip1.1cm
\begin{figure}[ht]
\begin{center}
\begin{picture}(0,0)%
\epsfig{file=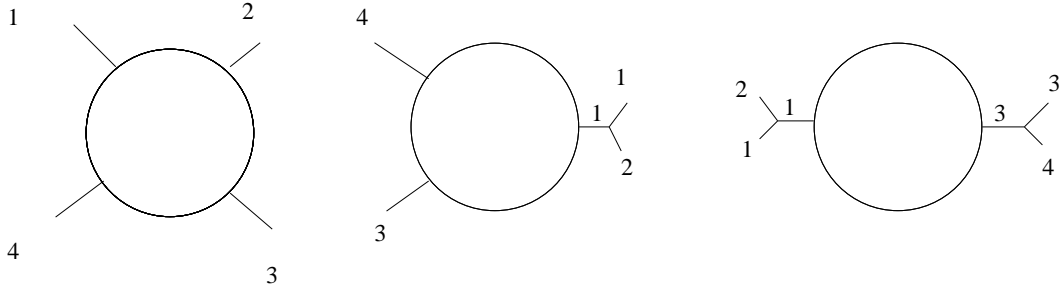}%
\end{picture}%
\setlength{\unitlength}{0.00087500in}%
\begingroup\makeatletter\ifx\SetFigFont\undefined
\def\x#1#2#3#4#5#6#7\relax{\def\x{#1#2#3#4#5#6}}%
\expandafter\x\fmtname xxxxxx\relax \def\y{splain}%
\ifx\x\y   
\gdef\SetFigFont#1#2#3{%
  \ifnum #1<17\tiny\else \ifnum #1<20\small\else
  \ifnum #1<24\normalsize\else \ifnum #1<29\large\else
  \ifnum #1<34\Large\else \ifnum #1<41\LARGE\else
     \huge\fi\fi\fi\fi\fi\fi
  \csname #3\endcsname}%
\else
\gdef\SetFigFont#1#2#3{\begingroup
  \count@#1\relax \ifnum 25<\count@\count@25\fi
  \def\x{\endgroup\@setsize\SetFigFont{#2pt}}%
  \expandafter\x
    \csname \romannumeral\the\count@ pt\expandafter\endcsname
    \csname @\romannumeral\the\count@ pt\endcsname
  \csname #3\endcsname}%
\fi
\fi\endgroup
\begin{picture}(7634,1536)(201,-850)
\end{picture}
\caption{\label{phi3diag} 
Diagrams in $\phi^3$ theory.}
\end{center}
\end{figure}
\vspace{5pt}
but excluding tadpoles, and diagrams where the loop is isolated on an
external leg:
\vskip10pt
\begin{figure}[ht]
\begin{center}
\begin{picture}(4000,0)%
\epsfig{file=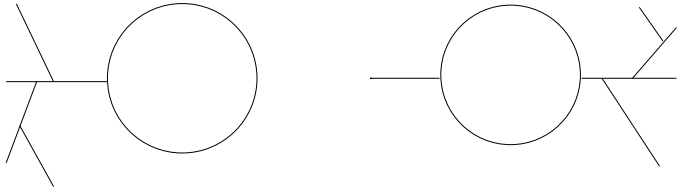}%
\end{picture}%
\setlength{\unitlength}{0.00087500in}%
\begingroup\makeatletter\ifx\SetFigFont\undefined
\def\x#1#2#3#4#5#6#7\relax{\def\x{#1#2#3#4#5#6}}%
\expandafter\x\fmtname xxxxxx\relax \def\y{splain}%
\ifx\x\y   
\gdef\SetFigFont#1#2#3{%
  \ifnum #1<17\tiny\else \ifnum #1<20\small\else
  \ifnum #1<24\normalsize\else \ifnum #1<29\large\else
  \ifnum #1<34\Large\else \ifnum #1<41\LARGE\else
     \huge\fi\fi\fi\fi\fi\fi
  \csname #3\endcsname}%
\else
\gdef\SetFigFont#1#2#3{\begingroup
  \count@#1\relax \ifnum 25<\count@\count@25\fi
  \def\x{\endgroup\@setsize\SetFigFont{#2pt}}%
  \expandafter\x
    \csname \romannumeral\the\count@ pt\expandafter\endcsname
    \csname @\romannumeral\the\count@ pt\endcsname
  \csname #3\endcsname}%
\fi
\fi\endgroup
\begin{picture}(4734,1536)(120,-850)
\end{picture}
\caption{\label{notadpoles} 
Diagrams to be omitted.}
\end{center}
\end{figure}

\noindent
The labels follow the cyclic ordering of the trace. One also attaches a label
to every internal line in the tree part of a diagram; for definiteness,
this  is taken to be the smallest one of the labels of the two lines into
which  the line splits to the outward.

\noindent
Every diagram $D_i$ contributes a parameter integral

\begin{equation}
D_i=\Gamma (m-{D\over 2})
\int_0^1 du_{i_1}
\int_0^{u_{i_1}}du_{i_{2}}\cdots\int_0^{u_{i_{m-2}}}du_{i_{m-1}}
{{P_i(u_{i_1},\ldots,u_{i_m})}\over
{{\bigl [-\sum_{r<s}^m G_{Bi_r i_s}K_{i_r}\cdot K_{i_s}\bigr ]}
^{m-{D\over 2}}}}
\label{D}
\end{equation}

\noindent
Here $D$ is the space-time dimension, $m$ is the number of legs directly
attached to the loop, and

\begin{equation}
G_{Bi_r i_s} = G_B(u_{i_r},u_{i_s}) = \mid u_{i_r} - u_{i_s}\mid
-{(u_{i_r}-u_{i_s})}^2 = (u_{i_r}-u_{i_s}) - {(u_{i_r}-u_{i_s})}^2
\label{G}
\end{equation}

\noindent
($u_{i_m}=0$).
$K_{i_r}$ denotes the sum of the external momenta flowing into
the tree which enters the loop at the point carrying the label
$i_r$.
$P_i$ is a  polynomial function of the loop parameters, and of the external
momenta and polarization vectors; it will be determined in steps 5
and 6. 

\begin{figure}[ht]
\begin{center}
\begin{picture}(7000,0)%
\epsfig{file=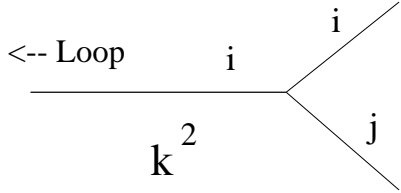}%
\end{picture}%
\setlength{\unitlength}{0.00087500in}%
\begingroup\makeatletter\ifx\SetFigFont\undefined
\def\x#1#2#3#4#5#6#7\relax{\def\x{#1#2#3#4#5#6}}%
\expandafter\x\fmtname xxxxxx\relax \def\y{splain}%
\ifx\x\y   
\gdef\SetFigFont#1#2#3{%
  \ifnum #1<17\tiny\else \ifnum #1<20\small\else
  \ifnum #1<24\normalsize\else \ifnum #1<29\large\else
  \ifnum #1<34\Large\else \ifnum #1<41\LARGE\else
     \huge\fi\fi\fi\fi\fi\fi
  \csname #3\endcsname}%
\else
\gdef\SetFigFont#1#2#3{\begingroup
  \count@#1\relax \ifnum 25<\count@\count@25\fi
  \def\x{\endgroup\@setsize\SetFigFont{#2pt}}%
  \expandafter\x
    \csname \romannumeral\the\count@ pt\expandafter\endcsname
    \csname @\romannumeral\the\count@ pt\endcsname
  \csname #3\endcsname}%
\fi
\fi\endgroup
\begin{picture}(3634,1536)(201,-850)
\end{picture}
\caption{\label{pinch} 
Removal of trees.}
\end{center}
\end{figure}

\vskip15pt
{\bf step 5}: {\bf tree replacement rules}
\vskip5pt
\noindent
Remove all trees, working from the outside of the diagram
toward the loop. 
If the diagram contains a vertex as shown in fig. (\ref{pinch})

\noindent
keep only those terms in $K_{\rm red}$
which contain exactly one $\dot G_{Bij}$.
In those, replace $\dot G_{Bij}$,
with $i<j$,
by ${2\over k^2}$, and replace all remaining $\dot G_{Bjr}$
by $\dot G_{Bir}$.
Repeat this procedure, till only the naked loop is left.

\vskip10pt
{\bf step 6}: {\bf loop replacement rules}
\vskip5pt
\noindent
It is only at this stage that one has to distinguish between the scalar, the
fermion, and the gluon loop.

\vskip.4cm\no
{\underline {Scalar loop:}}
Simply write out $K_{\rm red}$ (i. e. what became of $K_{\rm red}$ in step 5)
in terms of the integration variables, by substituting

\begin{equation}
\dot G_{Bij} \rightarrow  
{\rm sign}(u_i - u_j) 
-2(u_i - u_j)
\label{scalar}
\end{equation}
\no
Multiply by an overall factor of 2 if the scalar is complex.

\vskip.4cm\no
{\underline {Spinor loop:}}
Replace {\sl simultaneously}
every closed cycle 
$\dot G_{Bi_1i_2}
\dot G_{Bi_2i_3}\cdots\dot G_{Bi_ki_1}$
appearing in $K_{\rm red}$ (which may or may not follow
the ordering of the external legs) by

\begin{equation}
\dot G_{Bi_1i_2}
\dot G_{Bi_2i_3}\cdots\dot G_{Bi_ki_1}
- G_{Fi_1i_2}G_{Fi_2i_3}\cdots
G_{Fi_ki_1}
\label{fermion}
\end{equation}\no
An expression is considered
a cycle already if it
can be put into cycle form using the
antisymmetry of $\dot G_B$
(e.g. $\dot G_{Bij}\dot G_{Bij}
=-\dot G_{Bij}\dot G_{Bji}$). 

\noindent
Then replace all $\dot G_B$'s as in the scalar case,
and all $G_F$'s by

\begin{equation}
G_{Fij} \rightarrow {\rm sign}(u_i-u_j)
\label{Gf}
\end{equation}

\noindent
Multiply by an overall factor of 
\vskip.2cm

--4 \quad for a Dirac fermion

--2 \quad for a Weyl fermion

\vskip.4cm\no
\no
{\underline {Gluon loop:}}
In this case, there are two types of contributions,
which have to be summed:

\vskip.2cm\no
{\underline {type 1:}}
Replace $\dot G_{Bij}$ as above.

\vskip.2cm\no
{\underline {type 2:}}
For every closed cycle of $\dot G_{Bij}$'s appearing
in a term,
{\sl with the ordering of the indices following the 
ordering of the external legs}, 
write down {\sl one} additional contribution, obtained
in the following way: replace

\begin{eqnarray}
\dot G_{Bi_1i_2}
\dot G_{Bi_2i_1}&\rightarrow&
4\nonumber\\
\dot G_{Bi_1i_2}
\dot G_{Bi_2i_3}\cdots\dot G_{Bi_ki_1}
&\rightarrow&
2^{k-1}
\quad (k>2)\nonumber\\
\label{gluoncyclerule}
\end{eqnarray}

\noindent
and all remaining $\dot G_B$'s -- even those belonging
to other cycles -- as in the scalar case.

\noindent
Multiply by a factor of $2$ for both types.

\noindent
The expression obtained from $K_{\rm red}$ in this way
is the polynomial $P_i$ above.

\vskip10pt
{\bf step 7}
\vskip5pt
\noindent
Perform the parameter integrations. 
Techniques for their calculation may
be found in ~\cite{bedikopent,caglmi,biguhe}.
The four-point case is treated in appendix
\ref{boxint}, following \cite{bedikopent}.

{\bf step 8}
\vskip5pt
\noindent
Finally the
amplitude is given by a sum over all diagrams,
with an overall normalization factor:

\begin{equation}
\Gamma^{a_1\cdots a_N}
[k_1,\varepsilon_1;\ldots;k_N,\varepsilon_N]
 = (-ig)^N
{\rm tr}
(T^{a_1}\cdots T^{a_N})
{{{(4\pi\mu^2)}^{-{\epsilon\over 2}}}\over{32{\pi}^2}}
\sum_{{\rm diagrams}} D_i
\label{partialamplitude}
\end{equation}

\noindent
Here $\epsilon = D-4$, and
$\mu$ is the  usual unit of mass appearing
in the dimensional continuation. 
$T^{a_i}$ is a color matrix in the representation
of the loop particle.
We have specialized to a specific version of
dimensional regularization, the so-called
four-dimensional helicity scheme ~\cite{berkos:npb379}.
This is only the partial amplitude corresponding to
the considered fixed ordering of the external states.
To obtain the complete amplitude one must still
sum over all possible 
non-cyclic
permutations of the states,
so that
\be
\Gamma(\lbrace a_i,k_i,\varepsilon_i\rbrace )
=
\sum_{\pi \in S_N/Z_N}
\Gamma^{a_{\pi (1)}\cdots a_{\pi (N)}}
[k_{\pi (1)},\varepsilon_{\pi (1)};\ldots;
k_{\pi (N)},\varepsilon_{\pi (N)}]
\label{totalamplitude}
\ee\no
In chapter four we will explicitly apply these rules to the 
four-gluon case.

See \cite{bedush,dunnor} for the analogous rules for 
graviton scattering in quantum gravity.

\section{Worldline Path Integral Representations for Effective Actions}
\renewcommand{\theequation}{3.\arabic{equation}}
\setcounter{equation}{0}

In this chapter, we derive worldline path integral representations
for a number of one-loop effective actions 
involving some of the 
most basic interactions in quantum field theory. 
Those derivations are based on the fact that
one-loop effective actions can generally
be expressed in terms of the determinant of the kinetic
operator in field theory. Using the $ln(det)=tr(ln)$ -- formula
and the Schwinger proper-time representation, one obtains an
integral over the space of all closed trajectories of a
quantum mechanical particle moving in spacetime. 

Generally, to every such closed loop path integral one finds
associated a similar open-ended
path integral, representing the field theory propagator 
of the loop particle in the background field.
The propagator path integral 
has the same worldline Lagrangian, possibly with some
boundary terms added. 
It is to be performed 
over the space of trajectories
connecting two fixed points in spacetime, with appropriate
boundary conditions.

In the present review we will concentrate on the
effective action, or closed loop case, simply because almost all
explicit calculations which have been done so far
pertain to this case.  
By this we do not mean to imply that
propagator path integrals may not
play an important role in
future extensions of this formalism.

Our derivations are mostly formal. Only in
the spin -- $1$ loop case will we
touch upon
the subtle issues connected with the existence of
different discretization prescriptions
etc. Those have recently been
investigated in much detail 
for the case of curved backgrounds, 
a subject which will be shortly discussed in section
\ref{otherbackgrounds}.

\subsection{Scalar Field Theory}

Let us begin with the simplest case of a 
real massive scalar 
field $\phi$ 
with a self-interaction potential
$U(\phi)$.
According to standard quantum field theory
(see, e.g., \cite{itzzub})  the
Euclidean one-loop effective action for
this field theory can be written as
\footnote{
We work with relativistic
quantum field theory conventions, $\hbar = c =1$.
Functional traces are denoted by $Tr$, finite
dimensional traces by $tr$.}

\be
\Gamma[\phi]
=-{1\over 2}
{\rm Tr}
\ln
\biggl[
{{-\Box + m^2 + U''(\phi)}
\over
{-\Box + m^2}}
\biggr]
\label{scaleffact}
\ee
\no
We use the formula

\be
-\Tr\ln \Bigl(
{A\over B}
\Bigr)
=
\Tint
\Tr
\biggl
(
{\rm e}
^{-AT}
-
{\rm e}
^{-BT}
\biggr)
\label{lnformel}
\ee\no
valid for positive definite operators $A,B$,
delete the irrelevant $\phi$ -- independent
term,
and perform the functional trace in $x$ -- space.
This gives

\be
\Gamma[\phi]
=
{1\over 2}
\Tint
\int d^D x\,
\bigl\langle x\mid
\exp 
\biggl\lbrace
-T
\Bigl[
-\Box + m^2 + U''(\phi(x))
\Bigr]
\biggr\rbrace
\mid x\bigr\rangle
\label{Gammatrace}
\ee\no

Now compare this with Feynman's path integral
formula for the evolution operator in
non-relativistic quantum mechanics.
For a particle with mass $\tilde m$
moving in a time-independent
potential $\tilde V(x)$ this
formula reads
(see, e.g., \cite{ditreu:candq,kleinert}),

\be
\langle x'' \mid
{\rm e}^{-i(t''-t')H}
\mid x'\rangle
=
\int
_{x(t')=x'}
^{x(t'')=x''}
{\cal D}x(t)
\,
{\rm e}^{i\int_{t'}^{t''}
dt
\Bigl[
{\tilde m\over 2}
\dot x^2
-\tilde V(x)
\Bigr]
}
\label{feynqm}
\ee\no
We can therefore interpret our kinetic operator
above 
as the Hamilton operator $H$ for
a fictitious particle moving in 
$D$ dimensions,

\be
H={p^2\over 2\tilde m}
+ \tilde V(x)
\label{Hfict}
\ee\no
by identifying

\bear
\tilde V(x) &=& m^2 + U''(\phi(x)) \nonumber\\
\tilde m &=& {1\over 2} \nonumber\\
i(t''-t')&=& T\nonumber\\
\label{identify}
\ear\no
Without retracing the usual 
path integral discretization procedure
\cite{ditreu:candq,kleinert}
we can thus immediately write

\be
\langle x\mid
{\rm exp}
\biggl\lbrace
-T
\Bigl[
-\Box + m^2 + U''(\phi(x))
\Bigr]
\biggr\rbrace
\mid x\bigr\rangle
=
\int
_{x(0)=x}
^{x(T)=x}
{\cal D}x(\tau)
{\rm e}^{-\int_{0}^{T}
d\tau
\Bigl[
\fourth
\dot x^2
+ m^2
+
U''(\phi(x(\tau)))
\Bigr]
}
\label{scalpix}
\ee\no
($\tau = it$). Taking into account that

\be
\int d^Dx\,
\int_{x(0)=x(T)=x}{\cal D}x(\tau)
=
\int_{x(0)=x(T)}
{\cal D}x(\tau)
\label{joinzero}
\ee\no
we obtain the desired path integral representation for the
effective action,

\be
\Gamma[\phi]
=\half
\Tintm
\int_{x(T)=x(0)}{\cal D}x(\tau)\,
e^{-\int_0^T d\tau\Bigl(
\kinb
+U''(\phi(x(\tau)))
\Bigr)}
\label{scalftpi}
\ee\no

In a completely analogous way one derives the path integral
representation for the scalar propagator in the background
field $\phi$,

\bear
\langle x''\mid
{\biggl[ -\Box + m^2 + U''(\phi(x))
\biggr]
}^{-1}
\mid x' \rangle
&=&
\int_0^{\infty}
dT
\,\langle x''\mid
\exp
\biggl\lbrace
-T
\Bigl[
-\Box + m^2 + U''(\phi(x))
\Bigr]
\biggr\rbrace
\mid x'\rangle
\non\\
&=&
\int_0^{\infty}
dT\,
\e^{-m^2T}
\int_{x(0)=x'}^{x(T)=x''}
{\cal D}x(\tau)\,
e^{-\int_0^T d\tau\Bigl(
\kinb
+U''(\phi(x(\tau)))
\Bigr)}
\non\\
\label{scalftprop}
\ear\no

\subsection{Scalar Quantum Electrodynamics}

The path integral for a massive
(complex) scalar field minimally coupled to a
background Maxwell field can also be found simply by
recurring to quantum mechanics. The 
field theory kinetic operator now reads

\be
{(\partial+ieA)}^2-m^2
\label{scalqedkinop}
\ee\no
with
a fictitious Hamiltonian

\be
H={{(p+eA)}^2\over 2\tilde m}
+ m^2
\label{scalqedham}
\ee\no
This translates into

\bear
\Gamma_{\rm scal}[A]
&=&
-\half
{\rm Tr}
\ln
\biggl[
{{-{(\partial+ieA)}^2 + m^2}
\over
{-\Box + m^2}}
\biggr]
\non\\
&=&
\Tintm
\int_{x(T)=x(0)}{\cal D}x(\tau)\,
e^{-\int_0^T d\tau\Bigl(
\kinb
+ie\,\dot x\cdot A(x(\tau))
\Bigr)}
\label{scalarpichap3}
\ear\no
and analogously for the propagator.
Note that the global factor of $\half$ has disappeared,
since in taking the trace
we have to take the double number of degrees
of freedom of the complex scalar into account.

\subsection{Spinor Quantum Electrodynamics}
\label{sectionspinorqed}

For the spin $\half$ - particle
various worldline path integral
representations can be found in the 
literature.
The basic choice is between
bosonic 
~\cite{feynman:pr84,bardur}
and 
Grassmann 
~\cite
{fradkin,casalbuoninc,casalbuoniplb,bermar,bdzdh,brdiho,borcas,baboca,fragit,holten:npb457}
representations.
The latter can be derived using either coherent state methods
~\cite{ohnkas,hehoto,dhogag1,dhogag2}
or the ``Weyl symbol'' method
~\cite{bermar,henteibook}.

Our following treatment of the spin $\half$ -- case uses the
coherent state formalism.
We would like to find a path integral representation
for the Euclidean effective action
\footnote{Our Euclidean Dirac matrix conventions are
$\lbrace\gamma_{\mu},\gamma_{\nu}\rbrace = 2g_{\mu\nu}\Eins,
\gamma_{\mu}^{\dag} = \gamma_{\mu}, 
\gamma_5 = \gamma_1\gamma_2\gamma_3\gamma_4$.}

\be
\Gamma_{\rm spin}[A] = \ln {\rm Det}
[\slash p +e\slash A
-im 
]
\label{defEAeuc}
\ee\no
This time we will not be able to just take over results
from quantum mechanics; we have to construct our path integral
by brute force.

\no
Let us start with the well-known 
observation that we can rewrite

\be
(\slash p + e\slash A)^2 =
- (\partial_\mu 
+ i e A_\mu)^2 
-{i\over 2}\, e
\sigma^{\mu\nu}
F_{\mu\nu}
\label{1to2order}
\ee\no
($\sigma^{\mu\nu}=\half[\gamma^{\mu},\gamma^{\nu}]$).
Using the usual argument that

\be
{\rm Det}
\Bigl[(\slash p + e\slash A) - im\Bigr] = 
{\rm Det}
\Bigl[(\slash p + e\slash A) + im\Bigr] = 
{\rm Det}^{1/2}
\Bigl[(\slash p + 
e\slash A)^2 + m^2\Bigr]
\label{3gamma5trick}
\ee\no
we can then write
the effective action in the following form,

\be
\Gamma_{\rm spin}[A] = -\half\, \Tr\, \int_0^\infty \, 
\frac{dT}{T} \, \exp 
\Bigl\lbrace
- T \left[-
(\partial + ieA)^2 
-{i\over 2}
e\,
\sigma^{\mu\nu}
F_{\mu\nu}
+ m^2 \right]
\Bigr\rbrace\non\\
\label{Weucl}
\ee
Up to the global sign, this is 
formally identical with the effective action for
a scalar loop in a background containing, besides the
gauge field $A$, a potential term 

\be
V\equiv -{i\over 2} e\,
\sigma^{\mu\nu}
F_{\mu\nu}
\label{3defV}
\ee\no
We will now use the formalism developed in
\cite{ohnkas} 
to transform this functional trace into a quantum mechanical
path integral. Our treatment closely parallels the one in
\cite{dhogag1,dhogag2},
except that they work in six dimensions.
Define matrices $a_r^+$ and $a_r^-$, $r=1,2$, by

\be
a_1^{\pm} = \half(\gamma_1\pm i\gamma_3),
\quad
a_2^{\pm} = \half(\gamma_2\pm i\gamma_4)
\label{defa12}
\ee\no
Those satisfy Fermi-Dirac anticommutation rules

\be
\lbrace a_r^+,a_s^-\rbrace = \delta_{rs},
\quad
\lbrace a_r^+,a_s^+\rbrace = 
\lbrace a_r^-,a_s^-\rbrace = 
0
\label{arsanticomm}
\ee
Thus we can use $a_r^+$ and $a_r^-$ as creation and
annihilation operators for a Hilbert space with a
vacuum defined by

\be
a_r^- |0\rangle =
\leftvac a_r^+ =0
\label{defavacuum}
\ee\no
Next we introduce Grassmann variables $\eta_r$ and $\bar\eta_r$,
$r=1,2$,
which anticommute with one another and with the operators
$a_r^{\pm}$, and commute with the vacuum $\rightvac$.
The coherent states are then defined as

\bear
\langle\eta\mid\equiv i\langle0\mid
(\eta_1-a_1^-)
(\eta_2-a_2^-)
\qquad
&&
\mid\eta\rangle
\equiv
\exp (-\eta_1a_1^+ -\eta_2a_2^+)
\rightvac
\non\\
\langle\bar\eta\mid
\equiv
\leftvac
\exp (-a_1^-\bar\eta_1-a_2^-\bar\eta_2)
\qquad &&
\mid\bar\eta\rangle
\equiv
i
(\bar\eta_1-a_1^+)
(\bar\eta_2-a_2^+)
\rightvac
\non\\
\label{defcoherentstates}
\ear
It is easily verified that those satisfy the defining
equations for coherent states,

\bear
\langle\eta\mid a_r^- = \langle\eta\mid\eta_r
\qquad
a_r^-\mid\eta\rangle = \eta_r\mid\eta\rangle
\qquad
&&
\langle\eta\mid\bar\eta\rangle
=
\exp
(\eta_1\bar\eta_1
+\eta_2\bar\eta_2
)
\non\\
\langle\bar \eta\mid a_r^+ = \langle\bar \eta\mid\bar\eta_r
\qquad
a_r^+ \mid\bar\eta\rangle = \bar\eta_r\mid\bar\eta\rangle
\qquad
&&
\langle\bar\eta\mid\eta\rangle
=
\exp
(\bar\eta_1\eta_1
+\bar\eta_2\eta_2
)
\label{verifycoherentstates}
\ear
Also one introduces the corresponding Grassmann integrals,
defined by

\be
\int\eta_i d\eta_i
=
\int\bar\eta_i d\bar\eta_i
=i
\label{Grassmannetaint}
\ee
The $d\eta_r,d\bar\eta_r$ commute with one another and with the vacuum,
and anticommute with all Grassmann variables and the $a^{\pm}_r$.
This leads to the completeness relations

\be
\Eins = i
\int\mid\eta\rangle\langle\eta\mid
d^2\eta
= -i
\int d^2\bar\eta
\mid\bar\eta\rangle
\langle
\bar\eta\mid
\label{completenessrelations}
\ee
($d^2\eta=d\eta_2 d\eta_1,
\,\,
d^2\bar\eta
=d\bar\eta_1
d\bar\eta_2
$), and to the following representation for a 
trace in the Fock space generated by the
$a^{\pm}_r$,

\be
\Tr (U)
= i
\int
d^2\eta\,
\langle
-\eta
\mid
U\mid
\eta\rangle
\label{Diractracerepresentation}
\ee
We can now apply these fermionic
coherent states together with the usual complete
sets of coordinate states to rewrite the
functional trace in
(\ref{Weucl}) in the following way, 

\bear
\Tr
\e^{-T\Sigma}
&=& i
\int d^4x \int d^2\eta\,
\langle
x,-\eta\mid
\e^{-T\Sigma}
\mid
x,\eta\rangle
\non\\
&=& i^N
\int
\prod_{i=1}^N
\Bigl(
d^4x^id^2\eta^i
\langle
x^i,\eta^i\mid
\e^{-{T\over N}\Sigma}
\mid
x^{i+1},\eta^{i+1}
\rangle
\Bigr)
\label{rewritefunctionaltrace}
\ear
where 
$\Sigma = - (\partial 
+ ieA)^2 + V
$.
The boundary conditions on the $x$ and $\eta$ integrations
are
$(x^{N+1},\eta^{N+1})
=
(x^1,-\eta^1)$.
For the evaluation of this matrix element it will be useful to look
first at the matrix elements of products of
Dirac matrices. For the product of two $\gamma$'s one finds

\be
\langle
\eta^i
\mid
\gamma_{\mu}\gamma_{\nu}
\mid
\eta^{i+1}
\rangle
=
- i
\int d^2\bar\eta^{i,i+1}
\langle
\eta^i\mid
\bar\eta^{i,i+1}
\rangle
\langle
\bar\eta^{i,i+1}
\mid\eta^{i+1}
\rangle
2
\phantom{,}^i\psi_{\mu}
\psi_{\nu}^{i+1}
,\quad
\mu\ne \nu
\label{gammatopsi}
\ee
where

\bear
\psi_{1,2}^{i+1}
\equiv
{1\over\sqrt 2}
(\eta_{1,2}^{i+1}+\bar\eta_{1,2}^{i,i+1})
,\qquad
&&
\psi_{3,4}^{i+1}
\equiv
{i\over\sqrt 2}
(\eta_{1,2}^{i+1}
-\bar\eta_{1,2}^{i,i+1}
),
\non\\
\phantom{,}^i\psi_{1,2}
\equiv
{1\over\sqrt 2}
(\eta_{1,2}^i
+
\bar\eta_{1,2}^{i,i+1}
),\qquad
&&
\phantom{,}^i
\psi_{3,4}
\equiv
{i\over\sqrt 2}
(\eta_{1,2}^i -
\bar\eta_{1,2}^{i,i+1}
)
\label{etatopsi}
\ear
To verify this equation one
rewrites the Dirac matrices
in terms of the $a_r^{\pm}$
and then
inserts a complete set of coherent
states
$\mid\bar\eta^{i,i+1}\rangle$
in between them. 

With this information it is now easy to compute that

\bear
\langle
x^i,\eta^i\mid
\e^{-{T\over N}\Sigma[p,A,\gamma_{\mu}\gamma_{\nu}]}
\mid x^{i+1},\eta^{i+1}\rangle
&=&
-
{i\over {(2\pi)}^4}
\int
d^4p^{i,i+1}
d^2\bar\eta^{i,i+1}
\e^{i(x^i-x^{i+1})p^{i,i+1}
+(\eta^i-\eta^{i+1})_r
\bar\eta_r^{i,i+1}}
\non\\
&&\!\!\!\!\!\!\!\!\!\!\!\times
\biggl\lbrace
1-{T\over N}
\Sigma\Bigl[p^{i,i+1},A^{i,i+1},
2\phantom{,}^i\psi_{\mu}
\psi_{\nu}^{i+1}\Bigr]
+O\Bigl({T^2\over N^2}\Bigr)
\biggr\rbrace
\non\\
\label{matrixelement}
\ear
Here the superscript $\phi^{i,i+1}$ on a field denotes
the average of the corresponding fields with superscripts
$i$ and $i+1$. 
Inserting this result back into eq.(\ref{rewritefunctionaltrace})
one obtains, after symmetrizing the positions of the Grassmann
variables in the exponentials,

\bear
\Tr
\e^{-T\Sigma}
&=&
{1\over {(2\pi)}^{4N}}
\int
\prod_{i=1}^N
d^4x^i
d^4p^{i,i+1}
d^2\eta^i
d^2\bar\eta^{i,i+1}
\Bigl(
1-{T\over N}
\Sigma_i
+
{\rm O}
({T^2\over N^2})
\Bigr)
\non\\
&&\times
\exp
\biggl\lbrace
\sum_{i=1}^N
\Bigl[
i(x^i-x^{i+1})
p^{i,i+1}
+\half
(\eta_r^i
-\eta_r^{i+1})
\bar\eta_r^{i,i+1}
-\half
\eta_r^i
(\bar\eta_r^{i-1,i}-\bar\eta_r^{i,i+1})
\Bigr]
\biggr\rbrace
\non\\
\label{Tracediscretized}
\ear
Introducing an interpolating proper-time
$\tau$ such that $\tau_1=T$, $\tau^{N+1}=0$,
and $\tau^i-\tau^{i+1}  = {T\over N}$, and
taking the limit $N\to\infty$
in the usual naive way, we finally obtain the
following path integral representation,

\be
\Tr
\e^{-T\Sigma}
=
\int{\cal D}p
\int{\cal D}x
\int_A{\cal D}\eta{\cal D}\bar\eta
\exp
\biggl\lbrace
\int_0^Td\tau\,
\Bigl[
i\dot x\cdot p
+\half
\dot\eta_r\bar\eta_r
-\half
\eta_r
\dot{\bar\eta_r}
-\Sigma [p,A,2\psi_{\mu}\psi_{\nu}]
\Bigr]
\biggr\rbrace
\label{pxpi}
\ee
The ``A'' denotes the antiperiodic boundary conditions which we
have for $\eta,\bar\eta$. 

\no 
The continuum limits of eqs.(\ref{etatopsi})
are

\be
\psi_{1,2}(\tau)
=
{1\over\sqrt 2}
(\eta_{1,2}(\tau)+\bar\eta_{1,2}(\tau)),
\qquad
\psi_{3,4}(\tau)
=
{i\over\sqrt 2}
(\eta_{1,2}(\tau)-\bar\eta_{1,2}(\tau))
\label{etatopsicont}
\ee
This suggests a change of variables from
$\eta,\bar\eta$ to $\psi$, which we
complete by rewriting the fermionic kinetic term,

\be
\half
\dot\eta_r\bar\eta_r
-\half
\eta_r\dot{\bar\eta}_r
=
-\half
\psi^{\mu}\dot\psi_{\mu}
\label{etatopsikin}
\ee
The boundary conditions are now 
$(x(T),\psi(T))=(x(0),-\psi(0))$.
Finally, we note that the momentum path integral
is Gaussian, and perform it by a naive completion
of the square (for a less unscrupulous treatment 
of this point see again \cite{dhogag1,dhogag2}, 
as well as for the various normalization factors involved). 
This brings us to our following
final result
\footnote{Our definition of the Euclidean effective action
differs by a sign from the one used in \cite{dhogag1,dhogag2}.}

\bear
\Gamma_{\rm spin}[A]
&=& -\half\, \int_0^\infty \, 
\frac{dT}{T} 
\e^{-m^2T}
\int_{P}{\cal D}x
\int_{A}{\cal D}\psi
\, \e^
{
-\int_0^Td\tau\,
L_{\rm spin}
}\label{3spinorpi}\\
L_{\rm spin} &=&
\kinb
+\half\psi_{\mu}\dot\psi^{\mu}
+ie\dot x^{\mu}A_{\mu}
-ie\psi^{\mu}F_{\mu\nu}\psi^{\nu}
\
\label{3spinorlag}
\ear\no
which we already quoted in the introduction,
eq.(\ref{spinorpi}). Although in the present
review we will be exclusively concerned
with four-dimensional
field theories, it should be mentioned that
the obtained path integral representation
is valid for all even spacetime dimensions.
Note also that only the even subspace of the
Clifford algebra came into play in the above.
This is different in the case of an open fermion line,
and is the reason why the 
corresponding path integral representation for the
electron propagator in a background field
is significantly more complicated 
\cite{baboca,faimar,hehoto,fragit,holten:npb457,rescsc}.

In the introduction it was also mentioned that
the worldline Lagrangian (\ref{3spinorlag})
has a global supersymmetry, (\ref{wlsusy}).
One consequence of this is that we can 
make use of 
a one-dimensional superfield formalism.
Introducing

\begin{eqnarray}
X^{\mu} &=& x^{\mu} 
+ \sqrt 2\,\theta\psi^{\mu} 
\label{defX}\\
Y^{\mu} &=& X^{\mu}-x_0^{\mu}
\label{defY}\\
D &=& {\partial\over{\partial\theta}} - 
   \theta
{\partial\over{\partial\tau}} \label{defD}\\
\int d\theta\theta &=& 1 
\label{deftheta}
\end{eqnarray}
we can
combine the $x$ -- and $\psi$ -- path
integrals into the following
super path 
integral ~\cite{polbook,rajeev,andtse,ss3},

\begin{equation}
\Gamma_{\rm spin}
\lbrack A\rbrack   = - \half{\displaystyle\int_0^{\infty}}
{dT\over T}
e^{-m^2T}
{\displaystyle\int} {\cal D} X
\,{\rm e}^{-
\int_0^T d\tau\int d\theta \, \Bigl [-{1\over 4}X\cdot D^3 X 
- ieDX\cdot A(X)\Bigr ]}
\label{superpi}
\end{equation}
\no
Written in this way, the spinor path integral becomes formally
analogous to the scalar one, and can be considered as its
``supersymmetrization''.
Note, however, that  
the supersymmetry is broken by the
different periodicity
conditions
which we have for the 
coordinate and the Grassmann path integrals.
For a constant Grassmann parameter $\eta$ those are not
respected by the supersymmetry transformations
(\ref{wlsusy}).

\subsection{Non-Abelian Gauge Theory}

\subsubsection{Scalar Loop Contribution to the Gluon Effective Action}

The simplest non-abelian generalization which one can
consider is the
contribution to the gluon scattering amplitude due to
a scalar loop.
Retracing the above derivation of the
scalar path integral for photon scattering, 
one finds that the non-abelian nature
of the background field leads to the following changes
in eq.(\ref{scalarpi}):

\begin{enumerate}

\item
The trace now includes a global color trace.

\item
The corresponding 
quantum mechanical Hamilton operators at
different times need not commute any more, so that the
exponential must be taken path-ordered.

\end{enumerate}
\no
We have thus

\be
\Gamma_{\rm scal}[A]
=
{\rm tr}
\Tintm
\int{\cal D}x(\tau)\,
{\cal P}
e^{-\int_0^T d\tau\Bigl(
\kinb
+ig\dot x\cdot A(x(\tau))
\Bigr)}
\label{scalarpinonab}
\ee\no
where now $A_{\mu}=A_{\mu}^aT^a$.
$\cal P$ denotes the path ordering operator, and
$\rm tr$ the color trace.

\subsubsection{Spinor Loop Contribution to the Gluon Effective Action}

In addition to these two changes, in the spinor loop
case the $F_{\mu\nu}$ appearing in the
worldline Lagrangian (\ref{3spinorlag})
must now be taken to be the
full non-abelian field strength tensor, including the
commutator term 
$[A_{\mu},A_{\nu}]$
~\cite{bacalu:npb124,bssw:prd15}.  

One may wonder
how this commutator term is to be accommodated in 
the superfield formalism. 
As was shown in
~\cite{andtse}, a very convenient
way of doing so is to
introduce a super path ordering.
The ordinary path ordering can be 
defined by 

\be
{\cal P}\prod_{i=1}^N\int_0^T d\tau_i
\equiv
N!
\int_0^T\, d\tau_1
\int_0^{\tau_1}\, d\tau_2
\cdots
\int_0^{\tau_{N-1}}\, d\tau_N
=
N!
\int_0^T
d\tau_1
\cdots
\int_0^Td\tau_N
\prod_{i=1}^{N-1}
\theta(\tau_i-\tau_{i+1})
\label{defpathorder}
\ee\no
The super path ordering is obtained from
this simply by replacing the 
proper-time differences in the
arguments of the $\theta$ -- functions by
super-differences, 

\be
\hat\tau_{ij}\equiv \tau_i - \tau_j + \theta_i\theta_j
\label{defsuperdiff}
\ee\no
so that
\be
\hat{\cal P}\prod_{i=1}^N
\int_0^T
d\tau_i
\int
d\theta_i
\equiv
N!
\int_0^T
d\tau_1
\int
d\theta_1
\cdots
\int_0^T
d\tau_N
\int
d\theta_N
\prod_{i=1}^{N-1}
\theta(\hat\tau_{i(i+1)})
\label{defsuperpathorder}
\ee\no
(here and in the following we use the convention that
$\prod_{i=1}^N d\theta_i \equiv d\theta_1 d\theta_2\cdots d\theta_N$).
Then expanding the $\theta$ -- functions one finds

\be
\theta(\hat\tau_{i(i+1)})=
\theta(\tau_i-\tau_{i+1})+\theta_i\theta_{i+1}
\delta(\tau_i-\tau_{i+1})
\label{supertheta}
\ee\no
and the 
$\delta$ -- function terms
will generate precisely the commutator terms above.
With this definition of $\hat{\cal P}$,
we can thus generalize eq.(\ref{superpi})
to the non-abelian case as

\begin{equation}
\Gamma_{\rm spin}
\lbrack A\rbrack   = - \half{\rm tr}
{\displaystyle\int_0^{\infty}}
{dT\over T}
e^{-m^2T}
{\displaystyle\int} {\cal D} X
{\hat{\cal P}}\,{\rm e}^{-
\int_0^T d\tau\int d\theta \, \Bigl [-{1\over 4}X\cdot D^3 X 
- ig DX\cdot A(X)\Bigr ]}
\label{superactionnonab}
\end{equation}
\no
This remarkable interplay between worldline supersymmetry
and spacetime gauge symmetry has recently attracted
some attention ~\cite{migdal96}.

\subsubsection{Gluon Loop Contribution to the Gluon Effective Action}

We proceed to the 
much more delicate case of the gluon loop, i.e. we wish now 
to derive a
path integral describing a spin-1 particle coupled to a
spin-1 background. Here one would expect to run into difficulties.
It is well-known how to construct free path integrals for
particles of arbitrary spin (see, e.g., \cite{hppt,bucshv}). 
However,
the quantization of those path integrals usually leads to 
inconsistencies 
as soon as one tries to couple a path integral with
spin higher than $\half$ to a spin-1 background. 
In ~\cite{strassler1} this problem had been circumvented by the
introduction of auxiliary degrees of freedom, and we will
follow a similar approach here ~\cite{rescsc}.

We employ the background gauge fixing technique
so that the effective action $\Gamma[A^a_\mu]$ becomes a gauge invariant
functional of $A^a_\mu$ \cite{abbott,ditreu:seltop}.
The gauge fixed classical action reads, in $D$ dimensions,
\be\label{3.1}
S[a;A]=
-
\frac{1}{4} \int d^D x F^a_{\mu\nu} (A+a) F^{a\mu\nu} (A+a)
-
\frac{1}{2\alpha}\int d^D x\left( D^{ab\mu} [A]a^b_\mu\right)^2\ee
A priori, the background field $ A^a_\mu$ is unrelated to the
quantum field $a^a_\mu$. The kinetic operator of the gauge boson
fluctuations is obtained as the second functional
derivative of $S[a,A]$
with respect to $a^a_\mu$,  at fixed $A^a_\mu$.  
This leads to  the inverse  
propagator
\be\label{3.2}
{\cal D}^{ab}_{\mu\nu}=- D^{ac}_\rho D^{cb}_\rho \delta_{\mu\nu}-2ig
F^{ab}_{\mu\nu}\ee
and the effective action
\be\label{3.3}
\Gamma_{\rm glu}[A]=
-
\frac{1}{2}\ln \det({\cal D})
=\frac{1}{2}\int^{\infty}_0\frac{dT}{T} {\Tr}
\bigl( e^{-T{\D}}\bigr)\ee
In writing down eq.(\ref{3.2}) we have 
adopted the Feynman gauge $\alpha=1$.
The covariant derivative $D_\mu\equiv \partial_\mu+ig A^a_\mu  T^a$ and
the field strength $F^{ab}_{\mu\nu}\equiv F^c_{\mu\nu}(T^c)^{ab}$ are
matrices in the adjoint representation of the gauge group
\footnote{Our definition for the
non-abelian covariant derivative is
$D_\mu\equiv \partial_\mu+ig A^a_\mu  T^a$,
with $[T^a,T^b] = i f^{abc}T^c$. The adjoint
representation is given by $(T^a)^{bc}=-i f^{abc}$.}.
The full effective action is obtained by
adding the contribution of the Faddeev-Popov ghosts to eq. (\ref{3.3}).
The evaluation of the ghost determinant proceeds along the same lines as
scalar QED, and will be dealt with later on.

\no
In order to derive a path integral representation of the  heat-kernel

\begin{equation}
{\Tr}\bigl(\e^{-T{\D}}\bigr)  
\end{equation}
\noindent
we first look at a slightly more general problem. We
generalize the operator ${\D}$ to
\be\label{3.4}
\widehat h_{\mu\nu}\equiv - D^2 \delta_{\mu\nu}+M_{\mu\nu}(x)\ee
where $M_{\mu\nu}(x)$ is an arbitrary matrix in color space. In
particular,
we do not assume that the Lorentz trace $M_{\mu}^{\mu}$ is zero. 
Given
$M_{\mu\nu}$, we construct the following one-particle Hamilton operator:
\be\label{3.5}
\widehat H=(\widehat p_\mu+g A_\mu(\widehat x))^2-:
\widehat{\bar\psi^{\mu}}
M_{\nu\mu}(\widehat x)\widehat\psi^\nu:\ee
The system under consideration has a graded phase-space coordinatized
by $x_\mu,p_\mu$ and two sets of anti-commuting variables, $\psi_\mu$
and $\bar\psi_\mu$, which obey canonical anti-commutation relations:
\be\label{3.6}
\widehat\psi_\mu \widehat{\bar\psi}_\nu+
\widehat{\bar\psi}_\nu\widehat\psi_\mu
=\delta_{\mu\nu}\ee
For a reason which will become obvious in a moment we have adopted
the ``anti-Wick'' ordering in (\ref{3.5}): all $\bar\psi$'s are
on the right of all $\psi$'s, e.g.
\bear\label{3.7}
:\widehat\psi_\alpha\widehat{\bar\psi}_\beta:&=&\widehat\psi_\alpha
\widehat{\bar\psi}_\beta\nonumber\\
:\widehat{\bar\psi}_\beta \widehat\psi_\alpha:&=&-\widehat\psi_\alpha
\widehat{\bar\psi}_\beta\ear
We can represent the commutation relations on a space of wave functions
$\Phi(x,\psi)$ depending on $x_\mu$ and a set of classical Grassmann
variables $\psi_\mu$. The ``position'' operators $\widehat x_\mu=x_\mu,\
\widehat\psi_\mu=\psi_\mu$ act multiplicatively on $\Phi$, 
the conjugate
momenta as derivatives $\widehat p_\mu=-i\partial_\mu$ 
and $\widehat{\bar\psi}_\mu
=\partial/\partial\psi^\mu$. 
Thus the Hamiltonian becomes~\cite{reuter95}

\be\label{3.8}
\widehat H=-D^2+\psi^\nu M_{\nu\mu}(x)\frac{\partial}
{\partial \psi_\mu}\ee
The wave functions $\Phi$ have a decomposition of the form

\be\label{3.9}
\Phi(x,\psi)=\sum^D_{p=0}\frac{1}{p!}\ \phi^{(p)}_{\mu_1\dots \mu_p}
(x)\ \psi^{\mu_1}\dots\psi^{\mu_p}\ee\no

This suggests the interpretation of $\Phi$ as an inhomogeneous
differential form on  $\R^D$ with the fermions $\psi^\mu$ playing
the role of the differentials $dx^\mu$~\cite{goreth,gozreu}. 
The form-degree
or, equivalently, the fermion number is measured  by the operator

\be\label{3.10}
\widehat F=\widehat\psi^\mu\widehat{\bar\psi}_\mu
=\psi^\mu\frac{\partial}{\partial
\psi^\mu}\ee
We are particularly interested in one-forms:

\be\label{3.11}
\Phi(x,\psi)=\varphi_\mu(x)\psi^\mu\ee\no

The Hamiltonian (\ref{3.8}) acts on them according to
\be\label{3.12}
(\widehat H\Phi)(x,\psi)=(\widehat h_{\mu}^{\phantom{\mu}\nu}
\varphi_\nu)\psi^\mu \ee\no

We see that, when restricted to the one-form sector,
 the quantum system with the Hamiltonian (\ref{3.5})
is equivalent to the one defined by the bosonic matrix Hamiltonian
$\widehat h_{\mu\nu}$~\cite{reuter95,gozreu}.
\no
The Euclidean proper time
evolution of the wave functions $\Phi$ is implemented by the kernel
\be\label{3.13}
K(x_2,\psi_2, {\tau}_2|x_1,\psi_1,{\tau}_1)=\langle x_2,\psi_2| 
e^{-({\tau}_2-{\tau}_1)
\widehat H}
|x_1,\psi_1\rangle\ee
which obeys the Schr{\"o}dinger equation
\be\label{3.14}
\left(\frac{\partial}{\partial T}+\widehat H\right)
K(x,\psi,T|x_0,\psi_0,0)=0\ee
with the initial condition $K(x,\psi,0|x_0,\psi_0,0)=\delta(x-x_0)\delta
(\psi-\psi_0)$. It is easy to write down a path integral solution
to eq. (\ref{3.14}). For the trace of $K$ one obtains

\be\label{3.15}
{{\Tr}}\bigl( e^{-T\widehat H_W}\bigr) =\int_{\PP}{{\D}}  
x(\tau)\int_A{{\D}}\psi(\tau){{\D}}\bar\psi(\tau)
{\rm tr} {\PP} e^{-\int^T_0 d\tau L}\ee
with
\be\label{3.16}
L=\frac{1}{4}\dot x^2+ig \dot x^\mu A_{\mu}+\bar\psi^\mu
\bigl( \delta_{\mu\nu}{d\over d\tau} -M_{\nu\mu}\bigr)\psi^\nu\ee
(the subscript ``W'' for $H$ will be explained in a moment).
We have again periodic boundary conditions for $x^{\mu}(\tau)$,
and anti-periodic ones for $\psi^{\mu}(\tau)$.

At this point we have to be careful.
If we regard the Hamiltonian (\ref{3.5})
as
a function of the anti-commuting $c$-numbers $\psi_\mu$ and $\bar\psi_\mu$
it is related to the classical Lagrangian (\ref{3.16}) by a standard
Legendre transformation. As is well-known, 
the information about the operator ordering
is implicit in the discretization which is used for the definition of the
path-integral. 
Different operator orderings correspond to different
discretization prescriptions
(see, e.g., \cite{weinberg}). 
In our derivation of the worldline path integral representation
for the spinor loop effective action in the previous section
we used the so-called midpoint prescription~\cite{sakitabook}
for the discretization. The reason for this choice is that,
by Sato's theorem ~\cite{sato.m}, 
only in this case the familiar
path-integral manipulations are allowed. Those will be needed
to justify the naive one-dimensional perturbation
expansion which we have in mind.

It is known
~\cite{sakitabook,sato.m,berezin,mizrahi,gerjev} that, at the 
operator level, this is
equivalent to using the Weyl ordered Hamiltonian $\widehat H_W$.  This is
the reason why we wrote $\widehat H_W$ rather than $\widehat H$ on the
l.h.s. of eq. (\ref{3.15}).
In order to arrive at the relation (\ref{3.12})
we had to assume that the fermion operators in $\widehat H$ 
are ``anti-Wick''
ordered. Weyl ordering amounts to a symmetrization in $\bar\psi$ and
$\psi$
so that

\bear\label{3.18}
\widehat H_W&=&(\widehat p_\mu+g A_\mu(\widehat x))^2+\frac{1}{2}
M_{\nu\mu}(\widehat x)(\widehat\psi^\nu\widehat{\bar\psi^\mu}-
\widehat{\bar\psi^\mu}\widehat\psi^\nu)\nonumber\\
&=&\widehat H-\frac{1}{2} M_{\mu}^{\mu} (\widehat x)\ear
In the second line of (\ref{3.18}) we used (\ref{3.5})
and (\ref{3.6}). (With respect  to $\widehat x_\mu$ and $\widehat p_\mu$
Weyl ordering is used throughout.) If we employ (\ref{3.18}) in
(\ref{3.15})
we obtain the following representation for the partition function of
the anti-Wick ordered exponential:

\bear
{{\Tr}}\bigl( e^{-T\widehat H}\bigr) 
&=&\int_P {\D} x (\tau)\int_A{\D}\psi(\tau)
{\D}\bar\psi(\tau){\rm tr} 
{\cal P}\exp\left[-\int^T_0 d\tau\left\{ L(\tau)
+\frac{1}{2} M_{\mu}^{\mu}(x(\tau))\right\}\right]\non\\
\label{3.19}
\ear
Let us now calculate the partition function 
${{\Tr}}\bigl( \exp (-T\widehat h)\bigr)$
which is a generalization of the heat-kernel needed in eq. (\ref{3.3}).
By virtue of eq. (\ref{3.12}) we may write

\be\label{3.20}
{{\Tr}}\bigl( e^{-T\widehat h}\bigr) 
={{\Tr}}_1\bigl( e^{-T\widehat H}\bigr)\ee
where ``${\Tr}_1$'' denotes the trace in the one-form sector of the
theory which contains the worldline fermions. In order to perform
the projection on the one-form sector we identify $M_{\mu\nu}$ with

\be\label{3.21}
M_{\mu\nu}=C\delta_{\mu\nu}-2ig F_{\mu\nu}\ee
where $C$ is a real constant. As a consequence,

\be\label{3.22}
\widehat H=\widehat H_0+C\widehat F\ee
with

\be\label{3.23}
\widehat H_0\equiv(\widehat p_\mu+g A_\mu(\widehat x))^2-2ig F_{\nu\mu}
(\widehat x) \widehat\psi^\nu \widehat{\bar\psi^\mu}\ee
denoting the Hamiltonian which corresponds to the inverse propagator ${\D}$.
The fermion number operator $\widehat F\equiv\widehat\psi^\mu  
\widehat{\bar\psi}_\mu$
is anti-Wick ordered by definition. Its spectrum  consists of the
integers
$p=0,1,2,\dots ,D$. Note that $M_{\mu}^{\mu}=DC$, and that because of the
antisymmetry of $F_{\mu\nu}$ the Hamiltonian $\widehat H_0$ has no ordering
ambiguity  in its fermionic piece.
It will prove useful to apply eq. (\ref{3.19}) not to $\widehat H$ directly,
but to
$\widehat H-C=\widehat H_0+C(\widehat F-1)$.
This leads to

\bear
{{\Tr}}\bigl( e^{-CT(\widehat F-1)}e^{-T\widehat H_0}\bigr)
&=& \exp\Bigl[-CT(\frac{D}{2}-1)\Bigr]\int_P{\cal D}x(\tau)\int_A
{\cal D}\psi(\tau){\cal D}\bar\psi(\tau){\rm tr} 
{\PP}e^{-\int^T_0d\tau L_{\rm glu}}
\non\\
\label{3.24}
\ear
with
\be\label{3.25}
L_{\rm glu} =\frac{1}{4}\dot x^2+ig \dot x^\mu A_{\mu}
+\bar\psi^\mu
\Bigl[\delta_{\mu\nu}({d\over d\tau}-C)-2igF_{\mu\nu}\Bigr]
\psi^\nu\ee
After having performed the path integration in (\ref{3.24})
we shall send $C$ to infinity. While this has no effect in
the one-form sector, it leads to an exponential suppression
factor $\exp[-CT(p-1)]$ in the sectors with fermion
numbers $p=2,3,...,D$. Hence only the zero and the one forms
survive the limit $C\to\infty$. In order to eliminate the
contribution from the zero forms we insert the projector
$[1-(-1)^{\widehat F}]/2$ into the trace. It projects on the
subspace of odd form degrees, and is easily implemented
by combining periodic and anti-periodic boundary conditions
for $\psi_\mu$. In this way we arrive at the following 
representation
of the partition function of $\widehat H_0$, restricted to the
one-form sector:

\bear\label{oneformsector}
{{\Tr}}_1[e^{-T\widehat H_0}]
&=&\lim_{C\to\infty}{{\Tr}}\left[\frac{1}{2}(1-(-1)^{\widehat F})
e^{-CT(\widehat F-1)}e^{-T\widehat H_0}\right]\nonumber\\
&=&\lim_{C\to\infty}\exp\left[-CT\left(\frac{D}{2}-1\right)
\right]\int_P{\cal D}x(\tau)
\frac{1}{2}(\int_A-\int_P){{\D}}\psi(\tau){{\D}}\bar\psi(\tau)
{{\tr}}{\PP}e^{-\int_0^Td\tau L_{\rm glu}}\non\\
\ear\no
Because ${{\Tr}}\bigl( \exp(-T{{\D}})\bigr)
={{\Tr}}_1\bigl( \exp(-T\widehat H_0)\bigr)$,
eq. (\ref{oneformsector}) implies for the effective action
\cite{strassler1,rescsc}

\bear\label{gluonpi}
\Gamma_{\rm glu}[A]&=&\frac{1}{2}\lim_{C\to\infty}
\int^\infty_0\frac{dT}{T}\exp\Bigl[-CT(\frac{D}{2}-1)\Bigr]
\int_P{\cal D}x\
\frac{1}{2}(\int_A-\int_P){{\D}}\psi{\cal D}\bar\psi
\nonumber\\
&&\times 
{{\tr}}{\PP}
\exp\biggl[-\int^T_0
d\tau\left\{\frac{1}{4}\dot x^2+ig \dot x^\mu A_\mu
+\bar\psi^\mu\Bigl[\delta_{\mu\nu}({d\over d\tau}-C)
-2ig F_{\mu\nu}\Bigr]\psi^\nu\right\}
\biggr]
\nonumber\\
\ear
Note that, 
from the point of view of the worldline
fermions, $C$ plays the role of a mass.
The factor $\exp[-CTD/2]$ in
(\ref{gluonpi}) is due to the difference between the Weyl and
the anti-Wick-ordered Hamiltonian. It is crucial for obtaining
a finite result in the limit $C\to\infty$. In fact, for $D=4$ it
converts the prefactor $\e^{CT}$ to a decaying exponential
$\e^{-CT}$ 
\footnote
{This reordering factor was missing in
~\cite{strassler1}, where the change of the sign
in $D=4$ was instead erroneously attributed
to a difference between Minkowski and Euclidean spacetime.}.

A similar worldline path integral representation 
can also be written down for the gluon propagator in a background
Yang-Mills field \cite{rescsc}. This may be useful for future
extensions of the string -- inspired formalism.

\section{Calculation of One-Loop Amplitudes}
\label{loop}
\renewcommand{\theequation}{4.\arabic{equation}}
\setcounter{equation}{0}

We proceed to the evaluation of worldline path integrals
at the one-loop level. As was already mentioned
the method used is a very specific one, and analogous
to the techniques used in string perturbation theory.
All path integrals will be manipulated into
Gaussian form, which reduces their evaluation
to the calculation of worldline propagators and
determinants, and standard combinatorics.
Of course there exist many alternatives to this procedure
(see, e.g., 
\cite{friedbook,baboca,frgish,gitshv,stkakt,rossch,gitzla,kksw}).
Those will not be discussed here.

While most of the formalism developed here applies to an
arbitrary spacetime dimension, or at least to even dimensions,
in this review all of our applications will be to four dimensional
field theories (for some calculations in $D=2$ see
\cite{mnss2}, in $D=3$ \cite{cadhdu}).

\subsection{The N - point Amplitude in Scalar Field Theory}

At the one-loop level, the worldline formalism has been
used for a large variety of purposes.
Let us begin with the simplest possible case, the
one-loop $N$ - point amplitude in 
massive $\phi^3$ - theory.
Choosing 

\be
U(\phi)=
{\lambda\over 3!}
\phi^3
\label{defphito3}
\ee\no
in eq.(\ref{scalftpi}),
the path integral for 
the corresponding effective action
reads

\be
\Gamma[\phi]
=\half
\Tintm
\int_{x(T)=x(0)}{\cal D}x(\tau)\,
e^{-\int_0^T d\tau\Bigl(
\kinb
+\lambda\phi(x(\tau))
\Bigr)}
\label{scalphi3pi}
\ee\no
We intend to calculate this path integral using the
elementary Gauss formula

\be
\int\,
dx\,
\e^{-x \cdot A \cdot x+2b\cdot x} 
\sim
{(\det (A))}^{-\half}
\e^{\, b\cdot A^{-1}\cdot b}
\label{gauss}
\ee\no
As we already explained
in the introduction,
first
one has to deal with the zero-mode
contained in the coordinate path integral
$\int{\cal D}x(\tau)$, i.e.
the constant loops.
This is done by separating off
the integration over the 
loop center of mass $x_0$,
which reduces the coordinate 
path integral to an integral
over the relative coordinate $y$:
 
\begin{eqnarray}
{\dps\int}{\cal D}x &=&
{\dps\int}d x_0{\dps\int}{\cal D} y\nonumber\\
x^{\mu}(\tau) &=& x^{\mu}_0  +  
y^{\mu} (\tau )\nonumber\\
\int_0^T d\tau\,   y^{\mu} (\tau ) &=& 0\nonumber\\
\label{split}
\end{eqnarray}

\noindent
The effective action thereby gets expressed in
terms of an effective Lagrangian ${\cal L}_{eff}$,

\begin{equation}
\Gamma = \int d x_0\,{\cal L}_{eff}(x_0)
\label{defL}
\end{equation}

\noindent
and ${\cal L}_{eff}(x_0)$ is represented 
as an integral over the space of all loops
with fixed common center of mass $x_0$.

In the reduced Hilbert space without
the zero mode, the kinetic operator is invertible,
and the inverse is easily found using the
eigenfunctions of the derivative operator on the circle
with circumference $T$, $\lbrace
\e^{2\pi in{\tau\over T}},n\in {\bf Z}\backslash
\lbrace 0 \rbrace
\rbrace$:

\be
2\bigl\langle\tau_1\mid
{({d\over d\tau})}^{-2}
\mid\tau_2\bigr\rangle
= 2T
\sum_{{n=-\infty}\atop{n\ne 0}}^{\infty}
{\e^{2\pi in{{\tau_1-\tau_2}\over T}}
\over {(2\pi in)}^2}
=
\mid \tau_1-\tau_2\mid 
-{{(\tau_1-\tau_2)}^2\over T}
-{T\over 6}
\label{calcGchap4}
\ee
\no
($\tau_1-\tau_2 \in [-T,T]$).
It will be seen later on that
the constant $-T/6$ drops out
of all physical results,
so that we can delete it at the beginning.
The remainder is the ``bosonic'' Green's function which
we already introduced in eq.(\ref{defG}),

$$
G_B(\tau_1,\tau_2)=
\mid \tau_1-\tau_2\mid 
-{{(\tau_1-\tau_2)}^2\over T}
\, 
$$
\no
Note that it is continuous
as a function on $S^1\times S^1$. Its value depends
neither on the location of the zero
on the circle, nor on the choice of orientation.
This Green's function we will always use as the
correlator for the coordinate ``field'',

\begin{equation}
\langle y^{\mu}(\tau_1)y^{\nu}(\tau_2)\rangle
   = - g^{\mu\nu}G_B(\tau_1,\tau_2)
\nonumber\\
\label{wicky}
\end{equation}
\no
The zero mode fixing prescription, and consequently also
the form of this worldline correlator, are not unique
~\cite{fss1,fhss4,fhss1,schvan}.
This ambiguity is of some technical importance,
and will be discussed in chapter 7 in connection 
with the calculation of the effective action itself.

The only other information required for the execution of a
Gaussian path integral is the free path integral
determinant. With our conventions, the free coordinate
path integral at fixed proper-time $T$ is

\begin{equation}
{\dps\int} {\cal D} y\,
 \exp\Bigl [- \int_0^T d\tau
\fourth{\dot y}^2\Bigr ]
 =  {( 4\pi T)}^{-{D\over 2}}\nonumber\\
\label{normy}
\end{equation}
\no
Here the $T$ - dependence can be easily determined by, e.g.,
$\zeta$ - function regularization ~\cite{hehoto}, while
the factor ${\lbrack 4\pi \rbrack}^{-{D\over 2}}$
corresponds to the usual loop-counting factor in
quantum field theory.

How to continue now depends on whether we wish to
compute the effective action itself, or the corresponding
scattering amplitudes.
For the calculation of the effective action, one way to proceed 
after the expansion of the interaction exponential is to
Taylor-expand the external field at the loop center of mass
$x_0$,

\begin{equation}
\phi(x) =
e^{y \cdot \partial}\phi(x_0)
\label{taylor}
\end{equation}

\no
As we will see in chapter 7,
this leads to a highly efficient algorithm 
~\cite{ss1,fss1,fhss4,fhss1}
for calculating derivative expansions of effective actions.

At the moment we are interested in the
calculation of the
$N$ - point amplitude,
which proceeds
somewhat differently.
According to standard quantum field theory
(see, e.g., \cite{itzzub}), 
the (one-particle-irreducible)
$N$ -- point function 
can be obtained
from the one-loop
effective action $\Gamma[\phi]$ by 
a $N$ - fold functional differentiation
with respect to $\phi$.
In $x$ -- space, we can implement
this operation simply by
expanding the interaction exponential to
$N$ - th order, and 
inserting appropriate $\delta$ -- functions into the
path integral ~\cite{polbook}:

\bear
\Gamma_{\rm 1PI}[x_1,\ldots,x_N]
&=&\half
{(-\lambda)}^N
\Tintm
\int{\cal D}x
\nonumber\\
&&\times
\int_0^T
\prod_{i=1}^N
d\tau_i\,
\delta
\Bigl(
x(\tau_i)-x_i
\Bigr)
e^{-\int_0^T d\tau\bigl(
\kinb
\bigr)}
\label{Npointphi3pix}
\ear\no
Thus only trajectories running through the
prescribed points $x_1,\ldots,x_N$ will contribute
to the amplitude.
The subscript ``1PI'' stands for one-particle-irreducible,
and needs to be introduced here since, 
in contrast to the $N$ - photon amplitude
treated in the introduction, the $N$ - point 
function in $\phi^3$ - theory
has also one-particle-reducible contributions. 

Similarly, the $N$ -- point function in
momentum space can be obtained by specializing
the background to a sum of plane waves,

\be
\phi(x)=\sum_{i=1}^N{\rm e}^{ip_i\cdot x}
\label{pwbscalft}
\ee\no
Then one picks out the term containing every
$p_i$ precisely once
(compare eqs.(\ref{expandint}), (\ref{planewavebackground})).
This leads to 

\bear
\Gamma_{\rm 1PI}[p_1,\ldots,p_N]
&=&\half
{(-\lambda)}^N
\Tintm
\int_0^T
\prod_{i=1}^N
d\tau_i
\int dx_0
\int{\cal D}y
\nonumber\\
&&\times
{\rm exp}\Bigl[i\sum_{i=1}^Np_i\cdot x(\tau_i)\Bigr]
e^{-\int_0^T d\tau\bigl(
\kinb
\bigr)}
\label{Npointphi3pi}
\ear\no
Note that now every external leg is represented by
a scalar vertex operator, eq.(\ref{scalarvertop})
\footnote{In our present conventions momenta appearing in
vertex operators are {\sl ingoing}.}.
Since $x_i\equiv x(\tau_i)=x_0+y(\tau_i)$,
the $x_0$ - integral just gives momentum
conservation,

\be
\int d x_0\,
{\rm \exp}\Bigl[{ix_0\cdot \sum_{i=1}^N p_i}\Bigr]
= {(2\pi)}^D\delta\Bigl(\sum_{i=1}^N p_i\Bigr)
\label{momentumcons}
\ee\no
The $y$ - path integral is now
Gaussian, and
can be simply calculated by ``completing the square''.
One obtains the following parameter integral,

\begin{eqnarray}
\Gamma_{\rm 1PI}[p_1,\ldots,p_N] &=&
\half{(-\lambda)}^N
{(2\pi )}^D\delta (\sum p_i)
{\dps\int_{0}^{\infty}}{dT\over T}
{(4\pi T)}^{-{D\over 2}}e^{-m^2T} 
\nonumber\\&&\times
\prod_{i=1}^N \int_0^T 
d\tau_i
\exp\biggl[\half\sum_{i,j=1}^NG_B(\tau_i,\tau_j) p_i\cdot p_j\biggr]
\nonumber\\
\label{scalarmaster}
\end{eqnarray}
\no
This 
representation of the one-loop $N$-point amplitude
in $\phi^3$ -- theory
appears, as far as is known to the author, first
in ~\cite{polbook}. It is the simplest example of
a Bern-Kosower type formula.
Note that a constant added to $G_B$ would drop out 
immediately on account of momentum conservation.

Using momentum conservation 
this parameter integral can, for any given ordering of the
external legs along the loop, be readily transformed into
the corresponding
standard Feynman parameter integral 
\cite{berkos:npb379,berdun,vanholten:zpc66,dittrich}. 
\vspace{15pt}

\par
\begin{figure}[ht]
\vbox to 4.5cm{\vfill\hbox to 13.8cm{\hfill
\epsffile{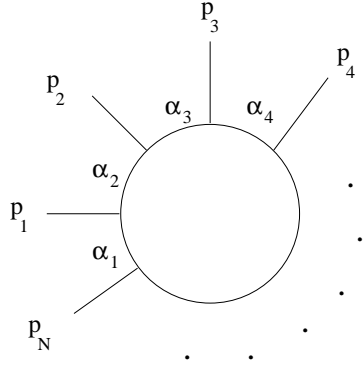}
\hfill}\vfill}
\caption[dum]{One-loop N-point 
diagram.\hphantom{xxxxxxxxxxxxxxxxxxxxxxxx}}
\label{1loopNpoint}
\end{figure}
\par

\no

To obtain the contribution of the
Feynman diagram with the standard ordering of the momenta
$p_1,\ldots,p_N$ (fig. \ref{1loopNpoint}), one simply restricts the
$\tau$ - integrations to the sector defined by
$T\geq\tau_1\geq\tau_2\geq\ldots\tau_N=0$, and transforms
from $\tau$ - parameters to $\alpha$ - parameters:
\vspace{-10pt}

\begin{eqnarray}
\alpha_1 &=& T-\tau_1\nonumber\\
\alpha_2 &=& \tau_1 -\tau_2\non\\
\ldots&&\ldots\nonumber\\
\alpha_N &=& \tau_{N-1} 
\label{trafotaualpha}
\end{eqnarray}

\no
Here we have made
use of the freedom to choose the zero
somewhere on the ``worldloop'' for setting $\tau_N = 0$.
This is always possible, since our 
worldline Green's function $G_B(\tau_1,\tau_2)$
is translation invariant in $\tau$. 
The complete parameter integral
represents the sum of that 
particular Feynman diagram
together with all the ``crossed'' ones. 

The case of $\phi^4$ -- theory is only marginally
different at the one-loop level. A
field theory potential of
$U(\phi)={\lambda\over 4!}\phi^4$
leads to a worldline interaction Lagrangian
of $L_{\rm int} = {\lambda\over 2}\phi^2$.
The formula
for the $2N$ -- point amplitude
analogous to
eq.(\ref{scalarmaster})
is

\bear
\Gamma_{\rm 1PI}[p_1,\ldots,p_{2N}]
&=&\half
{(-\lambda)}^N
\Tintm
\int_0^T
\prod_{i=1}^N
d\tau_i
\int dx_0
\int{\cal D}y
\nonumber\\
\!\!\!\!\!\!\!\!\!
&&\times
{\rm exp}\Bigl[i\sum_{i=1}^N
(p_{2i-1}+p_{2i})\cdot x(\tau_i)\Bigr]
e^{-\int_0^T d\tau\bigl(
\kinb
\bigr)}
+
{\rm permuted\,\, terms}
\non\\
\label{Npointphi4pi}
\ear\no
Here one must explicitly sum over all
possible ways of partitioning the $2N$ external
states into $N$ pairs.

\subsection{Photon Scattering in Quantum Electrodynamics}

Since the scalar loop contribution to the one-loop
$N$ -- photon scattering amplitude
has already been discussed
in the introduction, we 
immediately turn to the
spinor loop case. The appropriate
path integral representation
for the one-loop effective action 
was given in eq.(\ref{spinorpi}),

\begin{eqnarray}
\Gamma_{\rm spin}
\lbrack A\rbrack &  = &- \half {\displaystyle\int_0^{\infty}}
{dT\over T}
e^{-m^2T}
{\displaystyle\int}_P 
{\cal D} x
{\displaystyle\int}_A 
{\cal D}\psi\nonumber\\
& \phantom{=}
&\times
{\rm exp}\biggl [- \int_0^T d\tau
\Bigl ({1\over 4}{\dot x}^2 + {1\over
2}\psi\cdot\dot\psi
+ ieA\cdot\dot x - ie
\psi\cdot F\cdot\psi
\Bigr )\biggr ]
\label{spinorpichap4}
\end{eqnarray}

\noindent
Besides the periodic
coordinate functions $x^{\mu}(\tau )$ 
we now need to also integrate over
the $\psi^{\mu}(\tau )$'s, which are
anti-periodic Grassmann functions.
As was explained before, the scalar loop
case is obtained from this simply by 
discarding all Grassmann quantities.
As a consequence, in this formalism
all calculations performed in
fermion QED
include the corresponding scalar
QED results as a byproduct 
(as far as the calculation of the 
bare regularized amplitudes
is concerned). 

Thus the calculation of the $x$ - path integral
proceeds as before.
For the $\psi$ - path integral,
first note that there is no zero-mode
due to the anti-periodicity.
To find the appropriate
worldline Green's function,
we now need to invert the first 
derivative in the Hilbert space of
anti-periodic functions. This
yields 

\be
2\bigl\langle\tau_1\mid
{({d\over d\tau})}^{-1}
\mid\tau_2\bigr\rangle
= 2
\sum_{n=-\infty}^{\infty}
{\e^{2\pi i{(n+\half)}{{\tau_1-\tau_2}\over T}}
\over {2\pi i{(n+\half)}}}
=
{\rm sign}(\tau_1-\tau_2)
\equiv G_F(\tau_1,\tau_2)
\label{calcGF}
\ee
\no
($\tau_1-\tau_2\in [T,-T]$).
Thus we have now the following two Wick contraction
rules,

\begin{eqnarray}
\langle y^{\mu}(\tau_1)y^{\nu}(\tau_2)\rangle
   & = &- g^{\mu\nu}G_B(\tau_1,\tau_2)
    = \quad - g^{\mu\nu}\biggl[ \mid \tau_1 - \tau_2\mid -
{{(\tau_1 - \tau_2)}^2\over T}\biggr]\nonumber\\
\langle \psi^{\mu}(\tau_1)\psi^{\nu}(\tau_2)\rangle
   & = &{1\over 2}\, g^{\mu\nu} G_F(\tau_1,\tau_2)
   = \quad {1\over 2}\, g^{\mu\nu}{\rm sign}
(\tau_1 -\tau_2 )\nonumber\\
\label{wickrules}
\end{eqnarray}
\noindent
With our present conventions the free 
$\psi$ - path integral
is normalized as
\footnote{This convention differs from the
one used in \cite{ss1,ss2,rescsc}.}

\begin{eqnarray}
{\displaystyle\int} {\cal D} \psi\,
{\rm exp}\Bigl [- \int_0^T d\tau
{1\over2}\psi\cdot\dot\psi\Bigr ]
& = & 4\qquad \nonumber\\
\label{norm}
\end{eqnarray}
\noindent
This takes into account 
the fact that a Dirac spinor 
in four dimensions has four real
degrees of freedom.

One--loop scattering amplitudes are 
again obtained
by specializing the background to 
a finite sum of plane waves of definite
polarization. Equivalently one introduces
a photon vertex operator $V^A$
representing
an external photon of definite momentum and
polarization (compare eq.(\ref{gluonvertop})).
For the spinor loop case this vertex operator
is

\begin{equation}
V^A_{\rm spin}[k,\varepsilon]
=
\int_0^Td\tau
\Bigl[
\varepsilon\cdot \dot x
+2i
\varepsilon\cdot\psi
k\cdot\psi
\Bigr]
\,\e^{ik\cdot x}
\label{photonvertop}
\end{equation}
\noindent
The photon vertex operator $V^A_{\rm scal}[k,\varepsilon]$ for the
scalar loop is given by the same expression without
the Grassmann term.
We can then express the scalar and spinor QED 
$N$ - photon amplitudes in terms
of Wick contractions of vertex operators as follows,

\bear
\Gamma_{\rm scal}[k_1,\varepsilon_1;\ldots;k_N,\varepsilon_N]
&=&{(-ie)}^N\Tint\e^{-m^2T}
{(4\pi T)}^{-{D\over 2}}
\Bigl\langle
V_{{\rm scal},1}^A\cdots
V_{{\rm scal},N}^A
\Bigr\rangle
\non\\
\label{Nphotonwickscal}\\
\Gamma_{\rm spin}[k_1,\varepsilon_1;\ldots;k_N,\varepsilon_N]
&=&-2
{(-ie)}^N\Tint\e^{-m^2T}
{(4\pi T)}^{-{D\over 2}}
\Bigl\langle
V_{{\rm spin},1}^A
\cdots
V_{{\rm spin},N}^A
\Bigr\rangle
\non\\
\label{Nphotonwickspin}
\ear
\no
Here the zero-mode integration has already been
performed, and the
resulting factor (\ref{momentumcons}) been omitted. 
The normalization refers to the complex scalar 
and Dirac fermion cases.

\no
The bosonic Wick contractions may be performed using the
formal exponentiation as explained in the
introduction, eq.(\ref{formexp}), leading to
the Bern-Kosower master formula
eq.(\ref{scalarqedmaster}).
Alternatively, one may follow the following simple
general rules for the 
Wick contraction of expressions
involving both
elementary fields and exponentials:

\begin{enumerate}
\item
Contract fields with each other as usual, and
fields with exponentials according to

\be
\Bigl\langle y^{\mu}(\tau_1)
\,\e^{ik\cdot y(\tau_2)}\Bigr\rangle
= i\langle y^{\mu}(\tau_1)y^{\nu}(\tau_2)
\rangle k_{\nu}
\e^{ik\cdot y(\tau_2)}
\label{wickfieldexp}
\ee\no
(the field disappears, the exponential stays in the game).

\item
Once all elementary fields have been eliminated, 
the contraction of the remaining
exponentials yields a universal 
factor

\be
\Bigl\langle
\e^{ik_1\cdot y_1}
\cdots
\e^{ik_N\cdot y_N}
\Bigr\rangle 
=
\exp\biggl[
-\half
\sum_{i,j=1}^N
k_{i\mu}
\langle
y^{\mu}(\tau_i)
y^{\nu}(\tau_j)
\rangle
k_{j\nu}
\biggr]
\label{wickNexp}
\ee\no
\end{enumerate}
\no
Since the photon vertex operator eq.(\ref{photonvertop})
contains $\dot x$ we will also need to Wick-contract expressions
involving $\dot y$. It is always assumed that Wick-contractions
commute with derivatives. Therefore the first and
second derivatives of $G_B$ will appear,
\begin{eqnarray}
\dot G_B(\tau_1,\tau_2) &=& {\rm sign}(\tau_1 - \tau_2)
- 2 {{(\tau_1 - \tau_2)}\over T}\nonumber\\
\ddot G_B(\tau_1,\tau_2)
&=& 2 {\delta}(\tau_1 - \tau_2)
- {2\over T}\quad \nonumber\\
\label{GdGddchap4}
\end{eqnarray}
\noindent
Turning our attention to the $\psi$ -- path integral,
first note that its
explicit execution 
would be algebraically equivalent to the
calculation of the corresponding
Dirac traces in field theory.
For example, the correlator of
four $\psi$'s gives

\be
\Bigl\langle
\psi^{\kappa}(\tau_1)\psi^{\lambda}(\tau_2)
\psi^{\mu}(\tau_3)\psi^{\nu}(\tau_4)
\Bigr\rangle
=
\fourth
\biggl[
G_{F12}
G_{F34}
g^{\kappa\lambda}g^{\mu\nu}
-G_{F13}G_{F24}
g^{\kappa\mu}g^{\lambda\nu}
+G_{F14}
G_{F23}
g^{\kappa\nu}g^{\lambda\mu}
\biggr]
\label{4psicorr}
\ee\no
Choosing a definite ordering of the proper-time
arguments, e.g. $\tau_1>\tau_2>\tau_3>\tau_4$,
we have the familiar alternating sum of products
of metric tensors at hand which appears also in the
trace of the product of four Dirac matrices.

However, the
explicit computation of the
$\psi$ - integral can be avoided,
due to the following remarkable
feature of the Bern-Kosower
formalism.
After the evaluation of the $x$ -- path integral,
one is left with an integral over the
parameters $T,\tau_1,\ldots,\tau_N$, where
$N$ is the number of external legs. 
The integrand  
is an expression consisting of 
the ubiquitous exponential factor
${\rm exp}\Bigl[\half\sum_{i,j=1}^NG_B(\tau_i,\tau_j) k_i
\cdot k_j\Bigr]$
multiplied by a prefactor $P_N$ 
which is a polynomial function of the
$\dot G_{Bij}$'s and $\ddot G_{Bij}$'s, as well as
of the kinematic invariants.

As Bern and Kosower
have shown 
in appendix B of
~\cite{berkos:npb362},
all the $\ddot G_{Bij}$'s can 
be eliminated by suitable chains of
partial integrations in the parameters
$\tau_i$,
leading to an equivalent parameter
integral involving only the
$G_{Bij}$'s and $\dot G_{Bij}$'s.
According to the Bern-Kosower rules for
the spinor loop case,
all contributions
from fermionic Wick contractions 
may then 
be taken into account simply 
by simultaneously
replacing every closed 
cycle of $\dot G_B$'s appearing, say
$\dot G_{Bi_1i_2} 
\dot G_{Bi_2i_3} 
\cdots
\dot G_{Bi_ni_1}$, 
by its ``supersymmetrization'',

\begin{equation}
\dot G_{Bi_1i_2} 
\dot G_{Bi_2i_3} 
\cdots
\dot G_{Bi_ni_1}
\rightarrow 
\dot G_{Bi_1i_2} 
\dot G_{Bi_2i_3} 
\cdots
\dot G_{Bi_ni_1}
\nonumber\\
-
G_{Fi_1i_2}
G_{Fi_2i_3}
\cdots
G_{Fi_ni_1}\nonumber
\end{equation}
\noindent
(see eq.(\ref{fermion})).

The validity 
of this rule can be understood either in terms
of worldsheet ~\cite{berntasi}
or worldline ~\cite{strassler1} supersymmetry.
We will give a direct combinatorial proof
in appendix \ref{proof}, using the
worldline superfield formalism 
introduced in section \ref{sectionspinorqed}
and results from
section \ref{NphotonNgluon} below.
The rule reduces the transition from the
scalar to the spinor loop
case to a mere pattern matching problem.

Yet another option in the spinor loop calculation
is the explicit use of the superfield formalism.
If one takes the
super path integral representation of the
effective action, eq. (\ref{superpi}), as 
a starting point in the construction of
the photon scattering amplitude, one still
obtains
eq.(\ref{Nphotonwickspin}).
However
the photon vertex operator then appears rewritten as

\begin{equation}
V_{\rm spin}^A[k,\varepsilon]= \int_0^T d\tau \int d\theta \,\varepsilon
\cdot
DX{\rm exp}[ik\cdot X]
\label{supervertex}
\end{equation}
\noindent
This allows one to combine the two Wick contraction
rules eqs.(\ref{wickrules}) into a single one,

\begin{equation}
\langle Y^{\mu}(\tau_1,\theta_1)
Y^{\nu}(\tau_2,\theta_2)\rangle
    = - g^{\mu\nu}\hat G(\tau_1,\theta_1;\tau_2,\theta_2)
\label{superwick}
\end{equation}\no
with
a worldline superpropagator

\be
\hat G(\tau_1,\theta_1;\tau_2,\theta_2)
\equiv G_B(\tau_1,\tau_2) +
\theta_1\theta_2 G_F(\tau_1,\tau_2)
\label{superpropagator}
\ee\no
One can then write down a
master formula for $N$ -- photon scattering
\cite{strasslerthesis}
which is formally analogous
to the one for the scalar loop,
eq.(\ref{scalarqedmaster}),

\begin{eqnarray}
\Gamma_{\rm spin}
[k_1,\varepsilon_1;\ldots;k_N,\varepsilon_N]
&=&
-2
{(-ie)}^N
{(2\pi )}^D\delta (\sum k_i)
{\dps\int_{0}^{\infty}}{dT\over T}
{(4\pi T)}^{-{D\over 2}}e^{-m^2T}
\nonumber\\&&
\!\!\!\!\!\!\!\!\!\!\!\!\!\!\!\!\!\!\!\!
\!\!\!\!\!\!\!\!\!\!\!
\!\!\!\!\!\!\!\!\!\!\!\!\!\!\!\!\!\!\!\!
\!\!\!\!\!\!\!\!\!\!\!\!\!\!\!\!\!\!\!\!
\times
\prod_{i=1}^N \int_0^T 
d\tau_i
\int
d\theta_i
\exp\biggl\lbrace
\sum_{i,j=1}^N
\Biggl\lbrack
\half\hat G_{ij} k_i\cdot k_j
+iD_i\hat G_{ij}\varepsilon_i\cdot k_j
+\half D_iD_j\hat G_{ij}\varepsilon_i\cdot\varepsilon_j\Biggr]
\biggr\rbrace
\mid_{\varepsilon_1\ldots\varepsilon_N}
\nonumber\\
\label{supermaster}
\end{eqnarray}
\no
Here, as well as in (\ref{supervertex}),
we introduce the further convention
that also the polarization vectors 
$\varepsilon_1,\ldots,\varepsilon_N$
are to be treated as Grassmann variables.
Thus we have now all $\psi$'s, $\theta$'s, $d\theta$'s,
and $\varepsilon$'s anticommuting with each other.
The overall sign of the master formula refers to the
standard ordering of the polarization vectors
$\varepsilon_1\varepsilon_2\ldots\varepsilon_N$
\footnote{
With our conventions a Wick rotation 
$k_i^4\rightarrow -ik_i^0, T\rightarrow is$
yields the $N$ - photon amplitude in the
conventions of \cite{weinberg}.}.

The superfield formalism thus also avoids the
explicit execution of 
the Grassmann - Wick contractions.
Those are now replaced
by a number of Grassmann integrations, which have
to be performed at a later stage of the
calculation.
Ultimately the superfield
formalism leads to the same collection of
parameter integrals to be performed
as the component formalism, however
it is often useful for keeping
intermediate expressions compact.

\subsection{Example: QED Vacuum Polarization} 

As a first example, let us recalculate in detail
the one-loop vacuum polarization
tensors in scalar and spinor QED
\cite{strassler1}. 

\subsubsection{Scalar QED}

\no
According to the above
the one-loop two-photon amplitude
in scalar QED can be written as

\bear
\Gamma^{\mu\nu}_{\rm scal}[k_1,k_2]&=&
{(-ie)}^2\Tint\e^{-m^2T}
\Dx
\int_0^T d\tau_1
\int_0^T d\tau_2\non\\
&&\times
\dot x^{\mu}(\tau_1)
\e^{ik_1\cdot x(\tau_1)}
\dot x^{\nu}(\tau_2)
{\rm e}^{ik_2\cdot x(\tau_2)}
\freeexp
\label{scalarqed2point}
\ear\no
Separating off the zero mode according to
eqs.(\ref{split}),(\ref{momentumcons}),
one obtains

\bear
\Gamma^{\mu\nu}_{\rm scal}[k_1,k_2]&=&
{(2\pi )}^D \delta(k_1+k_2)
\,\Pi_{\rm scal}^{\mu\nu}(k_1)\non\\
\Pi_{\rm scal}^{\mu\nu}(k) &=&
-e^2\Tint\e^{-m^2T}
\dps
\int_0^T d\tau_1
\int_0^T d\tau_2
\non\\
&&\times
\int
{\cal D}y\,
\dot y^{\mu}(\tau_1)
\e^{ik\cdot y(\tau_1)}
\dot y^{\nu}(\tau_2)
\e^{-ik\cdot y(\tau_2)}
{\rm e}^{-\int_0^Td\tau {1\over 4}{\dot y}^2}
\non\\ 
\label{scalarqed2pointDy}
\ear\no
The Wick contraction of the two photon vertex operators
according to the above rules
produces two terms,

\be
\Bigl\langle
\dot y^{\mu}(\tau_1)
\e^{ik\cdot y(\tau_1)}
\dot y^{\nu}(\tau_2)
\e^{-ik\cdot y(\tau_2)}
\Bigr\rangle
=
\biggl\lbrace
g^{\mu\nu}
\ddot G_{B12}
-k^{\mu}k^{\nu}
\dot G_{B12}^2
\biggr\rbrace
\;\e^{-k^2G_{B12}}
\label{scalarqed2pointwick}
\ee\no
Now one could just write out $G_B$ and its derivatives,
and perform the parameter integrals.
It turns out to be useful, though, to first remove the
$\ddot G_{B12}$ appearing in the first
term by a partial integration in
the variable $\tau_1$ or $\tau_2$. The integrand
then turns into

\be
\biggl\lbrace
g^{\mu\nu}k^2
-k^{\mu}k^{\nu}
\biggr\rbrace
\dot G_{B12}^2
\e^{-k^2G_{B12}}
\label{scalarqed2pointpartint}
\ee\no
Note that this makes the transversality of the
vacuum polarization tensor manifest.
We rescale to the unit circle, 
$\tau_i = Tu_i, i = 1,2$, and use the translation
invariance in $\tau$ to fix the zero to 
be at the location of the second vertex operator,
$u_2=0, u_1=u$.
We have then

\bear
G_B(\tau_1,\tau_2)&=&Tu(1-u)\nonumber\\
\dot G_B(\tau_1,\tau_2)&=&1-2u
\nonumber\\
\label{scaledown}
\ear\no
Taking the free determinant factor eq.(\ref{normy})
into account, and
performing the global proper-time integration, one finds  

\begin{figure}
\begin{center}
\begin{picture}(28000,12000)(0,3000)

\put(4000,10000){\circle{5000}} 
\drawline\photon[\SW\REG](4000,8000)[3]
\drawline\photon[\SE\REG](4000,8000)[3]

\put(10500,9500){\Large$+$}

\put(20000,10000){\circle{5000}} 
\drawline\photon[\W\REG](17980,10000)[3]
\drawline\photon[\E\REG](21980,10000)[3]

\end{picture}
\caption{Scalar QED vacuum polarization diagrams.
\label{scalqedvp}}
\end{center}
\end{figure}
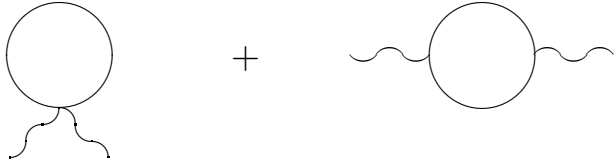

\bear
\Pi^{\mu\nu}_{\rm scal}(k)&=&
e^2
\Bigl[k^{\mu}k^{\nu}-
g^{\mu\nu}k^2
\Bigr]
\Tint\e^{-m^2T}
{(4\pi T)}^{-{D\over 2}}
T^2
\int_0^1du
{(1-2u)}^2
\e^{-Tu(1-u)k^2}
\non\\ 
&=&
{e^2\over {(4\pi )}^{D\over 2}}
\Bigl[k^{\mu}k^{\nu}-g^{\mu\nu}k^2\Bigr]
\Gamma\bigl(2-{D\over 2}\bigl)
\int_0^1du
(1-2u)^2
{\Bigl[
m^2 + u(1-u)k^2
\Bigr]
}^{{D\over 2}-2}
\non\\
\label{scalarvpresult}
\ear\no
The reader is invited to verify that this
agrees with the result reached by calculating,
in dimensional regularization,
the sum of the 
corresponding two Feynman diagrams
(fig. \ref{scalqedvp}).

\par
\par

\subsubsection{Spinor QED}

For the fermion loop the path integral for the two-photon amplitude
becomes, in the component formalism,

\vspace{-10pt}
\bear
\Gamma_{\rm spin}^{\mu\nu}[k_1,k_2]&=&
-\half {(-ie)}^2\Tint\e^{-m^2T}
\Dx\Dpsi
\int_0^T d\tau_1
\int_0^T d\tau_2\non\\
&&
\times
\Bigl(
\dot x^{\mu}_1
+2i\psi^{\mu}_1\psi_1\cdot k_1
\Bigr)
\e^{ik_1\cdot x_1}
\Bigl(
\dot x^{\nu}_2
+2i\psi^{\nu}_2\psi_2\cdot k_2
\Bigr)
\e^{ik_2\cdot x_2}
{\rm e}^{-\int_0^Td\tau
\Bigl(
\kinb +\kinf
\Bigr)
}
\non
\label{spinorqed2point}
\ear\no 
The calculation
of ${\cal D}x$ is identical with the scalar QED calculation.
Only the calculation of ${\cal D}\psi$ is new, and amounts to
a single Wick contraction,

\be
{(2i)}^2
\Bigl\langle
\psi^{\mu}_1\psi_1\cdot k_1
\psi^{\nu}_2\psi_2\cdot k_2
\Bigr\rangle
= G_{F12}^2
\Bigl[g^{\mu\nu}k_1\cdot k_2 -k_2^{\mu}k_1^{\nu}\Bigr]
\label{spinwick2point}
\ee\no
Adding this term to the bosonic result shows that, up to the global
normalization, 
the parameter integral for the spinor loop is obtained
from the one for the scalar loop simply by replacing, in
eq.(\ref{scalarqed2pointpartint}),

\be
\dot G_{B12}^2 \rightarrow
\dot G_{B12}^2 
- G_{F12}^2 
\label{subs2point}
\ee\no
This is in accordance with eq.(\ref{fermion}).
The complete change thus amounts
to supplying eq.(\ref{scalarvpresult})
with a global factor of $-2$, and replacing ${(1-2u)}^2$
by $-4u(1-u)$. This leads to

\be
\Pi_{\rm spin}^{\mu\nu}(k)
=
8{e^2\over {(4\pi )}^{D\over 2}}
\Bigl[k^{\mu}k^{\nu}-g^{\mu\nu}k^2\Bigr]
\Gamma\bigl(2-{D\over 2}\bigl)
\int_0^1du\,
u(1-u)
{\Bigl[
m^2 + u(1-u)k^2
\Bigr]
}^{{D\over 2}-2}
\label{spinorvpresult} 
\ee\no
again in agreement
with the result of the
standard textbook calculation.

\subsection{Scalar Loop Contribution to Gluon Scattering}
\label{scalarloopgluonscatter}

As we discussed in section 3.4.1, 
the path integral representing the effective action 
for the scalar loop in a gluon field differs from 
the photon case, eq.(\ref{scalarpi}),
only by the path-ordering of the exponentials, and the
addition of a global color trace. 
The gluon vertex operator 
eq.(\ref{gluonvertop}) therefore differs from the photon
vertex operator eq.(\ref{photonvertop}) only by
the additional $T^a$ factor, which denotes a gauge group generator
in the representation of the loop particle.
However, the path-ordering (= proper-time ordering = color ordering)
of the path integral has the effect that now those $N$ vertex operators
appear inserted on the worldloop in a fixed ordering. 
Thus the global color trace factors out, and the 
scalar loop Bern-Kosower
master formula eq.(\ref{scalarqedmaster}) generalizes to the
gluon scattering case as follows,

\begin{eqnarray}
\Gamma^{a_1\ldots a_N}[k_1,\varepsilon_1;\ldots;k_N,\varepsilon_N]
&=&
{(-ig)}^N
{\rm tr}
(T^{a_1}\cdots T^{a_N})
{(2\pi )}^D\delta (\sum k_i)
\non\\
&&
\!\!\!\!\!\!\!
\times
{\dps\int_{0}^{\infty}}dT
{(4\pi T)}^{-{D\over 2}}
e^{-m^2T}
\int_0^Td\tau_1\int_0^{\tau_1}d\tau_2\ldots \int_0^{\tau_{N-2}}
d\tau_{N-1}
\nonumber\\
&&
\!\!\!\!\!\!\!
\times
\exp\biggl\lbrace\sum_{i,j=1}^N 
\Bigl\lbrack \half G_{Bij} k_i\cdot k_j
-i\dot G_{Bij}\varepsilon_i\cdot k_j
+\half\ddot G_{Bij}\varepsilon_i\cdot\varepsilon_j
\Bigr\rbrack\biggr\rbrace
\mid_{\rm multi-linear}
\nonumber\\
\label{qcdmaster}
\end{eqnarray}
\no
Here we have already eliminated one integration by setting
$\tau_N =0$.
Note that it can now happen that
a $\delta(\tau_i-\tau_{i+1})$, generated by the
Wick contractions, appears multiplied by
a $\theta(\tau_i-\tau_{i+1})$, generated by
the path--ordering. Symmetry then dictates that just
one half of this
$\delta$ -- function should be allowed to contribute
to the color ordering under consideration.

The interpretation of this non-abelian master integral differs
from its abelian counterpart in two important ways.
Firstly, note that 
in the abelian case we can construct the complete amplitude
in either of two ways. We can
calculate the integral in the ordered sector
$\tau_1>\tau_2>\ldots >\tau_N=0$, and then generate the
``crossed'' contributions by explicit permutations of the
result. Alternatively, we can generate those permuted terms
by including the other ordered sectors in the integration.
The second option does not exist in the non-Abelian case,
since the crossed terms now generally have different color traces.

Secondly, in contrast to 
the one-loop photon scattering amplitudes the gluonic
amplitudes generally have one-particle reducible
contributions in addition to the irreducible ones.
Since our derivation of the above master formula
was based on a path integral representing the
one-loop effective action for the gluon field,
which is the generator for the
one-particle irreducible gluon correlators,
the master formula as it stands yields
precisely the contributions of all
one-particle irreducible graphs to the
amplitude in question. 
To complete the construction of the
full one - loop 
$N$ - gluon amplitude inside field theory
one would now have to generate the missing diagrams
by an explicit Legendre transformation,
amounting to sewing trees onto the one-particle
irreducible diagrams. 
While feasible \cite{berntasi}, this would to some extent spoil the
elegance of the string-inspired approach. Fortunately,
we have seen already in chapter two that,
as far as the on-shell amplitude is concerned,
the full set of string-derived rules
tells us how to bypass this procedure.
Steps 3 to 5 of the rules instruct us, starting from
the master formula, to remove all second derivatives
$\ddot G_B$'s, and then to apply the tree replacement
(``pinch'') rules. In this way all the 
one-particle-reducible terms are included automatically.
This is a remnant of the fact that the fragmentation
of the amplitude into
one-particle reducible and irreducible diagrams appears
only in the infinite string tension limit, when parts of the
string worldsheet get ``pinched off'' \cite{berkos:npb379,berntasi}.

\subsection{Spinor Loop Contribution to Gluon Scattering}

As we have already seen in the previous chapter,
the fermion loop case is more involved.
In the non-abelian case,
the worldline Lagrangian
eq.(\ref{spinorpi}) now contains a term
$\psi^{\mu}[A_{\mu},A_{\nu}]\psi^{\nu}$.
In the component formalism this
would force one to introduce,
besides the one-gluon vertex operator

\begin{equation}
V^A_{\rm spin}[k,\varepsilon,a]
=
T^a\int_0^Td\tau
\Bigl(
\varepsilon\cdot \dot x
+2i
\varepsilon\cdot\psi
k\cdot\psi
\Bigr)
{\rm exp}[ik\cdot x]
\label{onegluonvertop}
\end{equation}
\noindent 
an additional
two -- gluon vertex operator 
~\cite{strassler1,sato2}. 
This is not necessary in the superfield formalism, 
where  the single gluon super vertex operator 

\begin{equation}
V^A_{\rm spin}[k,\varepsilon,a]
=T^a\int_0^T d\tau d\theta 
\,\varepsilon
\cdot DX{\rm exp}[ik\cdot X]
\label{gluonsupervertex}
\end{equation}
\noindent
remains sufficient.
This is because, as 
explained in section 3.4,
the above commutator terms are 
then generated automatically by the
super $\theta$ -- functions 
implicit in the path-ordering.
Our eqs.(\ref{Nphotonwickscal}),(\ref{Nphotonwickspin})
for the N-point functions in the abelian case
thus generalize to the non-abelian case
simply as follows,

\bear
\Gamma_{\rm 1PI,scal}^{a_1\ldots a_N}
[k_1,\varepsilon_1;\ldots;k_N,\varepsilon_N]
&=&{(-ig)}^N
\tr\Tint\e^{-m^2T}
{(4\pi T)}^{-{D\over 2}}
\Bigl\langle
V_{{\rm scal},1}^A
\cdots
V_{{\rm scal},N}^A
\Bigr\rangle
\non\\&&\times
\delta({\tau_N\over T})
\prod_{i=1}^{N-1}
\theta(\tau_i-\tau_{i+1})
\non\\
\label{Ngluonwickscal}\\
\Gamma_{\rm 1PI,spin}^{a_1\ldots a_N}
[k_1,\varepsilon_1;\ldots;k_N,\varepsilon_N]
&=&-2 
{(-ig)}^N
\tr\Tint\e^{-m^2T}
{(4\pi T)}^{-{D\over 2}}
\Bigl\langle
V_{{\rm spin},1}^A
\cdots
V_{{\rm spin},N}^A
\Bigr\rangle
\non\\&&\times
\delta({\tau_N\over T})
\prod_{i=1}^{N-1}
\theta(\hat\tau_{i(i+1)})
\non\\
\label{Ngluonwickspin}
\ear
\no
Here it is understood that $V_{i}^A$ carries a
$T^{a_i}$. 

\subsection{Gluon Loop Contribution to Gluon Scattering}
\label{gluonloopgluonscatter}

In section 3.4 we derived the following path integral
representation of the one-loop effective action in
pure Yang-Mills theory, using the background field
method in Feynman gauge,
 
\bear
\Gamma_{\rm glu}[A]&=&\frac{1}{2}\lim_{C\to\infty}
\int^\infty_0\frac{dT}{T}
\exp\biggl[-CT(\frac{D}{2}-1)\biggr]
\int_P{\cal D}x\,
\frac{1}{2}\Bigl(\int_A-\int_P\Bigr)
{{\cal D}}\psi{\cal D}\bar\psi
\nonumber\\
&&\!\!\!\!\!\!\!\!\!\!\times 
{{\tr}}{\PP}
\exp\biggl\lbrace -\int^T_0
d\tau\biggl[\frac{1}{4}\dot x^2+ig \dot x^\mu A_{\mu}
+\bar\psi^\mu\Bigl[(\ddtau -C)
\delta_{\mu\nu}-2ig F_{\mu\nu}\Bigr]
\psi^\nu\biggr]
\biggr\rbrace \non
\label{gluonpichap4}
\ear
\no
Recall that
this path integral describes a whole
multiplet of $p$ -- forms, $p=0,\ldots,D$ circulating in
the loop; the role of the limit $C\rightarrow\infty$
is to suppress all contributions from $p\geq 2$, and the 
contributions from the zero form cancel out in the combination
$\int_A-\int_P$.

The Grassmann path integral now appears both with
anti-periodic and periodic boundary
conditions.
The worldline Green's function to be used for its evaluation
is ~\cite{rescsc}

\be\label{defGC}
{G}^C(\tau_1,\tau_2)
\equiv
\langle\tau_1|(\ddtau -C)^{-1}|\tau_2\rangle\ee
\no
and
reads for periodic and anti-periodic boundary conditions, 
respectively,
\bear\label{calcGCP}
&&{G}^C_P(\tau_1,\tau_2)=-\Bigl[\theta(\tau_2-\tau_1)+
\theta(\tau_1-\tau_2)e^{-CT}\Bigr]
\frac{e^{C(\tau_1-\tau_2)}}{1-e^{-CT}}\\
\label{calcGCA}
&&{G}^C_A(\tau_1,\tau_2)=
-\Bigl[\theta(\tau_2-\tau_1)-\theta(\tau_1-\tau_2)e^{-CT}\Bigr]
\frac{e^{C(\tau_1-\tau_2)}}{1+e^{-CT}}\ear
We observe that for $C\to\infty$ there is an increasingly
strong asymmetry between the forward and backward propagation
in the proper time. 
The derivation of these Green's functions
is given in appendix \ref{greendet}.

The representation (\ref{gluonpichap4}) of the effective action does
not coincide with the one used by Strassler \cite{strassler1}.
While he uses the same kinetic term in the fermionic worldline
Lagrangian, he modifies the interaction term according to
\be\label{psitochi}
\bar\psi^\mu F_{\mu\nu}\psi^\nu\to
\half\chi^\mu
F_{\mu\nu}\chi^\nu\equiv\bar\psi^\mu
F_{\mu\nu}\psi^\nu+\frac{1}{2} F_{\mu\nu}
(\psi^\mu\psi^\nu+\bar\psi^\mu\bar\psi^\nu)\ee
where
\be\label{defchi}
\chi^\mu(\tau)\equiv\psi^\mu(\tau)+\bar\psi^\mu(\tau)\ee
\no
After this modification,  Wick contractions involve the 2 -- point 
function of $\chi$, i.e.

\be
\langle \chi^{\mu}(\tau_1)
\chi^{\nu}(\tau_2)
\rangle
=
g^{\mu\nu} 
{G}^{\chi}(\tau_1,\tau_2)
\label{wickchi}
\ee\no
where

\be
{G}^\chi(\tau_1,\tau_2)
\equiv{G}^C(\tau_1,\tau_2)-{G}^C(\tau_2,\tau_1) 
\label{defGchi}
\ee\no
From
(\ref{calcGCP}), (\ref{calcGCA}) we obtain explicitly
\bear\label{calcGchi}
{G}^\chi_P(\tau_1,\tau_2)&=&
{\rm sign}(\tau_1-\tau_2)\frac{\sinh[C(\frac{T}{2}
-|\tau_1-\tau_2|)]}{\sinh[CT/2]}\nonumber\\
{G}^\chi_A(\tau_1,\tau_2)&=&
{\rm sign}(\tau_1-\tau_2)\frac{\cosh[C(\frac{T}{2}
-|\tau_1-\tau_2|)]}{\cosh[CT/2]}\ear\no
These Green's functions 
still do not quite coincide with the ones given by
Strassler \cite{strassler1}; however,  they become effectively equivalent
in the limit $C\to\infty$. The 
modified version is more convenient for the calculation
of scattering amplitudes.

Although the worldline Lagrangians for the
gluon and for the spinor loop are similar
in structure, there is no analogue of the
supersymmetry transformations eq.(\ref{wlsusy})
in the spin -- 1 case. As was noted in ~\cite{sato2},
it is useful nevertheless to introduce the superfield formalism
as a book-keeping device.
In complete analogy to the spinor loop case we introduce
new superfields
\vspace{-7pt}

\begin{eqnarray}
\tilde X^{\mu} &=& x^{\mu} 
+ \sqrt 2\,\theta\chi^{\mu} 
\nonumber\\
\tilde Y^{\mu} &=& \tilde X^{\mu}-x_0^{\mu}
\label{defsupergluon}
\end{eqnarray}
with the same super conventions as before.
The gluon vertex operator becomes
(compare eq.(\ref{gluonsupervertex}))

\begin{equation}
V^A_{\rm glu}[k,\varepsilon,a]
= T^a\int_0^T d\tau d\theta 
\,\varepsilon
\cdot D\tilde X{\rm exp}[ik\cdot\tilde X]
\label{gluonloopsupervertex}
\end{equation}
\noindent
(with $T^a$ in the adjoint representation).
The appropriate worldline super propagator is

\begin{equation}
\hat G^{\chi}_{P,A}
(\tau_1,\theta_1;\tau_2,\theta_2)
\equiv
G_B(\tau_1,\tau_2)
+2\theta_1\theta_2
{G}^{\chi}_{P,A}
(\tau_1,\tau_2)
\label{defsuperpropgluon}
\end{equation}\no
The super Wick contraction rule

\begin{equation}
\langle \tilde Y^{\mu}(\tau_1,\theta_1)
\tilde Y^{\nu}(\tau_2,\theta_2)\rangle_{P,A}
    = - g^{\mu\nu}\hat G^{\chi}_{P,A}
(\tau_1,\theta_1;\tau_2,\theta_2)
\label{superwickgluon}
\end{equation}\no
then correctly reproduces the
component field expressions. 
It allows us to take over 
all the conveniences of the superfield formalism
encountered before, and to write 
the one-particle-irreducible off-shell
$N$ -- gluon Green's function in a way analogous to
eq.(\ref{Ngluonwickspin}),

\bear
\Gamma_{\rm 1PI,glu}^{a_1\ldots a_N}
[k_1,\varepsilon_1;\ldots;k_N,\varepsilon_N]
&=& 
-
\fourth
{(-ig)}^N
\lim_{C\to\infty}
\tr\Tint
\e^{-CT}
{(4\pi T)}^{-{D\over 2}}
\non\\
&&\hspace{-30pt}
\times
\sum_{p=P,A}
\sigma_p
Z_p
\Bigl\langle
V_{{\rm glu},1}^A
\cdots
V_{{\rm glu},N}^A
\Bigr\rangle_{p}
\delta({\tau_N\over T})
\prod_{i=1}^{N-1}
\theta(\hat\tau_{i(i+1)})
\non\\
\label{Ngluonwickgluon}
\ear
\no
Here we have defined $\sigma_P=1$, $\sigma_A=-1$.
$Z_{A,P}$ are the fermionic determinant factors

\be\label{defZAP}
Z_{A,P}\equiv \Det{}_{A,P}
\biggl
[({d\over d\tau}-C)\delta_{\mu\nu}
\biggr]
=\left(\Det{}_{A,P}
\Bigl[{d\over d\tau}-C
\Bigr]\right)^D\ee
\no
Those can be easily calculated using the same basis of
circular eigenfunctions of the derivative operator
as was used in the computation of $G_B, G_F$ above.
The result is

\bear
Z_A&=&(2\cosh[CT/2])^D\nonumber\\
Z_P&=&(2\sinh[CT/2])^D
\label{calcZAP}
\ear\no
Note that we have already
set $D=4$ in the reordering factor, and the same will
be done for $Z_{A,P}$. This corresponds to the choice
of a certain dimensional reduction variant of
dimensional regularization, the four-dimensional
helicity scheme developed by Bern and Kosower
~\cite{berkos:npb379}
(compare chapter 2).

The super formalism allows us not only to do without an
additional two-gluon vertex operator, but also to
generalize the replacement rule
eq.(\ref{fermion}) to the gluon loop case. This means
that one can, for finite $C$ and a fixed choice of the
fermionic boundary conditions, first perform the bosonic Wick
contractions, then partially integrate away all
$\ddot G_{Bij}$'s, and finally include the terms from
the fermionic sector by replacing

\begin{equation}
\dot G_{Bi_1i_2} 
\dot G_{Bi_2i_3} 
\cdots
\dot G_{Bi_ni_1}
\rightarrow 
\dot G_{Bi_1i_2} 
\dot G_{Bi_2i_3} 
\cdots
\dot G_{Bi_ni_1}
\nonumber\\
-
2^n
{G}^{\chi}_{p\,i_1i_2}
{G}^{\chi}_{p\,i_2i_3}
\cdots
{G}^{\chi}_{p\,i_ni_1}
\nonumber\\
\label{gluonsubrule}
\end{equation}
The analysis of the $C\to\infty$ 
limit ~\cite{berkos:npb379,strassler1} 
shows, however, that in this limit all terms containing multiple
products of fermionic cycles get suppressed.
Moreover, of the terms containing precisely one cycle, only
those survive for which the ordering of the indices follows the
ordering of the external legs. 
For those, the $C$ -- dependence is isolated in a factor of

\be
z_n(C)\equiv
e^{-CT}
\half
\sum_{p=P,A}
\sigma_p
Z_p\,
{G}^{\chi}_{p\,i_1i_2}
{G}^{\chi}_{p\,i_2i_3}
\cdots
{G}^{\chi}_{p\,i_ni_1}
\label{defnp}
\ee\no
In the limit this expression
goes to a constant, namely

\be
\lim_{C\to\infty}
z_n(C) =
\left\{ \begin{array}{r@{\quad:\quad}l}
2 & n=2 \\
1 & n>2, \quad {\rm indices\,\, follow \,\, legs}\\
0 & n>2, \quad {\rm all \,\, other \,\, orderings} 
\end{array} \right.
\label{calcnp}
\ee\no
Here the sign in the second line refers to the 
descending ordering

\be
\tau_{i_1}>\tau_{i_2}>\ldots >\tau_{i_n}
\label{descending}
\ee\no
If the ordering of the indices follows the
ordering of the legs,
this can always be arranged for using the
cyclicity and the 
antisymmetry of $G^{\chi}_{A,P}$.
\no
This then leads just to the gluonic
cycle rule part of the Bern-Kosower rules,
eq.(\ref{gluoncyclerule}).
\no
For the remaining purely bosonic terms the 
$C$ -- dependence is trivial, and gives a factor

\be
\lim_{C\to\infty}
e^{-CT}
\half(Z_A-Z_P)
= 4
\label{trivialClimit}
\ee\no
The bosonic part alone will therefore 
yield four times the
contribution of a real scalar in the loop.
This corresponds just to the four
degrees of freedom of the gluon. 
But as we know from field theory the
ghost contribution to the amplitude is $-2$ times the
contribution of a real scalar. Therefore
it just subtracts 
the contribution of
the two unphysical degrees of
freedom of the gluon, and the whole ghost
contribution can be taken into account
simply by changing the above factor from
$4$ to $2$. This explains the factor $2$
which we had 
in the Bern-Kosower rules 
for the ``type 1'' contributions to the
gluon loop. 

\subsection{Example: QCD Vacuum Polarization}

With all this machinery in place, we can now easily
generalize our results for the QED vacuum polarization
tensors,
eqs.(\ref{scalarvpresult})
and (\ref{spinorvpresult}),
to the QCD case ~\cite{strassler1,sato2}.

For the QCD gluon vacuum polarization we have to take
the scalar, spinor, gluon, and ghost loops into
account. 
For the scalar and spinor loops, the replacement
of eqs.(\ref{Nphotonwickscal}),(\ref{Nphotonwickspin})
by their non-abelian counterparts 
eqs.(\ref{Ngluonwickscal}),(\ref{Ngluonwickspin})
has the sole effect that the
amplitude gets multiplied by a color
trace ${\rm tr}(T^{a_1}T^{a_2})$.
The $\delta$ -- function term contained in
$\theta(\tau_1 -\tau_2 +\theta_1\theta_2)$
after the $\theta$ -- integrations produces a
term proportional to 

$$\delta_{12}G_{F12}\,\e^{-k^2G_{B12}}$$

\no
however its $\tau$ -- integral 
vanishes since $G_{F12}(0) = 0$.

For the gluon loop, the coordinate part 
together with the ghost loop give the
same as a complex scalar.
Again the terms produced by the
$\delta$ -- function drop out due to the
antisymmetry of $G^{\chi}_{A,P}(\tau_1,\tau_2)$.
The only nontrivial new contribution comes
from the gluon spin in the loop.
This one is a two-cycle, and according to the above
is related to the scalar contribution by a
replacement of

$$\dot G_{B12}\dot G_{B21} \rightarrow 4$$

\no
If we assume the loop scalars and fermions to be
massless and in the adjoint representation,
the results can be combined into
the following single parameter integral
(compare eqs.(\ref{scaledown}),
(\ref{scalarvpresult}),(\ref{spinorvpresult})),

\bear
\Pi^{\mu\nu}_{\rm adj}
(k)&=&
\tr (T_{\rm adj}^{a_1}T_{\rm adj}^{a_2})
{g^2\over {(4\pi )}^{D\over 2}}
\Bigl[k^{\mu}k^{\nu}-g^{\mu\nu}k^2\Bigr]
\Gamma\bigl(2-{D\over 2}\bigl)
\int_0^1du
{\Bigl[
u(1-u)k^2
\Bigr]
}^{{D\over 2}-2}
\non\\
&&\times
\biggl[
({N_s\over 2}-N_f+1)
{(1-2u)}^2
+N_f -4
\biggr]
\label{gluonvpresult}
\ear\no
Here $N_s$ denotes the number of (real) scalars,
$N_f$ the number of Weyl fermions.
It should be remembered that, in field theory terms,
this result corresponds to a calculation using
the background field method and Feynman gauge. 
It is nice to verify
~\cite{berntasi,strassler1}
that the
second line vanishes for $N_s = 6$,
$N_f=4$, corresponding to the case of $N = 4$
Super Yang-Mills theory, which is a finite theory.
Note that the amplitude then vanishes
already at the integrand level. 

\subsection{$N$ Photon / $N$ Gluon Amplitudes}
\label{NphotonNgluon}

Before proceeding to higher numbers of external legs, let
us introduce some notation to keep the formulas manageable.
Writing out the exponential in the master formula
eq.(\ref{scalarqedmaster}) for a fixed number $N$ of
photons, one obtains an integrand

\be \exp\biggl\lbrace 
\biggr\rbrace \mid_{\rm multi-linear} \quad={(-i)}^N P_N(\dot
G_{Bij},\ddot G_{Bij}) \exp\biggl[\half \sum_{i,j=1}^N G_{Bij}k_i\cdot
k_j \biggr] \label{defPN} \ee\no 
with a certain polynomial $P_N$
depending on the various  $\dot G_{Bij}$'s, $\ddot G_{Bij}$'s, 
as well as on the kinematic invariants. 
To be able to apply the Bern-Kosower rules, we need to
remove all second derivatives
$\ddot G_{Bij}$ appearing  by suitable partial integrations
in the variables $\tau_1,\ldots,\tau_N$. 
This transforms $P_N$ 
into another polynomial $Q_N$ depending only
on the $\dot G_{Bij}$'s alone:

\be
P_N(\dot G_{Bij},\ddot G_{Bij})
\,\e^{\half\sum G_{Bij}k_i\cdot k_j}
\quad
{\stackrel{\sy{\rm part. int.}}{\longrightarrow}}
\quad
Q_N(\dot G_{Bij})
\,\e^{\half\sum G_{Bij}k_i\cdot k_j}
\label{partint}
\ee\no
As a result of the partial integration procedure
certain combinations of the kinematic
invariants are going to appear, the ``Lorentz cycles'' $Z_n$,

\bear
Z_2(ij)&\equiv&
\varepsilon_i\cdot k_j
\varepsilon_j\cdot k_i
-\varepsilon_i\cdot\varepsilon_j
k_i\cdot k_j
\non\\
Z_n(i_1i_2\ldots i_n)&\equiv&
{\rm tr}
\prod_{j=1}^n 
\Bigl[
k_{i_j}\otimes \varepsilon_{i_j}
- \varepsilon_{i_j}\otimes k_{i_j}
\Bigr]
\quad (n\geq 3)
\label{defZn}
\ear\no
Those generalize the transversal projector
which is familiar from the two-point case. 
(In the (abelian) effective action 
they would correspond to a ${\rm tr}(F^n)$.)
We also introduce the notion of a ``$\tau$ - cycle'',
which is a product of $\dot G_{Bij}$'s with the indices
forming a closed chain,

\be
\dot G_{Bi_1i_2} 
\dot G_{Bi_2i_3} 
\cdots
\dot G_{Bi_ni_1}
\label{deftaucycle}
\ee
\no
(It should be remembered from chapter 2 
that an expression is considered a cycle
already if it can be put into cycle form
using the antisymmetry of $\dot G_B$.) 
With these notations, the two-point
result is

\bear
Q_2 &=& Z_2(12)\dot G_{B12}\dot G_{B21}
\label{Q2}
\ear
In the three - point case one starts with

\bear
P_3&=& 
\dot G_{B1i}\varepsilon_1\cdot k_i
\dot G_{B2j}\varepsilon_2\cdot k_j
\dot G_{B3k}\varepsilon_3\cdot k_k
\non\\&&
- 
\Bigl[
\ddot G_{B12}\varepsilon_1\cdot\varepsilon_2
\dot G_{B3i}\varepsilon_3\cdot k_i\, 
+\, (1\rightarrow 2\rightarrow 3)\,
+\, (1\rightarrow 3\rightarrow 2)
\Bigr] 
\non\\
\label{P3}
\ear\no
Here and in the following the dummy indices
$i,j,k$
should be summed over
from $1$ to $N$, and one has
$\dot G_{Bii}=0$ by antisymmetry.
Removing all the $\ddot G_{Bij}$'s
by partial integrations one finds

\bear
Q_3&=&
\dot G_{B1i}\varepsilon_1\cdot k_i
\dot G_{B2j}\varepsilon_2\cdot k_j
\dot G_{B3k}\varepsilon_3\cdot k_k
\non\\
&&
+\half
\biggl\lbrace
\dot G_{B12}
\varepsilon_1\cdot\varepsilon_2
\biggl\lbrack
\dot G_{B3i}
\varepsilon_3\cdot k_i
\bigl(
\dot G_{B1j}k_1\cdot k_j
-\dot G_{B2j}k_2\cdot k_j
\bigr)
\non\\
&&
+
\bigl(
\dot G_{B31}
\varepsilon_3\cdot k_1
-\dot G_{B32}
\varepsilon_3\cdot k_2
\bigr)
\dot G_{B3j}
k_3\cdot k_j
\biggr\rbrack
+ 2\,\, {\rm permutations}
\biggr\rbrace
\non\\
&=& Q_3^3 + Q_3^2 \non\\
\label{Q3}
\ear\no
where

\bear
Q_3^3&=&
\dot G_{B12}\dot G_{B23}\dot G_{B31}
Z_3(123)\non\\
Q_3^2&=&
\dot G_{B12}\dot G_{B21}
Z_2(12)
\dot G_{B3i}\varepsilon_3\cdot k_i
+ (1\rightarrow 2\rightarrow 3)
+ (1\rightarrow 3\rightarrow 2)
\non\\
\label{Q3components}
\ear\no
Here we have decomposed the result of the partial
integration, $Q_3$, according to its ``cycle content'',
which is indicated by the upper index.
$Q_3^3$ contains a 3-cycle, while the terms in $Q_3^2$ 
have a
2-cycle.
Note that a $\tau$ - cycle comes multiplied with the corresponding
Lorentz cycle. This turns out to be true in general,
and motivates the further definition of a 
``bi-cycle'' as the product of the two,

\be
{\dot G}(i_1i_2\ldots i_n)
\equiv
\dot G_{Bi_1i_2}
\dot G_{Bi_2i_3}
\cdots
\dot G_{Bi_ni_1}
Z_n(i_1i_2\ldots i_n)
\label{defbicycle}
\ee\no
After all bi-cycles have been separated out
in a given term, whatever remains is called the
``tail'' of the term, or
``$m$-tail'', where $m$ denotes the
number of left-over $\dot G_{Bij}$'s.
In the case of  $Q_3^2$ we have
just a $1$-tail $\dot G_{B3i}\varepsilon_3\cdot k_i$.

In the abelian case the 3-point amplitude must vanish
according to Furry's theorem. In the present formalism this
can be immediately seen by noting that the integrand is odd
under the orientation-reversing transformation of
variables $\tau_i=T-\tau_i'$,
$i=1,2,3$.

In the three-point case,
$Q_3$ is still unique; all possible
ways of performing the partial integrations
lead to the same result.
The same is not true any more in the
four-point case, where the result
of the partial integration procedure
turns out to depend on the
specific chain of partial integrations
chosen. 
In \cite{menphoton} it was shown that this ambiguity
can be fixed
by requiring $Q_N$ to have, like $P_N$, the full permutation
symmetry in the external legs, and  
a particular algorithm for the partial integration
was given
which manifestly preserves this symmetry.
This algorithm is explained in detail in 
appendix \ref{q2to6}, where we also explicitly write down 
the resulting polynomials $Q_N$ up to the six-point case.  
For the four-point case the result can be written in the following form,

\bear
Q_4&=&Q_4^4+Q_4^3+Q_4^2+Q_4^{22}\nonumber\\
Q_4^4 &=& 
\dot G_{B12}
\dot G_{B23}
\dot G_{B34}
\dot G_{B41}
Z_4(1234)
+ 2 \,\, {\rm permutations}
\non\\
Q_4^3 &=&
\dot G_{B12}
\dot G_{B23}
\dot G_{B31}
Z_3(123)
\dot G_{B4i}
\varepsilon_4\cdot k_i
+ 3 \,\, {\rm perm.}
\non\\
Q_4^2 &=&
\dot G_{B12}\dot G_{B21}
Z_2(12)
{\sum}'
\biggl\lbrace
\dot G_{B3i}
\varepsilon_3\cdot k_i
\dot G_{B4j}
\varepsilon_4\cdot k_j
\non\\&&
+\half
\dot G_{B34}
\varepsilon_3\cdot\varepsilon_4
\Bigl[
\dot G_{B3i}
k_3\cdot k_i
-
\dot G_{B4i}
k_4\cdot k_i
\Bigr]
\biggr\rbrace
+ \, 5 \,\, {\rm perm.}
\non\\
Q_4^{22} &=&
\dot G_{B12}\dot G_{B21}
Z_2(12)
\dot G_{B34}\dot G_{B43}
Z_2(34)
+ 2 \,\, {\rm perm.}
\non\\
\label{Q4}
\ear\no
Here the terms in the partially integrated integrand
have already been grouped according to their cycle
content.
The $\sum'$ appearing in 
the two-tail of $Q_4^2$ means 
that in the summation over the dummy variables
$i,j$ the term with $i=4, j=3$ must be omitted, since for these values
an additional two-cycle would be present in the tail.
Thus our final representation for the four-photon
amplitude in scalar QED is the following,

\bear
\Gamma_{\rm scal}
[k_1,\varepsilon_1;\ldots ;k_4,\varepsilon_4]
&=&
{e^4\over (4\pi )^{D\over 2}}
{\dps\int_{0}^{\infty}}{dT\over T}
T^{4-{D\over 2}}
e^{-m^2T}
\non\\
&&\hspace{-30pt}\times
\int_0^1 
du_1 du_2 du_3 du_4\;
Q_4(\dot G_{Bij})
\exp\biggl\lbrace {T\over 2}\sum_{i,j=1}^4 
G_{Bij} k_i\cdot k_j
\biggr\rbrace
\label{4photonscalarqed}
\ear\no
Here we have already rescaled to the unit circle,
$\tau_i = Tu_i, 
G_{Bij} = \mid u_i - u_j \mid - (u_i-u_j)^2$.
Note that this is already the complete (off-shell) amplitude,
with no need to add ``crossed'' terms. The summation over
crossed diagrams which would have to be done in
a standard field theory calculation here is implicit
in the integration over the various ordered sectors.

As an unexpected bonus of the whole procedure it turns
out that this decomposition according to cycle content
coincides, for arbitrary $N$,
with a decomposition into gauge invariant
partial amplitudes. Every single one of the 16 terms
contained in the decomposition 
of $Q_4$
is individually
gauge invariant, i.e. it
either
vanishes or turns into a total derivative
if the replacement
$\varepsilon_i \rightarrow k_i$
is made for any of the external legs.
For example, if in
$Q_4^3$ we substitute
$k_4$ for $\varepsilon_4$ (in the un-permuted term)
we have the following total derivative at hand,

$$\partial_4
\biggl[
\dot G_{B12}
\dot G_{B23}
\dot G_{B31}
{\e}^{\half G_{Bij}k_i\cdot k_j}
\biggr]
$$
\no
These total derivatives become more and more
complicated with increasing lengths of the
tails \cite{menphoton}.
Note also that the partial integration has the
effect of homogeneizing the integrand;
every term in $Q_N$ has $N$ factors of $\dot G_{Bij}$
and $N$ factors of external momentum. In the
four-point case this has the additional 
advantage
of making the UV finiteness of the photon-photon
scattering amplitude manifest. As is well-known,
in a Feynman diagram calculation this 
property would be seen
only after adding up all diagrams.
Similarly, in the present approach the initial 
parameter integral still contains spurious
divergences, since $P_4$
has terms involving products of two $\ddot G_{Bij}$'s
which lead to a logarithmically divergent
$T$ -- integral. After the partial integration
the integrand is finite term by term
so that there is no necessity any more for an
UV regulator; as far as the QED case is
concerned, we can set $D=4$ in 
(\ref{4photonscalarqed}).

Applying the Bern-Kosower rules to the above
integrand we can immediately obtain the corresponding
parameter integrals for (off-shell) photon-photon scattering
in spinor QED, as well as for (on-shell) gluon-gluon scattering
in QCD. As was already mentioned, in the latter case
the presence of color factors forces one to restrict
the integrations to the standard ordered sector
$1\geq u_1\geq u_2\geq u_3\geq u_4 =0$, and to explicitly
sum over all non-cyclic permutations.

The same partial integration algorithm can also be used
for the fermion loop in the superfield formalism. In appendix
\ref{proof} this will be used for a simple proof of the
replacement rule (\ref{fermion}).

Finally, what does one gain by the partial integration
procedure in terms of the difficulty of the arising
parameter integrals? 
The partial integration increases the total
number of terms in the integrand, but decreases the number
of independent integrals. But then one must take into account
the fact that those fewer independent integrals have, on the
average, more complicated Feynman numerators
(this fact turned out to be of significance in the calculation
of the five -- gluon amplitude ~\cite{bernpc}).
It is therefore
difficult to say in general,
and to confuse matters more
we will see in section \ref{2loopeh}
that sometimes even taking a linear combination of both
integrands can be useful.

\subsection{Example: Gluon -- Gluon Scattering}

Let us now
have a look at a somewhat more substantial calculation,
namely the one-loop gluon -- gluon scattering amplitude
in massless QCD. While this amplitude still does not
present a challenge by
modern standards, and was obtained by Ellis and Sexton
many years ago \cite{ellsex}, the advantages of the
following recalculation over a standard Feynman diagram
calculation should be evident.

The calculation proceeds in two parts. The first step
is concerned with the reduction of this
amplitude to a collection of parameter integrals;
the second one with their explicit calculation.
As always one finds those integrals to be of the
same type as the corresponding Feynman parameter
integrals. 
At the 4 - point level those integrals are still
fairly easy. A convenient
method for their calculation is described in appendix
\ref{boxint}, following \cite{bedikopent}.

Gluon scattering amplitudes are usually calculated in a
helicity basis. A gluon has only two different
physical polarizations, which are chosen to be helicity
eigenstates ``+'' and ``$-$''. In QCD those amplitudes are
not all independent, since CP invariance implies that 
simultaneously flipping
all helicities is equivalent to changing all momenta from
ingoing to outgoing, and vice versa. 
Therefore in the 4 - point case there are only four amplitudes to consider,
$A(++++)$, $A(-+++)$, $A(--++)$, and $A(-+-+)$. 
As always in the non-abelian
case we have to fix the ordering of the external legs, which
we choose as the standard ordering (1234). 
In the end one must sum
over all non-cyclic permutations.

Quite generally it turns out that the calculation of the
$N$ gluon or photon amplitudes is easiest for the all ``+''
(or all ``$-$'')
cases, and most difficult for the completely ``mixed'' ones
\footnote{
It should be noted that $A(++++)$ does {\sl not} describe a
helicity conserving process, due to the convention used here
that all momenta are ingoing.}.
In the 4 - point case the
calculation of $A(++++)$ is similar to the one of $A(-+++)$, 
while $A(--++)$ is similar to $A(-+-+)$. 
We will therefore restrict ourselves
to the computation of $A(-+++)$ and $A(--++)$. 

\par
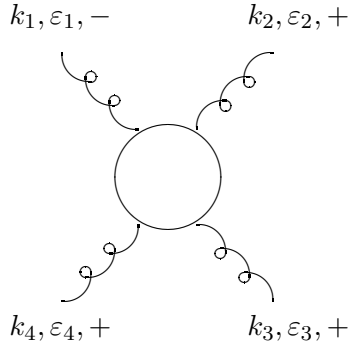
\begin{figure}
\begin{center}
\begin{picture}(20500,14800)(0,1900)

\put(10000,10000){\circle{7000}} 
\drawline\gluon[\SE\REG](11100,8200)[2]
\put(13000,4000){$k_3,\varepsilon_3,+$}
\drawline\gluon[\SW\REG](8900,8200)[2]
\put(4000,4000){$k_4,\varepsilon_4,+$}
\drawline\gluon[\NE\REG](11100,11800)[2]
\put(13000,15800){$k_2,\varepsilon_2,+$}
\drawline\gluon[\NW\REG](8900,11800)[2]
\put(4000,15800){$k_1,\varepsilon_1,-$}

\end{picture}
\caption{Gluon -- gluon scattering amplitude.
\label{gluongluon}}
\end{center}
\end{figure}
\par

Let us start with the easier one, $A(-+++)$ (fig. \ref{gluongluon}). 
A given
assignment of polarizations can be realized by many different
choices of polarization vectors, and it is desirable to make
this choice in such a way that the number of non-zero kinematic
invariants $\varepsilon_i\cdot\varepsilon_j$,
$\varepsilon_i\cdot k_j$ is minimized. 
A convenient way of finding such a set of polarization vectors
for a given polarization assignment is provided by the
spinor helicity technique (see, e.g., \cite{manpar}), which makes use of
the freedom to perform independent gauge transformations on all
external legs. While this technique is already very useful in the
corresponding field theory calculation, here its efficiency is further
enhanced by the fact that there is no loop momentum, which reduces
the number of kinematic invariants from the very beginning.
Appropriate sets of polarization vectors have been given 
in \cite{berkos:npb379,berntasi}.
For $A(-+++)$
they find that using a reference momentum $k_4$ for the first
gluon and $k_1$ for the other ones
makes all products of
polarization vectors vanish,

\be
\varepsilon_i\cdot\varepsilon_j = 0, \qquad i,j = 1,\ldots,4
\label{epsdoteps=0}
\ee\no
Moreover one has the following further relations,
\vspace{-10pt}

\bear
&&k_4\cdot\varepsilon_1=k_1\cdot\varepsilon_2=
k_1\cdot\varepsilon_3=k_1\cdot\varepsilon_4
=0\non\\
&&
k_3\cdot\varepsilon_1
=
-k_2\cdot\varepsilon_1,
\quad
k_4\cdot\varepsilon_2
=
-k_3\cdot\varepsilon_2,
\quad
k_4\cdot\varepsilon_3
=
-k_2\cdot\varepsilon_3,
\quad
k_3\cdot\varepsilon_4
=
-k_2\cdot\varepsilon_4
\non\\
\label{rel-+++}
\ear\no
Those allow one to express all non-zero invariants
in terms of $\varepsilon_1\cdot k_3,
\varepsilon_2\cdot k_4,
\varepsilon_3\cdot k_4,
\varepsilon_4\cdot k_3$.
Using them
in the representation eq.(\ref{Q4})
for $Q_4$
we find that they make most of the Lorentz cycles
$Z_n$ vanish; the surviving ones are

\bear
Z_3(234) &=& 2\epsk24\epsk34\epsk43\non\\
Z_2(23)  &=& \epsk24\epsk34\non\\
Z_2(24)  &=& -\epsk24\epsk43\non\\
Z_2(34)  &=& \epsk43\epsk34\non\\
\label{surviveZn-+++}
\ear\no
This leads to the vanishing of all pure cycle
terms,

\be
Q_4^4 = Q_4^{22} = 0
\label{purecyclesvanish}
\ee\no
The surviving structures $Q_4^3$ and $Q_4^2$ yield

\bear
Q_4
&=&
C_{-+++}
(\dot G_{B13}-\dot G_{B12})
\biggl\lbrace
2\Gp23\Gp34\Gp42
-\Gp23^2(\Gp43-\Gp42)
\non\\
&&
\qquad\qquad\qquad
+\Gp24^2(\Gp34-\Gp32)
-\Gp34^2(\Gp24-\Gp23)
\biggr\rbrace
\label{Q4-+++}
\ear\no
where

\be
C_{-+++} \equiv \epsk13\epsk24\epsk34\epsk43
\label{defC-+++}
\ee\no

\begin{figure}[ht]
\vbox to 6.5cm{\vfill\hbox to 15.8cm{\hfill
\epsffile{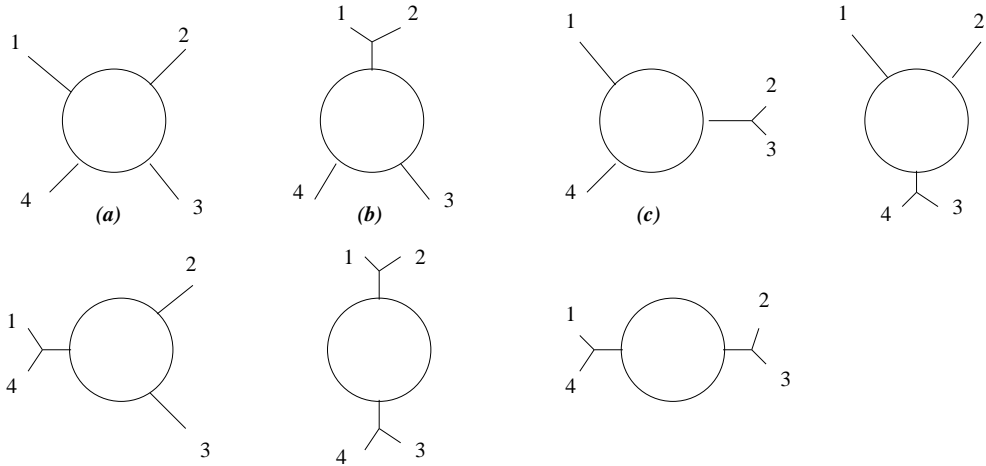}
\hfill}\vfill}
\vspace{10pt}
\caption[dum]{4 - point $\phi^3$ - diagrams\hphantom{xxxxxxxxxxxxx}}
\label{4pointphito3}
\end{figure}

According to step 4 of the Bern - Kosower rules we should
now search for possible pinch terms. 
The relevant $\phi^3$ diagrams at this order are shown
in fig. \ref{4pointphito3}.
\vspace{20pt}

\par
\par

Considering diagram (b) we note that it is indeed a candidate
for a pinch
term, since $Q_4$ contains $\Gp12$ linearly. 
Step 5 of the Bern-Kosower rules tells us that
to find its contribution
we have to replace $\Gp12$ by $2\over {(k_1+k_2)}^2$, and
replace the variable $u_2$ by $u_1$ in the remainder of the
expression. This transforms $Q_4$ 
into 

\bear
P_b
&=&
-2{C_{-+++}\over {(k_1+k_2)}^2}
\biggl\lbrace
2\Gp13\Gp34\Gp41
-\Gp13^2(\Gp43-\Gp41)
+\Gp14^2(\Gp34-\Gp31)
\non\\&&\hspace{70pt}
-\Gp34^2(\Gp14-\Gp13)
\biggr\rbrace
\label{Q4-+++pinch}
\ear\no
Diagram (c) looks like another candidate, however its pinch contribution
disappears since the first factor in $Q_4$ vanishes if $u_3$ is
replaced by $u_2$. Similarly the pinch contributions of all
remaining diagrams can be seen to be zero.

\no
Thus for the scalar loop we have to calculate the
following two
parameter integrals,

\bear
D_a &=& \Gamma(4-\Dhalf )
\int_0^1du_1\int_0^{u_1}du_2\int_0^{u_2}du_3
{P_a(u_1,u_2,u_3,u_4)\over
{{\bigl [-\sum_{i<j=1}^4 G_{Bij}k_i\cdot k_j\bigr ]}
^{4-{D\over 2}}}}
\non\\
D_b &=&
\Gamma(3-\Dhalf )
\int_0^1du_1\int_0^{u_1}du_3
{P_b(u_1,u_3,u_4)\over
{{\bigl [-\sum_{i<j=1,3,4} G_{Bij}k_i\cdot k_j\bigr ]}
^{3-{D\over 2}}}}
\non\\
\label{DaDb-+++}
\ear
(with $P_a$ the $Q_4$ of eq.(\ref{Q4-+++})).
Both integrals turn out to be finite, so that one can set $D=4$.
For their calculation
we introduce Mandelstam variables $s$,$t$ according to
eq.(\ref{mandelstam}) of appendix \ref{boxint}, and write out the
various $G_{Bij}$'s, $\dot G_{Bij}$'s for the standard ordering
of the external legs, $u_1>u_2>u_3>u_4=0$ (the
usual rescaling to the unit circle has already been done).
Then the first integral turns into

\be
D_a
=
-16C_{-+++}
\int_0^1du_1\int_0^{u_1}du_2\int_0^{u_2}du_3
{{(u_2-u_3)}^2u_3(1-u_2)
\over
{\Bigl[
s(u_2-u_3)(1-u_1)
+t(u_1-u_2)u_3
\Bigr]
}^2
}
\label{Daexplicit}
\ee\no
Beginning with $u_1$ all three integrations can be done elementarily, 
and one obtains

\be
D_a = -{8\over 3}{C_{-+++}\over st}
\label{Daresult}
\ee\no
Similarly the other diagram yields a

\be
D_b = 
-{8\over 3}{C_{-+++}\over s^2}
\label{Dbresult}
\ee\no 
To find the corresponding numerator polynomials for the spinor and gluon
loops, we have to apply the cycle replacement rules 
eqs.(\ref{fermion}),(\ref{gluoncyclerule}). 
For example, the first term inside the braces in
eq.(\ref{Q4-+++})
is a 3-cycle, and will give additional terms both for the
fermion and the gluon cases, and both for diagrams (a) and (b).
However, adding up all terms generated by the application of the
replacement rule, and writing them out in the $u_i$, one finds them
to cancel out exactly.
Thus for this particular helicity
component there are no further integrals to calculate, and
the amplitudes for the scalar, fermion and gluon loop
cases differ, apart from the group theory factor,
only by the global factors counting the
differences in statistics and degrees of freedom:

\bear
\Gamma_{\rm scal}^{a_1\cdots a_4}(-+++)&=&
-{g^4\over 12\pi^2}
{\rm tr}
(T^{a_1}_{\rm scal}\cdots T^{a_4}_{\rm scal})
{s+t\over s^2t}
C_{-+++}
\non\\
\Gamma_{\rm spin}^{a_1\cdots a_4}(-+++)&=&
{g^4\over 6\pi^2}
{\rm tr}
(T^{a_1}_{\rm spin}\cdots T^{a_4}_{\rm spin})
{s+t\over s^2t}
C_{-+++}
\non\\
\Gamma_{\rm glu}^{a_1\cdots a_4}(-+++)&=&
-{g^4\over 6\pi^2}
{\rm tr}
(T^{a_1}_{\rm adj}\cdots T^{a_4}_{\rm adj})
{s+t\over s^2t}
C_{-+++}
\non\\ 
\label{A-+++QCD}
\ear\no
Here the normalizations refer to a real scalar and to a
Weyl fermion, and it should be remembered 
that the gluon loop contribution includes the
contribution of its ghost. 
The factor $C_{-+++}$ can, using the spinor helicity method,
be expressed in terms of the
Mandelstam variables up to a complex phase
factor \cite{manpar,berntasi}.
Remember that this is not yet the complete amplitude, but must
still be summed over all non-cyclic permutations according
to (\ref{totalamplitude}).

As in the case of the vacuum polarization, things become 
particularly simple if the scalars and fermions are
also in the adjoint representation, which is the case
in $N=4$ Super-Yang-Mills theory. Here one has

\bear
\Gamma_{\rm spin}(-+++)&=&
-2 \Gamma_{\rm scal}(-+++)
\non\\
\Gamma_{\rm glu}(-+++)&=&
2 \Gamma_{\rm scal}(-+++)
\non\\ 
\label{A-+++SUSY}
\ear\no
These simple relations hold only for $A(-+++)$ and $A(++++)$,
and can be derived from the spacetime supersymmetry
\cite{berntasi}. 
In contrast to a standard field theory calculation here
they are visible already at the integrand level.

We proceed to the more substantial calculation of $A(--++)$.
Again following \cite{berkos:npb379,berntasi}
we choose reference momenta
$(k_4,k_4,k_1,k_1)$, which leads to the following relations:

\bear
\epseps12 &=& \epseps13 = \epseps14 = 
\epseps24 = \epseps34 = 0 \non\\
\epsk14 &=& \epsk24 = \epsk31 = \epsk41 = 0 \non\\
\epsk13 &=& -\epsk12, \quad
\epsk21 = -\epsk23, \quad
\epsk34 = -\epsk32, \quad
\epsk42 = -\epsk43 \non\\
\epseps23 &=& {\epsk23\epsk32\over \kk23}
\label{rel--++}
\ear\no
Using these relations in $Q_4$ one finds that this time
$Q_4^3$ is vanishing. The others all do contribute, 
yielding

\bear
Q_4 &=& C_{--++}
\biggl\lbrace
\Gp12^2\Gp34^2
+\Gp12^2\Bigl[\Gp23\Gp34-\Gp23\Gp24-\Gp24\Gp34\Bigr]
\non\\
&&
\qquad\qquad
+\Gp34^2\Bigl[\Gp12\Gp23-\Gp12\Gp13-\Gp13\Gp23\Bigr]
+\Gp12\Gp13\Gp24\Gp34
\non\\
&&
\qquad\qquad
+\Gp13\Gp14\Gp23\Gp24
-\Gp12\Gp14\Gp23\Gp34
\biggr\rbrace
\label{Q4--++}
\ear\no
with

\be
C_{--++}\equiv \epsk12\epsk21\epsk32\epsk42
\label{defC--++}
\ee\no
For this helicity component all pinch terms turn out to vanish.
For example, $Q_4$ contains four terms containing $\Gp12$ linearly,
however they cancel in pairs once the replacement $u_2\rightarrow u_1$
is made. 

On the other hand, this time the replacement rules will have an effect.
For example, for the spinor loop the first term 
in $Q_4$ 
has to be replaced by 

\be
\Gp12^2\Gp34^2 \rightarrow   
(\Gp12^2-\GF12^2)(\Gp34^2- \GF34^2)
\label{replacefirstspinor}
\ee\no
The analogous replacement has to be made for the
three 4 -- cycles appearing.
For the gluon loop case the first term should instead be replaced by

\be
\Gp12^2\Gp34^2 \rightarrow   
\Gp12^2\Gp34^2 - 4\Gp12^2 -4\Gp34^2
\label{replacefirstgluon}
\ee\no
Of the three 4 - cycles only one has the ordering of the indices
following the (standard) ordering of the external legs, and thus
needs to be replaced by

\be
\Gp12\Gp14\Gp23\Gp34
\rightarrow
\Gp12\Gp14\Gp23\Gp34
-8
\label{replacegluon3cycle}
\ee\no
Writing out the results in the variables $u_i$, and then
transforming to Feynman parameters according to 
eq.(\ref{trafoutoaint}) of appendix \ref{boxint},
one obtains the Feynman polynomials to be integrated.
For the case of the gluon loop one finds, taking the global factor
of $2$ into account,

\bear
P_a &=&
2C_{--++}
\Bigl(
8-12a_3-20a_1a_3-16a_2a_3+16a_1a_2a_3+16a_2^2a_3
-4a_3^2+32a_1a_3^2
\non\\&&\hspace{20pt}
+64a_2a_3^2-48a_1a_2a_3^2-48a_2^2a_3^2
+32a_3^3-32a_1a_3^3-48a_2a_3^3-16a_3^4
\Bigr)
\non\\
\label{Pa--++}
\ear\no
For this helicity component the parameter integrals turn out to
be divergent, so that their calculation is not elementary any more. 
In appendix \ref{boxint} we explain
a method \cite{bedikopent}
for the calculation of arbitrary on-shell massless four-point
tensor integrals in dimensional regularization.
For the numerator polynomial $P_a$ this yields the following,

\bear
D_a &=& 32{C_{--++}\over st}
{\Gamma(1-\epshalf)\Gamma^2(1+\epshalf)
\over
\Gamma(1+\eps)
}
\biggl\lbrace
{8\over\epsilon^2} +
{2\ln (s) +2\ln (t) -{11\over 3}\over\epsilon}
\non\\
&&\hspace{80pt}
+\ln(s)\ln(t)-{11\over 6}\ln(t)-{\pi^2\over 2}+{32\over 9}
+{\rm O}(\epsilon)
\biggr\rbrace
\label{Da--++}
\ear
The double pole in this 
expansion is due to infrared divergences alone,
while the simple pole comes from both infrared and ultraviolet
divergences. The infrared divergences will ultimately
cancel against contributions from the five-gluon tree
amplitude, however the ultraviolet divergence must be
removed by renormalization. 
The final result for the gluon loop contribution 
becomes 

\bear
\Gamma_{\rm glu}^{a_1\cdots a_4}(--++)&=&
{g^4\over 32\pi^2}
{\rm tr}
(T^{a_1}_{\rm adj}\cdots T^{a_4}_{\rm adj})
(4\pi)^{-{\epsilon\over 2}}
D_a^{\rm ren}
\label{A--++final}
\ear
where 

\bear
D_a^{\rm ren} &=&
32{C_{--++}\over st}
{\Gamma(1-\epshalf)\Gamma^2(1+\epshalf)
\over
\Gamma(1+\eps)
}
\biggl\lbrace
{8\over\epsilon^2} +
{2\ln \Bigl({s\over\mu^2}\Bigr) 
+2\ln \Bigl({t\over\mu^2}\Bigr) -{22\over 3}\over\epsilon}
\non\\
&&\hspace{80pt}
+\ln\Bigl({s\over\mu^2}\Bigr)
\ln\Bigl({t\over\mu^2}\Bigr)
-{11\over 6}\ln\Bigl({t\over\mu^2}\Bigr)-{\pi^2\over 2}+{32\over 9}
\biggr\rbrace
\label{Daren--++}
\ear
and $\mu$ is the renormalization scale.
 
As usual we have worked in the Euclidean; the analytic
continuation to physical momenta requires the use of
the appropriate $i\epsilon$ prescription for the
Mandelstam variables, $s\rightarrow s-i\epsilon$ etc.
\footnote{Note that, due to our use of the 
metric $(-+++)$,  our definition of the Mandelstam variables
differs by a sign from the one used in \cite{berkos:npb379,berntasi}.}

\subsection{Boundary Terms and Gauge Invariance}

So far we have completely disregarded possible boundary
terms in the partial integration procedure.
In the abelian case boundary terms are clearly absent,
since all integrations are over the complete circle,
and the integrand 
is written in terms of the worldline Green's functions,
which have the appropriate periodicity properties.
This is different in the
non-abelian case, where boundary terms will generally appear.
When using the string-derived rules for the calculation
of the gluon scattering amplitudes those can still be ignored,
since their contributions are automatically included
by the application of the pinch rules.
However, the validity of the original Bern-Kosower
pinch rules is restricted to the on-shell case
\footnote{Some steps towards 
an extension of those rules to the off-shell case
were taken in \cite{strassler1}.}. 
Therefore the boundary terms come into play if one
wishes to apply the partial integration procedure
to the calculation of the nonabelian effective action
itself, or to the corresponding 
one-particle-irreducible off-shell vertex function
$\Gamma_{\rm 1PI}[k_1,\ldots,\varepsilon_N]$.
Let us therefore 
investigate their structure for the simplest
case of a scalar loop (in this section we 
closely follow \cite{strassler2}). 

\begin{figure}[ht]
\vbox to 5.5cm{\vfill\hbox to 13.8cm{\hfill
\epsffile{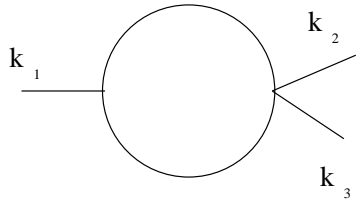}
\hfill}\vfill}
\vspace{-10pt}
\caption[dum]{Structure of three-point 
boundary term\hphantom{xxxxxxxxxxxxx}}
\label{boundarytermfig}
\end{figure}

Let us thus consider the standard low-energy expansion of the
one-loop effective action in gauge theory, to be discussed
at length in chapter \ref{revea} below.
In scalar QED, the first
non-trivial term in this expansion is proportional to
the Maxwell term $F_{\mu\nu}F_{\mu\nu}$. Its coefficient is given
by the zero-momentum limit of the vacuum polarization
tensor, eq.(\ref{scalarvpresult}).
This term must also appear, with the same coefficient,
in the scalar loop contribution to the
QCD effective action. 
However by gauge invariance it must now involve the
full non-Abelian field strength tensor

\be
F_{\mu\nu}=\partial_{\mu}A_{\nu}-\partial_{\nu}A_{\mu}
+ig[A_{\mu},A_{\nu}]
\label{defFnonabelian}
\ee\no
and thus be of the form

\be
\tr F_{\mu\nu}F_{\mu\nu}=
\tr F_{\mu\nu}^0F^0_{\mu\nu}
+2ig\tr F_{\mu\nu}^0[A_{\mu},A_{\nu}]+\ldots
\label{expandmaxwell}
\ee\no
Here we have defined 

\be
F^0_{\mu\nu}=\partial_{\mu}A_{\nu}-\partial_{\nu}A_{\mu}
\label{Fabelian}
\ee\no
as the ``abelian'' part of the non-Abelian field strength tensor.
Obviously, in the non-Abelian case the ``bulk'' integral will 
just produce the $F^0_{\mu\nu}F^0_{\mu\nu}$ - part. The 
additional terms involve the color commutator, and thus,
in field theory terms, a quartic vertex. But Feynman
diagrams involving quartic vertices have, compared to
those with only cubic vertices, a smaller number of internal
propagators, 
and thus of Feynman parameters. Since the boundary
terms appearing in the worldline partial integration have
fewer integrations but the same number of polarization vectors
as the main term, they have the right structure to represent
the missing color commutator terms, and a closer analysis of the
effective action reveals that this is indeed their role. 
Here we will be satisfied with seeing how the second term in
eq.(\ref{expandmaxwell}) makes its appearance.
Consider the three -- point integrand before partial integration,
the $P_3$ of eq.(\ref{P3}).
If we take just the term

\be
- 
\ddot G_{B12}\varepsilon_1\cdot\varepsilon_2
\dot G_{B3i}\varepsilon_3\cdot k_i 
\label{P3part}
\ee\no
multiply by the three-point exponential, and
partially integrate it in the variable $\tau_2$, 
we find a boundary contribution

\be
\int_0^Td\tau_1
\dot G_{B12}\varepsilon_1\cdot\varepsilon_2
\dot G_{B3i}\varepsilon_3\cdot k_i 
\e^{\half\sum G_{Bij}k_i\cdot k_j}
\mid_{\tau_2=\tau_3}^{\tau_2=\tau_1}
\quad
=
-\int_0^Td\tau_1
\dot G_{B13}
\varepsilon_1\cdot\varepsilon_2
\dot G_{B31}\varepsilon_3\cdot k_1 
\e^{G_{B13}k_1\cdot(k_2+k_3)}
\non\\
\label{boundaryterm}
\ee\no
($\tau_3 =0$; only the lower boundary contributes, since
$\dot G_{B12}=0$ for $\tau_1=\tau_2$).
We note that the new integrand
is, but for the Lorentz factors, identical 
with the one which we got in
the two-point case after the partial
integration, eq.(\ref{scalarqed2pointpartint}).
The momenta $k_2$ and $k_3$ now appear only in the
combination $k_2+k_3$, so that we may think of the
vertex operators $V_2$ and $V_3$ as having merged
to form a quartic vertex (fig. \ref{boundarytermfig}).

\par
\par

\no
In our comparison with the effective action we thus should
clearly identify leg number 1 with the $A$ - field appearing
in $F^0$, and indeed with this assignment the correct Lorentz
structure ensues. What still needs to be seen is the
color commutator.
So far we have just a global color factor
of $\tr(T^{a_1}T^{a_2}T^{a_3})$, but
we must remember that the
full amplitude is only obtained after summing over all non-cyclic
permutations of the result reached with the standard ordering.
In the three-point case we have already two 
non-equivalent orderings, and taking the other ordering
into account one finds a second boundary contribution which is
identical to the one above except that it has the reverse
color trace. Both traces can then be combined to a
$\tr(T^{a_1}[T^{a_2},T^{a_3}])$.

\no
We conclude that

\begin{enumerate}

\item
Boundary terms always involve color commutators, and thus
in the effective action picture contribute merely to the
``covariantization'' of the main term. 

\item
They lead to integrals which are already known from
lower-point calculations.

\end{enumerate}
The first fact may have been expected on general grounds; the
second one is non-trivial, and has been verified only to low orders.

\subsection{Relation to Feynman Diagrams} 
\label{comparefeynman}

Finally, how do the integrand polynomials $P_N$ relate
to the ones encountered in an ordinary Feynman parameter
calculation of the $N$ -- photon amplitude?
For the scalar QED case the connection is still very
direct ~\cite{berdun}. Consider again the
two Feynman diagrams for the scalar QED vacuum
polarization, fig. \ref{scalqedvp}.
The first one is a tadpole diagram involving the
seagull vertex. Now the result of the worldline
calculation before partial integration,
eq.(\ref{scalarqed2pointwick}), contained a
$\ddot G_{Bij}$, and thus a $\delta(\tau_1 -\tau_2)$.
This $\delta$ -- function also creates a quartic vertex,
and comparing the parameter integrals one finds that,
not surprisingly, its contribution to the amplitude
matches with the tadpole diagram.

This correspondence 
carries over to the $N$ -- point case, if 
one fixes the ordering of the external legs, and
transforms from $\tau$ - to $\alpha$ - parameters according
to eq.(\ref{trafotaualpha}).
The partially un-integrated Bern-Kosower integrand is thus
obtained from the Feynman parameter integrand 
by a transformation of variables, and a certain regrouping
of terms. This transformation has two effects. First, it allows
one to combine into one expression an individual Feynman
diagram and all the ones related to it by a permutation
of the external states. Second, by regrouping the
$\alpha$ -- parameter expressions in terms of
$G_{Bij}$,
$\dot G_{Bij}$, $\ddot G_{Bij}$, which are functions
well-adapted to the circle,
the integrand is 
brought into a form suitable for partial integration,
since now one needs, at least in the abelian case,
not to worry about possible boundary
terms.

In the spinor-loop case, comparison with the Feynman
calculation is not quite so straightforward. The
resulting parameter integrals obviously include
those from the scalar loop, and thus contain
contributions from diagrams including the seagull
vertex. Clearly they cannot correspond to the
parameter integrals obtained from the  
standard QED Feynman rules. 
It turns out that they 
correspond to a different break-up of those 
photon scattering amplitudes, a break-up
according to
a second-order formalism for fermions
~\cite{feygel,brown,tonin,hostler,berdun,strassler1,morgan} 
(this holds true
also for the more general theories considered in
~\cite{mnss1,mnss2,dimcsc}).  

The Feynman rules for (Euclidean) spinor QED in the second order formalism
(see \cite{morgan} and references therein) are, up to statistics
and degrees of freedom, the ones for scalar QED
with the addition of a third vertex 
(fig. \ref{Fig2ndOrderRules}).

\newlength{\fdwidth}
\setlength{\fdwidth}{1.1in}
\newlength{\frwidth}
\setlength{\frwidth}{.9in}
\begin{figure}[ht]
\centering
\begin{tabular}{@{\hspace{-1.7in}}r@{\hspace{.7in}}l}
\raisebox{0.2in}
{\epsfig{file=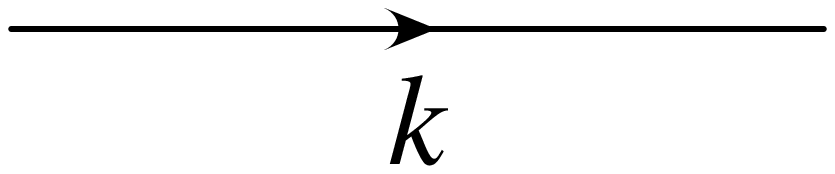, width=\fdwidth} } 
& \raisebox{0.65in}{\parbox{\frwidth}{\begin{displaymath} \frac{1}{k^2
+m^2} 
\end{displaymath}}} 
\vspace{-.21in}
\\
\epsfig{file=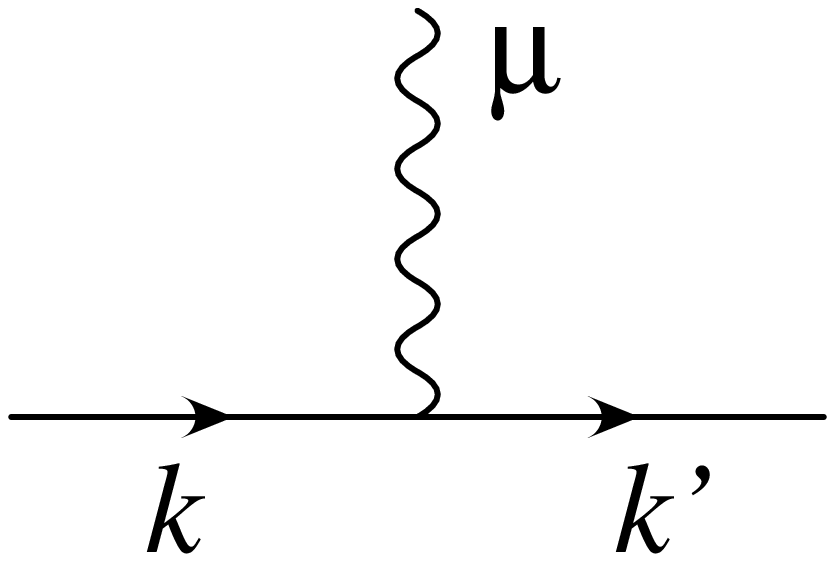, width=\fdwidth}   
& \raisebox{0.6in}{\parbox{\frwidth}{$$ e \left( k + k'\right)_\mu $$}} 
\\
\vspace{-.21in}
\epsfig{file=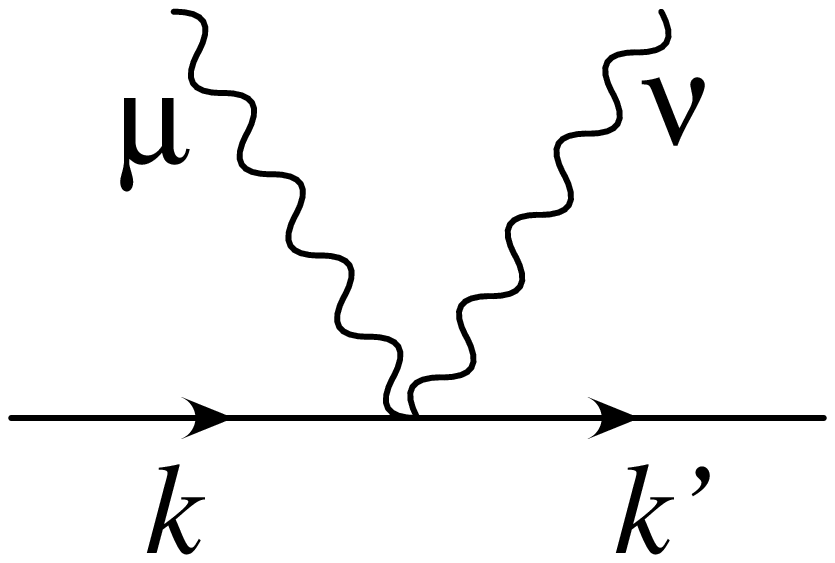,width=\fdwidth} 
& \raisebox{0.6in}{ \parbox{\frwidth}{ $$-2e^2g_{\mu\nu}$$ } }
\\\vspace{-.21in}
\epsfig{file=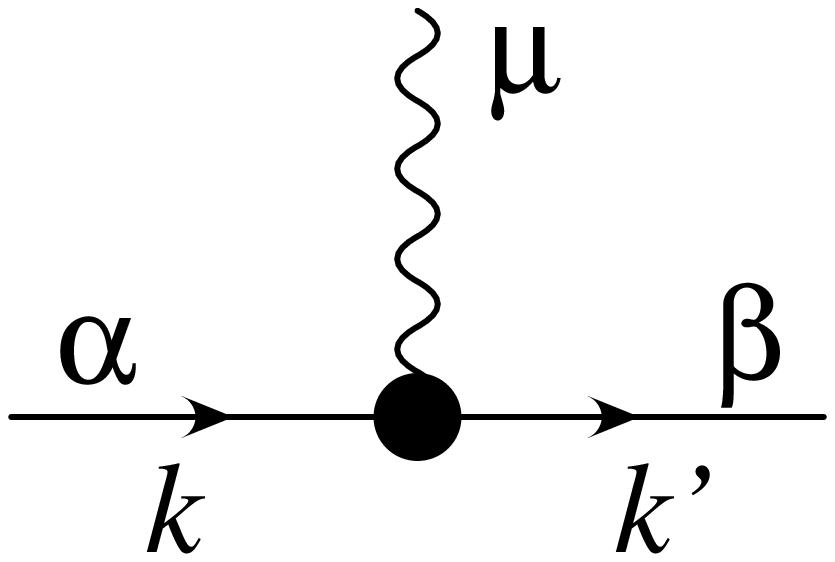,width=\fdwidth} 
& \raisebox{0.6in}{\parbox{\frwidth}{$$e (\sigma_{\mu\nu})_{\beta\alpha}\left
(k'-k\right )^\nu$$} }
\end{tabular}
\vspace{30pt}
\caption{Second order Feynman rules for spinor QED.
\label{Fig2ndOrderRules}}
\end{figure}

\no
The third vertex involves 
$\sigma^{\mu\nu}={1\over 2}[\gamma^{\mu},\gamma^{\nu}]$
and corresponds to the $\psi^{\mu}F_{\mu\nu}\psi^{\nu}$
-- term in the worldline Lagrangian $L_{\rm spin}$
(compare eq.(\ref{Weucl})).
For the details and for the non-abelian case see
~\cite{morgan}. There also an algorithm is given,
based on the Gordon identity,
which transforms the sum of Feynman (momentum)
integrals resulting from the first order rules
into the ones generated by the second order rules.

This explains the close relationship between scalar and spinor
QED calculations in the worldline formalism, which we already
noted before, and will encounter again at the multiloop
level.

\section{QED in a Constant External Field}
\renewcommand{\theequation}{5.\arabic{equation}}
\setcounter{equation}{0}
\label{qed}

An important role in quantum electrodynamics
is played by
processes involving constant
external fields.
An obvious physical reason is
that in many cases a general field can be treated as a constant
one to a good approximation. In QED this is expected to be the
case if the variation of the field is small on the scale of the
electron Compton wavelength.
Mathematically, the constant
field is distinguished by being 
one of the very few known 
field configurations for which the Dirac equation
can be solved exactly, allowing one to obtain results which are
nonperturbative in the field strength.  
For QED calculations in constant external fields it is possible 
and advantageous to take account of the 
field already at the level of the Feynman rules,
i.e. to absorb it into the free electron propagator.
Suitable formalisms have been developed 
many years ago ~\cite{geheniau,tsai,bks1,bks2}.
However, beyond the simplest special cases they lead to
extremely tedious and cumbersome calculations.
As we will see in the present chapter, 
in the string-inspired 
formalism the inclusion of constant external fields 
requires only relatively minor modifications
\cite{ss1,gussho1,gussho2,shaisultanov,rescsc}.
For this reason it has been extensively applied
to constant field processes in QED in four
\cite{ss1,gussho1,gussho2,shaisultanov,adlsch,rescsc,frss,korsch,shovkovy,ditsha,mevv}
as well as in three dimensions \cite{cadhdu}. 

\subsection{Modified Worldline Green's Functions
and Determinants}

Similar to the absorption of a constant field into the
electron propagator in standard field theory, in the
worldline approach we would like to absorb the field
into the basic worldline correlators.
Let us thus assume that we have, in addition to the
background field $A^{\mu}(x)$ we started with,
a second one, $\bar A^{\mu}(x)$, 
with constant field strength tensor
$\bar F_{\mu\nu}$. 
Using Fock--Schwinger gauge
centered at $x_0$ \cite{ss1} we may
take $\bar A^{\mu}(x)$ to be of the form

\begin{equation}
\bar A_{\mu}(x) = 
{1\over 2}y^{\nu}\bar F_{\nu\mu}\\
\label{fockschwinger}
\end{equation}

\noindent
The constant field contribution to the 
worldline Lagrangian (\ref{3spinorlag}) may then be written
as

\begin{equation}
\Delta L_{\rm spin} = {1\over 2}iey^{\mu}\bar F_{\mu\nu}
\dot y^{\nu} - ie\psi^{\mu}\bar F_{\mu\nu}\psi^{\nu}\\
\label{DeltaLkomp}
\end{equation}
\noindent
in components, or as

\begin{equation}
\Delta L_{\rm spin} = -{1\over 2}ieY^{\mu}\bar F_{\mu\nu}
DY^{\nu}\\
\label{DeltaLsuper}
\end{equation}
\noindent
in the superfield formalism.

Since it is still quadratic in the worldline
fields, we need not consider it as part
of the interaction Lagrangian; we can absorb it
into the free worldline propagators. 
This means that we need to 
replace the defining equations (\ref{calcG}) and
(\ref{calcGF})
for the worldline Green's functions
by

\begin{eqnarray}
2\bigl\langle\tau_1\mid
{\biggl(
{d^2\over {d\tau}^2}
-2ie\bar F {d\over d\tau}
\biggr)
}^{-1}
\mid\tau_2\bigr\rangle
&\equiv &
{\cal G}_{B}(\tau_1,\tau_2) 
\label{defcalGB}\\
2\bigl\langle\tau_1\mid
{
\biggl(
{d\over d\tau}
-2ie\bar F
\biggr)
}^{-1}
\mid\tau_2\bigr\rangle
&\equiv &
{\cal G}_{F}(\tau_1,\tau_2) 
\label{defcalGF}
\end{eqnarray}
\noindent
These inverses are calculated in appendix \ref{greendet},
with the result (deleting the ``bar'')

\begin{eqnarray}
{\cal G}_{B}(\tau_1,\tau_2) &=&
{T\over 2{({\cal Z})}^2}\biggl({{\cal Z}\over{{\rm sin}({\cal Z})}}
{\rm e}^{-i{\cal Z}\dot G_{B12}}
+i{\cal Z}\dot G_{B12} -1\biggr)
\nonumber\\
{\cal G}_{F}(\tau_1,\tau_2) &=&
G_{F12}
{{\rm e}^{-i{\cal Z}\dot G_{B12}}\over {\rm cos}({\cal Z})}
\nonumber\\
\label{calGBGF}
\end{eqnarray}
\noindent
where ${\cal Z} \equiv eFT$.
These expressions should be understood as power
series in the Lorentz matrix $\cal Z$
(note that eqs.(\ref{calGBGF})
do not require the field strength tensor
$F$ to be invertible).
Equivalent formulas have
been given for the magnetic field case
in ~\cite{cadhdu}, and for the general case in
~\cite{shaisultanov}.
Note also that the generalized Green's functions are still
translation invariant in $\tau$, and thus
functions of $\tau_1 - \tau_2$.
By writing them in terms of the ordinary Green's function
$G_B$ we have avoided an explicit case distinction
between $\tau_1 >\tau_2$ and $\tau_1 <\tau_2$ which would
become necessary otherwise \cite{shaisultanov}.
Note the symmetry properties

\bear
{\cal G}_B(\tau_1,\tau_2) = 
{\cal G}_B^{T}(\tau_2,\tau_1),
\quad
\dot{\cal G}_B(\tau_1,\tau_2) = 
-\dot{\cal G}_B^{T}(\tau_2,\tau_1),
\quad
{\cal G}_F(\tau_1,\tau_2) = 
-
{\cal G}_F^{T}(\tau_2,\tau_1)
\non\\
\label{symmcalGBF}
\ear\no
Those generalized Green's functions are, 
in general, nontrivial Lorentz matrices,
so that the Wick contraction rules eq.(\ref{wickrules})
have to be replaced by

\begin{eqnarray}
\langle y^{\mu}(\tau_1)y^{\nu}(\tau_2)\rangle
&=&
-{\cal G}_B^{\mu\nu}(\tau_1,\tau_2)\nonumber\\
\langle\psi^{\mu}(\tau_1)\psi^{\nu}(\tau_2)\rangle
&=&
\frac{1}{2}{\cal G}_F^{\mu\nu}(\tau_1,\tau_2)\nonumber\\
\label{exfieldGreen's}
\end{eqnarray}
\noindent\no
We also need the
generalizations of $\dot G_B,\ddot G_B$,
which are (see appendix \ref{greendet})

\begin{eqnarray}
\dot{\cal G}_B(\tau_1,\tau_2)
&\equiv&2\langle\tau_1\mid
{\bigl({d\over d\tau} -2ieF\bigr)}^{-1}
\mid\tau_2\rangle
=
{i\over {\cal Z}}\biggl({{\cal Z}\over{{\rm sin}({\cal Z})}}
{\rm e}^{-i{\cal Z}\dot G_{B12}}-1\biggr)
\nonumber\\
\ddot{\cal G}_{B}(\tau_1,\tau_2)
&\equiv&2\langle\tau_1\mid
{\Bigl(\Eins-2ieF{({d\over d\tau})}^{-1}\Bigr)}
^{-1}\mid\tau_2\rangle
= 2\delta_{12} -{2\over T}{{\cal Z}\over{{\rm sin}({\cal Z})}}
{\rm e}^{-i{\cal Z}\dot G_{B12}}\nonumber\\
\label{derivcalGB}
\end{eqnarray}
\noindent
Let us also write down the first few terms
in the expansion in $F_{\mu\nu}$ for 
all four functions,

\begin{eqnarray}
{\cal G}_{B12} &=& G_{B12}-{T\over 6}
-{i\over 3}
\dot G_{B12}G_{B12}TeF+({T\over 3}G_{B12}^2
-{T^3\over 90}){(eF)}^2+O(F^3)\nonumber\\
\dot{\cal G}_{B12} 
&=&\dot G_{B12}+2i\bigl(G_{B12}-{T\over 6}\bigr)eF
+{2\over 3}\dot G_{B12}G_{B12}T{(eF)}^2 +  O(F^3)
\nonumber\\
\ddot{\cal G}_{B12} 
&=& \ddot G_{B12}+2i\dot G_{B12}eF
-4\bigl(G_{B12}-{T\over 6}\bigr){(eF)}^2+O(F^3)\nonumber\\
{\cal G}_{F12}&=& G_{F12}-iG_{F12}\dot G_{B12}TeF
+2G_{F12}G_{B12}T{(eF)}^2+O(F^3)\nonumber\\
\label{GB(F)expand}
\end{eqnarray}
\noindent
(here we used the identity $\dot G_{B12}^2 = 1-{4\over T}G_{B12}$).
To lowest order in this expansion the
field dependent worldline Green's functions coincide,
of course, with their vacuum counterparts.

Contrary to the vacuum case,
in the constant field background 
one finds nonvanishing coincidence
limits not only for ${\cal G}_{B}$, but also for
$\dot{\cal G}_B$ and ${\cal G}_F$: 

\bear
{\cal G}_{B}(\tau,\tau)&=&
{T\over 2{({\cal Z})}^2}
\biggl({\cal Z}\cot({\cal Z})-1
\biggr)
\non\\
\dot {\cal G}_B(\tau,\tau) &=& i{\rm cot}({\cal Z})
-{i\over {\cal Z}}\non\\
{\cal G}_F(\tau,\tau) &=& -i\,{\rm tan}({\cal Z})\non\\
\label{coincalG}
\end{eqnarray}
\noindent
To correctly obtain this and other coincidence
limits, one has to apply the rules

\begin{equation}
\dot G_B(\tau,\tau)=0,\quad
\dot G_B^2(\tau,\tau)=1
\label{coincidencerules}
\end{equation}
\noindent
which follow from symmetry and continuity, respectively.

Again ${\cal G}_B$ and ${\cal G}_F$ may be assembled into
a super propagator,

\begin{equation}
\hat {\cal G}(\tau_1,\theta_1;\tau_2,\theta_2)
\equiv {\cal G}_B(\tau_1,\tau_2) +
\theta_1\theta_2 {\cal G}_F(\tau_1,\tau_2)\\
\label{calsuperpropagator}\nonumber
\end{equation}
\noindent
allowing one to generalize (\ref{superwick}) to

\begin{equation}
\langle Y^{\mu}(\tau_1,\theta_1)
Y^{\nu}(\tau_2,\theta_2)\rangle
    = - \hat {\cal G}^{\mu\nu}(\tau_1,\theta_1;\tau_2,\theta_2)
\label{superwickF}
\end{equation}\no

At first sight this definition would
seem not to accommodate the
non-vanishing
coincidence limit of ${\cal G}_F$
(which can{\sl not} be subtracted).
Nevertheless, 
comparison with the component field formalism shows
that the correct expressions are again
reproduced if one takes coincidence limits
{\sl after} superderivatives. For instance,
the correlator 
$\langle D_1X(\tau_1,\theta_1)X
(\tau_1,\theta_1)\rangle$
is evaluated by calculating  

\begin{equation}
\langle D_1X(\tau_1,\theta_1)
X(\tau_2,\theta_2)\rangle
=
\theta_1\dot{\cal G}_{B12}-\theta_2{\cal G}_{F12}
\label{wickDXX}
\end{equation}
\noindent
and then setting $\tau_2=\tau_1$.

This is almost all we need to know for computing one-loop
photon scattering amplitudes, or the corresponding
effective action, in a constant overall background field. 
The only further information required at the one--loop
level is the change in the free path integral determinants
due to the external field. 
As we will show in a moment, this
change is (\cite{ss1}; see also \cite{baboca,ckmw})

\bear
\!\!\!\!
{(4\pi T)} ^{-{D\over 2}}
&\rightarrow&
{(4\pi T)}^{-{D\over 2}}
{\rm det}^{-{1\over 2}}
\biggl[{\sin({\cal Z})\over {{\cal Z}}}
\biggr] \quad\qquad{\rm (Scalar\; QED)}
\label{scaldetext}\\
\!\!\!\!
{(4\pi T)}^{-{D\over 2}}
&\rightarrow&
{(4\pi T)}^{-{D\over 2}}
{\rm det}^{-{1\over 2}}
\biggl[{\tan({\cal Z})\over {{\cal Z} }}
\biggr] \quad\qquad{\rm (Spinor\; QED)}
\label{spindetext}
\ear\no
Since those determinants describe the vacuum amplitude in
a constant field one finds them to be, of course, just the
integrands of the well-known 
Euler-Heisenberg-Schwinger formulas.

\subsection{Example: 1-Loop Euler-Heisenberg-Schwinger Lagrangians}

\no
Let us
shortly retrace this calculation.
In the scalar QED case, we have to replace

\begin{equation}
{\displaystyle\int} {\cal D} y\,
 {\rm exp}\Bigl [- \int_0^T d\tau
{1\over 4}{\dot y}^2\Bigr ]
={\rm Det'}^{-{1\over 2}}_P\bigl[ -{d^2\over d\tau^2}\bigr]
=  (4\pi T)^{-{D\over 2}}\\
\label{bosnorm}
\end{equation}
\noindent
by
\begin{equation}
{\rm Det'}^{-{1\over 2}}_P
\Bigl[ -{d^2\over d\tau^2}
+2ieF{d\over d\tau}\Bigr]
=(4\pi T)^{-{D\over 2}}
{\rm Det'}_P^{-{1\over 2}}
\Bigl[\Eins -2ieF{({d\over d\tau})}^{-1}\Bigr]
\quad 
\label{Fbosnorm}
\end{equation}
\noindent
(as usual, the prime denotes the absence of the
zero mode in a determinant).
Application of the 
$ln\,det = tr\,ln$ identity yields
\footnote{Note that, although the determinants
considered here become formally identical with 
the ones appearing in (\ref{defZAP}) by
$2ieF_{\mu\nu}\rightarrow C\delta_{\mu\nu}$, 
in the periodic case
the results are not of the same form. The reason is that
here the zero-mode had to be excluded from the
determinant, while it needs to be included in the
calculation of $Z_P$.}

\begin{eqnarray}
{\rm Det'}_P^{-{1\over 2}}
\Bigl[\Eins -2ieF{({d\over d\tau})}^{-1}\Bigr]
&=&
{\rm exp}\biggl\lbrace
\frac{1}{2}\sum_{n=1}^{\infty}
{(2ie)^n\over n}
{\rm tr}[F^n]{\rm Tr}
\Bigl[{({d\over d\tau})}^{-n}\Bigr]
\biggr\rbrace
\nonumber\\
&=&
{\rm exp}\biggl[-
\frac{1}{2}\sum_{n=2\atop n\,{\rm even}}^{\infty}
{B_n\over n!n}{(2ieT)}^n
{\rm tr}[F^n]
\biggr]
\nonumber\\
&=&
{\rm det}^{-{1\over 2}}
\biggl[{{\rm sin}(eFT)\over eFT}\biggr]
\label{scaldetcomp}
\end{eqnarray}
\noindent
where the $B_n$ are the Bernoulli numbers.
In the second step eq.(\ref{tracechainP})
was used.
The analogous calculation
for the Grassmann path integral 
yields a factor

\begin{equation}
{\rm Det}_A^{+{1\over 2}}
\Bigl[\Eins -2ieF{({d\over d\tau})}^{-1}\Bigr]
=
{\rm det}^{1\over2}
\Bigl[\cos(eFT)\Bigr]
\label{grassfermcomp}
\end{equation}
For spinor QED we therefore find a 
total overall determinant factor of

\begin{equation}
(4\pi T)^{-{D\over 2}}
{\rm det}^{-{1\over 2}}
\biggl[{{\rm tan}(eFT)\over eFT}
\biggr]
\quad \\
\label{Ffermnorm}
\end{equation}
\noindent
Expressing these matrix determinants in terms of the
two standard Lorentz invariants of the Maxwell field
(see section \ref{exprep} 
below) and continuing to Minkowski space,
one obtains the well-known
Schwinger proper-time representation
of the (unrenormalized) Euler-Heisenberg-Schwinger
Lagrangians 
\cite{eulkoc,eulhei,weisskopf,schwinger51},

\bear
{\cal L}_{\rm scal}
&=& -{1\over 16\pi^2}\int_0^{\infty}{ds\over s}
\,\e^{-ism^2}\,
{e^2ab\over \sin(eas)\sinh(ebs)}
\label{eulheiscal}\\
{\cal L}_{\rm spin}
&=& {1\over 8\pi^2}\int_0^{\infty}{ds\over s}
\,\e^{-ism^2}\,
{e^2ab \over \tan(eas)\tanh(ebs)}
\label{eulheispin}
\ear\no
where $a^2-b^2 \equiv {\bf B}^2 - {\bf E}^2$,
$ab \equiv {\bf E}\cdot{\bf B}$.

\subsection{The $N$ - Photon Amplitude in a Constant Field}

Retracing our above calculation of the $N$ - photon path integral
with the external field included we arrive at the
following generalization of eq.(\ref{scalarqedmaster}),
representing the scalar QED $N$ - photon scattering amplitude 
in a constant field \cite{shaisultanov,rescsc}:

\begin{eqnarray}
&&\Gamma_{\rm scal}
[k_1,\varepsilon_1;\ldots;k_N,\varepsilon_N]
=
{(-ie)}^N
{(2\pi )}^D\delta (\sum k_i)
\non\\
&&\hspace{20pt}\times
{\dps\int_{0}^{\infty}}{dT\over T}
{(4\pi T)}^{-{D\over 2}}
e^{-m^2T}
{\rm det}^{-{1\over 2}}
\biggl[{{\rm sin}({\cal Z})\over {\cal Z}}\biggr]
\prod_{i=1}^N \int_0^T 
d\tau_i
\non\\
&&\hspace{20pt}\times
\exp\biggl\lbrace\sum_{i,j=1}^N 
\Bigl\lbrack \half k_i\cdot {\cal G}_{Bij}\cdot  k_j
-i\varepsilon_i\cdot\dot{\cal G}_{Bij}\cdot k_j
+\half
\varepsilon_i\cdot\ddot {\cal G}_{Bij}\cdot\varepsilon_j
\Bigr\rbrack\biggr\rbrace
\mid_{\rm multi-linear}\quad
\nonumber\\
\label{scalarqedmasterF}
\end{eqnarray}
\no
From this formula it is obvious that adding a constant 
Lorentz matrix
to ${\cal G}_B$ will have no effect due to momentum
conservation. As in the vacuum case, we 
can use this fact to get rid
of the coincidence limit of ${\cal G}_B$, (\ref{coincalG}).
Thus instead of ${\cal G}_B$ we will generally work with
the equivalent Green's function $\bar{\cal G}_B$,
defined by

\bear
\bar{\cal G}_B(\tau_1,\tau_2)
\equiv
{\cal G}_B(\tau_1,\tau_2) - {\cal G}_B(\tau,\tau)
= 
{T\over 2{\cal Z}}
\biggl(
{\e^{-i\dot G_{B12}{\cal Z}}-\cos({\cal Z})\over\sin({\cal Z})}
+i\dot G_{B12}\biggr)
\label{defbarcalGB}
\ear\no
No such redefinition is possible
for $\dot{\cal G}_B$ or ${\cal G}_F$.

The transition from scalar to spinor QED is done as in the
vacuum case, again with only some minor modifications.
The spinor QED integrand for a given number of
photon legs $N$ is obtained from the scalar QED
integrand by the following generalization
of the Bern-Kosower algorithm:

\begin{enumerate}

\item
{\it Partial Integration:}
After expanding out the exponential in the
master formula (\ref{scalarqedmasterF}), and 
taking the part linear in all $\varepsilon_1,\ldots,
\varepsilon_N$, remove all second derivatives
$\ddot{\cal G}_B$ appearing in the result by
suitable partial integrations in $\tau_1,\ldots,\tau_N$.

\item
{\it Replacement Rule:}
Apply to the resulting new integrand the
replacement rule (\ref{fermion}) with
$\dot G_B, G_F$ substituted by $\dot {\cal G}_B,
{\cal G}_F$.
Since the Green's functions ${\cal G}_B,{\cal G}_F$
are, in contrast to their vacuum counterparts,
non-trivial matrices in the Lorentz indices,
it must be
mentioned here that the 
cycle property is defined solely in terms of the
$\tau$ -- indices, irrespectively of what happens
to the Lorentz indices. For example, the expression

$$
\varepsilon_1\cdot\dot{\cal G}_{B12}\cdot k_2\,
\varepsilon_2\cdot\dot{\cal G}_{B23}\cdot\varepsilon_3\,
k_3\cdot\dot{\cal G}_{B31}\cdot k_1\,
$$

\noindent
would have to be replaced by

$$
\varepsilon_1\cdot\dot{\cal G}_{B12}\cdot k_2\,
\varepsilon_2\cdot\dot{\cal G}_{B23}\cdot\varepsilon_3\,
k_3\cdot\dot{\cal G}_{B31}\cdot k_1\,
-
\varepsilon_1\cdot {\cal G}_{F12}\cdot k_2\,
\varepsilon_2\cdot {\cal G}_{F23}\cdot\varepsilon_3\,
k_3\cdot {\cal G}_{F31}\cdot k_1
$$

\noindent
The only other difference
to the vacuum case is due to the 
non-vanishing coincidence limits
(\ref{coincalG})
of $\dot{\cal G}_B,{\cal G}_F$.
Those lead to an extension of the ``cycle
replacement rule'' to include one-cycles
\cite{rescsc}:

\begin{equation}
\dot{\cal G}_B(\tau_i,\tau_i)\rightarrow
\dot{\cal G}_B(\tau_i,\tau_i)
-{\cal G}_F(\tau_i,\tau_i)
\label{onecycle}
\end{equation}

\item

The scalar QED Euler-Heisenberg-Schwinger determinant factor
must be replaced by its spinor QED equivalent,

\be
{\rm det}^{-\half}
\biggl[{\sin({\cal Z})\over {{\cal Z}}}
\biggr] 
\rightarrow
{\rm det}^{-\half}
\biggl[{\tan({\cal Z})\over {{\cal Z} }}
\biggr] 
\label{detchange}
\ee\no

\item

Multiply by the usual factor of $-2$ for statistics and degrees
of freedom.

\end{enumerate}

\subsection{Explicit Representations of the Modified Worldline
Green's Functions}
\label{exprep}

For the result to be practically useful
it will be necessary to write
${\cal G}_B,{\cal G}_F$ in more explicit form.
This can be done by choosing some special Lorentz system,
such as the one where $\bf E$ and $\bf B$ are both pointing
along the $z$ - direction, and working with the explicit matrix
form of the worldline correlators, which becomes particularly
simple in such a system. This approach turns out to be
quite adequate for the case of a purely magnetic
(or purely electric) field \cite{rescsc,adlsch}.
However, it is also possible to directly express all
generalized worldline Green's functions
in terms of Lorentz
invariants, without specialization of the Lorentz frame.
This procedure is not only more elegant but appears also to be more
efficient computationally in the general case.

\subsubsection{Special Constant Fields}

\begin{enumerate}

\item
{\it Magnetic field case:}
With the $B$ -- field chosen
along the z -- axis, introduce matrices
$g_{\perp}$ and $g_{\parallel}$ projecting on
the $x,y$ -- and $z,t$ -- planes, so that
\vspace{10pt}

\begin{equation}
F =
\left(
\begin{array}{*{4}{c}}
0&B&0&0\\
-B&0&0&0\\
0&0&0&0\\
0&0&0&0
\end{array}
\right),
g_{\perp}\equiv
\left(
\begin{array}{*{4}{c}}
1&0&0&0\\
0&1&0&0\\
0&0&0&0\\
0&0&0&0
\end{array}
\right),
g_{\parallel}\equiv
\left(
\begin{array}{*{4}{c}}
0&0&0&0\\
0&0&0&0\\
0&0&1&0\\
0&0&0&1
\end{array}
\right)\nonumber\\
\label{defBmatrices}
\end{equation}
\vspace{10pt}

\no
We also introduce
$z=eBT$, and 
$\hat{F}={F\over B}$.
With these notations,
we can rewrite the determinant factors
eqs.(\ref{scaldetext}),(\ref{spindetext}) 
as

\bear
{\rm det}^{-{1\over 2}}
\biggl[{\sin({\cal Z})\over {{\cal Z}}}
\biggr]&=&
{z\over{\sinh(z)}}
\non\\
{\rm det}^{-{1\over 2}}
\biggl[{\rm tan ({\cal Z})\over {{\cal Z}}}
\biggr]&=&
{z\over{\tanh(z)} }
\non\\
\label{detextB}
\ear\no
The worldline correlators
eqs.(\ref{calGBGF}),(\ref{derivcalGB}),(\ref{defbarcalGB}) 
specialize to

\begin{eqnarray}
\bar{\cal G}_{B}(\tau_1,\tau_2) 
&=&G_{B12}\,{g_{\parallel}}
-{T\over 2}{\Bigl(\cosh(z\dot G_{B12})-\cosh(z)\Bigr)
\over z\sinh(z)}
{g_{\perp}}\nonumber\\
&&+{T\over{2z}}\biggl({\sinh(z\dot G_{B12})\over\sinh(z)}
-\dot G_{B12}\biggr)i{\hat{F}}\nonumber\\
\dot{\cal G}_{B}(\tau_1,\tau_2)
&=&\dot G_{B12}\,{g_{\parallel}}+{\sinh(z\dot G_{B12})\over\sinh(z)}
{g_{\perp}}
-\biggl({\cosh(z\dot G_{B12})\over \sinh(z)}-{1\over z}
\biggr)i{\hat{F}}\nonumber\\
\ddot{\cal G}_{B}(\tau_1,\tau_2)
&=& \ddot G_{B12}\,{g_{\parallel}}
+2\biggl(\delta_{12}-{z\cosh(z\dot G_{B12})\over T
\sinh(z)}\biggr){g_{\perp}}
+2{z\sinh(z\dot G_{B12})\over T\sinh(z)}i{\hat{F}}\nonumber\\
{\cal G}_{F}(\tau_1,\tau_2) &=&G_{F12}\,{g_{\parallel}}
+G_{F12}{\cosh (z\dot G_{B12})\over \cosh (z)}{g_{\perp}}
-G_{F12}{{\sinh (z\dot G_{B12})}\over{\cosh (z)}}i{\hat{F}}
\nonumber\\
\label{GB(F)pureB}
\end{eqnarray}
\noindent
Note that from ${\cal G}_B$ we subtracted already its its coincidence
limit, indicated by the ``bar''.
Not removable are the coincidence limits for $\dot{\cal G}_B$
and ${\cal G}_F$, 

\begin{eqnarray}
\dot {\cal G}_B(\tau,\tau) &=& -\biggl(\coth(z)
-{1\over z}\biggr)i{\hat{F}}\nonumber\\
{\cal G}_F(\tau,\tau) &=& -\tanh(z)i{\hat{F}}\non\\
\label{coincidencepureB}
\end{eqnarray}

\item
{\it Crossed field case:}
In a ``crossed field'', defined by 
${\bf E}\perp{\bf B}, E=B$, both invariants 
$B^2-E^2$ and ${\bf E}\cdot{\bf B}$
vanish. For such a field $F^3 =0$, so that the power series 
(\ref{calGBGF}) break off after their quadratic terms.
The worldline correlators thus get truncated to
those terms which were given in (\ref{GB(F)expand}).
The determinant factors are trivial,

\bear
{\rm det}^{-{1\over 2}}
\biggl[{\sin({\cal Z})\over {\cal Z}}
\biggr]
&=&
{\rm det}^{-{1\over 2}}
\biggl[{\tan({\cal Z})\over {\cal Z}}
\biggr] = 1
\label{detcrossed}
\ear\no

The importance of this case lies in the fact that a general
constant field can be well-approximated by
a crossed field at sufficiently high energies
(see, e.g., \cite{bokumi}). 

\end{enumerate}

\subsubsection{Lorentz Covariant Decomposition for a General Field}

Defining the Maxwell invariants

\bear
f &\equiv& \fourth F_{\mu\nu}F^{\mu\nu}
=
\half (B^2 - E^2)
\non\\
g&\equiv& \fourth F_{\mu\nu}\tilde F^{\mu\nu}
=
i {\bf E}\cdot {\bf B}
\non\\
\label{deffg}
\ear
we have the relations

\bear
F^2 + \tilde F^2
&=&
-2f\Eins
\label{eqF2Ftilde2}\\
F\tilde F
&=&
-g\Eins
\label{eqFFtilde}
\ear

\noindent
Define

\bear
F_{\pm} &\equiv& 
{N_{\pm}^2F-N_+N_-\tilde F\over N_{\pm}^2-N_{\mp}^2}
\label{defFpm}\\
N_{\pm} &\equiv&
\np \pm \nm
\label{defNpm}\\
n_{\pm}
&\equiv&
{\sqrt{f\pm g\over 2}}
\label{defnpm}
\ear
Then one has 

\bear
F &=& F_+ + F_-
\label{decompF}\\
F^2 F_{\pm} &=& -N_{\pm}^2 F_{\pm}
\label{FsquareFpm}\\
F_{+}F_{-}
&=&
0
\label{FpFmorth}
\label{propFpm}
\ear\no
With the help of these relations one easily derives
the following formulas, 

\bear
f_{\rm even}(F) &=&
f_{\rm even}(iN_{+}){F_{+}^2\over (iN_{+})^2}
+ f_{\rm even}(iN_{-}){F_{-}^2\over (iN_{-})^2}
\non\\&=& 
{1\over N_+^2-N_-^2}
\Bigl\lbrace
-f_{\rm even}
(iN_{+})\bigl[\Nm^2\Eins+F^2\bigr]
+f_{\rm even}(i\Nm)\bigl[\Np^2\Eins + F^2\bigr]\Bigr\rbrace
\non\\
f_{\rm odd}(F) &=&
f_{\rm odd}(iN_{+}){F_{+}\over iN_{+}}
+f_{\rm odd}(iN_{-}){F_{-}\over iN_{-}}
\non\\&=&
{i\over N_+^2-N_-^2}
\Bigl\lbrace
\bigl[\Nm f_{\rm odd}(i\Nm) -\Np f_{\rm odd}(i\Np)\bigr] F
\non\\
&&\hspace{50pt}
+
\bigl[\Nm f_{\rm odd}(i\Np)- \Np f_{\rm odd}(i\Nm)\bigr]\tilde F
\Bigr\rbrace
\non\\
\label{formfevenodd}
\ear\no
where $f_{\rm even}$ ($f_{\rm odd}$) are
arbitrary even (odd) functions in the field strength
matrix regular at $F=0$,

\bear
f_{\rm even}(F) = \sum_{n=0}^{\infty}
c_{2n}F^{2n},\qquad
f_{\rm odd}(F) = \sum_{n=0}^{\infty}
c_{2n+1}F^{2n+1}
\label{deffevvenodd}
\ear\no
Decomposing ${\cal G}_{B,F}$ 
into their even (odd) parts
${\cal S}_{B,F}$ (${\cal A}_{B,F}$),

\bear
{\cal G}_{B,F}
&=&
{\cal S}_{B,F}
+
{\cal A}_{B,F}
\label{decomposecalGBGF}
\ear\no
and applying the above formulas we obtain the 
following matrix decompositions
of ${\cal G}_B, \dot {\cal G}_B, \ddot {\cal G}_B, {\cal G}_F$,

\bear
{\cal S}_{B12}
&=&
{T\over 2}
\Bigl[
{A^+_{B12}\over z_+}\hat{\cal Z}_+^2
+{A^-_{B12}\over z_-}\hat{\cal Z}_-^2
\Bigr]
\non\\&=&
{T\over 2(z_+^2-z_-^2)}
\Bigl\lbrace
\Bigl[
{z_-^2\over z_+} A_{B12}^{+}
-{z_+^2\over z_-} A_{B12}^{-}
\Bigr]\Eins
+
\Bigl[ {A_{B12}^{+}\over z_+} - {A_{B12}^{-}\over z_-}
\Bigr]{\cal Z}^2
\Bigr\rbrace
\non\\
{\cal A}_{B12}
&=&
{iT\over 2}\Bigl[
(S_{B12}^+ -\dot G_{B12}){\hat{\cal Z_+}\over z_+}
+(S_{B12}^- -\dot G_{B12}){\hat{\cal Z_-}\over z_-}\Bigr]  
\non\\&=&
{iT\over 2(z_+^2-z_-^2)}
\Bigl\lbrace
\bigl[ S_{B12}^{+} - S_{B12}^{-}\bigr] {\cal Z}
+ \Bigl[ {z_+\over z_-} (S_{B12}^{-}-\dot G_{B12}) 
 -{z_-\over z_+} (S_{B12}^{+}-\dot G_{B12})  \Bigr]
{\tilde{\cal Z}}
\Bigr\rbrace
\non\\
\dot{\cal S}_{B12}
&=&
-S^+_{B12}\hat{\cal Z}_+^2
-S^-_{B12}\hat{\cal Z}_-^2
\non\\&=&
{1\over z_+^2-z_-^2}
\Bigl\lbrace
\bigl[
z_+^2 S_{B12}^{-} -z_-^2 S_{B12}^{+} \bigr]\Eins
+
\bigl[ S_{B12}^{-} - S_{B12}^{+}\bigr]{\cal Z}^2
\Bigr\rbrace
\non\\
\dot{\cal A}_{B12}
&=&
-i\Bigl[A_{B12}^-\hat{\cal Z}_-
+ A_{B12}^+\hat{\cal Z}_+\Bigr]
\non\\&=&
{i\over z_+^2-z_-^2}
\Bigl\lbrace
\bigl[ z_{-} A_{B12}^{-} - z_{+} A_{B12}^{+}\bigr] {\cal Z}
+ \bigl[z_- A_{B12}^{+}-z_+ A_{B12}^{-}\bigr] \tilde{\cal Z}
\Bigr\rbrace
\non\\
\ddot{\cal S}_{B12}
&=&
\ddot G_{B12}\Eins +{2\over T}
\Bigl[z_{+}A_{B12}^{+}\hat{\cal Z}_+^2
+ z_{-}A_{B12}^{-}\hat{\cal Z}_-^2\Bigr] 
\non\\
&=&
\ddot G_{B12}\Eins +{2\over T (z_+^2-z_-^2)}
\Bigl\lbrace
\Bigl[
z_-^2 z_+ A_{B12}^{+}
-z_+^2 z_- A_{B12}^{-}
\Bigr]\Eins
+
\Bigl[z_+ A_{B12}^{+} - z_-A_{B12}^{-}
\Bigr]{\cal Z}^2
\Bigr\rbrace
\non\\
\ddot{\cal A}_{B12}
&=&
{2i\over T}\Bigl[z_+S_{B12}^+\hat{\cal Z}_+
+ z_-S_{B12}^-\hat{\cal Z}_-\Bigr]
\non\\
&=&
{2i\over T(z_+^2-z_-^2)}
\Bigl\lbrace
\bigl[ z_+^2S_{B12}^{+} - z_-^2S_{B12}^{-}\bigr] {\cal Z}
+ z_+z_-\Bigl[S_{B12}^{-}-S_{B12}^+ \Bigr]
{\tilde{\cal Z}}
\Bigr\rbrace
\non\\
{\cal S}_{F12} &=&
-S^+_{F12}\hat{\cal Z}_+^2
-S^-_{F12}\hat{\cal Z}_-^2
\non\\&=&
{1\over z_+^2-z_-^2}
\Bigl\lbrace
\bigl[z_+^2 S_{F12}^{-} -z_-^2 S_{F12}^{+}\bigr]
\Eins + \bigl[ S_{F12}^{-} -S_{F12}^{+}\bigr] {\cal Z}^2
\Bigr\rbrace\non\\
{\cal A}_{F12} &=&
-i\Bigl[A_{F12}^-\hat{\cal Z}_- +
A_{F12}^+\hat{\cal Z}_+\Bigr]
\non\\&=&
{i\over z_+^2-z_-^2}
\Bigl\lbrace
\bigl[z_- A_{F12}^{-} -z_+ A_{F12}^{+}\bigr] {\cal Z}
+ \bigl[z_- A_{F12}^{+} -z_+A_{F12}^{-}\bigr] \tilde {\cal Z}
\Bigr\rbrace\non\\ 
\label{decompcalSA}
\ear\no
Here we have further introduced

\bear
z_\pm \equiv eN_\pm T,\quad 
\tilde {\cal Z}\equiv eT\tilde F,\quad
{\cal Z}_{\pm}\equiv
eTF_{\pm}=
{z_{\pm}^2{\cal Z}-z_{+}z_{-}\tilde{\cal Z}\over z_{\pm}^2-z_{\mp}^2}  
,\quad
\hat{\cal Z}_{\pm} \equiv {{\cal Z}_{\pm}\over z_{\pm}}
\non\\
\label{defzs}
\ear\no
Note that ${\cal Z}\tilde{\cal Z}=-z_+z_-\Eins$,
$\hat{\cal Z}_{\pm}^3=-\hat{\cal Z}_{\pm}$.
The scalar, dimensionless 
coefficient functions appearing in these formulas are given by

\bear
S_{B12}^{\pm} &=&
{\sinh(z_{\pm}\dot G_{B12})\over \sinh(z_{\pm})} 
\non\\
A_{B12}^{\pm} &=&
{\cosh(z_{\pm} \dot G_{B12})\over 
\sinh(z_{\pm})}-{1\over z_{\pm}}
\non\\
S_{F12}^{\pm} &=&
G_{F12}{\cosh(z_{\pm}\dot G_{B12})\over\cosh(z_{\pm})}
\non\\
A_{F12}^{\pm} &=&
G_{F12}{\sinh(z_{\pm}\dot G_{B12})\over \cosh(z_{\pm})}
\non\\
\label{defAB}
\ear\no
Note that $S^{\pm}_{B/F12}$ ($A^{\pm}_{B/F12}$)
are odd (even) in $\tau_1-\tau_2$. Thus
the non-vanishing coincidence limits are in 
$A_{B,F}^{\pm}$,

\bear
A_{Bii}^{\pm} &=&
\coth(z_{\pm})-{1\over z_{\pm}}
\non\\
A_{Fii}^{\pm} &=&
\tanh(z_{\pm})
\non\\
\label{coinAB}
\ear\no
In the string-inspired formalism, the functions 
(\ref{defAB}) are the basic
building blocks of parameter integrals for processes involving
constant fields. 
Let us also write down the first few terms of the weak field
expansions of these functions,

\bear
S_{B12}^{\pm} &=&
\dot G_{B12}\biggl[
1-{2\over 3}{G_{B12}\over T}z_{\pm}^2
+\Bigl({2\over 45}{G_{B12}\over T}+
{2\over 15}{G_{B12}^2\over T^2}\Bigr)z_{\pm}^4
+\, {\rm O}(z_{\pm}^6)
\biggr]
\non\\
A_{B12}^{\pm} &=&
\Bigl(\third -2{G_{B12}\over T}\Bigr)z_{\pm}
+\Bigl(-{1\over 45}+{2\over 3}
{G_{B12}^2\over T^2}\Bigr)z_{\pm}^3
+\,  {\rm O}(z_{\pm}^5)
\non\\
S_{F12}^{\pm} &=&
G_{F12}\biggl[
1-2{G_{B12}\over T}z_{\pm}^2+{2\over 3}\Bigl({G_{B12}\over T}
+{G_{B12}^2\over T^2}\Bigr)
z_{\pm}^4 +\, {\rm O}(z_{\pm}^6) \biggr]
\non\\
A_{F12}^{\pm} &=&
G_{F12}\dot G_{B12}
\biggl[z_{\pm}-\Bigl(\third + {2\over 3}{G_{B12}\over T}\Bigr)
z_{\pm}^3+\, {\rm O}(z_{\pm}^5)\biggr]
\non\\
\label{expandAB}
\ear\no
In the same way one finds
for the determinant factors 
(\ref{scaldetext}),(\ref{spindetext})

\bear
{\rm det}^{-{1\over 2}}
\biggl[{\sin({\cal Z})\over {{\cal Z}}}
\biggr] &=&
{z_+z_-\over \sinh(z_+)\sinh(z_-)},
\non\\
{\rm det}^{-{1\over 2}}
\biggl[{\tan({\cal Z})\over {{\cal Z}}}
\biggr]
&=&
{z_+z_-\over \tanh(z_+)\tanh(z_-)}
\non\\
\label{decompdet}
\ear\no
Using the above formulas we can obtain 
explicit results in a Lorentz covariant way.
Nevertheless, it will be useful
to write down these formulas also for
the Lorentz system where $\bf E$ and $\bf B$ 
are both pointing along the positive z - axis,
${\bf E} = (0,0,E), {\bf B} = (0,0,B)$.
(For this to be possible we have to assume that
${\bf E}\cdot{\bf B} > 0$.) 
In this Lorentz system $g=iEB$, so that

\bear
n_{\pm} = \half(B\pm iE), N_+=B, N_-=iE,
F_+ = Br_{\perp}, F_-= iEr_{\parallel},
\label{special}
\ear\no
where 

\begin{equation}
r_{\perp} \equiv
\left(
\begin{array}{*{4}{c}}
0&1&0&0\\
-1&0&0&0\\
0&0&0&0\\
0&0&0&0
\end{array}
\right),\qquad
r_{\parallel} \equiv
\left(
\begin{array}{*{4}{c}}
0&0&0&0\\
0&0&0&0\\
0&0&0&1\\
0&0&-1&0
\end{array}
\right)\nonumber\\
\label{defr}\nonumber
\non\\
\vspace{4mm}
\end{equation}
\vspace{4 mm}

Using those and the projectors $g_{\perp},g{\parallel}$ 
introduced in (\ref{defBmatrices}) 
the matrix decompositions (\ref{decompcalSA}) can 
be rewritten as follows,

\bear
{\cal S}_{B12}^{\mu\nu}
&=&
-{T\over 2}
\sum_{\alpha ={\perp},{\parallel}}
{A_{B12}^{\alpha}\over z_{\alpha}}\,g_{\alpha}^{\mu\nu}
\non\\
{\cal A}_{B12}^{\mu\nu}
&=&
{iT\over 2}
\sum_{\alpha ={\perp},{\parallel}}
{S_{B12}^{\alpha}-\dot G_{B12}\over z_{\alpha}}
\,r_{\alpha}^{\mu\nu}
\non\\
\dot{\cal S}_{B12}^{\mu\nu} &=&
\sum_{\alpha ={\perp},{\parallel}}
S_{B12}^{\alpha}\,g_{\alpha}^{\mu\nu}
\non\\
\dot{\cal A}_{B12}^{\mu\nu} &=& 
-i
\sum_{\alpha ={\perp},{\parallel}}
A_{B12}^{\alpha}\,r_{\alpha}^{\mu\nu}
\non\\
\ddot{\cal S}_{B12}^{\mu\nu} &=& \ddot G_{B12}g^{\mu\nu}
-{2\over T}
\sum_{\alpha ={\perp},{\parallel}}
z_{\alpha}A_{B12}^{\alpha}\,g_{\alpha}^{\mu\nu}
\non\\
\ddot{\cal A}_{B12}^{\mu\nu} &=& 
{2i\over T}
\sum_{\alpha ={\perp},{\parallel}}
z_{\alpha}S_{B12}^{\alpha}\,r_{\alpha}^{\mu\nu}
\non\\
{\cal S}_{F12}^{\mu\nu} &=&
\sum_{\alpha ={\perp},{\parallel}}
S_{F12}^{\alpha}\,g_{\alpha}^{\mu\nu}
\non\\
{\cal A}_{F12}^{\mu\nu} &=& 
-i
\sum_{\alpha ={\perp},{\parallel}}
A_{F12}^{\alpha}\,r_{\alpha}^{\mu\nu}
\non\\
\label{specialdecompcalSA}
\ear\no
with
$S/A^{\perp}_{B/F}\equiv S/A^+_{B/F}\,
(z_+=eBT\equiv z_{\perp}),\,
S/A^{\parallel}_{B/F}\equiv S/A^-_{B/F}\,
(z_-=ieET\equiv z_{\parallel})$.

\subsection{Example: The Scalar/Spinor
QED Vacuum Polarization Tensors in a Constant Field}

We now apply this formalism to
a calculation of the scalar and spinor QED
vacuum polarization tensors in a general
constant field.
For the 2-point case 
the master formula (\ref{scalarqedmasterF})
yields the following integrand,

\be
\exp\biggl\lbrace
\ldots
\biggr\rbrace
\mid_{\rm multi-linear}\quad
=
\Bigl\lbrack
\varepsilon_1\cdot\ddot {\cal G}_{B12}\cdot\varepsilon_2
-
\varepsilon_1\cdot\dot{\cal G}_{B1i}\cdot k_i
\,\varepsilon_2\cdot\dot{\cal G}_{B2j}\cdot k_j
\Bigr\rbrack
\e^{k_1\cdot \bar{\cal G}_{B12}\cdot k_2}
\label{P2withF}
\ee
where summation over $i,j = 1,2$ is understood.
Removing the second derivative in the first term by a
partial integration in $\tau_1$ 
this becomes

\be
\biggl\lbrack
-\varepsilon_1\cdot\dot {\cal G}_{B12}\cdot\varepsilon_2
\,k_1\cdot \dot {\cal G}_{B1j}\cdot k_j
-
\varepsilon_1\cdot\dot{\cal G}_{B1i}\cdot k_i
\,\varepsilon_2\cdot\dot{\cal G}_{B2j}\cdot k_j
\biggr\rbrack
\e^{k_1\cdot \bar{\cal G}_{B12}\cdot k_2}
\label{P2withFpint}
\ee
We apply the ``cycle replacement rule'' to this expression
and use momentum conservation, $k\equiv k_1 = -k_2$.
The content of the brackets then turns into
$\varepsilon_{1\mu} I^{\mu\nu}\varepsilon_{2\nu}$,
where

\bear
I^{\mu\nu} &=&
\dot{\cal G}^{\mu\nu}_{B12}k\cdot\dot{\cal G}_{B12}\cdot k
-{\cal G}^{\mu\nu}_{F12}k\cdot{\cal G}_{F12}\cdot k
\non\\&&\hspace{-15pt}
- \biggl[
\Bigl(\dot{\cal G}_{B11}
-{\cal G}_{F11}
-\dot{\cal G}_{B12}\Bigr)
^{\mu\lambda}
  \Bigl(\dot{\cal G}_{B21}
-\dot{\cal G}_{B22}
+{\cal G}_{F22}
\Bigr)^{\nu\kappa}
+
{\cal G}^{\mu\lambda}_{F12}
{\cal G}^{\nu\kappa}_{F21}
\biggr]
k^{\kappa}k^{\lambda}
\non\\
\label{substint}
\ear\no
Next we would like to use the fact that this integrand
contains many terms which integrate to zero
due to antisymmetry
under the exchange
$\tau_1\leftrightarrow\tau_2$. 
This we can do by decomposing 
${\cal G}_B$ 
and ${\cal G}_F$
as in (\ref{decomposecalGBGF}).
First note that only the Lorentz even part
of ${\cal G}_B$ 
contributes in the exponent,
\vspace{-8pt}

\bear
k_1\cdot \bar {\cal G}_{B12}\cdot k_2
&=&
k_1\cdot \bigl( {\cal S}_{B12}-{\cal S}_{B11}\bigr)\cdot k_2
\equiv
-Tk\cdot\Phi_{12}\cdot k
\label{defPhi}
\ear\no
$I^{\mu\nu}$ turns,
after decomposing all factors of $\dot {\cal G}_B, {\cal G}_F$ as above,
and deleting all $\tau$ - odd terms, into 

\bear
I^{\mu\nu}_{\rm spin}
&\equiv&
\biggl\lbrace
\Bigl(
{\dot{\cal S}}^{\mu\nu}_{B12}{\dot{\cal S}}^{\kappa\lambda}_{B12}
- {\dot{\cal S}}^{\mu\lambda}_{B12}{\dot{\cal S}}^{\nu\kappa}_{B12}
\Bigr)
-
\Bigl(
{{\cal S}}^{\mu\nu}_{F12}{{\cal S}}^{\kappa\lambda}_{F12}
- {{\cal S}}^{\mu\lambda}_{F12}{{\cal S}}^{\nu\kappa}_{F12}
\Bigr)
\non\\
&&
+
\Bigl(
{\dot{\cal A}}_{B12}-{\dot{\cal A}}_{B11}+{{\cal A}}_{F11}
\Bigr)^{\mu\lambda}
\Bigl(
{\dot{\cal A}}_{B12}-{\dot{\cal A}}_{B22}+{{\cal A}}_{F22}
\Bigr)^{\nu\kappa}
\non\\
&&
-{\cal A}^{\mu\lambda}_{F12}
{\cal A}^{\nu\kappa}_{F12}
\biggr\rbrace
k^{\kappa}k^{\lambda}
\non\\
\label{intfinal}
\ear\no
(here we used (\ref{symmcalGBF})).

In this way we obtain the following integral representations 
for the dimensionally regularized
scalar/spinor QED vacuum polarization
tensors \cite{mecorfu}, 

\bear
\Pi^{\mu\nu}_{\rm scal}(k)
&=&
-{e^2\over {(4\pi)}^{D\over 2}}
{\dps\int_{0}^{\infty}}{dT\over T}
{T}^{2-{D\over 2}}
e^{-m^2T}
{\rm det}^{-\half}\biggl[{\sin({\cal Z})\over {{\cal Z}}}
\biggr] 
\int_0^1 du_1
\,\,I^{\mu\nu}_{\rm scal}
\,\e^{-Tk\cdot\Phi_{12}\cdot k}
\non\\
\label{vpscalreg}\\
\Pi^{\mu\nu}_{\rm spin}(k)
&=&
2{e^2\over {(4\pi)}^{D\over 2}}
{\dps\int_{0}^{\infty}}{dT\over T}
{T}^{2-{D\over 2}}
e^{-m^2T}
{\rm det}^{-\half}\biggl[{\tan({\cal Z})\over {{\cal Z}}}\biggr]
\int_0^1 du_1
\,\,I^{\mu\nu}_{\rm spin}
\,\e^{-Tk\cdot\Phi_{12}\cdot k}
\non\\
\label{vpspinreg}
\ear\no
Here $I^{\mu\nu}_{\rm scal}$ is obtained simply by
deleting, in eq. (\ref{intfinal}), all quantities
carrying a subscript ``F''. As usual we have rescaled
to the unit circle and 
set $u_2 =0$. 

Note that again the transversality
of the vacuum polarization tensors is manifest at the
integrand level, 
$k_{\mu}I^{\mu\nu}_{\rm scal/spin}
=
I^{\mu\nu}_{\rm scal/spin}k_{\nu} =0$.

The constant field vacuum polarization tensors contain
the UV divergences of the ordinary vacuum polarization tensors
(\ref{scalarvpresult}),(\ref{spinorvpresult}), and thus require
renormalization.
As is usual in this context we perform the
renormalization on-shell, i.e. we impose the following
condition on the renormalized 
vacuum polarization tensor $\bar\Pi^{\mu\nu}(k)$ 
(see, e.g., \cite{ditreu}),

\bear
\lim_{k^2\rightarrow 0}\lim_{F\rightarrow 0}
{\bar \Pi}^{\mu\nu}(k) =0
\label{renormcond}
\ear\no
Counterterms appropriate to this
condition are easy to find from our
above results for the ordinary vacuum polarization
tensors. From the representations
eqs. (\ref{scalarvpresult}), (\ref{spinorvpresult}) 
for these tensors it is obvious that we can
implement (\ref{renormcond}) by
subtracting those same 
expressions with the last factor $\e^{-G_{B12}k^2}$ 
deleted. 

In this way we find for the
renormalized vacuum polarization tensors

\bear
{\bar \Pi}^{\mu\nu}_{\rm scal}(k)
&=&
\Pi^{\mu\nu}_{\rm scal}(k)
+{\alpha\over 4\pi}
\bigl(g^{\mu\nu}k^2 -k^{\mu}k^{\nu}\bigr)
{\dps\int_{0}^{\infty}}{dT\over T}
e^{-m^2T}
\int_0^1 du_1
\dot G_{B12}^2
\non\\
{\bar \Pi}^{\mu\nu}_{\rm spin}(k)
&=&
\Pi^{\mu\nu}_{\rm spin}(k)
-
{\alpha\over 2\pi}
\bigl(g^{\mu\nu}k^2 -k^{\mu}k^{\nu}\bigr)
{\dps\int_{0}^{\infty}}{dT\over T} e^{-m^2T}
\int_0^1 du_1
\bigl(\dot G_{B12}^2-G_{F12}^2\bigr)
\non\\
\label{vpren}
\ear\no
The remaining $u_1$ - integral can be brought
into a more standard form
by a transformation of variables
$v = \dot G_{B12} = 1-2u_1$.

Writing the integrands explicitly
using the formulas (\ref{decompcalSA})
and continuing to Minkoswki space
\footnote{For the Maxwell invariants this means
$f\rightarrow {\cal F}$, 
$g\rightarrow i{\cal G}$,
$N_+\rightarrow a$, $N_-\rightarrow ib$
(to be able to fix all signs we assume 
${\cal G} \geq 0$).
Note also that $r_{\perp}k\rightarrow \tilde k_{\perp},
r_{\parallel}k\rightarrow -i\tilde k_{\parallel}$.
}
we obtain our final result for these amplitudes
\cite{mevv},

\bear
{\bar \Pi}^{\mu\nu}_{\rm scal}(k)
&=&
-{\alpha\over 4\pi}
{\dps\int_{0}^{\infty}}{ds\over s}
\,e^{-ism^2}
\int_{-1}^1 {dv\over 2}
\Biggl\lbrace
{z_+z_-\over \sinh(z_+)\sinh(z_-)}
\non\\&&\times
{\rm exp}\biggl[
-i{s\over 2}\sum_{\alpha=+,-}
{A_{B12}^{\alpha}-A_{B11}^{\alpha}\over z_{\alpha}}\,
k\cdot \hat{\cal Z}_{\alpha}^2\cdot k
\biggr]
\non\\&&\times
\sum_{\alpha,\beta =+,-}
\biggl(
S_{B12}^{\alpha}
S_{B12}^{\beta}
\Bigl[
\bigl(\hat{\cal Z}_{\alpha}^2\bigr)^{\mu\nu}
k\cdot \hat{\cal Z}_{\beta}^2 \cdot k
-
\bigl(\hat{\cal Z}_{\alpha}^2k\bigr)^{\mu}
\bigl(\hat{\cal Z}_{\beta}^2k\bigr)^{\nu}
\Bigr]
\non\\&&\hspace{55pt}
-
(A_{B12}^{\alpha}-A_{B11}^{\alpha})
(A_{B12}^{\beta}-A_{B22}^{\beta})
\bigl(\hat{\cal Z}_{\alpha}k\bigr)^{\mu}
\bigl(\hat{\cal Z}_{\beta}k\bigr)^{\nu}
\biggr)
\non\\&&
- \bigl({\eta}^{\mu\nu}k^2 -k^{\mu}k^{\nu}\bigr)v^2
\biggr\rbrace
\label{vpscalfinal}\\
{\bar \Pi}^{\mu\nu}_{\rm spin}(k)
&=&
{\alpha\over 2\pi}
{\dps\int_{0}^{\infty}}{ds\over s}
\,e^{-ism^2}
\int_{-1}^1 {dv\over 2}
\Biggl\lbrace
{z_+z_-\over \tanh(z_+)\tanh(z_-)}
\non\\&&\times
{\rm exp}\biggl[
-i{s\over 2}\sum_{\alpha=+,-}
{A_{B12}^{\alpha}-A_{B11}^{\alpha}\over z_{\alpha}}\,
k\cdot \hat{\cal Z}_{\alpha}^2\cdot k
\biggr]
\non\\&&\times
\sum_{\alpha,\beta =+,-}
\biggl(
\Bigl[
S_{B12}^{\alpha}
S_{B12}^{\beta}
-
S_{F12}^{\alpha}
S_{F12}^{\beta}
\Bigr]
\Bigl[
\bigl(\hat{\cal Z}_{\alpha}^2\bigr)^{\mu\nu}
k\cdot \hat{\cal Z}_{\beta}^2 \cdot k
-
\bigl(\hat{\cal Z}_{\alpha}^2k\bigr)^{\mu}
\bigl(\hat{\cal Z}_{\beta}^2k\bigr)^{\nu}
\Bigr]
\non\\&&\hspace{-5pt}
-
\Bigl[ 
(A_{B12}^{\alpha}-A_{B11}^{\alpha}+A_{F11}^{\alpha})
(A_{B12}^{\beta}-A_{B22}^{\beta}+A_{F22}^{\beta})
-A_{F12}^{\alpha}A_{F12}^{\beta}
\Bigr]
\bigl(\hat{\cal Z}_{\alpha}k\bigr)^{\mu}
\bigl(\hat{\cal Z}_{\beta}k\bigr)^{\nu}
\biggr)
\non\\&&
- \bigl({\eta}^{\mu\nu}k^2 -k^{\mu}k^{\nu}\bigr)(v^2-1)
\biggr\rbrace
\label{vpspinfinal}
\ear\no
where now

\bear
z_+ &=& iesa \non\\
z_- &=& -esb \non\\
\hat{\cal Z}_+ &=& {aF-b\tilde F\over a^2+b^2},
\quad\;\;\;\;
{\hat{\cal Z}_+}^2 = {F^2-b^2\Eins\over a^2+b^2}
\non\\
\hat{\cal Z}_- &=& -i{bF+a\tilde F\over a^2 +b^2} 
,\quad 
{\hat{\cal Z}_-}^2 = -{F^2+a^2\Eins\over a^2+b^2}
\non\\
\label{defsmink}
\ear\no
Here $a,b$ denote 
the standard `secular'
invariants which we already introduced in 
eqs.(\ref{eulheiscal}),(\ref{eulheispin}). In terms of
the invariants ${\cal F},{\cal G}$ those read

\bear
a &\equiv& \sqrt{\sqrt{{\cal F}^2+{\cal G}^2}+{\cal F}}
\non\\
b &\equiv& \sqrt{\sqrt{{\cal F}^2+{\cal G}^2}-{\cal F}}
\label{defsecular}
\ear
(${\cal F} = \half (B^2-E^2), \quad {\cal G} =
{\bf E}\cdot{\bf B}$).

For fermion QED, the 
constant field vacuum polarization tensor was obtained before
by various authors 
\cite{batsha,urrutia,gies,ditgiebook}.
For the sake of comparison with their results, let us 
also specialize to the Lorentz system where 
${\bf E}=(0,0,E)$ and ${\bf B}=(0,0,B)$.
In this system $a=B, b=E$.
Denoting

\bear
k_{\parallel} &=& (k^0,0,0,k^3),\quad
k_{\perp} = (0,k^1,k^2,0)\non\\
\tilde k_{\parallel} &=& (k^3,0,0,k^0),\quad
\tilde k_{\perp} = (0,k^2,-k^1,0)
\non\\
\label{defkktilde}
\ear\no
our result can be written as follows,
 
\bear
{\bar \Pi}^{\mu\nu}_{\bigl({{\rm spin}\atop{\rm scal}}\bigr)}(k)
&=&
-{\alpha\over 4\pi}
\biggl({-2\atop 1}\biggr)
{\dps\int_{0}^{\infty}}{ds\over s}
\int_{-1}^1 {dv\over 2}
\Biggl\lbrace
{zz'\over \sin(z)\sinh(z')}
\biggl({\cos(z)\cosh(z')\atop 1}\biggr)
\non\\
&&\times
\e^{-is\Phi_0}
\sum_{\alpha,\beta =\perp ,\parallel}
\Bigl[
s^{\alpha\beta}_{\bigl({{\rm spin}\atop{\rm scal}}\bigr)}
({\eta}^{\mu\nu}_{\alpha}k_{\beta}^2-k^{\mu}_{\alpha}k^{\nu}_{\beta})
+a^{\alpha\beta}_{\bigl({{\rm spin}\atop{\rm scal}}\bigr)}
\tilde k_{\alpha}^{\mu}\tilde k_{\beta}^{\nu}\Bigr]
\non\\
&& -\e^{-ism^2}({\eta}^{\mu\nu}k^2-k^{\mu}k^{\nu})
\biggl({v^2-1\atop v^2}\biggr)
\Biggr\rbrace
\label{specialvpfinal}
\ear\no
where $z=eBs, z'=eEs$, and

\bear
\Phi_0 = m^2 +{k_{\perp}^2\over 2}{\cos(zv)-\cos(z)\over z\sin(z)}
-{k_{\parallel}^2\over 2}{\cosh(z'v)-\cosh(z')\over z'\sinh(z')}
\non\\
\label{Phi0}
\ear

\bear
s_{\rm scal}^{\perp\perp} &=& {\sin^2(zv)\over\sin^2(z)} \non\\
s_{\rm scal}^{\perp\parallel,\parallel\perp}
&=&
{\sin(zv)\sinh(z'v)\over\sin(z)\sinh(z')}\non\\
s_{\rm scal}^{\parallel\parallel} &=&
{\sinh^2(z'v)\over\sinh^2(z')}
\non\\
a_{\rm scal}^{\perp\perp} &=&
\Bigl({\cos(zv)-\cos(z)\over\sin(z)}\Bigr)^2 \non\\
a_{\rm scal}^{\perp\parallel,\parallel\perp}&=&
-{\cos(zv)-\cos(z)\over\sin(z)}{\cosh(z'v)-\cosh(z')\over\sinh(z')}
\non\\
a_{\rm scal}^{\parallel\parallel} &=&
\Bigl({\cosh(z'v)-\cosh(z')\over\sinh(z')}\Bigr)^2 \non\\
\label{coefffinalscal}
\ear

\bear
s_{\rm spin}^{\perp\perp} &=& {\sin^2(zv)\over\sin^2(z)} 
-{\cos^2(zv)\over\cos^2(z)}\non\\
s_{\rm spin}^{\perp\parallel,\parallel\perp}
&=&
{\sin(zv)\sinh(z'v)\over\sin(z)\sinh(z')}
-
{\cos(zv)\cosh(z'v)\over\cos(z)\cosh(z')}
\non\\
s_{\rm spin}^{\parallel\parallel} &=&
{\sinh^2(z'v)\over\sinh^2(z')}
-
{\cosh^2(z'v)\over\cosh^2(z')}
\non\\
a_{\rm spin}^{\perp\perp} &=&
\Bigl({\cos(zv)-\cos(z)\over\sin(z)}-\tan(z)\Bigr)^2 
-{\sin^2(zv)\over\cos^2(z)}\non\\
a_{\rm spin}^{\perp\parallel,\parallel\perp}&=&
-\Bigl({\cos(zv)-\cos(z)\over\sin(z)}-\tan(z)\Bigr)
\Bigl({\cosh(z'v)-\cosh(z')\over\sinh(z')}+\tanh(z')\Bigr)
\non\\&&
-{\sin(zv)\sinh(z'v)\over\cos(z)\cosh(z')}
\non\\
a_{\rm spin}^{\parallel\parallel} &=&
\Bigl({\cosh(z'v)-\cosh(z')\over\sinh(z')}+\tanh(z')\Bigr)^2 
-{\sinh^2(z'v)\over\cosh^2(z')}
\non\\
\label{coefffinalspin}
\ear

In this form it can be easily identified 
with the field theory results of 
\cite{bks1} (scalar QED) and 
\cite{urrutia,gies} (fermion QED).

\subsection{Example: Photon Splitting in a Constant Magnetic Field} 
\label{phosplit}

Photon splitting in a constant magnetic field is
a process of potential astrophysical interest.
Its first exact calculation, valid for an arbitrary magnetic
field strength and photon energies up to
the pair creation threshold, 
was performed
by Adler in 1971 ~\cite{adler71}.
This calculation amounts essentially to the calculation of the
QED one-loop three-photon amplitude in a constant
field. This amplitude is finite, so that one can set $D=4$.

\subsubsection{Scalar QED}

To obtain the photon splitting amplitude
for scalar QED, we have to 
use the correlators (\ref{GB(F)pureB})
for the Wick contraction of
three photon vertex operators $V_0$ and $V_{1,2}$, 

$$
V^A_{{\rm scal},i}[k_i,\varepsilon_i]
=\int_0^T d\tau_i\,
\varepsilon_{i}\cdot\dot x(\tau_i)
\,{\rm exp}\Bigl[ ik_i\cdot x(\tau_i)\Bigr] 
$$
\no
representing the
incoming and the two outgoing photons. 

The calculation is
greatly simplified by the peculiar kinematics of this process.
Energy--momentum conservation $k_0+k_1+k_2=0$
forces collinearity of all three four--momenta, so that,
writing $k_0\equiv k\equiv \omega n$,

\begin{equation}
k_1=-{\omega_1\over\omega}k,k_2=-{\omega_2\over\omega}k;\,\,
k^2=k_1^2=k_2^2=k\cdot k_1=k\cdot k_2=k_1\cdot k_2=0.
\label{kconstraints}
\end{equation}

\noindent
By a simple Lorentz invariance argument \cite{adler71}
one can assume $\bf k$
to be orthogonal to the magnetic field direction.
Moreover, in \cite{adler71} it was shown, using
CP invariance together with
an analysis of dispersive effects, that
there is only one 
non-vanishing polarization case. This is the case
where the magnetic vector 
$\hat {\bf k}\times\hat{\bf\varepsilon_0}$ 
of the incoming photon is
parallel to the plane
containing the external field and the direction of propagation
$\hat {\bf k}$, and those of the
outgoing ones are both perpendicular to this plane.
Taking the magnetic field in the $z$ - direction,
and choosing
\footnote{Note that we are still using Euclidean
conventions.}
$n=(1,0,0,-i)$, we can implement this case
by taking $\varepsilon_0 = (0,1,0,0)$ and
$\varepsilon_1=\varepsilon_2 = (0,0,1,0)$.
This 
leads to the further
vanishing relations

\begin{equation}
\varepsilon_{1,2}\cdot\varepsilon_0
=\varepsilon_{1,2}\cdot k
=\varepsilon_{1,2}\cdot F=0\quad 
\label{epsconstraints}
\end{equation}

\noindent
which leave us with the only a small
number of nonvanishing Wick contractions:

\begin{eqnarray}
\langle V_{{\rm scal},0}^AV_{{\rm scal},1}^AV_{{\rm scal},2}^A
\rangle&=&
\Bigl\langle
\prod_{i=0}^2\int_0^T
d\tau_i\,
\varepsilon_i
\cdot
\dot x_i
\,
{\rm exp}
\Bigl[
ik_i\cdot x(\tau_i)
\Bigr]
\Bigr\rangle
\nonumber\\
&=&
\Bigl\langle
\prod_{i=0}^2\int_0^T
d\tau_i\,
\varepsilon_i
\cdot
\dot x_i
\,
{\rm exp}
\Bigl[
i\, n\cdot
\sum_{j=0}^2
\bar\omega_j
x_j
\Bigr]
\Bigr\rangle
\nonumber\\
&=&
-i
\prod_{i=0}^2\int_0^T
\!\!\!d\tau_i
\exp\biggl[{1\over 2}
\!\!\sum_{i,j=0}^2
\bar\omega_i\bar\omega_j
n\cdot\bar{\cal G}_{Bij}\cdot n\biggr]\,
\varepsilon_1\cdot
\ddot{\cal G}_{B12}\cdot\varepsilon_2\,
\sum_{i=0}^2
\bar\omega_i\varepsilon_0\cdot
\dot{\cal G}_{B0i}\cdot n\nonumber\\
\label{psscalwick}
\end{eqnarray}

\vskip.3cm
\noindent
To keep the notation compact, we have defined
$\bar\omega_0=\omega, \bar\omega_{1,2}=-\omega_{1,2}$.
A number of
terms which vanish by the above relations have been
omitted. For example, to see that the term involving

$$\langle
\varepsilon_0\cdot \dot x_0\,
\varepsilon_1\cdot \dot x_1\rangle
= \varepsilon_0\cdot\ddot{\cal G}_{B01}
\cdot\varepsilon_1
$$
\no
vanishes, remember that $\ddot{\cal G}_B$ is a power series
in the matrix $F_{\mu\nu}$. The first term in this expansion
is proportional to the Lorentz identity
and gives zero since $\varepsilon_0
\cdot\varepsilon_1=0$; all remaining ones
give zero because $ F\cdot \varepsilon_1 = 0$.

Performing the Lorentz contractions,
and taking the determinant factor eq.(\ref{detextB})
into account, one obtains the
following parameter integral for the
three-point amplitude:

\begin{eqnarray}
\Gamma_{\rm scal}[k_0,k_1,k_2]&=&
{(-ie)}^3
\Tint {\rm e}^{-m^2T}
{(4\pi T)}^{-2}
{z\over\sinh (z)}
\langle V_{{\rm scal},0}^AV_{{\rm scal},1}^AV_{{\rm scal},2}^A\rangle
\nonumber\\
&=&
ie^3
\int_0^{\infty}\!dT\,{\rm e}^{-m^2T}
{(4\pi T)}^{-2}
{z\over {\rm sinh}(z)}
\int_0^T\!\!d\tau_1\,d\tau_2\,
\ddot G_{B12}
\sum_{i=0}^2
\bar\omega_i
{\cosh(z\dot G_{B0i})\over \sinh (z)}
\nonumber\\
&&
\times
{\rm exp}
\biggl\lbrace\!\!-{1\over 2}\!\sum_{i,j=0}^2\bar\omega_i
\bar\omega_j\Bigl[G_{Bij} + {T\over 2z}
{{\rm cosh}(z\dot G_{Bij})\over {\rm sinh}(z)}
\Bigr]\biggr\rbrace\nonumber\\
\label{3pointamp}
\end{eqnarray}
\noindent
($z=eBT$).
Translation invariance in $\tau$ has been used to
set the position $\tau_0$ of the
incoming photon equal to $T$. 
Normalizing the amplitude according to
eq.(25) in ~\cite{adler71}, the final result becomes

\begin{eqnarray}
C_{\rm scal}[\omega,\omega_1,\omega_2,B] &=&
{m^8\over 8 \omega\omega_1\omega_2}
\int_0^{\infty}\!dT\,T{{\rm e}^{-m^2T}
\over {z}^2{\rm sinh}^2(z)}
\int_0^T\!\!d\tau_1\,d\tau_2\,
\ddot G_{B12}
\non\\
&&
\!\!\!\!\!\!\!\!\!\!\!\!\!\!\!\!\!\!\!\!\!\!\!\!\!
\times
\biggl[
\sum_{i=0}^2
\bar\omega_i\cosh(z\dot G_{B0i})
\biggr]
\,{\rm exp}
\biggl\lbrace\!\!-{1\over 2}\!\sum_{i,j=0}^2\bar\omega_i
\bar\omega_j\Bigl[G_{Bij} + {T\over 2z}
{{\rm cosh}(z\dot G_{Bij})\over {\rm sinh}(z)}
\Bigr]\biggr\rbrace\nonumber\\
\label{psscalresult}
\end{eqnarray}
\noindent
($C_{\rm scal}$ corresponds to $C_2$ there). 

\subsubsection{Spinor QED}

For the fermion loop case let us, for a change,
use the superfield formalism rather than the cycle replacement rule. 
Using the superfield representation (\ref{supervertex})
of the photon vertex operator, 

$$
V^A_{\rm spin}[k,\varepsilon]
=\int_0^Td\tau d\theta\,
\varepsilon\cdot
DX
{\rm e}^{ik\cdot X}
\label{vertopsup}
$$\no
we can write the result of the
Wick contraction for the spinor loop in complete
analogy to the scalar loop result 
eq.(\ref{psscalwick}):

\begin{equation}
\langle V_{{\rm spin},0}^AV_{{\rm spin},1}^AV_{{\rm spin},2}^A
\rangle\!=\!i\!
\prod_{i=0}^2\int_0^T
\!\!\!d\tau_i\!\!\int\!d\theta_i\,
\exp\biggl[{1\over 2}
\!\!\sum_{i,j=0}^2
\bar\omega_i\bar\omega_j
n\cdot\bar{\hat{\cal G}_{ij}}\cdot n\biggr]\,
\varepsilon_1\cdot
D_1D_2\hat{\cal G}_{12}\cdot\varepsilon_2\,
\sum_{i=0}^2
\bar\omega_i\varepsilon_0\cdot
D_0\hat{\cal G}_{0i}\cdot n
\label{psspinwick}
\end{equation}
\no
Here $\hat {\cal G}$ denotes the constant
field worldline super propagator,
eq.(\ref{calsuperpropagator}).
The Lorentz contractions
are performed as before. The
only difference is in the additional
$\theta$ -- integrations, which are
easy to do. The
final result becomes

\begin{eqnarray}
&&C_{\rm spin}
[\omega,\omega_1,\omega_2,B] =
{m^8\over 4 \omega\omega_1\omega_2}
\int_0^{\infty}\!dT\,T{{\rm e}^{-m^2T}
\over {z}^2{\rm sinh}(z)}\nonumber\\
&&\times\int_0^T\!\!d\tau_1\,d\tau_2\,
\,{\rm exp}
\biggl\lbrace\!\!-{1\over 2}\!\sum_{i,j=0}^2\bar\omega_i
\bar\omega_j\Bigl[G_{Bij} + {T\over 2z}
{{\rm cosh}(z\dot G_{Bij})\over {\rm sinh}(z)}
\Bigr]\biggr\rbrace\nonumber\\
&&\times 
\Biggl\lbrace\biggl\lbrack
-\cosh (z)\ddot G_{B12}\!+\!\omega_1\omega_2
\Bigl( {\rm cosh}(z)-\cosh (z\dot G_{B12})\Bigr )\biggr\rbrack
\nonumber\\
&&\times\biggl\lbrack
{\omega\over\sinh(z)\!\cosh(z)}
-\omega_1
{{\rm cosh}(z\dot G_{B01})\over {\rm sinh}(z)}
-\omega_2
{{\rm cosh}(z\dot G_{B02})\over {\rm sinh}(z)}
\biggr\rbrack\nonumber\\
&& +{\omega\omega_1\omega_2
G_{F12}\over \cosh (z)}
\biggl\lbrack
\sinh (z\dot G_{B01})\Bigl (\cosh (z)\!-\!\cosh (z\dot G_{B02}) \Bigr )
- (1 \leftrightarrow 2) 
\biggr\rbrack \Biggl\rbrace\non\\
\label{psspinresult}
\end{eqnarray}

\noindent
A numerical analysis of this three-parameter
integral 
has shown ~\cite{adlsch,adlerhomepage}
it to be in complete agreement with 
other known integral representations of this
amplitude ~\cite{adler71,bks2,stoneham,bamish}.

See ~\cite{barhar}
for a more extensive analysis, 
as well as for a discussion
of the relevance of photon splitting for the
spectra of $\gamma$ - ray pulsars and soft
$\gamma$ - repeaters. 
\section{Yukawa and Axial Couplings}
\renewcommand{\theequation}{6.\arabic{equation}}
\setcounter{equation}{0}

Up to now we have been concentrating almost exclusively on
QED and QCD amplitudes. This reflects the present state-of-the-art,
since almost all nontrivial applications of the string-inspired
technique have been to these theories. 
In fact, until recently worldline path integral representations
for more general theories were not available in the literature.
It will be recalled that, with the exception of the
gluon loop case, the worldline path integrals  which we have
used for QED and QCD have essentially been known since the sixties.
It is thus quite surprising that some relatively straightforward
extensions of these formulas were apparently never considered
until the recent revival of this subject triggered by
the work of Bern and Kosower, and Strassler.

Obviously, to be able to treat arbitrary fermion loop processes
in the standard model one would need
a worldline representation for the coupling of
a spin $\half$ - loop to a more general background including
a
scalar field $\phi$, 
pseudoscalar field $\phi_5$, vector field $A$, and axialvector 
field $A_5$.
In (Minkowski space) field theory we would thus be dealing with 
the following action:

\be
S[\phi,\phi_5,A,A_5]
=
-\int dx\,
\bar\psi
\Bigl[
\slpartial
+\phi
+i\gamma^5\phi_5
+i\slash A
+i\gamma^5
{\slash A}_5
\Bigr]
\psi
\label{Sspinhalfmink}
\ee\no
Here we absorbed the coupling constants into the
background fields.
The corresponding Euclidean effective action is given by

\be
\Gamma_{\rm E}[\phi,\phi_5,A,A_5] = \ln {\rm Det}
\Bigl[{\slash p}_E -i\phi +\gamma_{E5}\phi_5 +{\slash A}_E
+\gamma_{E5}{\slash A}_{E5}\Bigr] 
\label{defEAeuc-annchiral}
\ee\no
In the following we work in Euclidean
space-time as usual and drop the subscript $E$.

In contrast to the QED or QCD effective action, which
develops an imaginary part only due to threshold or nonperturbative
effects, in the presence of axial vectors or pseudoscalars
the effective action can become imaginary already in
Euclidean perturbation theory. To be precise,
in Euclidean space-time Feynman graphs with an even (odd) number
of $\gamma_5$ - vertices contribute to $Re(\Gamma_{\rm E})$
($Im(\Gamma_{\rm E})$).

A worldline path integral representation for this effective action
was constructed in \cite{mnss1,mnss2} 
for the abelian case, though in a 
heuristic way. In \cite{dhogag1,dhogag2} 
a completely rigorous treatment was given
which also includes the antisymmetric tensor coupling and the
non-abelian case.
A detailed treatment of the general case would
be lengthy.
We will therefore restrict ourselves to two special
cases of particular interest, 
the scalar-pseudoscalar and the vector -- axialvector 
backgrounds. Moreover, we will take all fields to be
abelian, and refer again to \cite{dhogag1,dhogag2} for the nonabelian
generalization.

\subsection{Yukawa Couplings from Gauge Theory}

For a beginning, let us restrict our attention to the case of only
a scalar and a pseudoscalar field. Here it is possible to
give a simple and instructive 
solution of the problem \cite{mnss1}, using a dimensional
reduction procedure.

As usual in this formalism we try to  stay in line with
string theory. Now in string theory Yukawa couplings are
usually generated in the process of compactifying
some of the unphysical dimensions. 
Our ansatz for this worldline action is 
therefore
to take $\phi_5$ and
$\phi$ as the fifth and sixth components of 
a Yang-Mills field in six dimensions, with the
other four components vanishing,
$A=(0,0,0,0,\phi_5,\phi)$.
It was already mentioned that the path integral representation 
eq.(\ref{spinorpi}) for the 
coupling of the spinor loop to
a background gauge field remains valid
for any even spacetime dimension.
In six dimensions and for an 
$A$ -- field of the form above, the
worldline Lagrangian becomes

\bear
L
&=& \kinb + \fourth\dot x_5^2 + \fourth\dot x_6^2
+
\kinf +\half\psi_5\dot\psi_5 +\half\psi_6\dot\psi_6 
+ig\dot x_5\phi_5
+ig\dot x_6\phi
\non\\
&&
-2ig\psi^{\mu}\psi_5\partial_{\mu}\phi_5
-2ig\psi^{\mu}\psi_6\partial_{\mu}\phi
\non\\
\label{LY1}
\ear\no
where $g$ denotes the Yang-Mills coupling.
We assume
$\phi$ and $\phi_5$ to
depend only on the
four physical dimensions,
so that the index $\mu$ runs only from
$1$ to $4$.
The six-dimensional
path integral is then Gaussian in the
coordinate fields $x_5$, $x_6$. 
Integrating those out
we obtain a new Lagrangian,

\be
L
=
\kinb +\kinf
 +\half\psi_5\dot\psi_5 +\half\psi_6\dot\psi_6
+g^2\phi^2 
+g^2\phi_5^2
-2ig\psi^{\mu}\psi_5\partial_{\mu}\phi_5
-2ig\psi^{\mu}\psi_6\partial_{\mu}\phi
\non\\
\label{LY2}
\ee\no
So far the loop fermion was taken massless
(which implies, in particular, that we cannot yet 
distinguish between the scalar and the pseudoscalar fields).
To generate a mass term for the loop fermion we now
use the scalar field as a Higgs field,
i.e. we give it a non-vanishing vacuum expectation
value by shifting

\be
\phi\rightarrow \phi + {m\over g}
\label{shift}
\ee\no
Moreover, since we do not insist on gauge invariance,
we
can choose different couplings $\lambda, \lambda_5$ 
for $\phi, \phi_5$.
Our final result for the worldline Lagrangian
then becomes 

\bear
L_{\rm yuk}
&=&
m^2+
\kinb +\kinf
 +\half\psi_5\dot\psi_5 +\half\psi_6\dot\psi_6
+\lambda^2\phi^2
+2m\lambda\phi
+2i\lambda\psi_6\psi\cdot\partial\phi
\non\\
&&
+\lambda_5^2\phi_5^2
+2i\lambda_5\psi_5\psi\cdot\partial\phi_5
\non\\
\label{LYfinal}
\ear\no
This Lagrangian
contains two new worldline fields,
$\psi_5$ and $\psi_6$. The Wick
contractions for those are the same as for 
the other $\psi$ -- components,

\begin{equation}
\langle \psi_{5,6}(\tau_1) \, 
\psi_{5,6}(\tau_2) \rangle
= \frac{1}{2} \, G_F(\tau_1,\tau_2)
\label{correlationfn}\nonumber
\end{equation}
\no
Their free path integrals are normalized to unity.
Note also the presence of several nonlinear terms in 
this worldline Lagrangian. For momentum space amplitude
calculations those have to be treated in the
same way as in the case of $\phi^4$ theory above
(see section 4.1).

This Lagrangian looks certainly less compelling than its
gauge theory analogue. Nevertheless, 
precisely the same Lagrangian was also obtained by
D'Hoker and Gagn{\'e} in their more rigorous
derivation \cite{dhogag1,dhogag2,gagne,gagnethesis}.
For now, let us shortly explore its practical usefulness
for amplitude calculations.

\subsection{$N$ Scalar / $N$ Pseudoscalar Amplitudes}

Consider the one-loop one-particle-irreducible amplitude 
involving either $N$ massless 
scalars or $N$ massless pseudoscalars,
interacting with a fermion loop via Yukawa interactions.
Using the above worldline Lagrangian in the usual
procedure one finds that a master formula for this
amplitude can be obtained from the one for the
$N$ - photon amplitude, eq.(\ref{supermaster}), by
the following modifications:

\begin{enumerate}

\item
Write eq.(\ref{supermaster}) in $D=5$, but with the same
path integral determinant factor ${(4\pi T)}^{-2}$ as in
four dimensions.

\item
Take all polarization vectors $\tilde\varepsilon_i$ in the unphysical
dimension, $\tilde\varepsilon_i = (0,0,0,0,1)$, 
and all momenta in the physical dimensions, 
$\tilde k_i = (k_i,0)$.

\item
Delete the constant term in the second derivative of the
bosonic worldline Green's function,

\be
\ddot G_{Bij} = 2\delta_{ij} - {2\over T}
\quad\rightarrow\quad
2\delta_{ij}
\label{deleteconstant}
\ee\no

\end{enumerate}
\no
Since point 2 implies that all five - dimensional Lorentz products
$\tilde\varepsilon_i\cdot \tilde k_j$ vanish 
we can write

\bear
\Gamma_{\rm yuk}^{\phi_{(5)}}[k_1,\ldots,k_N]
&=&
-2
{(i\lambda_{(5)})}^N
{(2\pi )}^D\delta (\sum k_i)
{\dps\int_{0}^{\infty}}{dT\over T}
{(4\pi T)}^{-{D\over 2}}
\prod_{i=1}^N \int_0^T 
d\tau_i
\int
d\theta_i
\nonumber\\
&&
\times\exp\biggl\lbrace
\sum_{i,j=1}^N
\Bigl\lbrack
\half
(G_{Bij}+\theta_i\theta_j G_{Fij})
k_i\cdot k_j
-\half
(G_{Fij} + \theta_i\theta_j
2\delta_{ij}
)
\tilde\varepsilon_i\cdot\tilde\varepsilon_j\Bigr]
\biggr\rbrace
\mid_{\tilde\varepsilon_1\ldots\tilde\varepsilon_N}
\nonumber\\
\label{yukawamaster}
\ear
\no
Here the $\tilde\varepsilon_i\cdot\tilde\varepsilon_j$
are all equal to unity, but before making use of this
the exponential must be expanded, and the $\tilde\varepsilon_i$'s
anticommuted to the standard ordering
$\tilde\varepsilon_1\ldots\tilde\varepsilon_N$.
 
In the massive case we have to distinguish between the scalar and
pseudoscalar cases. The simpler one is the pseudoscalar case,
since according to eq.(\ref{LYfinal})
here the only difference between the massless and massive cases is
in the usual $m^2$ - term. Thus all that is needed to generalize
the master formula eq.(\ref{yukawamaster}) to the massive loop
case is to supply it with the usual proper-time exponential
$\e^{-m^2T}$:

\bear
\Gamma_{\rm yuk}^{\phi_5}[k_1,\ldots,k_N]
&=&
-2
{(i\lambda_5)}^N
{(2\pi )}^D\delta (\sum k_i)
{\dps\int_{0}^{\infty}}{dT\over T}
\,\e^{-m^2T}
{(4\pi T)}^{-{D\over 2}}
\prod_{i=1}^N \int_0^T 
d\tau_i
\int
d\theta_i
\nonumber\\
&&
\times\exp\biggl\lbrace
\sum_{i,j=1}^N
\Bigl\lbrack
\half
(G_{Bij}+\theta_i\theta_j G_{Fij})
k_i\cdot k_j
-\half
(G_{Fij} + \theta_i\theta_j
2\delta_{ij}
)
\tilde\varepsilon_i\cdot\tilde\varepsilon_j\Bigr]
\biggr\rbrace
\mid_{\tilde\varepsilon_1\ldots\tilde\varepsilon_N}
\nonumber\\
\label{yukawamasterpseudoscalar}
\ear
\no
In the scalar case we have the same mass term, but also the
term $2\lambda m\phi$. After the usual formal exponentiation 
it produces an additional term in the master exponent, so that
the massive master formula for the scalar case becomes
somewhat more complicated than the pseudoscalar one:

\bear
\Gamma^{\phi}_{\rm yuk} [k_1,\ldots,k_N]
&=&
-2
{(i\lambda)}^N
{(2\pi )}^D\delta (\sum k_i)
{\dps\int_{0}^{\infty}}{dT\over T}
\,\e^{-m^2T}
{(4\pi T)}^{-{D\over 2}}
\prod_{i=1}^N \int_0^T 
d\tau_i
\int
d\theta_i
\nonumber\\
&&\hspace{-60pt}\times
\exp\biggl\lbrace
\sum_{i,j=1}^N
\Bigl\lbrack
\half
(G_{Bij}+\theta_i\theta_j G_{Fij})
k_i\cdot k_j
-\half
(G_{Fij} + \theta_i\theta_j
2\delta_{ij}
)
\tilde\varepsilon_i\cdot\tilde\varepsilon_j\Bigr]
+2im\sum_{i=1}^N\tilde\varepsilon_i\theta_i
\biggr\rbrace
\mid_{\tilde\varepsilon_1\ldots\tilde\varepsilon_N}
\nonumber\\
\label{yukawamasterscalar}
\ear
\no

Let us look explicitly at the two-point cases. For 
the scalar case we get from 
eq.(\ref{yukawamasterscalar}) the parameter integral

\bear
\Gamma [k_1,k_2] &=&
-2\lambda^2
{(2\pi)}^D \delta(k_1+k_2)
\TintmD
\int_0^Td\tau_1
\int_0^Td\tau_2
\int d\theta_1
\int d\theta_2
\non\\
&&\times
\biggl[
\Bigl(1+\theta_1\theta_2 G_{F12}k_1\cdot k_2\Bigr)
\Bigl(
G_{F12}+\theta_1\theta_2
2\delta_{12}
\Bigr)
-4m^2\theta_1\theta_2
\biggr]
\e^{G_{B12}k_1\cdot k_2}
\label{scalar2point}
\ear\no
As usual we rescale to the unit circle, and set
$\tau_2 =0$.
Performing the $\theta$ - and $T$ - integrals,
and using energy-momentum conservation, 
we obtain

\be
\Gamma(k)=
2{(4\pi )}^{-{D\over 2}}
\lambda^2
\biggl\lbrace
2\Gamma(1-{D\over 2})
m^{D-2}
-(4m^2+k^2)
\Gamma(2-{D\over 2})
\int_0^1
du
{
\Bigl[
m^2
+u(1-u)k^2
\Bigr]
}^{{D\over 2}-2}
\biggr\rbrace
\label{yukawascalarresult}
\ee\no
The two - point function for the pseudoscalar case is
obtained simply by deleting the term proportional to
$4m^2$ (and replacing $\lambda$ by $\lambda_5$). 

In both cases the parameter integrals which we have 
at hand can be identified with the corresponding
field theory Feynman parameter integrals, albeit
only after suitable integrations by parts.
This generalizes, of course, to the $N$ - point case.
However, it should be noted that our parameter integrals
have a certain advantage insofar as they are already
of the scalar type. The master formula involves,
apart from the usual factor of $\e^{\half G_{Bij}k_i\cdot k_j}$
which in the $T$ - integration
turns into the Feynman denominator, 
only $\delta(\tau_i-\tau_j)$ and ${\rm sign}(\tau_i-\tau_j)$.
After restriction to an ordered sector the Feynman
numerators are therefore constants. This is not
the case in a straightforward Feynman parameter integral
calculation of these amplitudes, where one would
generally encounter non-trivial numerator polynomials.
For example, the above master formula allows one to 
write the, say, six - point amplitudes immediately
in terms of scalar triangle, box, pentagon, and hexagon
integrals, without the need to perform
a Passarino-Veltman type reduction.
Thus it seems that, for the scalar/pseudoscalar amplitudes, one should
use the master formula as it stands, without performing
partial integrations. 

At this point it must be observed that we have been cheating a bit
in the pseudoscalar case. 
From a Feynman diagram analysis it can
be easily seen that, in Euclidean space, the amplitude
with a massive fermion loop and any number of scalar and pseudoscalar
legs is real (imaginary) for an even (odd) number of
pseudo-scalars. Since our master formula 
(\ref{yukawamasterpseudoscalar}) obviously vanishes
for $N$ odd, 
we have seemingly lost the imaginary part
of the pseudoscalar amplitude.
This was to be expected, since in our heuristic derivation
we started from the gauge theory amplitude in six dimensions,
which is real in Euclidean space.
The missing imaginary part can also be represented on the
worldline, though in a somewhat less natural way 
~\cite{mnss1,dhogag1,mnss2,dhogag2}.
In \cite{haasch} the resulting path integral representation
was applied to the calculation of the
radiative decay of the axion into two photons in a constant
electromagnetic field, and moreover generalized to the finite
temperature case.
\vspace{5pt}
\subsection{The Spinor Loop in a Vector and Axialvector Background}
\vspace{4pt}

Another generalization of obvious interest is the inclusion of
axialvectors. 
Here we will not follow the approach taken in 
\cite{dhogag1,dhogag2}, based
on the introduction of auxiliary dimensions, but a more
direct construction, which was
proposed in \cite{mcksch} and further elaborated in 
\cite{dimcsc,dilkesthesis}. 
This will also allow us to avoid the separation into
the real and the imaginary part of the effective action which
was implied in the approach of \cite{dhogag1,dhogag2}.

Thus we would now like to find a path integral representation
for (in Euclidean space)

\be
\Gamma_{\rm spin}[A,A_5] = \ln {\rm Det}
[\slash p +\slash A
+\gamma_5{\slash A}_5
-im 
]
\label{defEAeucva}
\ee\no
The method is a straightforward generalization of the
one used in section 3.3 for the pure vector case, and we
will indicate only the necessary changes.

\no
First, eq. (\ref{1to2order}) can be generalized to

\be
(\slash p + \slash A +\gamma_5 \not\!\!A_5)^2 =
- (\partial_\mu 
+ i{\cal A}_\mu)^2 + V
\label{introAcal}
\ee\no
where

\bear
{\cal A}_{\mu} &\equiv& A_{\mu} -\gamma_5\sigma_{\mu\nu}A_5^{\nu}
\label{defcalA}\\
V &\equiv& -{i\over 2}
\sigma_{\mu\nu}
\Bigl(
\partial_{\mu}A_{\nu}-\partial_{\nu}A_{\mu}
\Bigr)
+i\gamma_5A_{5,\mu}^{\mu} + (D-2) A_5^2 
\label{defV}
\ear\no
Here we have used the four - dimensional Dirac algebra,
but dimensionally continued with an anticommuting $\gamma_5$.
Using
\vspace{4pt}

\bear
{\rm Det}
\Bigl[\slash p + \slash A +\gamma_5 \slash A_5 - im\Bigr] &=& 
{\rm Det}
\Bigl[\slash p + \slash A +\gamma_5 \slash A_5 + im\Bigr] = 
{\rm Det}^{1/2}
\Bigl[(\slash p + 
\slash A +\gamma_5 \slash A_5)^2 + m^2\Bigr]
\non\\
\label{gamma5trick}
\ear\no
one obtains

\be
\Gamma_{\rm spin}[A,A_5] = -\half\, \Tr\, \int_0^\infty \, 
\frac{dT}{T} \, \exp 
\Bigl\lbrace
- T \left[-
(\partial_\mu + i{\cal A}_\mu)^2 + V + m^2 \right]
\Bigr\rbrace\non\\
\label{Weuclva}
\ee

Up to a global factor, this is 
formally identical with the effective action for
a scalar loop in a background containing a 
(Clifford algebra valued)
gauge field ${\cal A}$ and a potential $V$.
Note that the exponent is not hermitian, which is the
price we have to pay for writing down the whole effective
action in one piece. However it is still positive for
weak background fields, which is sufficient
for our perturbative purposes.

Applying the coherent state formalism in the same way as in
section 3.3 we arrive at the 
following representation, corresponding to
(\ref{rewritefunctionaltrace}),

\bear
\Tr
\e^{-T\Sigma}
&=& i
\int d^4x \int d^2\eta\,
\langle
x,-\eta\mid
\e^{-T\Sigma}
\mid
x,\eta\rangle
\non\\
&=& i^N
\int
\prod_{i=1}^N
\Bigl(
d^4x^id^2\eta^i
\langle
x^i,\eta^i\mid
\e^{-{T\over N}\Sigma}
\mid
x^{i+1},\eta^{i+1}
\rangle
\Bigr)
\label{rewritefunctionaltraceea}
\ear
where now
$\Sigma = - (\partial_\mu 
+ i{\cal A}_\mu)^2 + V
$.
The only essential 
novelty is the presence of the $\gamma_5$ - matrix.
To take it into account, it is 
crucial to observe that, expressed in terms
of the $a_r^{\pm}$, it is identical to the fermion number
counter or ``G-parity operator''
${(-1)}^F$ \cite{alvarezgaume,gsw},

\be
\gamma_5= {(-1)}^F = (1-2F_1)(1-2F_2)
\label{gamma5=fnco}
\ee
where 

\be
F \equiv F_1 +F_2, \quad {\rm with} \quad F_i = a_i^+ a_i^-
\label{defFi}
\ee
From the identity

\be
\langle -\eta\mid{(-)}^F
=i\leftvac\prod_{r=1}^2
(-\eta_r-a_r^-)(1-2a_r^+a_r^-)
=i
\leftvac
\prod_{r=1}^2
(-\eta_r+a_r^-)
=\langle\eta\mid
\label{fncoaction}
\ee
it is clear that the presence of
${(-1)}^F$ can be taken into account
by switching the boundary conditions 
on the Grassmann path integral from antiperiodic
to periodic.
Thus (\ref{matrixelement}) generalizes to

\bear
\langle
x^i,\eta^i\mid
\e^{-{T\over N}\Sigma[p,A,A_5,\gamma_{\mu}\gamma_{\nu},\gamma_5]}
\mid x^{i+1},\eta^{i+1}\rangle
&=&
-
{i\over {(2\pi)}^4}
\int
d^4p^{i,i+1}
d^2\bar\eta^{i,i+1}
\e^{i(x^i-x^{i+1})p^{i,i+1}
+(\eta^i-\eta^{i+1})_r
\bar\eta_r^{i,i+1}}
\non\\
&&\!\!\!\!\!\!\!\!\!\!\!\hspace{-17pt}
\times
\biggl\lbrace
1-{T\over N}
\Sigma\Bigl[p^{i,i+1},A^{i,i+1},A_5^{i,i+1},
2\phantom{,}^i\psi_{\mu}
\psi_{\nu}^{i+1},{(-1)}^F\Bigr]
+O\Bigl({T^2\over N^2}\Bigr)
\biggr\rbrace
\non\\
\label{matrixelementva}
\ear\no
In the continuum limit this leads to

\bear
\Gamma_{\rm spin}[A,A_5]
&=& -\half\, \int_0^\infty \, 
\frac{dT}{T} 
\e^{-m^2T}
\int{\cal D}x
\int_A{\cal D}\psi
\, \e^
{
-\int_0^Td\tau\,
L_{\rm VA}
}\non\\
L_{\rm VA} &=&
\kinb
+\half\psi\cdot\dot\psi
+i\dot x^{\mu}A_{\mu}
-i\psi^{\mu}F_{\mu\nu}\psi^{\nu}
-2i\hat\gamma_5\dot x^{\mu}\psi_{\mu}\psi_{\nu}A_5^{\nu}
+i\hat\gamma_5\partial_{\mu}A^{\mu}_5
+(D-2)A_5^2
\non\\
\label{vapi}
\ear\no
The operator ${(-1)}^F$ has turned into an operator $\hat\gamma_5$
whose only raison d'{\^e}tre is to determine the boundary conditions
of the Grassmann path integral; after expansion of the interaction
exponential a given term will have to be
evaluated using antiperiodic (periodic) boundary conditions
on ${\cal D}\psi$, if it contains
$\hat\gamma_5$ at an even (odd) power. Once the boundary conditions
are determined $\hat\gamma_5$ can be replaced by unity.

The perturbative evaluation of this double path integral
can be done in the usual way. 
For the coordinate path
integral everything proceeds as before.
But for the Grassmann path integral one 
now has to proceed differently
depending on the boundary conditions.
In the antiperiodic case (``A'') there is 
again nothing new, we can compute it using the
by now familiar Green's function $G_F$.
In the periodic case (``P'') however 
one now encounters a fermionic zero mode.
As for the coordinate path integral we first must remove this
zero mode before executing the path integral.
Analogously to eq.(\ref{split}) we can do this
by factorizing the Hilbert space of periodic
Grassmann functions into the constant functions
$\psi_0$ and their orthogonal complement
$\xi(\tau)$,

\bear
\int_P{\cal D}\psi &=& \int d\psi_0\int {\cal D}\xi \non\\
\psi^{\mu}(\tau)&=&\psi_0^{\mu} + \xi^{\mu}(\tau)\non\\
\int_0^Td\tau\,\xi(\tau) &=& 0\non\\
\label{splitgrass}
\ear\no
The zero mode integration 
then produces the expected $\varepsilon$ - tensor via

\be
\int d^4\psi_0
\psi_0^{\mu}\psi^{\nu}_0\psi^{\kappa}_0\psi^{\lambda}_0 
=\varepsilon^{\mu\nu\kappa\lambda}
\label{zeromodeintegral}
\ee\no
and the $\xi$ - path integral can be performed using the correlator
\vspace{2pt}

\bear
\langle\xi^{\mu}(\tau_1)\, \xi^{\nu}(\tau_2)\rangle
&=& g^{\mu\nu}
\Bigl(
\half {\rm sign}(\tau_1 -\tau_2 )
-
{\tau_1 -\tau_2 \over T}  
\Bigr)
=g^{\mu\nu}\half\dot G_{B12}
\non\\
\label{xicorr}
\ear\no
The free $\xi$ - path integral
is, in four dimensions, normalized to unity.
Summarizing, in the vector-axialvector case we have the
Wick contraction rules

\begin{eqnarray}
\langle y^\mu(\tau_1)\,y^\nu(\tau_2)\rangle
&=& -g^{\mu\nu}{G}_{B}(\tau_1,\tau_2)\nonumber\\
\langle\psi^{\mu}(\tau_1)\, \psi^{\nu}(\tau_2)\rangle_A
&=& 
g^{\mu\nu}\half {G}_{F}(\tau_1,\tau_2)\nonumber\\
\langle\xi^{\mu}(\tau_1)\, \xi^{\nu}(\tau_2)\rangle_P
&=& 
g^{\mu\nu}\half {\dot{G}}_{B}(\tau_1,\tau_2)
\label{modcorrelators}
\end{eqnarray}\no
and the free path integral determinants

\bear
\int{\cal D}y
\,\e^{-\int_0^T d\tau\, \fourth
\dot y^2}
&=& {(4\pi T)}^{-{D\over 2}} \nonumber\\
\int_A{\cal D}\psi\, \e^{-\int_0^Td\tau \,\half\psi\cdot\dot\psi}
&=& 4 =: N_A\nonumber\\
\int_P{\cal D}\xi\, \e^{-\int_0^Td\tau \,\half\xi\cdot\dot\xi}
&=& 1 =: N_P
\label{freepi}
\ear\no

\subsection{Master Formula for the One-Loop Vector-Axialvector
Amplitudes}
\label{SecMaster}

To extract the scattering amplitude from the effective action, as usual
we must specialize the background fields to plane waves,
and then keep the part of the effective action which is linear
in all polarization vectors. 
As a preliminary step, it is convenient to linearize the
term quadratic in $A_5$ by introducing an auxiliary path
integration, writing

\bear
\label{linearizeA52}
\exp \Bigl[-(D-2)\int_0^Td\tau A_5^2\Bigr]
&=&
\int {\cal D}z \exp \Bigl[-\int_0^Td\tau 
\Bigl({z^2\over 4}+i\sqrt{D-2}\hat\gamma_5\,z\cdot A_5\Bigr)
\Bigr]
\ear\no
The Wick contraction rule for this auxiliary field is
simply

\bear
\langle z^{\mu}(\tau_1)z^{\nu}(\tau_2)\rangle
&=&
2g^{\mu\nu}\delta (\tau_1-\tau_2)
\label{wickz}
\ear\no
and its free path integral is normalized to unity.
This allows us to define an axial-vector vertex operator
as follows,

\bear
V^{A_5}[k,\varepsilon] &\equiv&
\hat\gamma_5
\int_0^Td\tau
\Bigl(i\varepsilon\cdot k 
+ 2\varepsilon\cdot\psi
\dot x\cdot\psi
+ \sqrt{D-2}\,\varepsilon\cdot z
\Bigr)
\,\e^{ik\cdot x}
\label{defaxvectvertop}
\ear\no
Before using this vertex operator for Wick-contractions, as usual
it is convenient to formally rewrite it as a
linearized exponential,

\bear
V^{A_5}[k,\varepsilon] 
&=&
\hat\gamma_5
\int_0^Td\tau
\int d\theta
\exp\Bigl\lbrack ik\cdot x + i\theta \varepsilon\cdot k
+\sqrt{2}\varepsilon\cdot\psi
+\sqrt{2}\theta\dot x\cdot\psi
\non\\&&\hspace{70pt}
+\sqrt{D-2}\,\theta\varepsilon\cdot z
\Bigr\rbrack
\Large\mid_{{\rm lin}(\varepsilon)}
\label{rewriteaxvectvertop}
\ear\no
Here $\theta$ is a Grassmann variable with $\int d\theta\theta =1$,
and $\varepsilon$ must now also be formally treated as Grassmann.
The vectors are represented by the usual photon vertex
operator (\ref{photonvertop})

\bear
V^A[k,\varepsilon]&=&
\int_0^Td\tau
\Bigl[
\varepsilon\cdot \dot x
+2i
\varepsilon\cdot\psi
k\cdot\psi
\Bigr]
\,\e^{ik\cdot x}
\non\\
&=&
\int_0^Td\tau\int d\theta
\exp\Bigl\lbrack
ik\cdot(x+\sqrt{2}\theta\psi)
+\varepsilon\cdot (-\theta\dot x +\sqrt{2}\psi)
\Bigr\rbrack
\Large\mid_{{\rm lin}(\varepsilon)}
\label{rewritephotonvertop}
\ear
\noindent\no
Those definitions allow us to represent the
one-loop amplitude with $M$ vectors and $N$ axialvectors
in the following way,

\bear
\Gamma[\lbrace k_i,\varepsilon_i\rbrace,
\lbrace k_{5j},\varepsilon_{5j}\rbrace]
&=&
-\half N_{A,P}(-i)^{M+N}
\Tintm 
\PITD
\non\\
&&\Mneg
\Mneg
\times
\Bigl\langle
V^{A}[k_1,\varepsilon_1]\ldots
V^{A}[k_M,\varepsilon_M]
V^{A_5}[k_{51},\varepsilon_{51}]\ldots
V^{A_5}[k_{5N},\varepsilon_{5N}]
\Bigr\rangle
\label{repMvectorNaxial}
\ear\no
where the global sign refers to the ordering
$\varepsilon_1\varepsilon_2\ldots\varepsilon_M
\varepsilon_{51}\varepsilon_{52}\ldots\varepsilon_{5N}$
of the polarization vectors.
It is then straightforward to perform the bosonic path integrations
\footnote{In \cite{dimcsc} the factor of 2 in front of the term involving
$(D-2)\delta(\tau_i-\tau_j)$ had been missing in eqs. (2.14),(2.19), and
(2.22).},

\bear
&&
\int {\cal D}x\int{\cal D}z\,
V^{A}[k_1,\varepsilon_1]\ldots
V^{A_5}[k_{5N},\varepsilon_{5N}]
\,\e^{-\int^T_0 d\tau \Bigl({{\dot x}^2\over 4} + {z^2\over 4}\Bigr)}
=
\non\\&&
\PITD 
\int_0^T d\tau_1 \cdots \int d\theta_M
\int_0^T d\tau_{51} \cdots \int d\theta_{5N}
\non\\&&\times
\exp\biggl\lbrace
\half
G_{BIJ}K_I\cdot K_J
-i\theta_i\dot G_{BiJ}\varepsilon_i\cdot K_J
-i\sqrt{2}\theta_{5i}\dot G_{BiJ}\psi_i\cdot K_J
\non\\&&
-\half \ddot G_{Bij}\theta_i\theta_j\varepsilon_i\cdot\varepsilon_j
-\sqrt{2}\ddot G_{Bij}\theta_i\theta_{5j}\varepsilon_i\cdot\psi_j
-\ddot G_{Bij}\theta_{5i}\theta_{5j}\psi_i\cdot\psi_j
\non\\&&
+\sqrt{2}(\varepsilon_i\cdot\psi_i +\varepsilon_{5j}\cdot\psi_j)
+\sqrt{2}i\theta_i k_i\cdot\psi_i
+ i\theta_{5i}\varepsilon_{5i}\cdot k_{5i} 
\non\\&&
-2(D-2)\delta(\tau_i-\tau_j)\theta_{5i}\theta_{5j}
\varepsilon_{5i}\cdot\varepsilon_{5j}
\biggr\rbrace
\Large\mid_{{\rm lin}(\lbrace\varepsilon_i\rbrace;
\lbrace\varepsilon_{5j}\rbrace)}
\label{resdxdz}
\ear\no
Here and in the following all lower case repeated indices run 
over either the vector or the axial vector indices,
depending on whether $\theta_i,\varepsilon_i$ or
$\theta_{5i},\varepsilon_{5i}$ are involved,
while capital repeated indices run over all of them
($\lbrace K_I\rbrace $ denotes the set of all external momenta).
We omit the momentum conservation factor (\ref{momentumcons}).
The remaining
$\psi$ - path integral is still Gaussian. For its performance
we must
now distinguish between even and odd numbers of
axialvectors. For the antiperiodic case, $N$ even,
there is no zero-mode, and the integration 
can still be done in closed form.
The only complication is the
existence of the term 
$-\ddot G_{Bij}\theta_{5i}\theta_{5j}\psi_i\cdot\psi_j$.
It modifies the worldline propagator $G_F$ to

\bear
G_{F12}^{(N)} 
&\equiv&
2\langle \tau_1\mid
\Bigl(\partial + 2 B^{(N)}\Bigl)^{-1}
\mid\tau_2\rangle
\label{defGN}
\ear\no
where $B^{(N)}$ denotes the operator with integral kernel

\bear
B^{(N)}(\tau_1,\tau_2) &=&
\delta(\tau_1 - \tau_i)\theta_{5i}
\ddot G_{Bij}
\theta_{5j}\delta(\tau_j-\tau_2)
\label{defBN}
\ear\no
($B^{(N)}$ acts trivially on the Lorentz indices, which we
suppress in the following).
Expanding the right hand side of eq.(\ref{defGN}) 
in a geometric series and resumming one obtains a
matrix representation for $G_{F12}^{(N)}$, 

\bear
G_{F}^{(N)} 
&=&
{G_{F}
\over
\Eins +\ddot\Theta G_F}
=
G_F -G_F\ddot\Theta G_F + \ldots
\label{GNexplicit}
\ear\no
Here $\ddot\Theta_{ij}$ is the antisymmetric
$N\times N$ matrix with entries
$\theta_{5i}\ddot G_{Bij}\theta_{5j}$ (no summation).
Moreover, the fermionic path integral determinant
changes by a factor 

\bear
\Det^{\half}\Bigl(\Eins + 2B^{(N)}\partial^{-1}\Bigr)
&=&
{\det}
\bigl(\Eins +\ddot\Theta G_F\bigr)^{{D\over 2}} 
\label{detfact}
\ear\no
as is easily seen using the $ln det = tr ln$ - formula
(note that on the left hand side we have a functional determinant,
on the right hand side the determinant of a $N\times N$ matrix).
Using these results the fermionic path integral can be
eliminated, yielding the following master formula for
this amplitude \cite{dimcsc},

\bear
\Gamma_{\rm even}[\lbrace k_i,\varepsilon_i\rbrace;
\lbrace k_{5j},\varepsilon_{5j}\rbrace]
&=&
-{N_A\over 2}(-i)^{M+N}
\Tintm \PITD
\non\\
&&\Mneg\Mneg\Mneg\times
\int_0^Td\tau_1\int d\theta_1
\cdots
\int_0^Td\tau_{5N}\int d\theta_{5N}
\,{\det}
\bigl(\Eins +\ddot\Theta G_F\bigr)^{{D\over 2}} 
\non\\
&&\Mneg\Mneg\Mneg\times
\exp\biggl\lbrace
\half
G_{BIJ}K_I\cdot K_J
-i\theta_i\dot G_{BiJ}\varepsilon_i\cdot K_J
-\half \ddot G_{Bij}\theta_i\theta_j\varepsilon_i\cdot\varepsilon_j
\non\\
&&\Mneg\Mneg
- {G_{Fij}^{(N)}\over 2}
\Bigl(\varepsilon_i+i\theta_i k_i
+\varepsilon_{5i}
-i\theta_{5i}\dot G_{BiR}K_R
+\ddot G_{Bir}\theta_{5i}\theta_r\varepsilon_r
\Bigr)
\non\\ && \Mneg\Mneg
\cdot
\Bigl(\varepsilon_j+i\theta_j k_j
+\varepsilon_{5j}
-i\theta_{5j}\dot G_{BjS}K_S
+\ddot G_{Bjs}\theta_{5j}\theta_s\varepsilon_s
\Bigr)
+i\theta_{5i}\varepsilon_{5i}\cdot k_{5i}
\non\\ && \Mneg\Mneg
-2(D-2)\delta(\tau_i-\tau_j)\theta_{5i}\theta_{5j}
\varepsilon_{5i}\cdot\varepsilon_{5j}
\biggr\rbrace
\mid_{\rm lin}(\lbrace\varepsilon_i\rbrace;
\lbrace\varepsilon_{5j}\rbrace)
\label{axialevenmaster}
\ear
Let us verify the correctness of this formula for the 
case of the massive 2-point axialvector function
in four dimensions. 
Expanding out the exponential as well as the 
determinant factor, and performing the 
two 
$\theta$ - integrals,
we obtain the following parameter integral,

\bear
\Gamma[k_1,\varepsilon_1;k_2,\varepsilon_2]
&=&
2\Tintm \PITD 
\int_0^Td\tau_1\int_0^T d\tau_2
\non\\&&\Mneg\times
\e^{G_{B12}k_1\cdot k_2}
\biggl\lbrace
2(D-2)\delta(\tau_1-\tau_2)\varepsilon_1\cdot\varepsilon_2
-(D-1)\ddot G_{B12}G_{F12}^2\varepsilon_1\cdot\varepsilon_2
\non\\&&\Mneg
-G_{F12}^2\dot G_{B12}^2
(\varepsilon_1\cdot\varepsilon_2 k_1\cdot k_2 -
\varepsilon_1\cdot k_1 \varepsilon_2\cdot k_2)
-\varepsilon_1\cdot k_1 \varepsilon_2\cdot k_2
\biggr\rbrace
\label{A5A5int}
\ear\no
As usual we rescale 
$\tau_{1,2}=T u_{1,2}$ , and use the
translation invariance in $\tau$ to
set $u_2=0$. 
Setting also $k=k_1=-k_2$ this leads to

\bear
\Gamma^{\mu\nu}(k)
&=&
2\Tintm \PITD 
\biggl\lbrace
2(D-2)Tg^{\mu\nu}
-2(D-1)Tg^{\mu\nu}
\non\\&&\hspace{-50pt}
+ \int_0^1 du
\,\e^{-Tu(1-u)k^2}
\Bigl[
2(D-1)Tg^{\mu\nu}
+(1-2u)^2T^2(g^{\mu\nu}k^2-k^{\mu}k^{\nu})
+T^2k^{\mu}k^{\nu}
\Bigr]
\biggr\rbrace
\non\\
\label{A5A5int2}
\ear\no
In the massless case
the first two terms in braces
do not contribute in dimensional regularization,
since they are of tadpole type. For the remaining
terms both integrations are elementary,
and the result is, using
$\Gamma$ - function identities,
easily identified with the
standard result for the
massless QED vacuum polarization.

A suitable integration by part verifies
the agreement with field theory also for
the massive case.
Here the tadpole terms do contribute,
and the comparison
shows that to get the precise $D$ - dependence of
the amplitude, appropriate to dimensional
regularization using an anticommuting $\gamma_5$,
it was essential to keep the explicit $D$ - dependence
of the $A_5^2$ - term in the worldline Lagrangian
$L_{\rm VA}$.

For an odd number of axial vectors, we need to go back to
eq.(\ref{resdxdz}) and replace $\psi$ by $\psi_0 +\xi$.
The ${\cal D}\xi$ - path integral is then executed in the
same way as before, but with the propagator $G_F$ changed
to $\dot G_B$. The final result becomes \cite{dimcsc}

\bear
\Gamma_{\rm odd}[\lbrace k_i,\varepsilon_i\rbrace;
\lbrace k_{5j},\varepsilon_{5j}\rbrace]
&=&
-{N_P\over 2}(-i)^{M+N}
\Tintm \PITD
\non\\
&&\Mneg\Mneg\times
\int_0^Td\tau_1\int d\theta_1
\cdots
\int_0^Td\tau_{5N}\int d\theta_{5N}
\,{\det}
\bigl(\Eins +\ddot\Theta \dot G_B\bigr)^{{D\over 2}} 
\int d^4\psi_0
\non\\
&&\Mneg\Mneg\times
\exp\biggl\lbrace
\half
G_{BIJ}K_I\cdot K_J
-i\theta_i\dot G_{BiJ}\varepsilon_i\cdot K_J
-\half \ddot G_{Bij}\theta_i\theta_j\varepsilon_i\cdot\varepsilon_j
\non\\
&&\Mneg\Mneg
- {\dot G_{Bij}^{(N)}\over 2}
\Bigl(\varepsilon_i+i\theta_i k_i
+\varepsilon_{5i}
-i\theta_{5i}\dot G_{BiR}K_R
+\ddot G_{Bir}\theta_{5i}\theta_r\varepsilon_r
+\sqrt{2}\ddot G_{Bir}\theta_{5i}\theta_{5r}\psi_0
\Bigr)
\non\\ && \Mneg\Mneg
\cdot
\Bigl(\varepsilon_j+i\theta_j k_j
+\varepsilon_{5j}
-i\theta_{5j}\dot G_{BjS}K_S
+\ddot G_{Bjs}\theta_{5j}\theta_s\varepsilon_s
+\sqrt{2}\ddot G_{Bjs}\theta_{5j}\theta_{5s}\psi_0
\Bigr)
\non\\ && \Mneg\Mneg
+i\theta_{5i}\varepsilon_{5i}\cdot k_{5i}
-2(D-2)\delta(\tau_i-\tau_j)\theta_{5i}\theta_{5j}
\varepsilon_{5i}\cdot\varepsilon_{5j}
-i\sqrt{2}\theta_{5i}\dot G_{BiJ}\psi_0\cdot K_J
\non\\ && \Mneg\Mneg
-\sqrt{2}\ddot G_{Bij}\theta_i\theta_{5j}\varepsilon_i\cdot\psi_0
+\sqrt{2}\bigl(\sum\varepsilon_i +\sum\varepsilon_{5j}\bigr)\cdot\psi_0
+\sqrt{2}i\theta_i k_i\cdot\psi_0
\biggr\rbrace
\mid_{\rm lin}(\lbrace\varepsilon_i\rbrace;
\lbrace\varepsilon_{5j}\rbrace)
\non\\
\label{axialoddmaster}
\ear\no
Here $\dot G_B^{(N)}$ is defined analogously to eq.(\ref{GNexplicit}),

\bear
\dot G_{B}^{(N)} 
&=&
{\dot G_{B}
\over
\Eins +\ddot\Theta \dot G_B}
\label{GdotNexplicit}
\ear\no
The integrand still depends on the zero-mode $\psi_0$, which
is to be integrated according to eq.(\ref{zeromodeintegral}).

\subsection{The $VVA$ Anomaly}

\begin{figure}[ht]
\begin{center}
\vbox to 4.5cm{\vfill\hbox to 15.8cm{\hfill
\epsffile{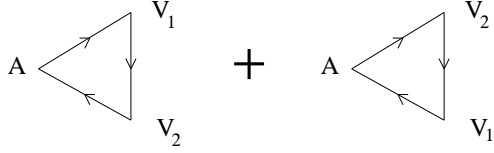}
\hfill}\vfill}\vskip-1.0cm
\end{center}
\caption[dum]{Sum of triangle diagrams in field theory.
\hphantom{xxxxxxxxxxxxxxx}}
\label{triangle}
\end{figure}

Any new formalism for calculations involving axialvectors must,
of course, be confronted with the existence of the chiral
anomaly \cite{adler,beljac}. Let us thus verify that our formulas
above correctly reproduce the anomaly for the $VVA$ case.
Calculation of the $VV\partial\cdot A$ amplitude,
either using the master formula eq.(\ref{axialoddmaster})
or a direct Wick-contraction of eq.(\ref{repMvectorNaxial}), 
yields the following parameter integral,

\bear
k_3^{\rho}
\langle A^{\mu}[k_1]A^{\nu}[k_2]A_5^{\rho}[k_3]\rangle
&=&
2
\varepsilon^{\mu\nu\kappa\lambda}k_1^{\kappa}k_2^{\lambda}
\Tint
{(4\pi T)}^{-2}
\prod_{i=1}^3\int_0^Td\tau_i
\non\\&&\hspace{-20pt}
\times \exp\Bigl\lbrack
\Bigl(G_{B12} -G_{B13}-G_{B23}\Bigr)k_1\cdot k_2 
-G_{B13}k_1^2 -G_{B23}k_2^2
\Bigr\rbrack
\non\\
&&\hspace{-50pt}\times
\biggl\lbrace
(k_1+k_2)^2 +
(\dot G_{B12} +\dot G_{B23} +\dot G_{B31})
(\dot G_{B13}-\dot G_{B23})
k_1\cdot k_2
-(\ddot G_{B13} +\ddot G_{B23})
\biggr\rbrace
\non\\
\label{divAAA5}
\ear
Here momentum conservation has been used to eliminate
$k_3$. 
It must be emphasized that this parameter integral
represents the complete three-point amplitude, and thus corresponds
to the sum of the two different triangle diagrams in field
theory, shown in fig. \ref{triangle}.

\par
\par

Removing the second derivatives
$\ddot G_{B13}$ ($\ddot G_{B23}$) by a
partial integration in $\tau_1$ ($\tau_2$),
the expression in brackets turns into

\bear
&&k_1\cdot k_2
\,
\biggl\lbrace
2 -(\dot G_{B12}+\dot G_{B23}+\dot G_{B31})^2
+ \dot G_{B12}^2 - \dot G_{B13}^2 - \dot G_{B23}^2
\biggr\rbrace
+k_1^2 (1-\dot G_{B13}^2) + k_2^2 (1-\dot G_{B23}^2)
\non\\&&
=
-{4\over T} \Bigl\lbrack
\Bigl(G_{B12} -G_{B13}-G_{B23}\Bigr)k_1\cdot k_2 
-G_{B13}k_1^2 -G_{B23}k_2^2
\Bigr\rbrack
\non\\
\label{rewritebraces}
\ear
In the last step we used the identities 
(\ref{Gdotsquare}),(\ref{Gdotsum}).
This is precisely the same expression which appears also in the
exponential factor in (\ref{divAAA5}). After 
performing the trivial $T$ - integral
we find therefore a complete cancellation between the Feynman
numerator and denominator polynomials, and obtain
without further integration the desired result,

\bear
k_3^{\rho}
\langle A^{\mu}A^{\nu}A_5^{\rho}\rangle
&=&
{8\over {(4\pi)}^2}
\,\varepsilon^{\mu\nu\kappa\lambda}k_1^{\kappa}k_2^{\lambda}
\non\\
\label{PCAC}
\ear
This is the usual expression for the divergence of the
axialvector current \cite{adler,beljac}. Note that in the present
formalism this divergence is unambiguously fixed to be at the
axialvector current. This can, in fact, be already seen at the
path integral
level, since 
the vectors are represented by the photon vertex operator
(\ref{rewritephotonvertop}) which, as is familiar from string theory,
turns into a total derivative when
contracted with its own momentum. This will lead to the
vanishing of the whole amplitude, independently of the
possible divergence of the global $T$ - integration.
Nothing analogous holds true for the axialvector vertex operator.
 
The behaviour of the present formalism with respect to the
chiral symmetry in the dimensional cointinuation is thus somewhat
unusual.
It should be remembered
that, in field theory, one has essentially a choice between
two evils. If one preserves the anticommutation
relation between $\gamma_5$ and the other Dirac matrices
\cite{chfuhi}
then the chiral symmetry is preserved for parity-even
fermion loops, but Dirac traces with an odd number
of $\gamma_5$'s are not unambiguously defined in general,
requiring additional prescriptions. The main alternative
is to use the
't Hooft-Veltman-Breitenlohner-Maison prescription
\cite{HV,Bremai}. In this case there are no ambiguities, but
the chiral symmetry is explicitly broken, so that 
in chiral gauge theories finite renormalizations
generally become
necessary to avoid violations of the gauge Ward identities
\cite{bonneau}.

Since our path integral representation was derived using an
anticommuting $\gamma_5$, we have not broken the chiral
symmetry. In particular, in the massless case the amplitude
with an even number of axialvectors 
should coincide with the
corresponding vector amplitude, and we have explicitly
verified this fact for the two-point case. 
(Even though the structure of the resulting Feynman numerators is quite
different from the equivalent ones derived from the ordinary
Bern-Kosower master formula.)
Nevertheless, we did not encounter any ambiguities 
even in the parity-odd case, not even in the anomaly
calculation. 
This property of the formalism may be useful for
applications to chiral gauge theories.

\subsection{Inclusion of Constant Background Fields}

As in the pure vector case, it is easy to take into account
an additional (vector) background field with constant
field strength tensor $F_{\mu\nu}$ \cite{meva}.
The presence of the background field modifies 
the Wick contraction rules (\ref{modcorrelators})
to 

\begin{eqnarray}
\langle y^{\mu}(\tau_1)y^{\nu}(\tau_2)\rangle
&=&
-{\cal G}_B^{\mu\nu}(\tau_1,\tau_2)\non\\
\langle\psi^{\mu}(\tau_1)\psi^{\nu}(\tau_2)\rangle
&=&
\frac{1}{2}{\cal G}_F^{\mu\nu}(\tau_1,\tau_2)\non\\
\langle\xi^{\mu}(\tau_1)\xi^{\nu}(\tau_2)\rangle
&=&
\frac{1}{2}{\dot{\cal G}}_B^{\mu\nu}(\tau_1,\tau_2)\non\\
\label{modcorrelatorsF}
\end{eqnarray}
\noindent
(compare chapter 5).
The Gaussian path integral determinants 
(\ref{freepi}) become

\bear
\int{\cal D}y
\,\e^{-\int_0^T d\tau\, \Bigl( \fourth
\dot y^2 +
{1\over 2} ie\,y^{\mu} F_{\mu\nu}
\dot y^{\nu} \Bigr)
}
&=& {(4\pi T)}^{-{D\over 2}} 
{\rm det}^{-{1\over 2}}
\biggl[{\sin({\cal Z})\over {{\cal Z}}}
\biggr]
\non\\
\int_A{\cal D}\psi\, \e^{-\int_0^Td\tau 
\Bigl(
\half\psi\cdot\dot\psi
- ie\,\psi^{\mu} F_{\mu\nu}\psi^{\nu}
\Bigr)
}
&=& 4\,
{\rm det}^{\half}\Bigl[\cos{\cal Z}\Bigr]
\non\\
\int_P{\cal D}\xi\, \e^{-\int_0^Td\tau 
\Bigl( \half\xi\cdot\dot\xi
- ie\,\xi^{\mu} F_{\mu\nu}\xi^{\nu}
\Bigr)
}
&=& 
{\rm det}^{1\over 2}
\biggl[{\sin({\cal Z})\over {{\cal Z}}}
\biggr]
\non\\
\label{freepiF}
\ear\no
Note that for the case of periodic
Grassmann boundary conditions the field dependence
of the determinant factors
cancels out between the coordinate and Grassmann
path integrals. This cancellation is
a consequence of the worldline supersymmetry (\ref{wlsusy})
\cite{berkos:npb362,berkos:npb379,strassler1,dimcsc}. 
It does not occur in the
antiperiodic case since here the supersymmetry is broken
by the boundary conditions
(see appendix C of \cite{dimcsc} for more on this point).

In the periodic case the external field must, moreover, be
also taken into account in the zero-mode integration,
as shown in the following example.

\subsection{Example: Vector - Axialvector Amplitude
in a Constant Field}

As an explicit example, we calculate the
vector -- axialvector two-point function
in a constant field. This amplitude is relevant, for example,
for photon -- neutrino processes at low photon energies
(see, e.g., \cite{raffelt,ioaraf,mecorfu}).
According to the above we can represent this amplitude as follows,

\bear
\langle
A_{\mu}(k_1)
A_{5\nu}(k_2)
\rangle
&=&
{1\over 2}
\Tintm
\Dx
\int_P{\cal D}\psi
\non\\
&&\!\!\!\!\!\!\!\!\!\!\!\!\!\!\!\!
\times\exp\biggl\lbrace
-\int_0^Td\tau
\Bigl[
\kinb
+{1\over 2}
\psi\cdot\dot\psi
+{i\over 2}e\, x\cdot F\cdot\dot x
-ie\, \psi\cdot F\cdot\psi
\Bigr]
\biggr\rbrace\non\\
&&\!\!\!\!\!\!\!\!\!\!\!\!\!\!\!\!
\times
\int_0^Td\tau_1
\Bigl(
\dot x_{\mu}(\tau_1)+2i\psi_{\mu}(\tau_1)
k_1\cdot\psi(\tau_1)
\Bigr)
\e^{ik_1\cdot x_1}
\non\\&&\!\!\!\!\!\!\!\!\!\!\!\!\!\!\!\!\times
\int_0^Td\tau_2
\Bigl(
ik_{2\nu}+2\psi_{\nu}(\tau_2)\dot x(\tau_2)\cdot\psi(\tau_2)
\Bigr)
\e^{ik_2\cdot x_2}
\non\\
\label{AA5withF}
\ear

This amplitude is finite, so that we can set $D=4$ in
its evaluation.
As a first step, the zero-modes of both path integrals
are separated out according to 
eqs.(\ref{split}),(\ref{splitgrass}),
and the Grassmann zero mode integrated out using 
eq.(\ref{zeromodeintegral}).
All terms which do not contain all four zero mode components
precisely once give zero. To explicitly perform this
integration we
note that by eq.(\ref{splitgrass}) we can rewrite, in the exponent
of eq.(\ref{AA5withF}),

\be
\int_0^Td\tau\, \psi(\tau)\cdot F\cdot\psi(\tau)
=
T
\psi_0\cdot F\cdot\psi_0
+
\int_0^Td\tau\,
\xi(\tau)\cdot F\cdot\xi(\tau)
\label{splitpsiFpsi}
\ee\no
Thus for 
the case at hand the Grassmann zero mode integral can appear
in the following three forms,

\bear
\int d^4\psi_0 
\,\e^{ieT\psi_0\cdot F\cdot\psi_0}
&=&
-{(eT)^2\over 2}
\varepsilon_{\mu\nu\kappa\lambda}
F_{\mu\nu}F_{\kappa\lambda}
=
-(eT)^2 F\cdot\tilde F\non\\
\int d^4\psi_0 
\,\e^{ieT\psi_0\cdot F\cdot\psi_0}
\psi_{0\mu}\psi_{0\nu}
&=&
ieT\varepsilon_{\mu\nu\kappa\lambda}F_{\kappa\lambda}
= 2ieT\tilde F_{\mu\nu}
\non\\
\int d^4\psi_0
\,\e^{ieT\psi_0\cdot F\cdot\psi_0}
\psi_{0\mu}\psi_{0\nu}\psi_{0\kappa}\psi_{0\lambda} 
&=&
\varepsilon_{\mu\nu\kappa\lambda}
\non\\
\label{zeromodeintegrals}
\ear\no
In the next step, both path integrations are performed
using the field-dependent Wick contraction rules
eqs. (\ref{modcorrelatorsF}). 
This results in the following parameter integral 
representation for the vector -- axialvector
vacuum polarization tensor
\footnote{
Since $G_F,{\cal G}_F$ do not occur for the periodic
case we delete the subscript ``B'' in the 
remainder of this section.}

\bear
\Pi_5^{\mu\nu}(k)
&=&
{e e_5\over 8\pi^2}
\int_0^{\infty}{dT\over T^3}\,\e^{-m^2T}
\int_0^Td\tau_1 d\tau_2
\,\,
J_5^{\mu\nu}(\tau_1,\tau_2)
\,\e^{-k\cdot \bar{\cal G}_{12}\cdot k}\non\\
J_5^{\mu\nu}(\tau_1,\tau_2)&=&
\Bigl[ 
\ddot{\cal G}_{12}^{\mu\alpha}
-
(
\dot{\cal G}_{21}^{\alpha\beta} 
-
\dot{\cal G}_{22}^{\alpha\beta} 
)
(
\dot{\cal G}_{11}^{\mu\rho} 
-
\dot{\cal G}_{12}^{\mu\rho} 
)
k_{\beta}k_{\rho}
\Bigr]
\Bigl(
i\tilde{\cal Z}_{\nu\alpha}
-{{\cal Z}\cdot\tilde{\cal Z}\over 4}
\dot{\cal G}_{22}^{\nu\alpha}
\Bigr)
\non\\
&&\hspace{-65pt}
+
{{\cal Z}\cdot\tilde{\cal Z}\over 4}
(
\dot{\cal G}_{11}^{\mu\rho} 
-
\dot{\cal G}_{12}^{\mu\rho} 
)
k_{\rho}k_{\nu}
+k_{\nu}k_{\rho}
\Bigl(
i\tilde{\cal Z}_{\mu\rho}
- {{\cal Z}\cdot\tilde{\cal Z}\over 4}
\dot{\cal G}_{11}^{\mu\rho}
\Bigr)
+k_{\rho}k_{\sigma}
(
\dot{\cal G}_{21}^{\alpha\rho}
-
\dot{\cal G}_{22}^{\alpha\rho}
)
\non\\
&&\hspace{-65pt}\times
\Bigl[
\varepsilon_{\mu\sigma\nu\alpha}
+i\bigl(
\dot{\cal G}_{22}^{\nu\alpha}
\tilde{\cal Z}_{\mu\sigma}
-
\dot{\cal G}_{12}^{\sigma\alpha}
\tilde{\cal Z}_{\mu\nu}
+
\dot{\cal G}_{12}^{\sigma\nu}
\tilde{\cal Z}_{\mu\alpha}
+
\dot{\cal G}_{12}^{\mu\alpha}
\tilde{\cal Z}_{\sigma\nu}
-
\dot{\cal G}_{12}^{\mu\nu}
\tilde{\cal Z}_{\sigma\alpha}
+
\dot{\cal G}_{11}^{\mu\sigma}
\tilde{\cal Z}_{\nu\alpha}
\bigr)
\non\\
&&\hspace{-65pt}
- {{\cal Z}\cdot\tilde{\cal Z}\over 4}
\bigl(
\dot{\cal G}_{11}^{\mu\sigma}\dot{\cal G}_{22}^{\nu\alpha}
-\dot{\cal G}_{12}^{\mu\nu}\dot{\cal G}_{12}^{\sigma\alpha}
+\dot{\cal G}_{12}^{\mu\alpha}\dot{\cal G}_{12}^{\sigma\nu}
\bigr)
\Bigr]
\non\\
\label{Pi5}
\ear\no
where
$k=k_1=-k_2, \tilde {\cal Z} \equiv eT\tilde F$.
As usual it is useful
to perform a partial integration on the one term 
involving $\ddot{\cal G}_{12}$, leading to the replacement

\be
\ddot{\cal G}_{12}^{\mu\alpha}
\rightarrow
\dot{\cal G}_{12}^{\mu\alpha}
k\cdot\dot{\cal G}_{12}\cdot k
\label{partintddotG}
\ee\no
By this partial integration, and the removal of some
terms which cancel against each other,
$J_5^{\mu\nu}(\tau_1,\tau_2)$ gets replaced 
by 

\bear
&&\hspace{-20pt}
k^{\rho}k^{\sigma}
\Bigl[
\dot {\cal G}^{\mu\alpha}_{12}
\dot{\cal G}_{12}^{\rho\sigma}
+
(
\dot{\cal G}_{21}^{\alpha\sigma} 
-
\dot{\cal G}_{22}^{\alpha\sigma} 
)
\dot{\cal G}_{12}^{\mu\rho} 
\Bigr]
\Bigl(
i\tilde{\cal Z}^{\nu\alpha}
-{{\cal Z}\cdot\tilde{\cal Z}\over 4}
\dot{\cal G}_{22}^{\nu\alpha}
\Bigr)
-
{{\cal Z}\cdot\tilde{\cal Z}\over 4}\dot{\cal G}_{12}^{\mu\rho} 
k^{\rho}k^{\nu}
\non\\
&&\hspace{-20pt}
+ik^{\nu}k^{\rho}
\tilde{\cal Z}^{\mu\rho}
+k^{\rho}k^{\sigma}
(
\dot{\cal G}_{21}^{\alpha\rho}
-
\dot{\cal G}_{22}^{\alpha\rho}
)
\Bigl[
\varepsilon^{\mu\sigma\nu\alpha} 
+ {{\cal Z}\cdot\tilde{\cal Z}\over 4}
\bigl(
\dot{\cal G}_{12}^{\mu\nu}\dot{\cal G}_{12}^{\sigma\alpha}
-\dot{\cal G}_{12}^{\mu\alpha}\dot{\cal G}_{12}^{\sigma\nu}
\bigr)
\non\\
&&\hspace{-20pt}
+i\bigl(
\dot{\cal G}_{22}^{\nu\alpha}
\tilde{\cal Z}^{\mu\sigma}
-
\dot{\cal G}_{12}^{\sigma\alpha}
\tilde{\cal Z}^{\mu\nu}
+
\dot{\cal G}_{12}^{\sigma\nu}
\tilde{\cal Z}^{\mu\alpha}
+
\dot{\cal G}_{12}^{\mu\alpha}
\tilde{\cal Z}^{\sigma\nu}
-
\dot{\cal G}_{12}^{\mu\nu}
\tilde{\cal Z}^{\sigma\alpha}
\bigr)
\Bigr]\non\\
\label{J5munusimpler}
\ear\no
As in the vector -- vector case, we decompose ${\cal G}_{ij}$ 
as

\bear
{\cal G}_{ij}
&=&
{\cal S}_{ij}
+
{\cal A}_{ij}
\label{decomposecalGBGFva}
\ear\no
where ${\cal S}$ (${\cal A}$) are its parts
even (odd) in $F$. 
We can then delete all terms
odd in $\tau_1 -\tau_2$ since they vanish
upon integration.
After using the identity $F\tilde F = -g\Eins$
and some combining of terms, 
$J_5$ finally turns into the following,
nicely symmetric expression $I_5$,

\bear
I_5^{\mu\nu}(\tau_1,\tau_2) &=&
i\biggl\lbrace
\tilde{\cal Z}^{\mu\nu}k{\cal U}_{12}k
+\Bigl[
(\tilde{\cal Z} k)^{\mu}({\cal U}_{12}k)^{\nu}
+ (\mu\leftrightarrow\nu )
\Bigr]
\non\\
&&\quad
-(\tilde{\cal Z}{\dot{\cal S}}_{12})^{\mu\nu}k{\dot{\cal S}}_{12}k
-\Bigl[
(\tilde{\cal Z} {\dot{\cal S}}_{12}k)^{\mu}({\dot{\cal S}}_{12}k)^{\nu}
+ (\mu\leftrightarrow\nu )
\Bigr]
\biggr\rbrace
\non\\
&&\hspace{-20pt}
+{{\cal Z}\cdot\tilde{\cal Z}\over 4}
\biggl\lbrace
-{\dot{\cal A}}_{12}^{\mu\nu}k{\cal U}_{12}k
-\Bigl[
({\dot{\cal A}}_{12}k)^{\mu}({\cal U}_{12}k)^{\nu}
+ (\mu\leftrightarrow\nu )
\Bigr]
\non\\
&&\quad
+({\dot{\cal A}}_{22}
{\dot{\cal S}}_{12}
)^{\mu\nu}
k{\dot{\cal S}}_{12}k
+\Bigl[
({\dot{\cal A}}_{22}{\dot{\cal S}}_{12}k)^{\mu}
({\dot{\cal S}}_{12}k)^{\nu}
+ (\mu\leftrightarrow\nu)
\Bigr]
\biggr\rbrace
\non\\
\label{defI5munu}
\ear\no
Here in addition to ${{\cal A}}$ and ${{\cal S}}$ 
we have introduced the combination
${\cal U}$,

\bear
{\cal U}_{12} &=& {\dot{\cal S}}_{12}^2 - ({\dot{\cal A}}_{12}
-{\dot{\cal A}}_{22})\bigl({\dot{\cal A}}_{12}+{i\over {\cal Z}}\bigr)
   =  {1-\cos({\cal Z}\dot G_{12})\cos({\cal Z})
   \over \sin^2({\cal Z})} \non\\
\label{U}
\ear\no
Defining also

\bear
\hat{\cal A} \equiv \dot{\cal A} + {i\over {\cal Z}}
\label{defhatA}
\ear\no
this expression can be further compressed to

\bear
I_5^{\mu\nu}(\tau_1,\tau_2) &=&
{{\cal Z}\cdot\tilde{\cal Z}\over 4}
\biggl\lbrace
-{\hat{\cal A}}_{12}^{\mu\nu}k{\cal U}_{12}k
-\Bigl[
({\hat{\cal A}}_{12}k)^{\mu}({\cal U}_{12}k)^{\nu}
+ (\mu\leftrightarrow\nu )
\Bigr]
\non\\
&&\quad
+({\hat{\cal A}}_{22}
{\dot{\cal S}}_{12}
)^{\mu\nu}
k{\dot{\cal S}}_{12}k
+\Bigl[
({\hat{\cal A}}_{22}{\dot{\cal S}}_{12}k)^{\mu}
({\dot{\cal S}}_{12}k)^{\nu}
+ (\mu\leftrightarrow\nu)
\Bigr]
\biggr\rbrace
\non\\
\label{defI5munusimp}
\ear\no
We can now use the matrix decompositions of
${\cal S},\dot{\cal S},\dot{\cal A}$, given 
in eq.(\ref{decomposecalGBGF}), to write the integrand
in explicit form. In this we have a choice between
the matrix bases $\lbrace\hat{\cal Z}_{\pm},
\hat{\cal Z}_{\pm}^2\rbrace$ 
or 
$\lbrace\Eins, F,\tilde F, F^2\rbrace$. We will use the former
one here since it leads to a somewhat more compact
expression.
After the usual rescaling to the unit circle,
a transformation of variables $v=\dot G_{12}$, and
continuation to Minkowski
space, we obtain our final result for the vector --
axialvector amplitude in a constant field
\cite{meva}
,

\bear
\Pi_5^{\mu\nu}(k)
&=&
{e^3e_5\over 8\pi^2}
{\cal G}
\int_0^{\infty}
ds\,s\,\e^{-ism^2}
\int_{-1}^1{dv\over 2}
\,{\rm exp}\biggl[
-i{s\over 2}\sum_{\alpha=+,-}
{\hat A_{B12}^{\alpha}
-\hat A_{B11}^{\alpha}\over z_{\alpha}}\,
k\cdot \hat{\cal Z}_{\alpha}^2\cdot k
\biggr]
\non\\&&\hspace{-15pt}\times
\, \sum_{\alpha,\beta=+,-}
\biggl[
\hat A_{12}^{\alpha}
\Bigl(
(\hat A_{12}^{\beta}-\hat A_{22}^{\beta})
\hat A_{12}^{\beta}
-
(S_{12}^{\beta})^2
\Bigr)
+\hat A_{22}^{\alpha}
S_{12}^{\alpha}S_{12}^{\beta}
\biggr]
\non\\&&\hspace{25pt}\times
\Bigl[\hat{\cal Z}_{\alpha}^{\mu\nu}k\hat{\cal Z}_{\beta}^2k
+(\hat{\cal Z}_{\alpha}k)^{\mu}(\hat{\cal Z}_{\beta}^2k)^{\nu}
+(\hat{\cal Z}_{\alpha}k)^{\nu}(\hat{\cal Z}_{\beta}^2k)^{\mu}
\Bigr]
\non\\
\label{axvpfinal}
\ear
where 

\bear
S_{12}^{\pm} &=&
{\sinh(z_{\pm}v)\over \sinh(z_{\pm})} 
\non\\
\hat A_{12}^{\pm} &=&
{\cosh(z_{\pm} v)\over 
\sinh(z_{\pm})}\non\\
\hat A_{ii}^{\pm} &=&
\coth(z_{\pm}) \non\\
\label{defsminkva}
\ear\no
and $z_{\pm},\hat{\cal Z}_{\pm},a,b$ are the same as in
(\ref{defsmink}),(\ref{defsecular}).
As in the vector -- vector case, 
this expression becomes somewhat more
transparent if one specializes to the Lorentz
system where $\bf E$ and $\bf B$ 
are both pointing along the positive z - axis,
${\bf E} = (0,0,E), {\bf B} = (0,0,B)$.
Here one obtains
\bear
\Pi_5^{\mu\nu}(k)
&=&i
{e^2e_5\over 8\pi^2}
\int_0^{\infty}
ds
\int_{-1}^1{dv\over 2}\,
\e^{-is\Phi_0}
\sum_{\alpha,\beta = \perp,\parallel}
c^{\alpha\beta}
\Bigl[
\tilde F_{\alpha}^{\mu\nu}k_{\beta}^2
+ (\tilde F_{\alpha}k)^{\mu}k_{\beta}^{\nu}
+ (\tilde F_{\alpha}k)^{\nu}k_{\beta}^{\mu}
\Bigr]
\non\\
\label{axvpfinalspecial}
\ear\no
where $z=eBs, z'=eEs$, $k_{\perp}=(0,k^1,k^2,0)$, 
$k_{\parallel} = (k^0,0,0,k^3)$,

\begin{equation}
(\tilde F_{\parallel})^{\mu\nu} \equiv
\left(
\begin{array}{*{4}{c}}
0&0&0&B\\
0&0&0&0\\
0&0&0&0\\
-B&0&0&0
\end{array}
\right),\qquad
(\tilde F_{\perp})^{\mu\nu} \equiv
\left(
\begin{array}{*{4}{c}}
0&0&0&0\\
0&0&-E&0\\
0&E&0&0\\
0&0&0&0
\end{array}
\right)\nonumber\\
\label{deftildeFperpparallel}\nonumber
\vspace{7 mm}
\end{equation}

\bear
\Phi_0 = m^2 +{k_{\perp}^2\over 2}{\cos(zv)-\cos(z)\over z\sin(z)}
-{k_{\parallel}^2\over 2}{\cosh(z'v)-\cosh(z')\over z'\sinh(z')}
\non\\
\label{Phi0va}
\ear

\bear
c^{\perp\perp}
&=&
z{\cos(zv)-\cos(z)\over\sin^3(z)}
\non\\
c^{\perp\parallel}
&=&
{z\cos(zv)\over\sin(z)}
{\cosh(z')\cosh(z'v)-1\over\sinh^2(z')}
-
{z\cos(z)\sin(zv)\over\sin^2(z)}
{\sinh(z'v)\over\sinh(z')}
\non\\
c^{\parallel\perp}
&=&
-{z'\cosh(z'v)\over\sinh(z')}
{\cos(zv)\cos(z)-1\over\sin^2(z)}
-{z'\cosh(z')\sinh(z'v)\over\sinh^2(z')}
{\sin(zv)\over\sin(z)}
\non\\
c^{\parallel\parallel}
&=&
-z'{\cosh(z'v)-\cosh(z')\over\sinh^3(z')}
\non\\
\label{calphabeta}
\ear\no
This result can still be slightly simplified using the
relations

\be
\tilde F^{\mu\nu}_{\alpha} k_{\alpha}^2 = 
(\tilde F_{\alpha}k)^{\mu}
k_{\alpha}^{\nu}
-
(\tilde F_{\alpha}k)^{\nu}
k_{\alpha}^{\mu}
\label{perpparallelidentity}
\ee\no
$(\alpha = \perp, \parallel)$.
It agrees, even at the integrand level, 
with the recent field theory result of
\cite{shaisultanovva}.

\no
We remark that via the axial Ward identity

\bear
k_2^{\nu}\langle A_{\mu}(k_1)A_{5\nu}(k_2)\rangle
&=&
-2im \langle A_{\mu}(k_1)\phi_5(k_2)\rangle
\label{axialward}
\ear
from $\Pi_5^{\mu\nu}$ one can also immediately obtain
the vector -- pseudoscalar amplitude in a constant
field. This amplitude leads, for example, to a
field -- induced effective axion -- photon interaction
\cite{mipava}.
More generally, from the Ward identity it is clear that
the Lorentz contraction $V^{A_5}[k,k]$ of the axialvector
vertex operator (\ref{defaxvectvertop}) can effectively
serve as a pseudoscalar vertex operator.

\section{Effective Actions and their Inverse Mass Expansions}
\label{revea}
\renewcommand{\theequation}{7.\arabic{equation}}
\setcounter{equation}{0}

In the previous chapters we have derived
worldline path integrals representing
effective actions, but applied them mainly to
the calculation of scattering amplitudes.
In this chapter, we calculate the effective action
directly in $x$ -- space, in a higher derivative expansion.
The method used is a pure $x$ -- space version of the
one used by Strassler in ~\cite{strassler2}, made
manifestly gauge invariant by the use of Fock-Schwinger
gauge.

\subsection{The Inverse Mass Expansion for Non-Abelian Gauge Theory}

\def\del{\partial}
\def\deli{\partial_{\kappa}}
\def\delj{\partial_{\lambda}}
\def\delk{\partial_{\mu}}
\def\delij{\partial_{\kappa\lambda}}
\def\delik{\partial_{\kappa\mu}}
\def\deljk{\partial_{\lambda\mu}}
\def\delki{\partial_{\mu\kappa}}
\def\delkl{\partial_{\mu\nu}}
\def\delijk{\partial_{\kappa\lambda\mu}}
\def\deljkl{\partial_{\lambda\mu\nu}}
\def\delikl{\partial_{\kappa\mu\nu}}
\def\delijkl{\partial_{\kappa\lambda\mu\nu}}
\def\delijklm{\partial_{\kappa\lambda\mu\nu o}}
\def\O(#1){O($T^#1$)} 
\def\O2{O($T^2$)}
\def\O3{O($T^3$)}
\def\O4{O($T^4)}
\def\O5{O($T^5$)}
\def\dA{\partial^2}
\def\DA{\sqsubset\!\!\!\!\sqsupset}
\def\eins{  1\!{\rm l}  }

\newcommand{\Vka}{V_{\kappa}}
\newcommand{\Vla}{V_{\lambda}}
\newcommand{\Vmu}{V_{\mu}}
\newcommand{\Vnu}{V_{\nu}}
\newcommand{\Vro}{V_{\rho}}
\newcommand{\Vkala}{V_{\kappa\lambda}}
\newcommand{\Vkamu}{V_{\kappa\mu}}
\newcommand{\Vkanu}{V_{\kappa\nu}}
\newcommand{\Vlamu}{V_{\lambda\mu}}
\newcommand{\Vlanu}{V_{\lambda\nu}}
\newcommand{\Vlaka}{V_{\lambda\kappa}}
\newcommand{\Vmunu}{V_{\mu\nu}}
\newcommand{\Vmuka}{V_{\mu\kappa}}
\newcommand{\Vnuro}{V_{\nu\rho}}
\newcommand{\Vkalamu}{V_{\kappa\lambda\mu}}
\newcommand{\Vkalanu}{V_{\kappa\lambda\nu}}
\newcommand{\Vkalaro}{V_{\kappa\lambda\rho}}
\newcommand{\Vkamunu}{V_{\kappa\mu\nu}}
\newcommand{\Vlamunu}{V_{\lambda\mu\nu}}
\newcommand{\Vmunuro}{V_{\mu\nu\rho}}
\newcommand{\Vkalamunu}{V_{\kappa\lambda\mu\nu}}
\newcommand{\Fkala}{F_{\kappa\lambda}}
\newcommand{\Fkanu}{F_{\kappa\nu}}
\newcommand{\Flaka}{F_{\lambda\kappa}}
\newcommand{\Flamu}{F_{\lambda\mu}}
\newcommand{\Fmunu}{F_{\mu\nu}}
\newcommand{\Fnumu}{F_{\nu\mu}}
\newcommand{\Fnuka}{F_{\nu\kappa}}
\newcommand{\Fmuka}{F_{\mu\kappa}}
\newcommand{\Fkalamu}{F_{\kappa\lambda\mu}}
\newcommand{\Flamunu}{F_{\lambda\mu\nu}}
\newcommand{\Flanumu}{F_{\lambda\nu\mu}}
\newcommand{\Fkamula}{F_{\kappa\mu\lambda}}
\newcommand{\Fkanumu}{F_{\kappa\nu\mu}}
\newcommand{\Fmulaka}{F_{\mu\lambda\kappa}}
\newcommand{\Fmulanu}{F_{\mu\lambda\nu}}
\newcommand{\Fmunuka}{F_{\mu\nu\kappa}}
\newcommand{\Fkalamunu}{F_{\kappa\lambda\mu\nu}}
\newcommand{\Flakanumu}{F_{\lambda\kappa\nu\mu}}

The higher derivative expansion 
is a standard tool
for the approximative calculation of one-loop
effective actions, and considerable work has
gone into the determination of its
coefficients 
for various theories
(see ~\cite{ball,avra} and refs. therein).

This expansion exists in several versions, which
differ by the grouping of terms. The one which we will consider
here is the ``inverse mass expansion'', which is 
just the
expansion in powers of 
the proper-time parameter $T$. 
This groups together
terms of equal mass dimension. Up to partial
integrations in space-time, it coincides with the
(diagonal part of the)
``heat kernel expansion'' for the
second order differential operator in
question. In particular, every coefficient
in this expansion 
is separately gauge invariant. 

Alternatively, one may calculate the same series 
up to a fixed
number of derivatives, but with an arbitrary number of
fields or potentials 
~\cite{zuk1a,zuk1b,chan,carsal,cadhdu}.
In QED this corresponds, to zeroeth order, to the
approximation of the effective Lagrangian by the
Euler-Heisenberg Lagrangian (see chapter 5).
See \cite{gussho1,gussho2} for a
calculation of the first gradient correction to the
Euler-Heisenberg Lagrangian.

Yet another option is to keep the number of
external fields fixed, and sum up the derivatives
to all orders. This leads to the notion of
Barvinsky-Vilkovisky form factors ~\cite{barvis1,barvis2,barvis3}.
For the calculation of some such form factors in the 
string-inspired formalism see \cite{ss1}.

Individual terms in this expansion are also 
relevant for the determination of
counterterms in the corresponding field
theories, defined at a spacetime dimension
$D$ which is related to the mass dimension of
the term considered ~\cite{vandeven}.

We consider a background consisting of a scalar
field and/or a gauge field, both 
possibly non-abelian.
In this background, the scalar loop path integral
(\ref{scalarpi}) generalizes to

 \begin{equation}
\Gamma_{\rm scal}\lbrack A,V\rbrack   =  {\dps\int_0^{\infty}}
{dT\over T}\,
e^{-m^2T}
{\rm tr}{\dps\int} {\cal D} x
\exp\Bigl [- \int_0^T d\tau
\Bigl (\fourth{\dot x}^2 
+ ig\dot x\cdot A
+ V
\Bigr )\Bigr ]
\label{avpi}
\end{equation}
\no
Here we have rewritten $V(x)\equiv U''(\phi(x))$,
where $U(\phi)$ is the field theory
interaction potential of chapter 3.
The path integral is path-ordered except if both A and V
are abelian.
in most applications of the heat kernel expansion the loop spin
can be taken into account by an appropriate choice of the scalar
part $V$ (see, e.g., ~\cite{carmcl,hekrsc}).
In this context we will therefore restrict ourselves to a treatment of
the scalar loop case, although the 
evaluation technique outlined below extends
to the spinor and gluon path integrals in an obvious way;
see \cite{ss1,gussho1,gussho2} for the case of the 
fermion loop in QED.
For QED this approach to the calculation of the
effective action has also been generalized to the finite
temperature case \cite{shovkovy}.

As always we separate out the ordinary integral
over the loop center of mass $x_0$, 
which reduces the effective action to the
effective Lagrangian,

$$
\Gamma [A,V] = \int dx_0 \,{\cal L}[A,V](x_0)\quad
$$

\noindent
To obtain the higher derivative expansion,
we Taylor--expand both $A$ and $V$ at $x_0$, 

\begin{eqnarray}
V(x) &=& e^{y\cdot\partial}V(x_0)\nonumber\\
\dot x^{\mu}A_{\mu}(x) &=&
{\dot y^{\mu}}e^{y\cdot \partial}A_{\mu}(x_0)\quad 
\nonumber\\
\label{taylorchap6}
\end{eqnarray}

\no
The path-ordered interaction 
exponential is then expanded to yield

\begin{eqnarray}
\Gamma_{\rm scal}\lbrack A,V\rbrack &=&  
{\rm tr}{\dps\int_0^{\infty}}
{dT\over T}
e^{-m^2T}
{\dps\int}d^D x_0\,\,
{\dps\sum_{n=0}^{\infty}}{{(-1)}^n\over n}
T\int_0^Td\tau_1\int_0^{\tau_{1}} d\tau_2
\ldots\int_0^{\tau_{n-2}} d\tau_{n-1}\nonumber\\
&\times&
{\dps\int} {\cal D} y
\biggl[ig\,
{\dot y^{\mu_1}}(\tau_1)
e^{y({\scriptscriptstyle\tau_1})
\partial_{(1)}}A^{(1)}_{\mu_1}(x_0)
+ e^{y({\scriptscriptstyle\tau_1})\partial_{(1)}}
V^{(1)}(x_0)\biggr]
\ldots\nonumber\\
&\times&
\biggl[ig\, 
{\dot y^{\mu_n}}(\tau_n )e^{y({\scriptscriptstyle\tau_n})
\partial_{(n)}}A^{(n)}_{\mu_n}(x_0)
+ e^{y({\scriptscriptstyle\tau_n})\partial_{(n)}}
V^{(n)}(x_0)\biggr]
{\rm exp} \Bigl [- \int_0^T d\tau
  {{\dot y}^2\over 4}\Bigr ]\nonumber\\
\label{expand}
\end{eqnarray}
   
\no
Here we have labelled the background fields, and
fixed $\tau_n=0$. This is also the
origin of the factor of $1\over n$.
We then use the Wick contraction rules for
evaluating the
individual terms in this expansion, e. g.,

\begin{eqnarray}
\langle e^{y({\sy\tau_1})\partial_{(1)}}
e^{y({\sy\tau_2})\partial_{(2)}}\rangle &=&
e^{{\scriptstyle -G_B}({\sy\tau_1},
{\sy\tau_2})\partial_{(1)}\cdot\partial_{(2)}}\nonumber\\
\langle \dot y^{\mu}({\tau_1})
e^{y({\sy\tau_1})\partial_{(1)}}
e^{y({\sy\tau_2})\partial_{(2)}}\rangle &=&
-\dot G_B({\tau_1},{\tau_2})\partial_{(2)}^{\mu}
e^{{\scriptstyle -G_B}({\sy\tau_1},
{\sy\tau_2})\partial_{(1)}\cdot\partial_{(2)}}
\nonumber\\
\label{expcon}
\end{eqnarray}
\no
As in our earlier constant background
field calculations,
we can enforce manifest gauge invariance by
choosing Fock-Schwinger
gauge centred at $x_0$.
The gauge condition is  

\begin{equation}
y^\mu A_{\mu}(x_0+y(\tau))\equiv 0
\label{deffs}
\end{equation}

\no
In this gauge,

\begin{equation}
A_{\mu}(x_0+y) = y^{\rho}
\int_0^1 d\eta\eta F_{\rho\mu}(x_0+\eta y)
\label{AtoF}
\end{equation}

\no
and 
$F_{\rho\mu}$ and V can be {\sl covariantly}
Taylor-expanded as (see, e.g., \cite{shifman})

\begin{eqnarray}
F_{\rho\mu}(x_0 + \eta y)&=& 
e^{\eta y\cdot D}F_{\rho\mu}(x_0)\nonumber\\
V(x_0 + y) &=&
e^{y\cdot D}V(x_0)\nonumber\\
\label{expFV}
\end{eqnarray}

\no
This leads also to a 
covariant Taylor expansion for A:

\begin{equation}
A_\mu(x_0+y)=\int^1_0 d\eta\eta\, y^\rho e^{\eta y\cdot D}
F_{\rho\mu}(x_0)
=\half y^\rho F_{\rho\mu}+{1\over 3}y^{\nu}y^{\rho} D_\nu
F_{\rho\mu}+...
\label{Aexpand}
\end{equation}
 
\no
Using these formulas, we obtain the following manifestly
covariant version of eq.(\ref{expand}):

\begin{eqnarray}
\Gamma_{\rm scal}\lbrack F,V\rbrack &=&
{\rm tr}{\dps\int_0^{\infty}}
{dT\over T}
e^{-m^2T}
{\dps\int}d^D x_0\,\,
{\dps\sum_{n=0}^{\infty}}{{(-1)}^n\over n}
T
\int_0^Td\tau_1\int_0^{\tau_1} d\tau_2
\ldots\int_0^{\tau_{n-2}} d\tau_{n-1}\nonumber\\
&& \hspace{-40pt}
\times
{\dps\int} 
{\cal D} y\,
{\rm exp} \Bigl [- \int_0^T d\tau
  {{\dot y}^2\over 4}\Bigr ]
\prod_{j=1}^n \!\Bigl[\;{\rm e}^{ y(\tau_j)D_{(j)}} 
V^{(j)}(x_0)\!+\! ig \dot{y}^{\mu_j}(\tau_j) y^{\rho_j}(\tau_j) 
\int_0^1 d\eta_j \eta_j {\rm e}^{\eta_j y(\tau_j) D_{(j)}}
F_{\rho_j\mu_j}^{(j)}(x_0)  \Bigr]
\nonumber\\
\label{masterchap6}
\end{eqnarray}

\no
From this master formula, the inverse mass 
expansion to some
fixed order N,

\begin{equation}
\Gamma_{\rm scal}[F,V] = \int_0^\infty \!{dT\over T} \; 
\frac{{\rm e}^{-m^2 T}}{(4\pi T)^{D/2}} \; 
{\rm tr} \; \int \! dx_0 \; \sum_{n=1}^N \; 
\frac{(-T)^n}{n!} \; O_n[F,V] \,,
\end{equation}\no
is obtained in three steps:

\vskip10pt
{\sl Wick contractions:} Truncate the master 
formula to $n=N$, 
and the covariant Taylor expansion
eq.(\ref{Aexpand})
accordingly. Perform the Wick contractions.
Alternatively, one may also first Wick contract
the complete expression for $n=N$, and truncate 
the Taylor expansion afterwards.
(This procedure is preferable in the
pure scalar field case.)

\vskip10pt
{\sl Integrations:}
Perform the $\tau$ -- integrations.
The integrand is a polynomial in
the worldline Green's function $G_B$, $\dot G_B$,
and
$\ddot G_B$. As usual, the $\tau$ -- integrals
can be rescaled to the unit circle, $\tau_i=Tu_i$.
The $\delta$-function in $\ddot G_B(u_i,u_j)$ only contributes
if $u_i$ and $u_j$ are neighbouring points on the loop.
(Note that this includes the case $\ddot G_B(1,u_n)$.) 
In the non-Abelian case the 
coefficient $2$ in front of the $\delta$-function has to be omitted, 
since only half of the $\delta$-function contributes to the ordered 
sector under consideration
(in the scattering amplitude context this rule was already stated in
section \ref{scalarloopgluonscatter}).

\vskip10pt
{\sl Reduction to a minimal basis:}
The result of this procedure is the effective Lagrangian at
the required order, albeit in redundant form.
To be maximally
useful for numerical applications, it still
needs to be reduced
to a minimal set of invariants, using all available symmetries.
Those are

\begin{enumerate}
\item
Cyclic invariance under the trace.
\item
Bianchi identities.
\item
For real representations of the gauge group
one has an additional symmetry
under transposition (up to a sign for every factor
of $F_{\mu\nu}$). 
\end{enumerate}
\noindent
Usually those symmetry operations would have to be
combined with judiciously chosen partial integrations
performed on the effective Lagrangian. It is a remarkable
property of the present calculational scheme
that the reduction of our result for the effective action
to
a minimal basis of invariants can be achieved without
any such partial integrations. In particular, for the
pure scalar case the reduction process amounts to 
nothing more than
the identification of cyclically equivalent terms
(in fact, 
for this special case the whole procedure can
be condensed into a purely 
combinatorial formula ~\cite{muellerformel}).
In the general case, the reduction to a 
minimal basis of invariants
is much more involved. The method adopted here
follows a proposal by M{\"u}ller ~\cite{muellerbasis1,muellerbasis2}, 
which is also explained in ~\cite{fhss4}.

Let us give the explicit result of this procedure
up to order $O(T^4)$
(absorbing the coupling 
constant $g$ into the fields, 
and abbreviating
$\Fkalamunu\equiv D_\kappa D_\lambda 
F_{\mu\nu}$ etc.):

\begin{eqnarray*}
O_1 &=& V \,\\
O_2 &=& V^2 + {1\over 6} \Fkala \Flaka \,\\
O_3 &=& V^3 + {1\over 2} \Vka \Vka 
          + {1\over 2} V \Fkala \Flaka 
          - {2\over 15} \,i\,\Fkala\Flamu\Fmuka
          + {1\over 20} \Fkalamu\Fkamula \,\\
O_4 &=& V^4 + 2 V\Vka\Vka + {1\over 5} \Vkala\Vlaka
          + {3\over 5} V^2 \Fkala\Flaka 
          + {2\over 5} V\Fkala V\Flaka \\
    &-& {4\over 5} \,i\, \Fkala\Vla\Vka 
          - {8\over 15} \,i\, V\Fkala\Flamu\Fmuka 
          + {1\over 5} V\Fkalamu\Fkamula 
          - {2\over 15} \Fkala\Flamu\Vmuka \\ 
    &+& {1\over 3}\Fkala\Fmulaka\Vmu 
          + {1\over 3} \Fkala\Vmu\Fmulaka 
          + {2\over 35} \Fkala\Flaka\Fmunu\Fnumu 
          + {4 \over 35} \Fkala\Flamu\Fkanu\Fnumu \\
    &-& {1\over 21} \Fkala\Flamu\Fmunu\Fnuka 
          - {8\over 105} \,i\,\Fkala\Flamunu\Fkanumu 
          - {6\over 35} \,i\, \Fkala\Fmulanu\Fmunuka \\
    &+& {11\over 420} \Fkala\Fmunu\Flaka\Fnumu  
          + {1\over 70} \Fkalamunu\Flakanumu \,
\end{eqnarray*}
\no
With the present method, a complete calculation of all
coefficients was achieved to order $O(T^6)$ in the
general case
\footnote{See the recent \cite{kwlemi} for an application
of this result to the approximate calculation of the
one-loop contribution 
by massive quarks
to the QCD vacuum tunneling amplitude.},
and to order $O(T^{12})$ in the case with
only a scalar field ~\cite{fss1,fhss1,fhss4}.

With conventional methods this 
expansion was previously obtained to order
O($T^5$) in the general case ~\cite{vandeven},
and only recently to order O($T^7$) in the
scalar case ~\cite{lanyov}.
\no
Detailed comparisons with other methods 
\cite{nepo,chan,vandeven}
of calculating the
higher derivative expansion 
have been made in
\cite{fss1,fhss4}.
We consider here only the Onofri-Zuk method,
which is the one most closely related
to the worldline technique.

In Onofri's work ~\cite{onofri},
the Baker-Campbell-Hausdorff formula
was employed to represent the coefficients for the
pure scalar case by Feynman diagrams in a 
one-dimensional auxiliary field theory.
Those Feynman diagrams are calculated using
the Green's function

\begin{equation}
G^0(\tau_1,\tau_2) = \mid \tau_1 - \tau_2\mid - 
(\tau_1 + \tau_2 ) + {2\over T}\tau_1\tau_2\quad 
\label{defG0}     
\end{equation}

\no
which is the kernel for the second derivative
operator
on an interval of length $T$
appropriate
to the boundary conditions

\begin{equation}
x(0) = x(T) = 0
\label{G0bc}
\end{equation}

\no
This representation was then used by Zuk
to calculate the effective Lagrangian for the pure
scalar case up to the terms with four derivatives
~\cite{zuk1a,zuk1b}.
This author further generalized the method to the 
gauge field case, and also used Fock-Schwinger gauge
to enforce manifest gauge invariance ~\cite{zuk2}.
The same Green's function is used in the ``Quantum
Mechanical Path Integral Method'' 
~\cite{mckeon:cjp70,mckeon:ap224,mckreb,mckshe},
which may be considered as an extension of
the Onofri-Zuk formalism.

To see the connection to our formalism, first
note that the Green's function 
which we used for the evaluation of the
reduced path integral $\int {\cal D}y(\tau)$,
eq.(\ref{defG}),
is by no means unique.
Since the naive defining equation

\begin{equation}
{1\over 2}
{\partial^2\over\partial\tau_1^2}
G(\tau_1,\tau_2)
=\delta(\tau_1 -\tau_2)
\label{laplaceequ}
\end{equation}
\noindent
has no periodic solutions, it needs to be modified by the
introduction of a background charge, leading to
\cite{strassler1}

\begin{equation}
{1\over 2}
{\partial^2\over\partial\tau_1^2}
G^{\rho}(\tau_1,\tau_2)
=\delta(\tau_1 -\tau_2)
-\rho(\tau_1)
\label{laplacemod}
\end{equation}
\noindent 
The distribution of the background charge $\rho$
along the circle is arbitrary, except that
it should integrate to unity,

\begin{equation}
\int_0^Td\tau
\rho(\tau)
=1
\label{normrho}
\end{equation}
\noindent
That this is necessary can be seen
by integrating eq.(\ref{laplacemod})
in the first variable.

If one further
requires the Green's
function $G^{\rho}$
to be symmetric in its
both arguments,
periodicity 
determines it up to
an irrelevant constant.
The solution can be expressed
in terms of the standard Green's function $G_B$,
eq.(1.16),
as follows, 

\begin{eqnarray}
G^{\rho}(\tau_1,\tau_2)
&=&
G_B(\tau_1,\tau_2)
-\int_0^T\, d\sigma
\rho(\sigma)
G_B(\sigma,\tau_2)
-\int_0^T\, d\sigma
G_B(\tau_1,\sigma)
\rho(\sigma)
\nonumber\\
&&
+
\int_0^T\,d\sigma_1
\int_0^T\, d\sigma_2
\,\rho(\sigma_1)
G_B(\sigma_1,\sigma_2)
\rho(\sigma_2)
\label{calcGrho}
\end{eqnarray}
\noindent
Any such $G^{\rho}$ can be used as a Green's function
for the evaluation of the path integral 
eq. (\ref{scalarpi}).
Different choices of $\rho$ 
must lead to the same effective action
or scattering amplitude.

This is easily verified by the following little
argument well-known from string perturbation theory.
Let us consider the scalar field theory case first.
For the 
$\phi^3$ scattering amplitude, eq.(\ref{scalarmaster})
becomes, if a general $\rho$ is used,

\begin{eqnarray}
\Gamma[p_1,\ldots,p_N] &=&
\half{(-\lambda)}^N
{(2\pi )}^D\delta (\sum p_i)
{\dps\int_{0}^{\infty}}{dT\over T}
{(4\pi T)}^{-{D\over 2}}e^{-m^2T} 
\nonumber\\&&\times
\prod_{i=1}^N \int_0^T 
d\tau_i
\exp\biggl[\half\sum_{i,j=1}^NG^{\rho}
(\tau_i,\tau_j) p_i\cdot p_j\biggr]
\nonumber\\
\label{scalarmasterchap6}
\end{eqnarray}
\no
Using eq.(\ref{calcGrho}) and momentum conservation
it is immediately seen that all $\rho$ - dependence
drops out in the exponent.

For the effective action the equivalence works in
almost the same way.
If only $V$ is present, 
eq.(\ref{expand}) 
after Wick contraction
turns into

\bear
\Gamma_{\rm scal}[V]
&=&\int_0^\infty {dT\over T}(4\pi T)^{-D/2}
{\rm e}^{-m^2T} {\rm tr} \int d^Dx_0
{\sum_{n=0}^\infty} {(-1)^n\over n} T \int_0^T
\! d\tau_1 \int_0^{\tau_1} d\tau_2
\;...\; \int_0^{\tau_{n-2}} \!\!\!\!\!\! d\tau_{n-1} \non\\
&&\times {\rm exp} \Bigl[ - \half\sum_{i,j=1}^n G^{\rho}(\tau_i,\tau_j)
\del_{(i)} \cdot
\del_{(j)} \Bigr] V^{(1)}(x_0) ... V^{(n)}(x_0)
\label{scalarrhocontract}
\ear\no
Using eq.(\ref{calcGrho}) we may rewrite the exponent
as follows,
 
\begin{eqnarray}
-{1\over 2}\sum_{i,j=1}^n
G^{(\rho)}(\tau_i,\tau_j)
\partial_{(i)}\cdot\partial_{(j)}
&=&
-{1\over 2}\sum_{i,j=1}^n
G_B(\tau_i,\tau_j)
\partial_{(i)}\cdot\partial_{(j)}
\non\\&&
+
\int_0^Td\sigma \rho(\sigma)
\biggl(\sum_{i=1}^n\partial_{(i)}
\cdot
\sum_{j=1}^n
G_B(\sigma,\tau_j)\partial_{(j)}
\biggr)
\nonumber\\
&&
-{1\over 2}
\int_0^T\,d\sigma_1
\int_0^T\, d\sigma_2
\,\rho(\sigma_1)
G_B(\sigma_1,\sigma_2)
\rho(\sigma_2)
\sum_{i,j=1}^n
\partial_{(i)}
\cdot
\partial_{(j)}
\non\\
\label{rewriteunivexp}
\end{eqnarray}
\noindent
This shows that all $\rho$ -- dependent
terms in the effective Lagrangian carry at least
one factor of $\sum_{i=1}^n\partial_{(i)}$.
They
are thus total derivative terms and
will disappear in the final $x_0$ -- integration
(under appropriate boundedness conditions
on the background field at infinity).
The same applies to the dependence
on a constant which one could always add to
$G^{\rho}$.

This argument easily carries over to the
gauge theory case, if one uses the
Bern-Kosower
master formula eq.(\ref{scalarqedmaster})
and its effective action analogue.
Different admissible Green's functions
will thus in general produce different 
effective Lagrangians, but the
same effective action and scattering amplitudes.

However, from string theory it is also known
that the choice of the background charge
can have some technical significance
at intermediate stages of calculations
\cite{dhopho,dhogid}.
The Green's function $G_B$ usually
used in the string--inspired 
approach corresponds to the
choice of a constant $\rho$,

\begin{equation}
\rho (\tau) = {1\over T}\quad 
\label{rhoconst}
\end{equation}

\no
This is the only choice leading to
a translation-invariant Green's function.
If one chooses the function

\begin{equation}
\rho (\tau) = \delta (\tau)
\label{rhodelta}
\end{equation}

\no
instead,
eq.(\ref{calcGrho})
yields
just the one used by Onofri, eq.(\ref{defG0}).
Expansion of the path integral as in 
eq.(\ref{masterchap6}) then 
precisely generates
Zuk's one-dimensional
Feynman rules.
The final result after integrating out $x_0$
will be the same.
However, the 
difference in the choice
of $\rho$
turns out to have some 
nontrivial technical consequences:

With our choice of the worldline Green's function
partial integrations never become necessary in the
reduction process. This is not true for the
Onofri-Zuk approach, 
a fact which becomes particularly conspicuous
in the pure scalar case ~\cite{fss1}.
Here the effective Lagrangian resulting from
our method is already minimal after identification
of cyclically equivalent terms, while large
numbers of partial integrations turn out
to be necessary to further minimize 
Zuk's result. 
 
Moreover, due to the translational invariance of
the worldline Green's function eq.(\ref{defG})
cyclically equivalent terms always come with
the same numerical coefficient. This considerably
facilitates the cyclic identification process.
Again, this property does not hold true 
if one uses the Green's function eq. (\ref{defG0});
for example, of the 
three cyclically equivalent terms
$V_{\mu}V_{\mu}V$, $VV_{\mu}V_{\mu}$
and $V_{\mu}VV_{\mu}$ appearing in the
scalar effective action at O($T^4$)
the first two then get assigned the same 
coefficient, while the coefficient of the
third one is different. 

Finally, let us to note that the ambiguity
in the choice of $\rho$ 
has an interpretation already 
at the path integral level.
From eq.(\ref{calcGrho}) it can be read off that
$G^{\rho}$ fulfills

\begin{equation}
\int_0^T\,d\sigma\rho(\sigma)
G^{\rho}(\sigma,\tau_2)
=\int_0^T\,d\sigma
G^{\rho}(\tau_1,\sigma)
\rho(\sigma)
=0
\label{Grhocond}
\end{equation}
\noindent
(for any $\tau_1,\tau_2$).
This indicates that
a given background charge $\rho$ corresponds to
the following 
generalization of the zero mode fixing
eqs.(\ref{split}),

\begin{eqnarray}
{\dps\int}{\cal D}x &=&
{\dps\int}d x_0{\dps\int}{\cal D} y\nonumber\\
x^{\mu}(\tau) &=& x^{\mu}_0  +
y^{\mu} (\tau )\nonumber\\
\int_0^T\,d\tau\rho(\tau)y^{\mu}(\tau)&=&0\nonumber\\
\label{gensplit}
\end{eqnarray}
\no
In particular, for $\rho(\tau)=\delta(\tau)$
one recovers the boundary conditions
eq.(\ref{G0bc}), $y(0)=y(T)=0$.

In the ``string-inspired'' formalism,
the effective Lagrangian 
${\cal L} (x_0)$ is 
obtained as
a path integral over the space of all loops
having $x_0$ as their common center of 
mass; in the Onofri-Zuk formalism, 
as a path
integral over the space of all loops
intersecting in $x_0$. 
And indeed, for Onofri's original formalism
precisely this path integral representation was
already
provided by Fujiwara et al. \cite{fow}.
Clearly
the center of mass choice is more ``symmetric'',
so that it is intuitively reasonable that it
should lead to a more compact form for the
effective Lagrangian.

\subsection{Other Backgrounds}
\label{otherbackgrounds}

The above procedure can be extended to the 
case of a mixed vector -- axialvector background
without difficulties; see \cite{mcksch} for the
computation of some heat kernel coefficients
for this background along the above lines.
To the contrary, the inclusion of gravitational
backgrounds poses new and interesting conceptual problems.
Here the problems 
connected to the existence of
different ordering prescriptions for the
quantum mechanical Hamiltonian, which we
already briefly encountered in our
discussion of the gluon loop,
are of a more serious nature.
Classically, the worldline
Lagrangian for a scalar point particle
coupled to a background gravitational
field could be taken as

\be
L^{\rm cl} =\half \dot x^{\mu}g_{\mu\nu}(x)\dot x^{\nu}
\label{Lgrav}
\ee\no
Quantum mechanically, the ambiguity in the factor ordering of the
Hamiltonian leads to the 
possible appearance of further terms in the
path integral action, which are
of order $\hbar^2$. 
According to the theorem by Sato
~\cite{sato.m} 
mentioned earlier our naive way of evaluating Gaussian path
integrals requires the use of the Weyl-ordered
Hamiltonian, which in turn is equivalent to using the
mid-point rule in the standard time-slicing definition
of the path integral 
\cite{sakitabook,sato.m,berezin,mizrahi,gerjev}.
This procedure 
leads to the following ``quantum'' Lagrangian ~\cite{mizrahi},

\be
L^{\rm qu}_{\rm TS}
=
\half \dot x^{\mu}g_{\mu\nu}\dot x^{\nu}
+
{\hbar^2\over 8}
\Bigl(
R+
g^{\mu\nu}\Gamma^{\kappa}_{\mu\lambda}\Gamma^{\lambda}_{\nu\kappa}
\Bigr)
\label{LTS}
\ee
where $\Gamma^{\kappa}_{\mu\lambda}$
denotes the Christoffel symbol.

Moreover, in a curved background the path integral measure
also becomes nontrivial. Since, as usual, we wish to
evaluate the path integral in terms of one-dimensional
Feynman diagrams, it is natural to absorb this measure into
the action by the introduction of appropriate ghost terms
into the action, in the spirit of Lee and Yang \cite{leeyan}.
In the present context this can be done in two slightly different
ways \cite{bpsv:npb446,dilmck}.  

To further complicate matters, a careful
analysis performed over the past
few years by van Nieuwenhuizen, Bastianelli, and their coworkers
\cite{basvan,bpsv:npb446,bpsv:npb459,bascva,schvan,bascor} 
has established that the coefficients of
the above $\hbar^2$-terms have no absolute meaning; they depend
on the choice of the regularization prescription which is implicit in the
definition of the path integral. For example, another plausible
definition would be to expand all worldline fields, including 
the ghosts, in a sine expansion about the classical trajectories,
and then integrate over the Fourier coefficients. Regulating the
resulting expressions by a universal cutoff on these Fourier mode
sums one arrives at the so-called ``mode regularization''. The
explicit computation \cite{bascva,bascor}
shows that, if one wishes to reproduce in this scheme the known results for the
heat kernel in curved space, then the above
Lagrangian must be replaced by

\be
L^{\rm qu}_{\rm MR}
=
\half \dot x^{\mu}g_{\mu\nu}\dot x^{\nu}
+
{\hbar^2\over 8}
\Bigl(
R-{1\over 3}g^{\mu\nu}g^{\kappa\lambda}
g_{\rho\sigma}\Gamma^{\rho}_{\mu\kappa}\Gamma^{\sigma}_{\nu\lambda}
\Bigr)
\label{LMR}
\ee
This ambiguity has also a natural interpretation 
in terms of the one-dimensional quantum field
theory defined by the path integral. 
This field theory constitutes a super-renormalizable 
nonlinear sigma model, which by power counting
has superficial 
ultraviolet divergences at the one- and two-loop levels,
but not at higher loop orders
\footnote{Here the loop counting refers to the one-dimensional
worldline field theory; in terms of the four-dimensional
field theory our discussion is, of course, at the one-loop
level.}. 
Those ultraviolet divergences cancel out in the sum of terms,
but, as is usual in such cases, leave a finite ambiguity
embodied by the above $\hbar^2$ - terms. 
Their coefficients cannot be determined 
inside the one-dimensional field theory without further input. 
Requesting the result of the worldline perturbation series
to reproduce the usual heat kernel expansion for the
space-time field theory provides such an input, and suffices to fix
them completely.
Once this has been done, the coefficient
of the $R$ - term turns out to be universal for the regularization
schemes considered in the works quoted above.
Since the heat kernel is a covariant
quantity, the appearance of the other, noncovariant terms in the
worldline Lagrangians (\ref{LTS}),(\ref{LMR}) is clearly connected
to the fact that both regularizations used, time-slicing and
mode regularization, break the covariance; 
the role of the explicit noncovariant terms in the
Lagrangians is to compensate for this. 
This leads to the question whether some regularization
can be found which would avoid this covariance breaking.
Very recently, Kleinert and Chervyakov \cite{kleche1,kleche2,kleche3} 
have claimed
that one-dimensional dimensional regularization provides such a
scheme, and they verified the absence of any $\hbar^2$ - terms in a simpler
model with a one-dimensional target space. 
In \cite{baconi1,baconi2} it was
then shown that, in the four-dimensional case, this scheme   
is indeed free of noncovariant counterterms, although the 
$R$ - term is still necessary:

\be
L^{\rm qu}_{\rm DR}
=
\half \dot x^{\mu}g_{\mu\nu}\dot x^{\nu}
+
{\hbar^2\over 8}
R
\label{LDR}
\ee
This scheme therefore seems
to be the most promising one
for future applications of the string-inspired
formalism to curved-space calculations
\footnote{As a necessary preliminary step to any such
application in \cite{baconi2} it was shown how to
extend dimensional regularization to a compact time intervall.} 
.

To use any of these worldline actions for the calculation of
the higher derivative expansion of the 
gravitational effective action, one 
would now wish to choose a Riemann normal
coordinate system centered at the loop center of mass
$x_0$. This is the gravitational analogue of
Fock-Schwinger gauge, and allows one to rewrite
the Taylor expansion of the metric at $x_0$
in terms of covariant derivatives of the
curvature tensor \cite{eisenhart,msv},

\be
g_{\mu\nu}(x)
=
\delta_{\mu\nu}
+
\third
y^{\alpha}y^{\beta}
R_{\mu\alpha\beta\nu}
+
\sixth
y^{\gamma}y^{\alpha}y^{\beta}
\nabla_{\gamma}
R_{\mu\alpha\beta\nu}
+
\cdots
\label{riemann}
\ee\no
Individual terms in the higher derivative expansion can
then again be calculated by the application of appropriate Wick contraction
rules 
\cite{basvan,bpsv:npb446,dilmck,bpsv:npb459,bascva,schvan,bascor,dilkesthesis}.

However, here another rather subtle complication 
appears in the choice of the worldline propagator.
As we remarked in the previous section for the
gauge theory case, and explicitly demonstrated for
scalar field theory, there exists a large family
of admissible worldline propagators, and the
effective actions computed with different such propagators
differ by total derivative terms. While this statement
remains true in the curved space case, an explicit computation
at the two-loop level has revealed \cite{schvan} that those total
derivative terms are in general {\sl not covariant}.
If one wishes to reproduce precisely the 
standard heat kernel expansion, which is 
manifestly covariant, then
the Onofri-Zuk propagator (\ref{defG0}) must be
used, corresponding to the boundary conditions
$y(0)=y(T)=0$ on the path integral after the
zero mode fixing. The use of the standard worldline
Green's function $G_B$, on the other hand, 
leads to a result which differs
from this by noncovariant total derivative terms. 
This poses no problems in principle, but in practice,
since it invalidates the application of Riemann
normal coordinates, which are useful only
for the computation of covariant quantities
\footnote{To be precise, the application of such coordinates
to the computation of a noncovariant quantitity will
produce an {\sl apparently} covariant result which is correct 
in that particular coordinate system but not in others.}.
Thus the present state of affairs is that, in the
application of the worldline technique to curved space
effective actions, one either has to forgo the convenience
of using the translation invariant worldline propagator
$G_B$ or, worse, of the use of Riemann normal coordinates
\footnote{It should be noted that, a priori, the analogous
problem could have appeared also in the gauge theory case
in the form of non -- gauge invariant  total derivative terms. Our
use of Fock-Schwinger coordinates in the previous section
was possible only because the compatibility of the
standard propagator $G_B$ with gauge invariance 
is known on general grounds (see chapter 4).}
.

\section{Multiloop Worldline Green's Functions}
\renewcommand{\theequation}{8.\arabic{equation}}
\setcounter{equation}{0}

While one could clearly construct multiloop
formulas of the Bern-Kosower type starting
with formulas 
such as eqs.(\ref{scalarmaster}),(\ref
{scalarqedmaster}),(\ref{supermaster}),
and then sewing together pairs of external
legs, 
such a procedure turns out to be 
unnecessarily cumbersome.
In the spirit of the Bern-Kosower
formalism, we would rather like to 
completely avoid the
appearance of internal momenta. 
This is not only for aesthetic reasons; it is
the absence of internal momenta which, in the
Bern-Kosower formalism, reduces the number of
independent kinematic invariants from the very
beginning, thereby rendering the spinor helicity
formalism even more useful than usual.

We will rather follow the example of string theory,
where multiloop amplitudes can be
represented as path integrals over 
Riemann surfaces of higher genus, 
embedded into some target space.
The Green's function of the
Laplacian (or some other kinetic operator)
for those surfaces  
~\cite{gsw,dhopho} is then
the basic quantity needed
for their calculation.
In the infinite string tension limit, the Riemann
surfaces correspond to graphs.
If we wish to preserve the analogy with string
theory, we ought to find out
how to calculate path integrals over
graphs embedded into spacetime. 

This is by no means a new idea.
For the example
of $\phi^3$ -- theory, it has been
repeatedly pointed out that it should
be possible to construct multiloop
amplitudes in terms of path integrals
over graphs 
~\cite{cagolo,tseytlin86,hehoto,kontsevich,dijkgraaf}.
The proper-time lengths of the propagators
making up those graphs would then just
correspond to the moduli parameters
in string theory.
However, 
none of those authors provided
an explicit computational
prescription for the evaluation
of this kind of path integral.
In ~\cite{ss2}, M.G. Schmidt and the
present author proposed such a prescription,
based on the concept of Green's functions defined
on graphs.
Let us therefore begin with explaining 
how to construct such 
``multiloop worldline Green's functions''
at the two-loop level.

\subsection{The 2-Loop Case}
   
At first glance, this looks like an ill-defined problem.
In contrast to the circle, a general graph is not
a differentiable manifold, and it is a priori
not obvious how to define the second derivative 
operator at the node points. 

\begin{figure}[ht]
\begin{center}
\vbox to 4.5cm{\vfill\hbox to 15.8cm{\hfill
\epsffile{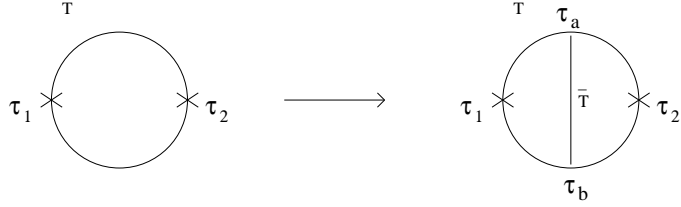}
\hfill}\vfill}\vskip-.4cm
\end{center}
\caption[dum]{Change of the one-loop Green's function
by a propagator insertion.
\hphantom{xxxxxxxxxxxxxxx}}
\label{2loopG}
\end{figure}

Instead, we will pose the following simple
question.
How does the Green's function 
$G_B(\tau_1,\tau_2)$ 
between two fixed points $\tau_1,\tau_2$ on
the circle change, if we insert, between
two other points $\tau_a$ and $\tau_b$,
a (scalar) propagator of fixed proper-time length
$\bar T$ (fig. \ref{2loopG})? 

\par
\par

\no
To answer this question, let us start with the
worldline-path integral representation for the
one-loop two-point -- amplitude in 
$\phi^3$ -- theory,
and sew together the two external
legs. The result is, of course,
the vacuum path
integral with a propagator insertion:

\begin{equation}
\Gamma^{(2)}_{\rm vac} =
\int_0^{\infty}{{dT}\over T}{\rm e}^{-m^2T}
{\dps\int}
{\cal D} x
\int_0^T d\tau_a \int_0^T d\tau_b
\,\langle \phi(x(\tau_a))\phi(x(\tau_b))
\rangle
\,\exp\biggl[- \int_0^T d\tau
{{\dot x}^2\over 4}
\biggr]
\label{pi+prop}
\end{equation}
\no
Here
$\langle \phi(x(\tau_a))\phi(x(\tau_b))
\rangle$
is the x--space scalar propagator in $D$
dimensions, which, if we specialize to the
massless case for a moment, would read

\begin{equation}
\langle \phi(x(\tau_a))\phi(x(\tau_b))
\rangle = {{\Gamma ({D\over 2}-1)}\over
{4{\pi}^{D\over 2}
{\Bigl[{(x_a-x_b)}^2\Bigr]}^{({D\over 2} -1)}}}
\label{masslessscalprop}
\end{equation}
\no
Clearly
this form of the propagator is not suitable for
calculations in our auxiliary
one-dimensional field theory.
The 
approach based on worldline
Green's functions which we have in mind
will work nicely only if we can manipulate
all our path integrals into Gaussian form.
To obtain a Gaussian path integral,
we therefore make further use of the
Schwinger proper-time representation
to exponentiate the
propagator insertion,

\begin{equation}
\langle\phi(x(\tau_a))\phi(x(\tau_b))\rangle
=\int_0^{\infty} d\bar T 
e^{-m^2\bar T}
{(4\pi \bar T)}^{-{D\over 2}}
{\rm exp}\Biggl
[-{{(x(\tau_a)-x(\tau_b))}^2\over 4\bar T}
\Biggr ]\, 
\label{intprop}
\end{equation}
\no
where the propagator mass was also reinstated.
We have then the following path integral
representation of the two -- loop
vacuum amplitude:

\begin{eqnarray}
\Gamma_{\rm vac}^{(2)}&=&
\int_0^{\infty}{{dT}\over T}{\rm e}^{-m^2T}
\int_0^{\infty} d\bar T{(4\pi \bar T)}^{-{D\over 2}} 
e^{-m^2\bar T}
\int_0^T d\tau_a \int_0^T d\tau_b
\nonumber\\
&&\times
{\dps\int}
{\cal D} x
\,\exp\Biggl [- \int_0^T d\tau\,
{{\dot x}^2\over 4}-
{{(x(\tau_a)-x(\tau_b))}^2\over 4\bar T}
\Biggr ]\nonumber\\
\label{pi+B}
\end{eqnarray}
\no
The propagator insertion has, for fixed parameters
$\bar T,\tau_a,\tau_b$, just produced an additional
contribution to the original free worldline action. 
Moreover, this term is quadratic 
in $x$, so that we can hope to absorb it into the
free worldline Green's function. For this
purpose, it is useful to introduce
an integral operator $B_{ab}$
with integral kernel

\begin{equation}
B_{ab}(\tau_1,\tau_2) = 
\Bigl[\delta(\tau_1 - \tau_a)-\delta(\tau_1-\tau_b)\Bigr]
\Bigl[\delta(\tau_a - \tau_2)-\delta(\tau_b-\tau_2)\Bigr]
\quad 
\label{defB}
\end{equation}
\no
($B_{ab}$ acts trivially on Lorentz indices).
We may then rewrite

\begin{equation}
{(x(\tau_a)-x(\tau_b))}^2
= 
\int_0^T d\tau_1 \int_0^T d\tau_2\,
x(\tau_1)B_{ab}(\tau_1,\tau_2)x(\tau_2)
\label{useB}
\end{equation}
\no
Obviously, the presence of the
additional term corresponds
to changing the defining equation for $G_B$,
eq.(\ref{calcG}),
to
\begin{equation}
G_B^{(1)}(\tau_1,\tau_2)=2
\bigl\langle\tau_1\mid
{\Bigl({d^2\over {d\tau}^2}-{B_{ab}\over \bar T}\Bigl)}^{-1}
\mid\tau_2\bigr\rangle 
\label{defG(1)}
\end{equation}
\no
After eliminating the zero-mode
as before,
this modified propagator can be constructed simply
as a geometric series:

\begin{equation}
{\Bigl({d^2\over {d\tau}^2}-{B_{ab}\over \bar T}\Bigl)}^{-1}
= {\Bigl({d\over {d\tau}}\Bigr)}^{-2}
+
{\Bigl({d\over {d\tau}}\Bigr)}^{-2}{B_{ab}\over \bar T}
{\Bigl({d\over {d\tau}}\Bigr)}^{-2}
+
{\Bigl({d\over {d\tau}}\Bigr)}^{-2}{B_{ab}\over \bar T}
{\Bigl({d\over {d\tau}}\Bigr)}^{-2}{B_{ab}\over \bar T}
{\Bigl({d\over {d\tau}}\Bigr)}^{-2}
+\cdots
\label{sumseries}
\end{equation}
\no
Noting that

\be
B_{ab}{({d\over d\tau})}^{-2}B_{ab}=
-G_{Bab}B_{ab}
\label{B2ident}
\ee\no
we can explicitly sum this series,
and obtain ~\cite{ss2}

\bear
G_B^{(1)}(\tau_1,\tau_2)=
G_B(\tau_1,\tau_2) + \half
{{[G_B(\tau_1,\tau_a)-G_B(\tau_1,\tau_b)]
[G_B(\tau_a,\tau_2)-G_B(\tau_b,\tau_2)]}
\over
{{\bar T} + G_B(\tau_a,\tau_b)}}
\non\\
\label{G(1)} 
\ear
\no
The worldline Green's function between points
$\tau_1$ and $\tau_2$ is 
thus simply the one-loop Green's function plus  
one additional piece, which takes the effect
of the insertion
into account. Observe that this piece
can still be written in terms of the various
one-loop Green's functions $G_{Bij}$. 
However it is not a function of
$\tau_1-\tau_2$ any more, nor is its
coincidence limit a constant
(for alternative derivations of this expression see
~\cite{ss2,dashsu,sato1}).

Knowledge of this Green's function is not yet sufficient
for performing two-loop calculations. We also need to know
how the path integral determinant is changed by
the propagator insertion. 
Using the $ln\,det=tr\,ln$ -- formula
this can be easily calculated, and yields

\begin{equation}
{{
{\dps\int}
{\cal D} y
\,\exp\Bigl [- \int_0^T d\tau
{{\dot y}^2\over 4}-
{{(y(\tau_a)-y(\tau_b))}^2\over 4\bar T}
\Bigr]
}
\over 
{
{\dps\int}
{\cal D} y
\,\exp\Bigl [- \int_0^T d\tau
{{\dot y}^2\over 4}
\Bigr]
}
}
= {
{{{\rm Det_P'}({d^2\over {d\tau}^2} - 
{B_{ab}\over \bar T})}^{-{D\over 2}}}
\over
{{{\rm Det_P'}({d^2\over {d\tau}^2})}^{-{D\over 2}}}
}
= {\Bigl(1+{G_{Bab}\over{\bar T}}\Bigr)}^{-{D\over 2}}
\label{normchange}
\end{equation}
\no

To summarize, the insertion of a scalar propagator into
a scalar loop can, for fixed values of the proper-time
parameters, be completely taken into account by
changing the path integral normalization, and
replacing $G_B$ by $G_B^{(1)}$.
\no
The vertex operators remain unchanged.

\begin{figure}[ht]
\vbox to 6.0cm{\vfill\hbox to 15.8cm{\hfill
\epsfysize=5.5cm
\epsffile{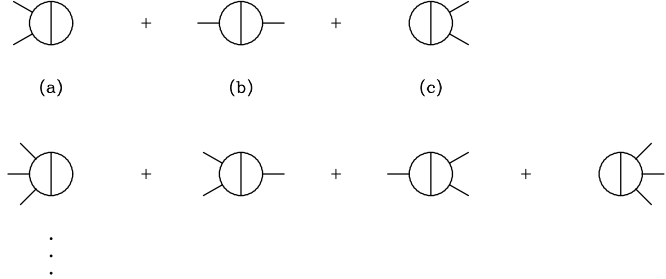}
\hfill}}
\caption[dum]{Summation of diagrams with N legs on the loop.
\hphantom{xxxxxxxxxxx}}
\label{sum2loopdiag}
\end{figure}

\no
In this way we arrive at the following two-loop
generalization of
eq. (\ref{scalarmaster}), 
 
\begin{eqnarray}
& &{\dps\int_0^{\infty}}
{dT\over T}
 {\dps\int_0^{\infty}}
d{\bar T}
e^{-m^2(T+\bar T)}
{(4\pi)}^{-D}
{\dps\int_0^T}d\tau_a
{\dps\int_0^T}d\tau_b
\,{[T\bar T + T
G_B(\tau_a,\tau_b)]}^{-{D\over 2}}\nonumber\\
&&\times\prod_{i=1}^N
\int_0^Td\tau_i\,
\exp\biggl[\sum_{k,l=1}^N G_B^{(1)}(\tau_k,\tau_l) k_k\cdot k_l
\biggr]
\label{2loopscalarmaster}
\end{eqnarray}
\no
For fixed $N$, this integral represents a 
certain linear combination of 
two-loop
diagrams in $\phi^3$ -- theory which have 
$N$ 
legs on the loop, and no 
leg on the internal line (fig. \ref{sum2loopdiag}).
 
\par
\par
\no
Observe that we have diagonal terms 
in the exponential,
since $G_B^{(1)}$ has
$G_B^{(1)}(\tau,\tau)\ne 0$ in general.
Self-contractions of vertex operators must therefore be taken into
account. As always, momentum conservation allows one to
absorb the diagonal terms into the non-diagonal ones,
however
in contrast to our previous experiences
the coincidence limit of $G_B^{(1)}$ is not constant.
To ensure that 
the subtracted Green's function 
has a zero coincidence limit, it
must now be defined
in the following way
\footnote
{We remark that an analogous ambiguity appears in the
electric circuit approach to Feynman parameter
integration ~\cite{bjorken,lamleb,lamqed}. In the terminology
of ~\cite{lamqed}
the condition $\bar G_B^{(1)}(\tau,\tau)=0$ corresponds
to the choice of a ``level zero scheme''.}

\begin{equation}
\bar G_B^{(1)}(\tau_1,\tau_2) \equiv
G_B^{(1)}(\tau_1,\tau_2)
-\half G_B^{(1)}(\tau_1,\tau_1)
-\half G_B^{(1)}(\tau_2,\tau_2)\, 
\label{renorm}
\end{equation}
\no
Explicitly this gives

\begin{figure}[ht]
\vbox to 4.5cm{\vfill\hbox to 15.8cm{\hfill
\epsffile{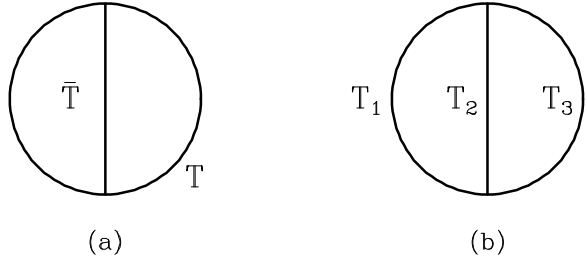}
\hfill}\vfill}\vskip-.4cm
\caption[dum]{Two different parametrizations 
of the two-loop diagram.
\hphantom{xxxxxxxxxx}}
\label{wlparametrizations}
\end{figure}

\bear
\bar G_B^{(1)}(\tau_1,\tau_2)&=&G_{B12}-{1\over 4}
{{\Bigl(G_{B1a}-G_{B1b}-G_{B2a}+G_{B2b}\Bigr)
}^2
\over
\bar T
+ G_{Bab}}
\non\\
\label{G2loopsub}
\ear
\no 
We note the following properties of this ``subtracted
two-loop worldline Green's function'':

\begin{enumerate}

\item
As one would expect $\bar G_B^{(1)}$ reduces to
$G_B$ in the limit where the proper-time $\bar T$ of
the inserted propagator becomes infinite.

\item
For the case that both points $\tau_{1,2}$ are 
located on the
same side of the propagator insertion
we can rewrite 
$\bar G_B^{(1)}$ as
the one-loop Green's function
$G_B$ with a modified global proper-time
$T\rightarrow T'$. In the parametrization
of fig. (\ref{wlparametrizations}b) this new
proper-time is given by
$T' = T_1 + {T_2 T_3\over T_2+T_3}$
\footnote{This property also has a direct analogue
in the electric circuit formalism.}.

\item
$\bar G_{B}^{(1)}$ is {\sl not} translation invariant, i.e. the equation

\bear
\Bigl(
{\partial\over\partial \tau_1} +
{\partial\over\partial \tau_2}
\Bigr)
\bar G_{B}^{(1)}(\tau_1,\tau_2)
&=&
0
\label{trala2loop}
\ear\no
is not true in general. It {\sl does} hold,
however, for $\tau_{1,2}$ on the same side of the propagator
insertion, as follows directly from the previous property.

\end{enumerate}

So far we are restricted to inserting vertex operators on the
loop only.
Due to the symmetry of the diagram, this restriction is
easily removed. Obviously, for any two points on the
two-loop vacuum graph we may regard those to be on
the loop, and the remaining branch -- or one of the remaining
branches -- to be the inserted line. We can thus
always use our formula eq.(\ref{G(1)}) up to
a reparametrization. In contrast to the string-theoretic
worldsheet Green's function the worldline Green's function
is a ``bi-scalar'', i.e. it transforms trivially under
reparametrizations.

\par

\par

\no
We therefore now find it convenient to
switch to the parametrization of fig. \ref{wlparametrizations}b, which is
symmetric with regard to the three branches. We
have then three ``moduli parameters'' $T_1,T_2$ and
$T_3$, and the location of a vertex operator on
branch $i$ will be denoted by a parameter
$\tau_i$ running from $0$ to $T_i$.

If we now  fix the
number of external legs on branch $i$ to be $n_i$, 
carrying momenta 
$k_i^{\sy (i)},\ldots,k_{n_i}^{\sy (i)}$, 
we obtain the following obvious generalization
of 
eq. (\ref{2loopscalarmaster}):
 
\begin{eqnarray}
 & &{\dps\prod_{a=1}^3}{\dps\int_0^{\infty}}{dT_a}
e^{-m^2(T_1+T_2+T_3)}
{(4\pi)}^{-D}
{(T_1T_2+T_1T_3+T_2T_3)}^{-{D\over 2}}\nonumber\\
&&\times\prod_{i=1}^{n_1}
\prod_{j=1}^{n_2}
\prod_{m=1}^{n_3}
\int_0^{T_1} d\tau_i^{\sy (1)}\,
\int_0^{T_2} d\tau_j^{\sy (2)}\,
\int_0^{T_3} d\tau_k^{\sy (3)}
\non\\&&\times
\exp\Biggl[
\sum_{r<s}
\sum_{k=1}^{n_r}
\sum_{l=1}^{n_s}
G_{Brs}^{\rm sym}(\tau_k^{\sy (r)},\tau_l^{\sy (s)})
k_k^{\sy (r)}\cdot k_l^{\sy (s)}
+
\sum_{r=1}^3
\half\sum_{k,l=1}^{n_r}
G_{Brr}^{\rm sym}(\tau_k^{\sy (r)},
\tau_l^{\sy (r)})k_k^{\sy (r)}\cdot k_l^{\sy (r)} 
\Biggr]\nonumber\\
\label{2loopsym}
\end{eqnarray}
\no
Here the 
$G_{B11}^{\rm sym},G_{B33}^{\rm sym},G_{B13}^{\rm sym}$ are
related to
$G_B^{(1)}$ by the mentioned reparametrization.
Again it is convenient to 
absorb the diagonal coincidence terms from
the beginning via eq.(\ref{renorm}).
After this subtraction, the
two-loop worldline Green's function in
symmetric parametrization becomes

\begin{eqnarray}
\bar G_{B11}^{\rm sym}(\tau_1^{\sy (1)},
\tau_2^{\sy (1)}) &=& \Delta
\mid \tau_1^{\sy (1)} - \tau_2^{\sy (1)}\mid\biggl
[(T_1 - \mid\tau_1^{\sy (1)} - \tau_2^{\sy (1)}
\mid )(T_2+T_3) + T_2 T_3\biggr]
\nonumber\\
&=& \mid \tau_1^{\sy (1)} - \tau_2^{\sy (1)}\mid
-\,\Delta (T_2+T_3)
{\bigl({\tau_1^{\sy (1)} - 
\tau_2^{\sy (1)}}\bigr)}^2\nonumber\\
\bar G_{B12}^{\rm sym}(\tau^{\sy (1)},
\tau^{\sy (2)}) &=&
\Delta\Bigl [
T_3(\tau^{\sy (1)}+\tau^{\sy (2)})\bigl[T_1+T_2-
(\tau^{\sy (1)}+\tau^{\sy (2)})\bigr]\nonumber\\
&&
+\, \tau^{\sy (2)}(T_2-\tau^{\sy (2)})T_1 
+\tau^{\sy (1)}(T_1-\tau^{\sy (1)})T_2\Bigr ]
\nonumber\\
&=&\tau^{\sy (1)}+\tau^{\sy (2)}
-\,\Delta\Bigl[{\tau^{\sy (1)}}^2 T_2 
+{\tau^{\sy (2)}}^2 T_1 
+{(\tau^{\sy (1)} + \tau^{\sy (2)})}^2 T_3\Bigr]\nonumber\\
\Delta &=& {(T_1T_2+T_1T_3+T_2T_3)}^{-1}\nonumber\\
\label{Gsymfin}
\end{eqnarray}
\no
plus permuted $\bar G_{Bij}$'s.

\subsection{Comparison with Feynman Diagrams}

\begin{figure}[ht]
\vbox to 5.7cm{\vfill\hbox to 15.8cm{\hskip1cm\hfill
\epsffile{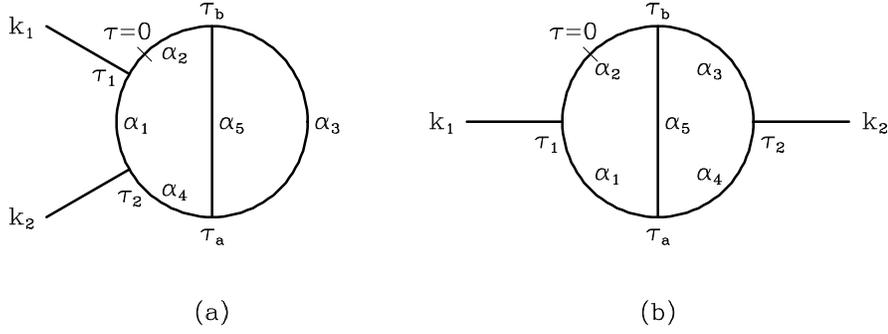}
\hfill}\vfill}\vskip-1cm
\caption[dum]{Reparametrization of the 
two-point two-loop diagrams.
\hphantom{xxxxxxxxxxxxxxxx}}
\label{2reparametrize}
\end{figure}

Let us look more closely at the two-point case, and
compare our approach
with the corresponding Feynman 
diagram calculation.
For $N=2$, eq.(\ref{2loopscalarmaster})
reads

\begin{eqnarray}
& &{\dps\int_0^{\infty}}
{dT\over T}
 {\dps\int_0^{\infty}}
d{\bar T}
e^{-m^2(T+\bar T)}
{(4\pi)}^{-D}
{\dps\int_0^T}d\tau_a
{\dps\int_0^T}d\tau_b
\,{[T\bar T + T
G_B(\tau_a,\tau_b)]}^{-{D\over 2}}\nonumber\\
&&\times
\int_0^Td\tau_1\,\int_0^Td\tau_2
\exp\biggl[\Bigl(
{1\over 2}G_B^{(1)}(\tau_1,\tau_1)
+{1\over 2}G_B^{(1)}(\tau_2,\tau_2)
-G_B^{(1)}(\tau_1,\tau_2) 
\Bigr)k^2\biggr]
\label{2loop2point}
\end{eqnarray}
\no
This should
correspond to the sum of graphs
(a), (b) and (c) of fig. \ref{sum2loopdiag}.
A straightforward Feynman parameter calculation of
graph (a) results in
($k=k_1=-k_2$)
\footnote{$P^{(a)}$ was misprinted in \cite{ss2}.}
 
\begin{equation}
{\dps\int_{0}^{\infty}}{d\hat T}\,
{(4\pi)}^{-D}e^{-m^2\hat T}
\,{\dps\prod_{i=1}^{5}}{\dps\int}d\alpha_i
\,\delta (\hat T - \sum_{i=1}^5\alpha_i)
\,{\bigl[ P^{(a)}(\alpha_i)\bigr ]}^{-{D\over 2}}
\,{\exp\bigl[-Q^{(a)}(\alpha_i)k^2\bigr]}
\label{feyn(a)}
\end{equation}
\no
with
\begin{eqnarray}
P^{(a)} &=& \alpha_5 (\alpha_1+\alpha_2+\alpha_3+\alpha_4)
+\alpha_3 (\alpha_1 +\alpha_2 +\alpha_4)
     \nonumber\\
P^{(a)}Q^{(a)} &=& 
\alpha_1[\alpha_5(\alpha_2+\alpha_3+\alpha_4)
   +\alpha_2\alpha_3 + \alpha_3\alpha_4]\nonumber\\ 
\label{PQ(a)}
\end{eqnarray}
\no
The analogue of the transformation eq. (\ref{trafotaualpha}) 
can be 
directly read off fig. \ref{2reparametrize}a
 
\par
\par
 
\no
\begin{eqnarray}
\alpha_1 &=& \tau_1-\tau_2\nonumber\\
\alpha_2 &=& T-\tau_1+\tau_b\nonumber\\
\alpha_3 &=& \tau_a-\tau_b\nonumber\\
\alpha_4 &=& \tau_2-\tau_a\nonumber\\
\alpha_5 &=& \bar T\nonumber\\
\end{eqnarray}
\no
As expected, it leads to the identifications
\begin{eqnarray}
P^{(a)} &=& T\bar T
{\bigl[1+{1\over {\bar T}}
G_B(\tau_a,\tau_b)\bigr]}\nonumber\\
Q^{(a)} &=& \bar G_B^{(1)}(\tau_1,\tau_2)\, 
\nonumber\\
\label{ident}
\end{eqnarray}
The Feynman calculation for diagram (b) yields polynomials
 
\begin{eqnarray}
P^{(b)} &=& \alpha_5 (\alpha_1+\alpha_2+\alpha_3+\alpha_4)
     + (\alpha_1+\alpha_2)(\alpha_3+\alpha_4)\nonumber\\
P^{(b)}Q^{(b)} &=& 
\alpha_5(\alpha_2+\alpha_3)(\alpha_1+\alpha_4)
    +\alpha_1\alpha_2(\alpha_3+\alpha_4)
    +\alpha_3\alpha_4(\alpha_1+\alpha_2)\nonumber\\
\label{PQ(b)}
\end{eqnarray}
\no
which are different as functions of the 
variables $\alpha_i$, but
after the corresponding transformation
 
\begin{eqnarray}
\alpha_1 &=& \tau_1 -\tau_a\nonumber\\
\alpha_2 &=& \tau_b + T - \tau_1\nonumber\\
\alpha_3 &=& \tau_2-\tau_b\nonumber\\
\alpha_4 &=& \tau_a-\tau_2\nonumber\\
\alpha_5 &=& \bar T\nonumber\\
\end{eqnarray}
\no
identify with the {\sl same} expressions ~(\ref{ident}).
It should be noted that this becomes apparent only 
after
everything has been expressed in terms of $G_B$, due to the
absolute sign contained in that function.
 
Our worldline formula
eq.~(\ref{2loop2point})
thus indeed 
unifies the three $\alpha$ -- parameter integrals,
arising in the calculation of diagrams (a), (b) and (c),
in a single $\tau$ -- parameter integral.
This correspondence has also been checked for 
a number of diagrams
with more external legs, as well as for diagrams with
legs on all three branches.

This is remarkable since in field theory
diagrams (a) and (b) have very different properties.
Both in the massless and in the massive
cases the integrals arising from topology (b) are
less elementary than the ones from (a). To
understand how this comes about we need only
remember a property of $\bar G_B^{(1)}$ stated
above, namely that it is translation invariant
for (a) but not for (b). Therefore for (a) the 
parameter integral has a redundancy, and
eq.(\ref{trala2loop}) can be used to
reduce the number of integrations by one.
 
Quite obviously we have found here a universality
property which is not visible in ordinary Feynman
parameter calculations, and constitutes a field theory
relic of the fact mentioned in the
introduction, namely that string perturbation theory
does not suffer from the usual proliferation of terms
due to the existence of many different topologies.
  
\subsection{Higher Loop Orders}

The whole procedure generalizes without difficulty 
to the case
of $m$ propagator insertions, resulting in an integral
representation combining into one expression
all diagrams with $N$ legs on the
loop, and $m$ inserted propagators:

\begin{eqnarray}
&&{\dps\int_{0}^{\infty}}{dT\over T}
T^{-{D\over2}}
(4\pi)^{-(m+1){D\over 2}}
\prod_{j=1}^m\int_0^{\infty}d\bar T_j
e^{-m^2(T+\sum_{j=1}^m\bar T_j)}
\int_0^Td\tau_{a_j}\int_0^Td\tau_{b_j}\nonumber\\
&&\times\prod_{i=1}^N
\int_0^Td\tau_i\,
{N^{(m)}}^{{D\over 2}}
\exp\Bigl[\half\sum_{k,l=1}^N G_B^{(m)}
(\tau_k,\tau_l) k_k\cdot k_l\Bigr]
\label{kinsert}
\end{eqnarray}
\no
where
\begin{eqnarray}
N^{(m)} &=& {\rm Det}(A^{(m)})           \nonumber\\
G_B^{(m)}(\tau_1,\tau_2) &=& G_B(\tau_1,\tau_2)
\nonumber\\
&&+\half {\dps \sum_{k,l=1}^m}
\bigl[G_B(\tau_1,\tau_{a_k})-G_B(\tau_1,\tau_{b_k})\bigr]
A^{(m)}_{kl}
\bigl[G_B(\tau_2,\tau_{a_l})-G_B(\tau_2,\tau_{b_l})\bigr]
           \nonumber\\
\label{nkgk}
\end{eqnarray}
\no
and the symmetric $m\times m$ -- matrix $A^{(m)}$ is defined by

\vspace{-1pt}
\begin{eqnarray}
A^{(m)} &=& {\Bigl[\bar T - {C\over 2}\Bigr]}^{-1}
\nonumber\\
\bar T_{kl} &=& \bar T_k \delta_{kl}\nonumber\\
C_{kl} &=& G_B(\tau_{a_k},\tau_{a_l})
- G_B(\tau_{a_k},\tau_{b_l})
- G_B(\tau_{b_k},\tau_{a_l})
+ G_B(\tau_{b_k},\tau_{b_l})\, \nonumber\\
\label{defA}
\end{eqnarray}
\noindent
Here
$\bar T_1,\ldots,\bar T_m$ denote the proper-time lengths
of the inserted propagators.

The coincidence terms can be subtracted as in the
two-loop case, leading to ``subtracted'' Green's
functions $\bar G^{(m)}_B$ related to the 
$G^{(m)}_B$ via the same eq.(\ref{renorm}).
Both choices lead to the same scattering
amplitudes.

Of course
one is free to choose the masses of the
inserted propagators to be different from each other, and from
the loop mass $m$. 
To give different masses to the individual field theory
propagators making up the loop is 
also possible, albeit only if one fixes the ordering
\be
\tau_{i_1}>\tau_{i_2}>\ldots
\label{fixordering}
\ee\no
of the interaction points around the loop. 
In this case, instead of the global factor
of $e^{-m^2T}$ one has to insert one factor of

\be
e^{-m_j^2(\tau_{i_j}-\tau_{i_{j+1}})}
\label{givemass}
\ee\no
for every massive propagator, where $m_j$ denotes the
mass for the propagator connecting $\tau_{i_j}$
and $\tau_{i_{j+1}}$.

Note that the formula above gives the 
worldline correlator only between
points on the loop. Beyond the
two -- loop level, the construction
of the correlators involving points
on the inserted propagators cannot
be achieved by symmetry arguments any more.
This extension was studied
by Roland and Sato \cite{rolsat:npb480,rolsat:npb515},
who obtained explicit formulas similar to
eqs.(\ref{nkgk}),(\ref{defA}) for the
Green's function between arbitrary points on
the same class of graphs.
This knowledge then
is sufficient
to write down
worldline representations for all $\phi^3$ graphs 
which have the topology of a loop with insertions. 
According to graph theory ~\cite{barnette},
the set of such graphs is surprisingly large.
For the first few orders of perturbation theory
such a loop, or ``Hamiltonian circuit'', can always
be found; all
trivalent graphs with less than 34 vertices do have this property
\footnote{This statement assumes that we disallow insertions of
the trivial one-loop propagator bubble graph.
(I thank D. Kreimer for pointing this out to me.)
}.
This would, of course, do for most practical
purposes, but still poses a problem in theory.

But there is a more bothersome
problem, which so far we have swept under
the carpet.
In checking the correspondence to Feynman graph calculations,
we verified that the correct integrands were produced,
but left aside the global statistical and symmetry factors
for the individual diagrams. 
As it turns out those do not work out
in the case of $\phi^3$ -- theory. Even in the simple example
analyzed above the complete $\tau$ -- integral contains
diagram (a) and (b) in a ratio of 2 : 1, while in field
theory one would have a ratio of 1 : 1.

Both these problems can be solved at the same time by 
an appropriate reformulation of the theory at the field
theory level \cite{satsch1}. 
This can be done in various ways. In
$\lambda\phi^3$ -- theory, the basic idea is to
rewrite the generating functional 

\be
Z[\eta]=\int{\cal D}\phi(x)\,
{\rm exp}
\biggl \lbrace
\int\,dx
\Bigl[-
\half 
\phi(x)
( -\Box + m^2)
\phi(x)
+\eta(x)\phi(x)
-\lambda
\phi^3(x)
\Bigr]
\biggr\rbrace
\label{defZ(eta)}
\ee\no
in the following way,

\be
Z[\eta]
=\int
{\cal D}A(x){\cal D}\phi (x)
\,\delta(A-\phi)
\,{\rm exp}
\biggl \lbrace
\int\,dx
\Bigl[-
\half 
\phi(x)
( -\Box + m^2)
\phi(x)
+\eta(x)\phi(x)
-\lambda
A(x)\phi^2(x)
\Bigr]
\biggr\rbrace
\label{rewriteZ}
\ee\no
After a Fourier transformation of the functional
$\delta$ -- function one finds that the new
scalar field theory obtained does not suffer from
the above problems, since the
interaction between $\phi$ and the
auxiliary field $A$ is of the Yukawa type.
For this theory the whole S-matrix can be exhausted
by Hamiltonian graphs, and moreover
letting ``legs slide around loops''
generates the correct statistical factors.
A similar reformulation exists for Yang-Mills theory \cite{satsch2}.
However the practical value of this procedure has not been
established yet, and we will not pursue this matter further here.

It is interesting to compare these difficulties in the
correct generation of the set of all Feynman diagrams to
the situation in string perturbation 
theory. There the apparent advantage
of being able to write down the full amplitude ``in one piece''
offered by the Polyakov path integral approach may, at higher loop
orders, become increasingly illusory due to the absence of a
convenient global parametrization of the corresponding moduli
spaces. The cell decomposition of moduli space provided by
second quantized string field theory \cite{zwiebach} may then turn into 
an advantage.

\subsection{Connection to String Theory}

Roland and Sato \cite{rolsat:npb480} 
provided a link back to string theory by 
analyzing the infinite string tension limit of
the Green's function $G_B^{RS(m)}$ of the corresponding 
Riemann surface, and identifying 
$\bar G_B^{(m)}$ with the
leading order term of $G_B^{RS(m)}$
in the ${1\over \alpha'}$ -- expansion:

\be
G_B^{RS(m)}(z_1,z_2)\quad 
{\stackrel{\alpha' \rightarrow 0}{\longrightarrow}}
\quad {1\over \alpha'}
\bar G_B^{(m)}(\tau_1,\tau_2) + {\rm finite}\quad
\label{Gasympt}
\ee\no
Note that their derivation automatically leads to the
``subtracted'' version of the multiloop Green's functions.

\subsection{Example: 3-Loop Vacuum Amplitude}
\label{3loopexample}

Finally, let 
us have a look at
the simplest example of a three -- loop
parameter integral calculation in this formalism
\cite{rheinproc}.
This is the one where
the integrand consists just of the 
bosonic three-loop determinant factor
${(N^{(2)})}^{D\over 2}$ (see eq.(\ref{kinsert})).
In dimensional regularization it reads

\be
{\Gamma}^{(3)}_{\rm vac}(D)=
(4\pi)^{-{3\over 2}D}\Tint \e^{-m^2 T}
T^{6-{3\over 2}D}I(D)
\label{defI3}
\ee\no

\begin{equation}
I(D)=\int_0^{\infty}\!\!\! d\hat T_1\, d\hat T_2
\int_0^1 da\, db\, dc\, dd\,
{\biggl[(\hat T_1+G_{Bab})(\hat T_2+G_{Bcd})-{C^2\over 4}\biggr]}
^{-{D\over 2}}
\label{defID}
\end{equation}

\noindent
Here $\hat T_{1,2}={T_{1,2}\over T}$ denote the proper-time lengths of the
two inserted propagators in units of $T$, and
$C\equiv G_{Bac}-G_{Bad}-G_{Bbc}+G_{Bbd}$.

In the following we will show how to compute
the ${1\over\epsilon}$ - pole of this amplitude,
which is the quantity needed for applications
to the calculation of renormalization group functions.

\begin{figure}[ht]
\begin{center}
\begin{picture}(5000,0)%
\epsfig{file=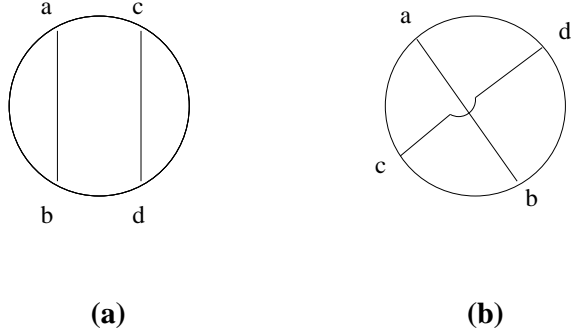}%
\end{picture}%
\setlength{\unitlength}{0.00087500in}%
\begingroup\makeatletter\ifx\SetFigFont\undefined
\def\x#1#2#3#4#5#6#7\relax{\def\x{#1#2#3#4#5#6}}%
\expandafter\x\fmtname xxxxxx\relax \def\y{splain}%
\ifx\x\y   
\gdef\SetFigFont#1#2#3{%
  \ifnum #1<17\tiny\else \ifnum #1<20\small\else
  \ifnum #1<24\normalsize\else \ifnum #1<29\large\else
  \ifnum #1<34\Large\else \ifnum #1<41\LARGE\else
     \huge\fi\fi\fi\fi\fi\fi
  \csname #3\endcsname}%
\else
\gdef\SetFigFont#1#2#3{\begingroup
  \count@#1\relax \ifnum 25<\count@\count@25\fi
  \def\x{\endgroup\@setsize\SetFigFont{#2pt}}%
  \expandafter\x
    \csname \romannumeral\the\count@ pt\expandafter\endcsname
    \csname @\romannumeral\the\count@ pt\endcsname
  \csname #3\endcsname}%
\fi
\fi\endgroup
\begin{picture}(3634,2000)(201,0)
\end{picture}
\vspace{5pt}
\caption{\label{3looptop} 
Planar and non-planar sectors.}
\end{center}
\end{figure}

In writing eq.(\ref{defI3}) we have already rescaled to
the unit circle, and separated off the
electron proper-time integral.
This integral decouples, and just yields
an overall factor of

\begin{equation} 
\int_0^{\infty}{dT\over T}{\rm e}^{-m^2T}T^{6-{3\over 2}D}
=\Gamma(6-{3\over2}D) \, m^{3D-12}
\sim -{2\over 3\epsilon}\quad 
\end{equation}
($\epsilon = D-4$).
The nontrivial integrations are 
$\int_0^1da\,db\,dc\,dd
\equiv\int_{abcd}$, 
representing the four propagator
end points 
moving around the 
loop (fig. \ref{3looptop}). 

\noindent
This fourfold
integral decomposes into
24 ordered sectors, of which 16 
constitute the planar (P)
(fig. \ref{3looptop}a) and 8 the non-planar (NP)
sector (fig. \ref{3looptop}b). Due to the symmetry
properties of the integrand, all sectors
of the same topology give an equal
contribution.  
The integrand has a trivial invariance
under the operator
${{\partial}\over{\partial} a}+
{{\partial}\over{\partial} b}+
{{\partial}\over{\partial} c}+
{{\partial}\over{\partial} d}
$,
which just shifts the location of the
zero on the loop.

As a first step in the calculation 
of $I(D)$, it is useful to
add and subtract the same integral
with $C=0$ and rewrite 

\be
I(D)=I_{\rm sing}(D)+I_{\rm reg}(D)
\label{splitI(D)}
\ee\no
\begin{eqnarray}
&& I_{\rm sing}(D) =
\int_0^{\infty}
d\hat T_1\,d\hat T_2
\int\limits_{\hspace{4mm} abcd}
{\Bigl[(\hat T_1+G_{Bab})(\hat T_2+G_{Bcd})\Bigr]}
^{-{D\over 2}}
\label{defIsing}
\end{eqnarray}

\noindent
$I_{\rm sing}(D)$ 
factorizes into two 
identical three-parameter
integrals, which are elementary:

\begin{equation}
I_{\rm sing}(D)=
{\biggl\lbrace
\int_0^{\infty}
dT\int_0^1 du
{\Bigl[
T+u(1-u)
\Bigr]}^{-{D\over 2}}
\biggr\rbrace}^2=
\biggl[{2B(2-{D\over 2},2-{D\over 2})\over D-2}\biggr]^2
\end{equation}

\noindent
The point of this split is that the remainder
$I_{\rm reg}(D)$ is finite. To see this,
set $D=4$, 
expand the original integrand in 
${C^2\over G_{Bab}G_{Bcd}}$, and note
that for all terms but the first one
the zeroes of $G_{Bab}$ ($G_{Bcd}$)
at $a\sim b$ ($c\sim d$) are 
neutralized
by
zeroes of $C^2$.
Since we want only the $1\over\epsilon$ -- pole  
of $\Gamma_{\rm vac}^{(3)}$
we can set $D=4$ 
in the calculation of $I_{\rm reg}(D)$.
The integrations over $\hat T_1,\hat T_2$ are then elementary,
and we are left with

\begin{eqnarray}
I_{\rm reg}(4) &=& 
\int\limits_{\hspace{4mm}abcd}
\biggl[-{4\over C^2}{\rm ln}
\Bigl(1-{C^2\over 4G_{Bab}G_{Bcd}}\Bigr)
-{1\over G_{Bab}G_{Bcd}}\biggr]
\label{Ireg(4)}
\end{eqnarray}

\noindent
For the calculation of this integral, observe
the following simple behaviour of the function $C$
under the operation
$D_{ab}\equiv {{\partial}\over{\partial}\tau_a}
+{{\partial}\over{\partial}\tau_b}$:

\begin{equation}
D_{ab}C= \pm 2\chi_{NP},\;
D_{ab}^2C= 2
\Bigl(
\delta_{ac}-\delta_{ad}-\delta_{bc}+\delta_{bd}
\Bigr)
\label{DC}
\end{equation}

\noindent
where 
$\chi_{NP}$ denotes the 
characteristic function
of the non-planar sector
(i.e. it is zero on the planar and
one on the non-planar sector).
From these identities and the symmetry properties 
one can easily derive the
following projection identities, which effectively
integrate out the variable $C$:

\begin{eqnarray}
\int_P f(C,G_{Bab},G_{Bcd})
&=&4\int_0^1da\int_0^adc(a-c)
f\Bigl(-2c(1-a),a-a^2,c-c^2\Bigr)\non\\
\int_{NP}f(C,G_{Bab},G_{Bcd})&=&
-4\int_0^1da\int_0^adc\int_0^{-2c(1-a)}
dCf\Bigl(C,a-a^2,c-c^2\Bigr)
\non\\
\label{project}
\end{eqnarray}
\noindent
Here $f$ is an arbitrary function in the variables
$G,G_{Bab},G_{Bcd}$, and 
$\int_0^C dC f$ denotes the integral of this
function in the variable $C$, with the other variables fixed.
The integrals on the left hand side are restricted to the
sectors indicated.
For $f$ the integrand of our formula eq.~(\ref{Ireg(4)}),
we have

\begin{eqnarray}
\int_0^C dC f & = &-{C\over G_{Bab}G_{Bcd}}+
{4\over C} \ln \Bigl( 1 -{C^2\over4 G_{Bab}G_{Bcd}} \Bigr) 
\nonumber\\
& & \hspace{1cm} + {4\over\sqrt{G_{Bab}G_{Bcd}}
} \, \mbox{arctanh} \Bigl( {1\over2} {C\over
\sqrt{G_{Bab}G_{Bcd}}} \Bigr)
\label{integrand}
\end{eqnarray}

\noindent
Inserted in the second equation of (\ref{project}) this
leaves us with three two--parameter integrals, of
which the first one is elementary.
Applying the substitution

\begin{equation}
y={c(1-a)\over a(1-c)}  
\end{equation}
\noindent
to the second integral, and

\begin{equation}
y^2={c(1-a)\over a(1-c)}
\end{equation}
\no
to the third integral, those are
transformed into known standard integrals,
tabulated for instance in ~\cite{devduk}.
The result is

\begin{equation}
\int_{NP}f=12\zeta(3) -8\zeta(2) \quad 
\end{equation}

\noindent
The calculation in the planar sector is elementary, and we just
give the result,

\begin{equation}
\int_Pf = 4\zeta(2)-4 \quad 
\end{equation}

\noindent
Putting the pieces together, we have,
up to terms of order $O(\epsilon^0)$,

\begin{equation}
{\Gamma}^{(3)}_{\rm vac}(D)=
m^{3D-12}(4\pi)^{-{3\over 2}D}
\Gamma(6-{3\over 2}D)\,
\biggl\lbrace
\Bigl[{2B(2-{D\over 2},2-{D\over 2})\over D-2}\Bigr]^2
+12\zeta(3)-4\zeta(2)-4
\biggr\rbrace
\end{equation}
\vskip-.2cm

\begin{figure}[ht]
\begin{center}
\begin{picture}(5000,0)%
\epsfig{file=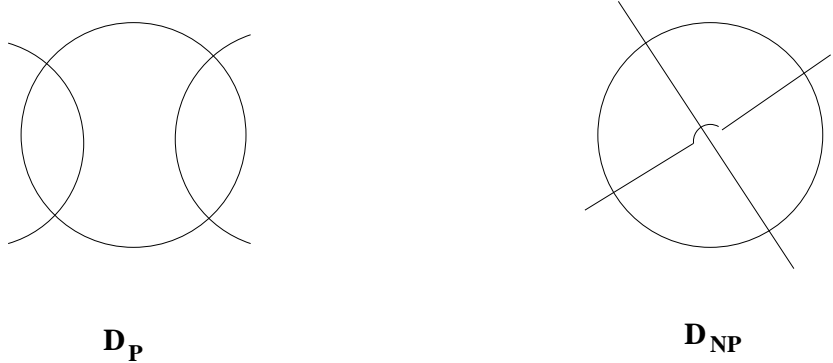}%
\end{picture}%
\setlength{\unitlength}{0.00087500in}%
\begingroup\makeatletter\ifx\SetFigFont\undefined
\def\x#1#2#3#4#5#6#7\relax{\def\x{#1#2#3#4#5#6}}%
\expandafter\x\fmtname xxxxxx\relax \def\y{splain}%
\ifx\x\y   
\gdef\SetFigFont#1#2#3{%
  \ifnum #1<17\tiny\else \ifnum #1<20\small\else
  \ifnum #1<24\normalsize\else \ifnum #1<29\large\else
  \ifnum #1<34\Large\else \ifnum #1<41\LARGE\else
     \huge\fi\fi\fi\fi\fi\fi
  \csname #3\endcsname}%
\else
\gdef\SetFigFont#1#2#3{\begingroup
  \count@#1\relax \ifnum 25<\count@\count@25\fi
  \def\x{\endgroup\@setsize\SetFigFont{#2pt}}%
  \expandafter\x
    \csname \romannumeral\the\count@ pt\expandafter\endcsname
    \csname @\romannumeral\the\count@ pt\endcsname
  \csname #3\endcsname}%
\fi
\fi\endgroup
\begin{picture}(3624,2300)(214,-100)
\end{picture}
\vspace{15pt}
\caption{\label{3loopsum} 
3-loop $\phi^4$ vertex diagrams.}
\end{center}
\end{figure}

This calculation method 
generalizes in an obvious
way to the tensor integrals which appear in worldline
calculations of three-loop renormalization group functions
in other abelian theories such as QED or the Yukawa
model.
The basic scalar integral considered
here appears in
the calculation of the 3~-~loop $\beta$~--~function
for $\phi^4$ - theory. This permits an easy check
of the above calculation against a Feynman diagram
calculation. 
In diagrammatic terms, the integral
${\Gamma}^{(3)}_{\rm vac}(D)$ corresponds to a weighted sum of the
two scalar 3-loop vertex diagrams depicted
in fig. \ref{3loopsum},
calculated at zero external momentum, with
massive propagators along the loop, and
massless propagator insertions.

\noindent
It is not difficult to verify that, with
an appropriate normalization, the relation

\begin{equation}
{\Gamma}^{(3)}_{\rm vac}(D)
 = 16D_P + 8D_{NP} 
\end{equation}
\vskip-.2cm
\noindent
indeed 
holds true for the singular parts of the $1\over\epsilon$ -- expansions.

\section{The QED Photon S-Matrix}
\renewcommand{\theequation}{9.\arabic{equation}}
\setcounter{equation}{0}

We generalize this ``multiloop worldline formalism''
to the case of photon scattering in quantum electrodynamics
~\cite{ss3,rescsc}.

\subsection{The Single Scalar Loop}

We begin with studying scalar electrodynamics at the two-loop level,
i.e. a scalar loop with an internal photon correction. 
A photon insertion in
the worldloop may, in Feynman gauge, be represented in terms of
the following current-current interaction term 
inserted into
the one-loop path integral,

\begin{equation}
-{e^2\over 2}
{{\Gamma (\lambda )}\over {4{\pi}^{\lambda +1}}}
\int_0^T d\tau_a \int_0^T d\tau_b
{{\dot x(\tau_a)\cdot\dot x(\tau_b)}\over
{\Bigl({[x(\tau_a) - x(\tau_b)]}^2\Bigr)}^{\lambda}}\quad 
\label{cci}
\end{equation}

\noindent
($\lambda = {D\over2} -1$).
This is essentially still Feynman's formula
eq.(\ref{feynform}), except that we
have rewritten it in $D$ dimensions,
and Euclidean conventions.

As in the case of the scalar propagator,
we can exponentiate the offending
``non-Gaussian'' denominator,

\begin{equation}
{\Gamma(\lambda )\over {4{\pi}^{\lambda +1}}
{\Bigl({[x(\tau_a) - x(\tau_b)]}^2\Bigr)}^{\lambda}}
=\int_0^{\infty} d\bar T 
{(4\pi \bar T)}^{-{D\over 2}}
{\rm exp}\Biggl
[-{{\Bigl(x(\tau_a)-x(\tau_b)\Bigr)}^2\over 4\bar T}
\Biggr ]\, 
\label{intpropchap8}
\end{equation}
\no
and absorb it into the worldline Green's function.
We obtain then, of course, the same 2-loop worldline
Green's function eq.(\ref{G(1)}) and
determinant factor eq.(\ref{normchange})
as before. 
The numerator $\dot x_a\cdot\dot x_b$
remains, and will participate in the Wick
contractions. 

This treatment of the photon propagator may appear somewhat
unnatural, but will be seen to work quite well in practice.
Moreover, only this procedure will enable us to use the
same universal Green's functions both for scalar field theory
and gauge theory calculations. 

As in the scalar field theory case, the
generalization from one-loop to two-loop
calculations of photon amplitudes
in scalar QED 
requires no changes of the formalism itself,
but only of the Green's functions used, and of
the global determinant factor. 

The generalization to an arbitrary fixed number of photon insertions
is obvious.
To obtain a parameter integral
representation for the sum of all diagrams with one
scalar loop and fixed numbers of photons, N external
and m internal, we have to Wick contract N 
photon vertex operators, together
with m factors of
$\int_0^Td\tau_a  \int_0^T d\tau_b\,
\dot x_a\cdot\dot x_b$, using the 
$(m+1)$ -- loop Green's function $G^{(m)}$:

\bear
\Gamma_{\rm scal}^{(m+1)}
[k_1,\varepsilon_1;\ldots;k_N,\varepsilon_N]
&=&{(-ie)}^N
{(-{e^2\over 2})}^m
\Tint\e^{-m^2T}
T^{-{D\over 2}}
{(4\pi)}^{-(m+1){D\over 2}}
\prod_{j=1}^m\int_0^{\infty}d\bar T_j
\nonumber\\
&&
\!\!\!\!\!\!\!\!\!\!\!\!
\times
\int_0^Td\tau_{a_j}\int_0^Td\tau_{b_j}
{N^{(m)}}^{{D\over 2}}
\Bigl\langle
\prod_{j=1}^m
\dot x_{a_j}\cdot\dot x_{b_j}
V_{{\rm scal},1}^A\cdots
V_{{\rm scal},N}^A
\Bigr\rangle
\non\\
\label{Nphotonwickscal(m)}
\ear\no
This is our $(m+1)$ -- loop generalization of
the one -- loop 
photon scattering formula
eq.(\ref{Nphotonwickscal})
(to avoid a further complication of
nomenclature, it should 
simply be understood in the following that this is
only the ``quenched'' part of the amplitude).
As in the one-loop case, one could 
immediately translate this into a
master formula of the type
eq.(\ref{scalarqedmaster}).

It is important to note that precisely the same
integral representation could be obtained starting
from the one-loop formula
eq.(\ref{Nphotonwickscal})
with $(N + 2m)$ external photons, and then
sewing together $m$ pairs of them, using
Feynman gauge. Of course this would be much more
laborious.
This also explains why the multiloop Green's
functions can be rewritten in terms of the
one-loop Green's function $G_B$.
It has the nontrivial
consequence that one is still
allowed to use the one-loop replacement rule;
after writing out the result of the Wick -- contractions
in terms of the one-loop Green's function
$G_{Bij}$, one can generate the corresponding
spinor loop integrand by the usual partial
integration routine, and use of
eq.(\ref{fermion}).

While this multiloop construction is done most simply using
Feynman gauge for the propagator insertions, other
gauges can be implemented as well (the gauge freedom was also
discussed in ~\cite{dashsu}).
In an arbitrary covariant gauge, the photon insertion term
eq.(\ref{cci}) would read

\begin{eqnarray}
-{e^2\over 2}
{1\over {4{\pi}^{{D\over2}}}}
\int_0^T d\tau_a \int_0^T d\tau_b
\Biggl\lbrace
{{1+\alpha}\over 2}\Gamma\Bigl({D\over 2}-1\Bigr)
{{\dot x_a\cdot\dot x_b\over
{\Bigl[{(x_a - x_b)}^2\Bigr]}^{{D\over 2}-1}}\quad
}\nonumber\\
+(1-\alpha)\Gamma\Bigl({D\over 2}\Bigr)
{\dot x_a\cdot(x_a-x_b)
(x_a-x_b)\cdot\dot x_b\over
{\Bigl[{(x_a - x_b)}^2\Bigr]}^{D\over 2}\quad
}
\Biggr\rbrace
\label{ccigengauge}
\end{eqnarray}
\noindent
Here $\alpha = 1$ corresponds to Feynman gauge,
$\alpha = 0$ to Landau gauge.
The integrand may also be written as

\begin{equation}
\Gamma\Bigl({D\over 2}-1\Bigr)
{{\dot x_a\cdot\dot x_b\over
{\Bigl[{(x_a - x_b)}^2\Bigr]}^{{D\over 2}-1}}\quad
}\nonumber\\
-{1-\alpha\over 4}
\Gamma\Bigl({{D\over 2}-2}\Bigr)
{\partial\over\partial\tau_a}
{\partial\over\partial\tau_b}
{\Bigl[{(x_a-x_b)}^2\Bigr]}
^{2-{D\over2}}
\label{ccigengaugetotdir}
\end{equation}
\noindent
This shows that, on the worldline, gauge transformations
correspond to the addition of total derivative
terms. This form of the
photon insertion is also the more practical one
for actual calculations.
The power of ${(x_a-x_b)}^2$ appearing in
the second term is then again to be exponentiated.

Before leaving this section, let us mention
that sometimes it can be useful to
exponentiate also the numerator of
the inserted Feynman propagator, 
rewriting

\be
\dot x_a\cdot \dot x_b
=
\half
\lim_{\tau_a'\to \tau_a}
\lim_{\tau_b'\to \tau_b}
\lim_{\alpha\to 0}
{\partial\over \partial\alpha}
{\partial\over\partial \tau_a'}
{\partial\over\partial \tau_b'}
\,\e^{-\alpha{(x_{a'}-x_{b'})}^2}
\label{dettrick}
\ee\no
This little 
point-splitting trick turns out to be
surprisingly useful for 
the computerization of the algorithm
~\cite{dennythesis}.
For example, it allows one
to generate the
integrand for the 3-loop 
scalar QED vacuum amplitude
by differentiations performed
on the 5-loop determinant
factor $N^{(4)}$, instead of 
Wick -- contractions at the
3 -- loop level.

\subsection{The Single Electron Loop}

As in the one-loop case, the transition to spinor
electrodynamics is most simply accomplished by 
supersymmetrization.
According to the supersymmetrization rules, the
photon insertion eq.(\ref{cci})
generalizes to the
spinor loop as follows,

\begin{equation}
{e^2\over 2} 
{{\Gamma (\lambda )}\over {4{\pi}^{\lambda +1}}}
\int_0^Td\tau_a \, d\theta_a
\int_0^Td\tau_b \, d\theta_b
{{DX_a\cdot D{X_b}}\over
{({(X_a - X_b)}^2)}^{\lambda}}\quad 
\label{ccisup}
\end{equation}

\noindent
The simplest way to verify the correctness of
this expression
is to write the one-loop two-photon amplitude in the
super-formalism, and then sewing together the external legs
to create an internal photon, using Feynman gauge.

Just as a demonstration of the usefulness of the
superfield formalism, let us rewrite the double integral in
components:

\begin{eqnarray}
&&
\int_0^Td\tau_a\int_0^Td\tau_b
\Biggl\lbrace
-{{\dot x_a^{\mu}\dot {x_b}_{\mu}}\over
{({(x_a-x_b)}^2)}^{\lambda}}
-4\lambda{{(x_a^{\mu}-x_b^{\mu})
(\psi_b^{\mu}\psi_b^{\nu}\dot {x_a}_{\nu}
-
\psi_a^{\mu}\psi_a^{\nu}\dot {x_b}_{\nu})}
\over {({(x_a-x_b)}^2)}^{\lambda+1}}
\nonumber\\
&&
+8\lambda{{(\psi_a^{\mu}{\psi_b}_{\mu})}^2
\over {({(x_a-x_b)}^2)}^{\lambda+1}}
-16\lambda (\lambda +1) {(x_a^{\mu}-x_b^{\mu})
(x_a^{\nu}-x_b^{\nu})
{\psi_a}_{\mu}{\psi_b}_{\nu}
\psi_a^{\kappa}{\psi_b}_{\kappa}\over
{({(x_a-x_b)}^2)}^{\lambda+2}}
\Biggr\rbrace\quad \nonumber\\
\label{ccicomp}
\end{eqnarray}

\noindent
The denominator of eq.(\ref{ccisup}) being bosonic, we can
again use the proper-time representation 
eq.(\ref{intpropchap8})
to get it into the exponent, and then absorb 
this exponent into the
worldline superpropagator. The algebra is completely
identical to the scalar case, and leads to
modified superpropagators $\hat G^{(m)}$
which are given by the same
formulas as in eqs.(\ref{G(1)}) and (\ref{nkgk}),
with all the one-loop Green's functions appearing on the
right-hand sides replaced by the corresponding
one-loop superpropagators eq.(\ref{superpropagator}).
The same applies to the determinant factor 
${(N^{(m)})}^{D\over 2}$.
The generalization of 
the
$(m+1)$ -- loop
$N$ -- photon scattering formula
eq.(\ref{Nphotonwickscal(m)})
to the spinor loop case is equally
trivial, and there is no point in writing
it down here.

Again what we have at hand is a parameter integral
combining into one formula all Feynman 
diagrams with one electron loop,
and fixed numbers of external and internal
photons. For instance, for $N=m=2$ this just corresponds to the
diagrams of fig. \ref{3loopbetadiag}
which we discussed in the introduction.

Finally, note that
the formulas (\ref{ccigengauge}),(\ref{ccigengaugetotdir})
for an arbitrary covariant gauge also carry
over to the spinor loop case
{\sl mutatis mutandis}.

\subsection{The General Case}

The general case of a multiple product of
scalar or
electron loop path integrals coupled by photon insertions
requires only two new considerations.

Firstly, every scalar/electron path integral has its own zero-mode
integral, which must be separated off, and yields
momentum conservation for the photon momenta
entering that particular loop. Total
momentum conservation is obtained only after all
zero mode integrals are performed.

Secondly, since we now have to Wick -- contract
vertex operators attached to the
different loops,
the multiloop worldline Green's function
becomes a matrix in the space of loops.
For instance, in the case of just two scalar/electron
 loops
one has a two by two matrix of Green's functions
$G_{B}^{\alpha\beta}$. The matrix element
$G_{B}^{11}$ has to be used for
the Wick contraction of two 
photon vertex operators
both on loop 1,
$G_{B}^{12}$ for the contraction of
one vertex operator on loop 1
and one on
loop 2, etc. The explicit expressions
for these Green's functions can be
found by the same procedure which we
described in the previous chapter. This leads to
formulas similar to 
eqs.  
(\ref{nkgk}),(\ref{defA}), which we will
not write down here (the simplest case
of two loops connected by a single propagator
was also considered in ~\cite{dashsu}).

Note that in the QED case
we encounter neither of the
two problems discussed in the last chapter,
which motivated the introduction of an auxiliary
field formalism.
First, Feynman's formula eq.(\ref{feynform}) 
and its supersymmetrization allow one to
neatly
exhaust the complete photon S-matrix in terms of
scalar/electron path integrals connected by photon
insertions. The question of non-Hamiltonian graphs
therefore does not arise. Moreover,
from the same formula it follows that
the contributions
of individual Feynman diagrams are always generated
with the appropriate statistical factors.

Note that we do not consider here amplitudes involving
external electrons; the corresponding formalism
has not yet been 
sufficiently developed. For a treatment of external
scalars in scalar QED along the present lines see
~\cite{dashsu}.

\subsection{Example: The 2-Loop QED $\beta$ -- Functions}
\label{2loopbeta}

As an illustration, we will use this calculus
for a re-derivation
of the two-loop QED $\beta$-function,
both for scalar and for spinor electrodynamics
~\cite{ss3,zako}.

As usual, matters much simplify if one is interested only
in the $\beta$-function contribution, as
opposed to a calculation of the complete
2-loop vacuum polarization amplitude. 
Our strategy here will be to use the effective action
formalism with a constant background field
$F_{\mu\nu}$,
and read off the $\beta$-function from the
coefficient of the induced $F_{\mu\nu}
F^{\mu\nu}$-term.   
Standard dimensional regularization will be used for
the treatment of the UV divergences.

Just for setting the stage,
let us first redo the one-loop calculation.
As always in the constant field case
we choose Fock-Schwinger gauge centered
at $x_0$, so that
$A_{\mu}={1\over 2} y^{\rho}F_{\rho\mu}$.
Using this $A$ -- field in the 
spinor-loop path-integral
eq.(\ref{spinorpi}), expanding the interaction 
exponential
to second order, 
and performing the Wick contractions, 
one obtains   

\begin{eqnarray}
\Gamma^{(1)}_{\rm spin}\lbrack F\rbrack &  = &
- \half {\displaystyle\int_0^{\infty}}
{dT\over T}
e^{-m^2T}
{\displaystyle\int} {\cal D} x{\cal D}\psi
\,{\rm exp}\biggl [- \int_0^T d\tau
\Bigl ({1\over 4}{\dot x}^2 + {1\over
2}\psi\dot\psi
\Bigr )\biggr ]
\nonumber\\
& \phantom{=}&\times
(-{e^2\over 2})\int_0^T d\tau_1 \int_0^T d\tau_2
\Bigl[{1\over 4} \dot x^{\mu}_1F_{\mu\nu}x^{\nu}_1
\dot x^{\alpha}_2F_{\alpha\beta}x^{\beta}_2
+ \psi_1^{\mu}F_{\mu\nu}\psi^{\nu}_1
\psi_2^{\alpha}F_{\alpha\beta}\psi^{\beta}_2\Bigr ] 
\nonumber\\
&=& 
{e^2\over 2} {\displaystyle\int_0^{\infty}}
{dT\over T}
(4\pi T)^{-{D\over 2}}
e^{-m^2T}
\int_0^T d\tau_1 \int_0^T d\tau_2
\Bigl({\dot {G_B}_{12}}^2
-{{G_F^2}_{12}}\Bigr)\int dx_0
F_{\mu\nu}F^{\mu\nu}\quad \nonumber\\
\label{oneloop}
\end{eqnarray}

\noindent
The parameter integral gives

\begin{equation}
\int_0^T d\tau_1 \int_0^T d\tau_2\,
\Bigl({\dot G_B}^2(\tau_1,\tau_2)
-G_F^2(\tau_1,\tau_2)\Bigr)
=  -{2\over3}\,T^2    \quad 
\label{parint}
\end{equation}
The singular part of the one-loop
effective action becomes

\begin{equation}
\Gamma^{(1)}_{\rm spin}[F] \sim {2\over 3}{(4\pi )}^{-2}
{1\over \epsilon}e^2 
\int dx_0 F_{\mu\nu}F^{\mu\nu}\quad 
\label{Gamma1}
\end{equation}
($\epsilon= D-4$).
From this one can read off the 
one-loop photon wave-function
renormalization factor 

\begin{equation}
{(Z_3 -1)}^{(1)} = {2\over {3\epsilon}}
{\alpha\over\pi}\quad 
\label{Z3}
\end{equation}

\noindent
leading to the usual value for
the one-loop spinor QED $\beta$ -- function,

\begin{equation}
\beta^{(1)}_{\rm spin}(\alpha) = {2\over3}{{\alpha^2}\over\pi}
\quad 
\label{beta1}
\end{equation}

\noindent
($\alpha = {e^2\over{4\pi}})$.    

The corresponding result for scalar QED is simply obtained
by omitting, in eq.(\ref{oneloop}), the
term involving $G_F$, and the global factor of $-2$.
This yields

\begin{equation}
\beta^{(1)}_{\rm scal}(\alpha) = {1\over6}{{\alpha^2}\over\pi}
\quad 
\label{beta1scal}
\end{equation}\no

\begin{figure}[ht]
\vbox to 4.5cm{\vfill\hbox to 15.8cm{\hfill
\epsffile{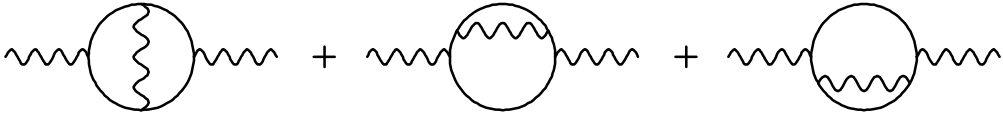}
\hfill}\vfill}\vskip-.4cm
\caption[dum]{Diagrams contributing to the
two-loop 
vacuum polarization.
\hphantom{xxxxxxxxxxxxxxx}}
\label{2loopvpdiag}
\end{figure}

Now let us tackle the two-loop calculation.
In the corresponding Feynman diagram calculation
(see, e.g., \cite{itzzub}), one would have to
separately calculate the three diagrams of fig. \ref{2loopvpdiag},
and then extract their $1\over\epsilon$ -- poles. 
Cancellation of the $1\over{{\epsilon}^2}$ -- poles
would be found in the sum of the results, 
indicating a 
cancellation of subdivergences
due to gauge invariance.
\vspace{6pt}

\par
\par

\vspace{6pt}
\noindent
Let us begin with the purely bosonic contributions, which
correspond to the scalar QED calculation. Those
are obtained by inserting 
the worldline current-current interaction term
eq.(\ref{cci}) into the bosonic one-loop
path-integral. After exponentiation
of the denominator
and absorption
into the worldline Green's function,
this results in

\begin{eqnarray}
{\Gamma}^{(2)}_{\rm bos}[F] &=& -2{\Gamma}^{(2)}_{\rm scal}
[F]\nonumber\\
&=&
-2{(4\pi )}^{-D}
\int_0^{\infty}{dT\over T}e^{-m^2T}T^{-{D\over 2}} 
\int_0^{\infty}d\bar T 
\int_0^T d\tau_a
\int_0^T d\tau_b
\,{\bigl[\bar T + G_{Bab}\bigr ]}
^{-{D\over 2}}\nonumber\\
&\phantom{=}&\times
{({-e^2\over 2})}^2
\int_0^T d\tau_1
\int_0^T d\tau_2\,
\int dx_0
{1\over 4}\langle
 \dot y^{\mu}_1F_{\mu\nu}y^{\nu}_1
\dot y^{\alpha}_2F_{\alpha\beta}y^{\beta}_2
\dot y^{\lambda}_a\dot y_{b\lambda}\rangle
\quad 
\label{bosecont}
\end{eqnarray}

\noindent
Note the appearance of the two-loop determinant
factor 
${\bigl[\bar T + G_B(\tau_a,\tau_b)\bigr ]}^{-{D\over 2}}$.
The Wick contraction of

\begin{equation}
\langle
 \dot y^{\mu}_1y^{\nu}_1
\dot y^{\alpha}_2y^{\beta}_2
\dot y^{\lambda}_a\dot y_{b\lambda}\rangle
\label{wick}
\end{equation}

\noindent
has now to be done, using the two-loop Green's function
eq.(\ref{G(1)}). Due to the symmetries of the problem 
there are only two nonequivalent
contraction possibilities, namely

\bear
\langle\dot y^{\mu}_1
y^{\beta}_2\rangle
\langle y^{\nu}_1
\dot y^{\alpha}_2\rangle
\langle\dot y^{\lambda}_a\dot y_{b\lambda}\rangle
&=&
-Dg^{\mu\beta}g^{\nu\alpha}
\partial_1G_{B12}^{(1)}
\partial_2G_{B12}^{(1)}
\partial_a\partial_bG_{Bab}^{(1)}
\non\\
\langle\dot y^{\mu}_1
\dot y^{\alpha}_2\rangle
\langle y^{\nu}_1\dot y^{\lambda}_a\rangle
\langle y^{\beta}_2
\dot y_{b\lambda}\rangle
\label{2wicks}
&=&
-g^{\mu\alpha}g^{\nu\beta}
\partial_1\partial_2 G_{B12}^{(1)}
\partial_a G_{B1a}^{(1)}
\partial_b G_{B2b}^{(1)}
\ear
Those occur with multiplicities $2$ and $8$, respectively.
Care must be taken with Wick
contractions involving $\dot y_a,\dot y_b$, as
the derivatives should not act on the
$\tau_a$,$\tau_b$ explicitly appearing in that
Green's function.
The result is written out in terms of the bosonic
one-loop Green's function and its derivatives. As in the
one-loop calculation, one next eliminates
all factors of $\ddot G_B$ appearing by partial
integrations with respect to $\tau_1,\tau_2,\tau_a,\tau_b$.
As the next step, all fermionic
contributions are included
by applying the one-loop replacement rule
(\ref{fermion}).
For example, one replaces

\begin{equation}
{\dot G_{B12}}\dot G_{B21}\dot G_{Bab}\dot G_{Bba}\rightarrow
(\dot G_{B12}\dot G_{B21}-G_{F12}G_{F21})
(\dot G_{Bab}\dot G_{Bba}- G_{Fab}G_{Fba})\quad 
\label{replace}
\end{equation}

\noindent
etc. 

\begin{figure}[ht]
\vbox to 5.7cm{\vfill\hbox to 15.8cm{\hskip1cm\hfill
\epsffile{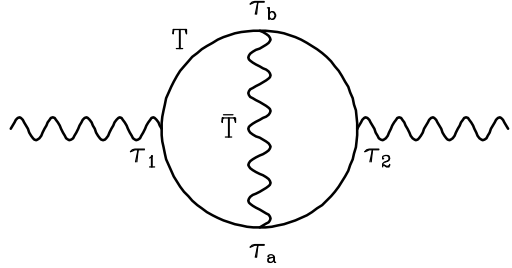}
\hfill}\vfill}\vskip-1cm
\caption[dum]{Definition of the six
integration parameters.
\hphantom{xxxxxxxxxxxxxxxx}}
\label{2loopvpint}
\end{figure}

At this stage, we have the desired contribution
to the two-loop effective
action in the form of a sixfold
integral (see fig. \ref{2loopvpint}),

\begin{eqnarray}
{\cal L}^{(2)}_{\rm spin}[F]&=& -2
{(4\pi)}^{-D}
{e^4\over 16}
\int_0^{\infty}{dT\over T}e^{-m^2T}T^{-{D\over2}}
\int_0^{\infty}d\bar T\nonumber\\
&\phantom{=}&\times 
\int_0^Td\tau_ad\tau_bd\tau_1d\tau_2
\,P(T,\bar T,\tau_a,\tau_b,\tau_1,\tau_2)
F_{\mu\nu}F^{\mu\nu}\quad \nonumber\\
\label{sixint}
\end{eqnarray}

\noindent
The integrand function P 
is a polynomial in the various
$G_{Bij},\dot G_{Bij},{G_F}_{ij}$, 
multiplied by powers of
$\gamma\equiv
{\Bigl[\bar T + G_B(\tau_a,\tau_b)\Bigr ]}^{-1}$.
Let us just write down its  
purely bosonic part $P_{\rm bos}$, which is
\footnote{In \cite{ss3} this integrand was
given incorrectly.}

\begin{eqnarray}
P_{\rm bos}&=& {\gamma}^{D\over 2}\Bigl\lbrace
D(D-1) \gamma \dot G_{Bab}^2\dot G_{B12}^2 + 
8D \gamma \dot G_{Bab}\dot G_{B12}\dot G_{B1a}\dot G_{B2b}
\nonumber\\
&&+ 8\gamma\dot G_{B1a}\dot G_{Bab}\dot G_{B12}
[\dot G_{B2a}-\dot G_{B2b}]\nonumber\\
&&- 4\gamma \dot G_{B1a}\dot G_{B2b}
[\dot G_{B1a}-\dot G_{B1b}]
[\dot G_{B2a}-\dot G_{B2b}]
\nonumber\\
&&+(D+2)(D-1) \gamma^2\dot G_{Bab}^2\dot G_{B12}
[\dot G_{B1a}-\dot G_{B1b}]
[G_{B2a}-G_{B2b}]\Bigr\rbrace\quad \nonumber\\
\label{Pbosexpl}
\end{eqnarray}
\no
In writing this polynomial,
we have used the symmetry with regard to interchange
of $\tau_1$ and $\tau_2$ to combine some terms, 
and omitted some terms
which are total derivatives with respect to $\int d\tau_1$ or
$\int d\tau_2$ (those terms are easy to identify at an
early stage of the calculation).

\par
\par

\noindent
It is convenient to begin with
the integrations over $\tau_1,\tau_2$.
Those
are polynomial, and easily performed
using a set of relations of
the type

\begin{eqnarray}
\int_0^1du_2 \dot G_{B12}\dot G_{B23} &=& 2{G_B}_{13}
-{1\over 3}\nonumber\\
\int_0^1du_2 {G_B}_{12}{G_B}_{23} &=&
-{1\over 6} {G_B^2}_{13}+{1\over 30}\nonumber\\
\vdots&&\vdots\nonumber\\
\label{relations}
\end{eqnarray}

\noindent
All those relations 
may be derived from the 
following master identities
proven in appendix \ref{greendet},

\begin{eqnarray}
\int_0^1 du_2\ldots du_n
\dot G_{B12}\dot G_{B23}\ldots\dot G_{Bn(n+1)} &=&
-{2^n\over n!}B_n(\vert u_1-u_{n+1}\vert)
{\rm sign}^n(u_1-u_{n+1})\nonumber\\
\int_0^1 du_2\ldots du_n
G_{F12}G_{F23}\ldots G_{Fn(n+1)} &=&
{2^{n-1}\over{(n-1)!}}
E_{n-1}(\vert u_1-u_{n+1}\vert)
{\rm sign}^n(u_1-u_{n+1})\quad \nonumber\\
\label{master}
\end{eqnarray}

\noindent
In writing these identities, we have scaled down
to the unit circle again.
$B_n$ denotes the $n^{th}$ Bernoulli-polynomial, and
$E_n$ the $n^{th}$ Euler-polynomial. Due to
the fact that those polynomials can be rewritten as

\begin{eqnarray}
B_n(x)  &=& P_n(x^2-x)
\phantom{(x-{1\over 2} )}\quad{\rm (n\quad even)}
\nonumber\\
B_n(x)  &=& P_n(x^2-x)(x-{1\over 2} ) 
\quad {\rm (n\quad odd)}    \nonumber\\
\label{berneuler}
\end{eqnarray}

\noindent
with another set of 
polynomials $P_n(x)$ (the same property holds true
for $E_n(x)$), 
the right hand sides can
always be re-expressed in terms of $G_B,\dot G_B$ and
$G_F$, so that explicit $u_i$'s will never appear
in those relations. Those integrals needed for the
present calculation are listed 
in appendix \ref{formulas}.

\noindent
Next we perform the $\bar T$ -- integration, which is
trivial:

\begin{equation}
\int_0^{\infty}d\bar T
{[\bar T + G_{Bab}]}^{-{D\over 2}-k}
= {G_{Bab}^{1-{D\over 2}-k}
\over {{D\over 2}}+k-1}
\quad\quad (k = 1, 2)
\label{Tbarint}
\end{equation}

\noindent
Collecting terms, and using (\ref{Gdotsquare}), we get

\begin{eqnarray}
\int_0^{\infty}d\bar T \int_0^T d\tau_1\int_0^T
d\tau_2\, P(T,\bar T,\tau_a,\tau_b,\tau_1,\tau_2)
&=& \nonumber\\
{16\over 3D}\biggl\lbrace 
(D-4)(D-1)
G_{Bab}^{1-{D\over 2}}T
&+& (D-2)(D-7)
G_{Bab}^{2-{D\over 2}}\biggr\rbrace
\quad  \nonumber\\
\label{spinorintegrand}
\end{eqnarray}
 
\noindent
The corresponding expression for
scalar QED is obtained by 
using only the bosonic part $P_{\rm bos}$
of the function $P$:

\begin{eqnarray}
\int_0^{\infty}d\bar T \int_0^T d\tau_1\int_0^T
d\tau_2\, 
P_{\rm bos}(T,\bar T,\tau_a,\tau_b,\tau_1,\tau_2)
= 
{2\over 3}(D-1)G_{Bab}^{-{D\over 2}}
T^2\nonumber\\
+(D-1)({32\over 3D}-4)
G_{Bab}^{1-{D\over 2}}T
+ {16\over 3D}
(D-2)(D-7)
G_{Bab}^{2-{D\over 2}}
\quad \nonumber\\
\label{scalarintegrand}
\end{eqnarray}

\noindent
Setting $\tau_a = 0$, the 
integration over $\tau_b$  
produces a couple of Euler Beta-functions,

\begin{equation}
\int_0^Td\tau_a\int_0^Td\tau_b\,
G_{Bab}^{k-{D\over 2}}
=B\Bigl(k+1-{D\over 2},k+1-{D\over 2}\Bigr)
T^{2+k-{D\over 2}}
\label{Eulerbeta}
\end{equation}

\noindent
As in the one-loop case,
the remaining electron proper-time integral 
just gives a $\Gamma$~--~function:

\begin{equation}
\int_0^{\infty}{dT\over T}{\rm e}^{-m^2 T}T^{4-D}
= \Gamma(4-D)m^{2(D-4)}
\label{2loopglobal}
\end{equation}

\noindent
Combining terms and performing 
the $\epsilon$ -- expansions
for the effective Lagrangians, we obtain

\begin{eqnarray}
{\cal L}^{(2)}_{\rm scal}[F] &\sim &
{1\over {2 \epsilon}}e^4{(4\pi)}^{-4}
F_{\mu\nu}F^{\mu\nu}
+ O({\epsilon}^0) \nonumber\\
{\cal L}^{(2)}_{\rm spin}[F] &\sim &
-{3\over \epsilon}e^4{(4\pi)}^{-4}
F_{\mu\nu}F^{\mu\nu}
+ O({\epsilon}^0)\nonumber\\
\label{Gexpansion}
\end{eqnarray}

\noindent

So far this is a calculation of the bare
regularized effective action. What about 
renormalization? 
The counterdiagrams
due to electron wave function and
vertex renormalization need not be taken into
account, since they cancel by 
the QED Ward identity
($Z_1=Z_2$). However, we have used the electron
mass as an infrared regulator for the electron 
proper-time integral eq.(\ref{2loopglobal});
mass renormalization must therefore be dealt with.

Since our
calculation corresponds to a field theory calculation
in dimensional regularization,
we need to know the corresponding one-loop
mass renormalization counterterms, both for
scalar and spinor QED. This is a simple textbook
calculation, of which we give the result only:

\begin{eqnarray}
{\delta m^2_{\rm scal} \over m^2_{\rm scal}}&=& 
{6\over \epsilon}e^2{(4\pi)}^{-2}\nonumber\\
{\delta m_{\rm spin}\over m_{\rm spin}} &=& 
{6\over \epsilon}e^2{(4\pi)}^{-2}\nonumber\\
\label{masscount}
\end{eqnarray}

\noindent
We insert those counterterms 
into the one-loop
path integrals, and obtain
the following additional contributions
to the two-loop effective Lagrangians,

\begin{eqnarray}
\Delta {\Gamma}^{(2)}_{\rm scal}[F]
&=& \delta m^2_{\rm scal}
{\partial\over \partial m^2}
\Gamma^{(1)}_{\rm scal}[F]\nonumber\\
&\sim& 
{1\over {2 \epsilon}}e^4{(4\pi)}^{-4}
\int dx_0 F_{\mu\nu}F^{\mu\nu}
+ O({\epsilon}^0) \nonumber\\
\Delta \Gamma^{(2)}_{\rm spin}[F]
&=& \delta m_{\rm spin}
{\partial\over \partial m}
\Gamma^{(1)}_{\rm spin}[F]\nonumber\\
&\sim &
{4\over \epsilon}e^4{(4\pi)}^{-4}
\int dx_0 F_{\mu\nu}F^{\mu\nu}
+ O({\epsilon}^0)
\quad \nonumber\\
\label{Gadd}
\end{eqnarray}

\noindent
The
extraction of the $\beta$ -- function coefficients
proceeds in the usual way. 
From the total effective Lagrangians

\begin{eqnarray}
{\cal L}^{(2)}_{\rm scal}[F] +
\Delta {\cal L}^{(2)}_{\rm scal}[F]
&\sim& 
{1\over {\epsilon}}e^4{(4\pi)}^{-4}
F_{\mu\nu}F^{\mu\nu}
\nonumber\\
{\cal L}^{(2)}_{\rm spin}[F] + 
\Delta {\cal L}^{(2)}_{\rm spin}[F]
&\sim&
{1\over {\epsilon}}e^4{(4\pi)}^{-4}
F_{\mu\nu}F^{\mu\nu}
\nonumber\\
\label{Gtotal}
\end{eqnarray}

\noindent
one obtains the two-loop photon wave-function
renormalization factors, and from those
the standard results for the two-loop
$\beta$ -- function coefficients
~\cite{joslut,bialynicka65},

\begin{equation}
\beta^{(2)}_{\rm scal}(\alpha )= 
\beta^{(2)}_{\rm spin}(\alpha ) 
=  {{\alpha}^3\over{2{\pi}^2}}
\quad 
\label{2loopcoeff}
\end{equation}

Observe that in the spinor-loop case, the
integrand after performance of the first three
integrations, eq.(\ref{spinorintegrand}), 
has only one term which would
be divergent for $D=4$
when integrated over $\tau_b$.
Moreover, the coefficient of this term vanishes for
$D=4$. This suggests that 
this calculation can be further simplified 
by using some
four-dimensional regularization scheme.
And indeed, if we do the spinor-loop
calculation in four dimension,
then instead of
eq.(\ref{spinorintegrand})
we find simply

\begin{equation}
\int_0^{\infty}d\bar T \int_0^T d\tau_1\int_0^T
d\tau_2\, P(T,\bar T,\tau_a,\tau_b,\tau_1,\tau_2)
= -8\quad
\label{threeint}
\end{equation}

\noindent
This time there is no dependence on $\tau_a,\tau_b$
left,
so that one immediately gets

\begin{equation}
{\cal L'}^{(2)}_{\rm spin}[F] = {(4\pi)}^{-4}
e^4
\int_0^{\infty}{dT\over T}e^{-m^2T}
F_{\mu\nu}F^{\mu\nu}
\label{final}
\end{equation}

It is only the final electron proper-time integral
that now needs to be regularized. This can be done
by introducing a proper-time cutoff $T_0$ at the
lower integration limit, which replaces
eq.(\ref{2loopglobal}) by

\begin{equation}
\int_{T_0}^{\infty}{dT\over T}{\rm e}^{-m^2 T}
\sim -{\rm ln}(m^2T_0)\qquad
\label{2loopcutoff}
\end{equation}

\noindent
(Pauli-Villars regularization could be used as 
well, although proper-time regularization appears
more natural in the worldline formalism).
With this regulator,
the two-loop effective
Lagrangian becomes

\begin{equation}
{\cal L'}^{(2)}_{\rm spin}[F] \sim 
-{\rm ln}(m^2T_0){(4\pi)}^{-4}
e^4
F_{\mu\nu}F^{\mu\nu}
+ {\rm finite} \quad 
\label{log}
\end{equation}

\noindent
In spite
of the  manifest suppression
of subdivergences, there is again a
contribution from mass
renormalization, which can be
determined by comparison with the
corresponding Feynman calculation. 
On-shell renormalization of spinor QED using a
proper-time cutoff has been 
studied in
\cite{tsai,ditreu}. 
It leads to a
one-loop mass renormalization 
counterterm 

\begin{equation}
{\delta {m}\over m} = 
 3{\rm ln}(m^2T_0)e^2{(4\pi)}^{-2}
+ {\rm finite}
\quad 
\label{masscountpt}
\end{equation}

\noindent
Insertion of this counterterm
into the one-loop path integral gives

\begin{eqnarray}
\Delta {\Gamma'}^{(2)}_{\rm spin}[F]
&=& \delta m
{\partial\over \partial m}
{\Gamma'}^{(1)}[F]\nonumber\\
&\sim &
2{\rm ln}(m^2T_0){(4\pi)}^{-4}
e^4\int dx_0
F_{\mu\nu}F^{\mu\nu}
+ {\rm finite} 
\nonumber\\
\label{Gmasspt}
\end{eqnarray}

\noindent
so that mass renormalization now
just amounts to a sign change
for the effective Lagrangian:

\begin{equation}
{\cal L'}^{(2)}_{\rm spin}[F] 
+ \Delta {\cal L'}^{(2)}_{\rm spin}[F]
\sim {\rm ln}(m^2T_0){(4\pi)}^{-4}
e^4
F_{\mu\nu}F^{\mu\nu}
\quad 
\label{G'total}
\end{equation}

\noindent
The extraction of the 
(still scheme-independent) 
$\beta$ -- function coefficient
$\beta^{(2)}_{\rm spin}(\alpha )$
is again standard ~\cite{ditreu}, and leads back to
eq.(\ref{2loopcoeff}).

\noindent
Let us summarize the properties of this calculation:

\begin{enumerate}
\item
Neither momentum integrals nor Dirac traces had to be 
calculated.

\item
The three diagrams of fig. \ref{2loopvpdiag} were combined into
one calculation (in fact, in this formalism it is somewhat
{\sl easier} to compute the sum than any single one of them).

\item
In the spinor-loop case, 
we have managed to obtain the correct 2-loop
coefficient without performing any
nontrivial integrals.

\end{enumerate}
\no
We will see later on that the 
cancellations which led to the
last property are not
accidental, but a direct consequence of renormalizability.

\subsection{Quantum Electrodynamics in a Constant External Field}
\label{qedmext}

As in the one-loop case,
it takes only minor modifications to
extend this formalism to the case when
an additional constant external field
$F_{\mu\nu}$ is present ~\cite{rescsc}.

Again we begin with the simplest case,
which is scalar QED at the 2-loop level.
Combining 
the contributions
to the quadratic part of the
worldline Lagrangian due
to the external field
eq.(\ref{DeltaLkomp}) 
and to the propagator insertion
eq.
(\ref{intpropchap8}),
we obtain the new total 
bosonic kinetic operator 

\begin{equation}
{d^2\over d\tau^2}
-2ieF{d\over d\tau}
-{B_{ab}\over\bar T}
\label{kinFB}
\end{equation}
\noindent
The inverse of this operator
can still be constructed as a 
geometric series,

\bear
{\Bigl({d^2\over d\tau^2} -2ieF{d\over d\tau}
-{B_{ab}\over\bar T} \Bigl)}^{-1}
&=& {\Bigl({d^2\over d\tau^2} -2ieF{d\over d\tau}\Bigr)}^{-1}
\non\\
&&
\!\!\!\!\!\!\!
+
{\Bigl({d^2\over d\tau^2} -2ieF{d\over d\tau}\Bigr)}^{-1}
{B_{ab}\over \bar T}
{\Bigl({d^2\over d\tau^2} -2ieF{d\over d\tau}\Bigr)}^{-1}
+\cdots
\label{sumserieschap8}
\ear
\no
Summing this series one obtains
the following
Green's function:

\begin{equation}
{\cal G}_B^{(1)}(\tau_1,\tau_2)=
{\cal G}_B(\tau_1,\tau_2) + {1\over 2}
{{[{\cal G}_B(\tau_1,\tau_a)-{\cal G}_B(\tau_1,\tau_b)]
[{\cal G}_B(\tau_a,\tau_2)-{\cal G}_B(\tau_b,\tau_2)]}
\over
{{\bar T} -{1\over 2}
{\cal C}_{ab}}}\,
\label{calG(1)} 
\end{equation}
\noindent
were we have defined

\begin{eqnarray}
{\cal C}_{ab}&\equiv& 
{\cal G}_B(\tau_a,\tau_a)
-{\cal G}_B(\tau_a,\tau_b)
-{\cal G}_B(\tau_b,\tau_a)
+{\cal G}_B(\tau_b,\tau_b)
\nonumber\\
&=&
T
{\cos ({\cal Z})-\cos ({\cal Z}\dot G_{Bab})
\over ({\cal Z})\sin ({\cal Z})}
\; \nonumber\\
\label{defCab}
\end{eqnarray}
\noindent
This is almost but not quite identical with what one would
obtain from the ordinary bosonic
two-loop Green's function, eq.(\ref{G(1)}),
by simply replacing all $G_{Bij}$'s appearing there
by the corresponding
${\cal G}_{Bij}$'s. The more complicated structure of the
denominator is due to the fact that
the ${\cal G}_{Bij}$'s are not any more symmetric 
under the interchange $i\leftrightarrow j$,
rather we have
${\cal G}_{Bij}={\cal G}_{Bji}^T$, and moreover
have non-vanishing coincidence limits.
The denominator is now in general a nontrivial
Lorentz matrix, and must be interpreted as
a matrix inverse (of course, all matrices
appearing here commute with each other).

The free Gaussian path integral is again
easily calculated using the
$ln\,det = tr\,ln$ -- identity,
yielding

\begin{eqnarray}
{\rm Det'}_P
\biggl[ -{d^2\over d\tau^2}
+2ieF{{d\over d\tau}}+{B_{ab}\over\bar T}\biggr]
&=&
{\rm Det'}_P
\Bigl[-{d^2\over d\tau^2}\Bigr]
{\rm Det'}_P
\biggl[{\Eins}-2ieF{({d\over d\tau})}^{-1}
\biggr]
\nonumber\\
&&\times
{\rm Det'}_P
\biggl[
{\Eins}-{B_{ab}\over\bar T}
{\Bigl(
{d^2\over d\tau^2}-2ieF{{d\over d\tau}}\Bigr)
}^{-1}
\biggr]
\nonumber\\
&=&
(4\pi T)^D
{\rm det}\biggl[{\sin({\cal Z})\over{{\cal Z}}}\biggr]
{\rm det}
\biggl[{{\Eins}
-{1\over {2\bar T}}
{\cal C}_{ab}
}\biggr]
\quad \nonumber\\
\label{FBbosnorm}
\end{eqnarray} 
\noindent
Thus we have now a product of two Lorentz matrix determinants.
The first one is
identical with the by now familiar
Euler-Heisenberg integrand, eq.(\ref{scaldetext}),
while the second one generalizes 
the two-loop determinant factor eq.(\ref{normchange})
to the external field case.

As in the case without a background field,
the whole procedure goes through essentially unchanged
for the fermion loop, if the superfield 
formalism is used.
As a consequence,
one finds the same 
close relationship 
as before between the parameter
integrals for the same amplitude 
calculated for the scalar and for the
fermion loop: They differ only
by a replacement of all ${\cal G}_B$'s
by $\hat{\cal G}$'s, and 
by the additional $\theta$ -- 
integrations.
Of course, one must also replace
the scalar QED one-loop Euler-Heisenberg
factor eq.(\ref{scaldetext})
by its spinor QED equivalent
eq.(\ref{spindetext}),
and as always the global factor of $-2$ must
be taken into account. 

The generalization to an arbitrary fixed number of photon insertions
is straightforward, and leads to
formulas for the generalized (super-) Green's functions 
and (super-) determinants identical with the ones 
given above for the vacuum case, up to a replacement of all
$G_B$'s ($\hat G$'s) by ${\cal G}_B$'s
($\hat{\cal G}$'s). The only point to be mentioned here
is that care must now be taken in writing the indices of
the ${\cal G}_{Bij}$'s appearing. For instance, the 
(un-subtracted) bosonic three-loop
Green's function, eq.(\ref{nkgk}) with $m=2$,
must be written as

\begin{eqnarray}
{\cal G}^{(2)}_B(\tau_1,\tau_2)
=
{\cal G}_B(\tau_1,\tau_2)
\!
+\frac{1}{2}
\sum_{k,l=1}^2
\Bigl[
{\cal G}_B(\tau_1,\tau_{a_k})
-{\cal G}_B(\tau_1,\tau_{b_k})
\Bigr]
A_{kl}^{(2)}
\Bigl[
{\cal G}_B(\tau_{a_l},\tau_2)
-{\cal G}_B(\tau_{b_l},\tau_2)
\Bigr]
\nonumber\\
\label{3loopGreen's}
\end{eqnarray}
\noindent
The matrix $A$ appearing here
is the inverse of the matrix

\begin{equation}
\left(
\begin{array}{*{2}{c}}
T_1-{1\over 2}
\Bigl({\cal G}_{Ba_1a_1}-{\cal G}_{Ba_1b_1}-{\cal G}_{Bb_1a_1}+{\cal G}_{Bb_1b_1}\Bigr)
&
-\half\Bigl({\cal G}_{Ba_1a_2}-{\cal G}_{Ba_1b_2}-{\cal G}_{Bb_1a_2}+{\cal G}_{Bb_1b_2}\Bigr)
\\
-\half
\Bigl({\cal G}_{Ba_2a_1}-{\cal G}_{Ba_2b_1}-{\cal G}_{Bb_2a_1}+{\cal G}_{Bb_2b_1}\Bigr)
&
T_2-\half
\Bigl({\cal G}_{Ba_2a_2}-{\cal G}_{Ba_2b_2}-{\cal G}_{Bb_2a_2}+{\cal G}_{Bb_2b_2}\Bigr)\\
\end{array}
\right)\\
\label{3loopA}
\end{equation}
\noindent
and
$T_1,T_2$ denote the proper-time lengths
of the two inserted propagators.

\subsection{Example: The 2-Loop Euler-Heisenberg Lagrangians}
\label{2loopeh}

As an application of the constant field
formalism at the two-loop level, in this
section we will
calculate the 
first radiative corrections to
the Euler-Heisenberg-Schwinger formulas
eqs.(\ref{eulheiscal}),(\ref{eulheispin})
\cite{rescsc,frss}.

\subsubsection{Scalar QED}

According to the above, for the scalar QED
case we can write this
effective Lagrangian in the form

\begin{eqnarray}
{\cal L}^{(2)}_{\rm scal}
[F]&=&
{(4\pi )}^{-D}
\Bigl(-{e^2\over 2}\Bigr)
\int_0^{\infty}{dT\over T}e^{-m^2T}T^{-{D\over 2}} 
\int_0^{\infty}d\bar T 
\int_0^T d\tau_a
\int_0^T d\tau_b
\nonumber\\
&\phantom{=}&\times
{\rm det}^{-{1\over 2}}
\biggl[{\sin({\cal Z})\over {{\cal Z}}}
\biggr]
{\rm det}^{-{1\over 2}}
\biggl[
\bar T 
-{1\over 2}
{\cal C}_{ab}
\biggr]
\langle
\dot y_a\cdot\dot y_b\rangle
\quad \nonumber\\
\label{Gamma2scal}
\end{eqnarray}
\noindent
In this fourfold parameter integral, 
$T$ and $\bar T$ represent the scalar
and photon proper-times, and $\tau_{a,b}$ the
endpoints of the photon insertion moving around
the scalar loop. 

A single Wick contraction is to be
performed
on the ``left-over'' numerator of the
photon insertion, using the modified worldline
Green's function eq.(\ref{calG(1)}). This yields

\begin{equation}
\langle
\dot y_a\cdot\dot y_b\rangle
=
{\rm tr}
\biggl[
\ddot{\cal G}_{Bab}+{1\over 2}
{(\dot {\cal G}_{Baa}-\dot {\cal G}_{Bab})
(\dot {\cal G}_{Bab}-\dot {\cal G}_{Bbb})
\over
{\bar T -{1\over 2}{\cal C}_{ab}}}
\biggr]\; 
\label{wickscal}
\end{equation}
\noindent
After replacing the 
one-loop Green's functions
$\dot {\cal G}_{Bij}$'s 
as well as ${\cal C}_{ab}$
by the explicit expressions given in eqs.(\ref{derivcalGB})
and eq.(\ref{defCab}),
we have already a parameter integral representation for
the bare dimensionally regularized effective Lagrangian.

Alternatively one may
remove $\ddot {\cal G}_B$ by a
partial integration with respect to $\tau_a$ or $\tau_b$.
Using the formula

\begin{equation}
d\det(M)=\det(M){\rm tr}(dMM^{-1})
\label{ddet}\nonumber
\end{equation}
\noindent
and $\dot {\cal G}_{Bab}=-\dot{\cal G}_{Bba}^T$, 
one obtains the equivalent parameter integral

\begin{eqnarray}
{\cal L}^{(2)}_{\rm scal}[F]&=&
{(4\pi )}^{-D}
\Bigl(-{e^2\over 2}\Bigr)
\int_0^{\infty}{dT\over T}e^{-m^2T}T^{-{D\over 2}} 
\int_0^{\infty}d\bar T 
\int_0^T d\tau_a
\int_0^T d\tau_b
\nonumber\\
&\phantom{=}&\times
{\rm det}^{-{1\over 2}}
\biggl\lbrack{\sin({\cal Z})\over {{\cal Z}}}
\biggr\rbrack
{\rm det}^{-{1\over 2}}
\biggl[
\bar T -{1\over 2}
{\cal C}_{ab}
\biggr\rbrack\nonumber\\
&\phantom{=}&\times
{1\over 2}
\Biggl\lbrace
{\rm tr}\dot{\cal G}_{Bab}
{\rm tr}
\biggl\lbrack
{\dot{\cal G}_{Bab}
\over
{\bar T -{1\over 2}{\cal C}_{ab}}}
\biggr\rbrack
+{\rm tr}
\biggl[
{(\dot {\cal G}_{Baa}-\dot {\cal G}_{Bab})
(\dot {\cal G}_{Bab}-\dot {\cal G}_{Bbb})
\over
{\bar T -{1\over 2}{\cal C}_{ab}}}
\biggr]
\Biggr\rbrace\; 
\nonumber\\
\label{Gamma2scalpI}
\end{eqnarray}
\noindent

To facilitate the further evaluation 
and renormalization of this Lagrangian,
we specialize the constant
field $F$ to a pure magnetic 
field. 
It will be instructive to 
do this calculation in two different regularizations,
proper-time and dimensional regularization.
The renormalization will be performed on-shell
in both cases.

Let us begin with the proper-time regularized version.
This regularization keeps the integrations fairly simple,
and was used in all previous calculations
of two-loop Euler-Heisenberg Lagrangians
~\cite{ritusscal,lebedev,ginzburg,ritusspin,ditreu}.
It means that in the following we set
$D=4$, and instead
introduce proper-time UV cutoffs for
the various proper-time integrals later on.

As in the photon-splitting calculation
of section \ref{phosplit}, we choose the field in
the $z$ -- direction. 
For this case the generalized worldline Green's 
functions and determinants were given in
(\ref{detextB}),(\ref{GB(F)pureB}). The combination
$C_{ab}$ becomes

\begin{equation}
{\cal C}_{ab}=-2G_{Bab}g_{\parallel}
-2G_{Bab}^zg_{\perp} 
\label{simplifyCab}
\end{equation}
\noindent
where

\begin{equation}
G_{Bab}^z\equiv {T\over 2}
{\Bigl(\cosh (z)-\cosh(z\dot G_{ab})\Bigr)\over
z\sinh (z)}
=G_{Bab}-{1\over 3T}G_{Bab}^2z^2+O(z^4)
\label{abbrzGz}
\end{equation}
\noindent
We will also use the derivative of this expression,

\bear
\dot G_{Bab}^z&=&{\sinh(z\dot G_{Bab})\over\sinh(z)}
\non\\
\label{abbrzdotGz}
\ear\no
Introducing the further abbreviations

\bear
\gamma&\equiv&{(\bar T +G_{Bab})}^{-1}\nonumber\\
\gamma^z&\equiv&{(\bar T +G_{Bab}^z)}^{-1}\nonumber\\
\label{defzgamma}\nonumber
\end{eqnarray}
\noindent
we can then rewrite the various Lorentz traces
and determinants 
appearing in eqs.~(\ref{Gamma2scal}),
(\ref{Gamma2scalpI})
as

\begin{eqnarray}
{\rm det}^{-{1\over 2}}
\biggl[{\sin({\cal Z})\over {{\cal Z}}}{\bigl(\bar T 
-{1\over 2}
{\cal C}_{ab}
\bigr )}
\biggr]
&=&
{z\over\sinh(z)}
\gamma\gamma^z\nonumber\\
{\rm tr}\Bigl[\ddot{\cal G}_{Bab}\Bigr]
&=&
8\delta_{ab}-4-4{z\cosh(z\dot G_{Bab})\over\sinh(z)}
\nonumber\\
{1\over 2}{\rm tr}\dot{\cal G}_{Bab}
{\rm tr}\biggl[
{\dot{\cal G}_{Bab}\over
{\bar T -{1\over 2}{\cal C}_{ab}}}
\biggr]
&=& 2
\biggl[
\dot G_{Bab}+{\sinh(z\dot G_{Bab})\over\sinh(z)}
\biggr]
\biggl[
\dot G_{Bab}\gamma
+{\sinh(z\dot G_{Bab})\over \sinh(z)}
\gamma^z
\biggr]
\nonumber\\
{1\over 2}{\rm tr}
\biggl[
{(\dot {\cal G}_{aa}-\dot {\cal G}_{ab})
(\dot {\cal G}_{ab}-\dot {\cal G}_{bb})
\over
{\bar T -{1\over 2}{\cal C}_{ab}}}
\biggr]
&=&
-\gamma^z
{\sinh^2(z\dot G_{Bab})
+{\Bigl[\cosh(z\dot G_{Bab})-\cosh(z)\Bigr]}^2
\over
\sinh^2(z)}
-\dot G_{Bab}^2\gamma\nonumber\\
\label{simplifyremainder}
\end{eqnarray}
\noindent
As usual
we rescale to the unit circle,
$\tau_{a,b}=Tu_{a,b}$,
and use translation invariance in $\tau$ to
set $\tau_b=0$, so that

\begin{eqnarray}
G_B(\tau_a,\tau_b)&=&TG_B(u_a,u_b)=T(u_a-u_a^2)\nonumber\\
\dot G_B(\tau_a,\tau_b)&=&\dot G_B(u_a,u_b)
=1-2u_a
\; \nonumber\\
\label{scaleG}\nonumber
\end{eqnarray}
\noindent
After performance of the $\bar T$ -- integration,
which is finite and elementary,
eq.(\ref{Gamma2scalpI})
turns into 

\begin{equation}
{\cal L}^{(2)}_{\rm scal}
[B]=-
{(4\pi )}^{-4}
{e^2\over 2}
\int_{0}^{\infty}{dT\over T^3}e^{-m^2T}
{z\over \sinh(z)}
\int_0^1 du_a
\,A(z,u_a)
\; 
\label{Gamma2scalB}
\end{equation}
\noindent
with 

\begin{eqnarray}
A&=&
\Biggl\lbrace
A_1
{\ln ({G_{Bab}/ G_{Bab}^z})
\over{(G_{Bab}-G_{Bab}^z)}^2}
+{A_2\over
(G^z_{Bab})(G_{Bab}-G^z_{Bab})}
+{A_3\over
(G_{Bab})(G_{Bab}-G^z_{Bab})}
\Biggr\rbrace\quad \nonumber\\
A_1&=&
4
\Bigl[
G_{Bab}^z z\coth(z)-G_{Bab}
\Bigr]
\quad  \nonumber\\
A_2&=&1+
2\dot G_{Bab}
\dot G_{Bab}^z
-4G^z_{Bab}z\coth(z)
\quad \nonumber\\
A_3&=&
-\dot G_{Bab}^2-
2\dot G_{Bab}
\dot G_{Bab}^z
\;  
\nonumber\\
\label{A1A2A3}
\end{eqnarray}
\noindent
All Green's functions now refer to the
unit circle, $T=1$.
Here and in the following we often 
use the identity
$\dot G_{Bab}^2=1-{4\over T}G_{Bab}$ to
eliminate $\dot G_{Bab}$ in favour of $G_{Bab}$.

Renormalization must
now be addressed, and will be performed
in close analogy to the discussion in ~\cite{ditreu}.
The integral in eq.(\ref{Gamma2scalB}) suffers from two
kinds of divergences: 

\begin{enumerate}

\item
An overall divergence of
the loop proper-time integral $\int_0^{\infty}dT$
at the lower integration limit (which is already
familiar from the $\beta$ -- function calculation
of section \ref{2loopbeta}).

\item
Divergences of
the $\int_0^1\,du_a$ parameter integral 
at the points $0,1$ where the
endpoints of the photon propagator
become coincident, $u_a=u_b$.

\end{enumerate}
\noindent
The first one will be removed by 
one- and two-loop photon wave function
renormalization, the second one by the one-loop 
renormalization of the scalar mass.
As was already mentioned, vertex renormalization 
and scalar self energy renormalization
need not be considered in this type
of calculation, since they must cancel 
on account of the QED Ward identity.

By power counting, an overall divergence can exist only for
the terms in the effective Lagrangian which are of order
at most quadratic in the external field $B$. 
Expanding the integrand of eq.(\ref{Gamma2scalB}),
$K(z,u_a)\equiv {z\over\sinh(z)}A(z,u_a)$,
in the variable $z$, we find 

\begin{equation}
K(z,u_a)=
\biggl[
 {3\over 
{G_{Bab}}^2}-{12\over G_{Bab}}
\biggr]
+\biggl[
-{1\over 2}{1\over G_{Bab}^2}
+{1\over G_{Bab}}
+2
\biggr]
z^2
+O(z^4)
\label{scalintexp}
\end{equation}
\noindent
The complicated singularity appearing here at
the point $u_a=u_b$ indicates that this
form of the parameter integral is not yet
optimized for the purpose of
renormalization. In particular, it 
shows a spurious singularity 
in the coefficient of the
induced Maxwell term $\sim z^2$.
This comes not unexpected as the
cancellation of subdivergences implied
by the Ward identity has,
in a general gauge, 
no reason to be
manifest at the parameter
integral level.

We could improve on this either
by switching to Landau gauge, or
by performing a suitable partial
integration on the integrand.
The latter procedure is less systematic,
but easy enough to implement for the simple
case at hand:
Inspection of the two versions which we
have of this parameter integral, 
the original one eq.(\ref{Gamma2scal})
and the partially integrated
one eq.(\ref{Gamma2scalpI}),
shows that we can optimize the integrand
by choosing a certain linear combination of
both versions, namely

\begin{equation}
{\cal L}^{(2)}_{\rm scal}[B]=
{3\over 4}\times
{\rm eq.}
(\ref{Gamma2scal})
+{1\over 4}\times
{\rm eq.}
(\ref{Gamma2scalpI})
.
\label{Gamma2scaloptim}
\end{equation}
\noindent
(taking the photon insertion in Landau
gauge would yield a similar simplification,
though the resulting parameter integrals
are not identical 
\footnote{We remark that they {\sl would}
be identical for the special case of a 
self-dual field.}).
After integration over $\bar T$, this leads to
another version of eq.(\ref{Gamma2scalB}),

\begin{equation}
{\cal L}^{(2)}_{\rm scal}
[B]=-
{(4\pi )}^{-4}
{e^2\over 2}
\int_{0}^{\infty}{dT\over T^3}e^{-m^2T}
{z\over \sinh(z)}
\int_0^1 du_a
\,A'(z,u_a)
\; 
\label{Gamma2scalBprime}
\end{equation}
\noindent
with a different integrand

\begin{eqnarray}
A'&=&
\Biggl\lbrace
A'_0
{\ln ({G_{Bab}/ G_{Bab}^z})
\over{(G_{Bab}-G_{Bab}^z)}}
+
A'_1
{\ln ({G_{Bab}/ G_{Bab}^z})
\over{(G_{Bab}-G_{Bab}^z)}^2}
\nonumber\\
&&
+{A'_2\over
(G^z_{Bab})(G_{Bab}-G^z_{Bab})}
+{A'_3\over
(G_{Bab})(G_{Bab}-G^z_{Bab})}
\Biggr\rbrace\quad ,\nonumber\\
A'_0&=&
3\biggl[
2z^2G^z_{Bab}-{z\over\tanh(z)}-1
\biggr]
\quad  \nonumber\\
A'_1 &=&
A_1-{3\over 2}
\Bigl[
\dot G_{Bab}^2 -\dot G_{Bab}^{z2}
\Bigr]
\quad  \nonumber\\
A'_2&=&
A_2-{3\over 2}
\Bigl[
\dot G_{Bab}\dot G_{Bab}^z
+\dot G_{Bab}^{z2}
\Bigr]
\quad \nonumber\\
A'_3&=&
A_3+{3\over 2}
\Bigl[
\dot G_{Bab}^2
+\dot G_{Bab}\dot G_{Bab}^z
\Bigr]
\;  
\label{A1A2A3prime}
\end{eqnarray}
\noindent
We have not yet taken into account here the term involving
$\delta_{ab}$, stemming from $\ddot{\cal G}_{Bab}$, which was
contained in the integrand of 
eq.(\ref{Gamma2scalBprime}). 
This term 
corresponds, in diagrammatic terms, to a tadpole insertion,
and could therefore be safely deleted. However, it will be
quite instructive to keep it and check explicitly that
it is taken care of by the renormalization procedure.
It leads to an integral $\int_0^{\infty} 
{d\bar T\over {\bar T}^2}$ which we regulate by introducing
an ultraviolet cutoff for the photon proper-time,

\be
\int_{{\bar T}_0}^{\infty}
{d\bar T\over {\bar T}^2}
={1\over {\bar T}_0}
\; 
\label{calctadpole}
\ee
\no
It gives then a further contribution $E({\bar T}_0)$ to
${\cal L}^{(2)}_{\rm scal}[B]$,

\begin{equation}
E({\bar T}_0)=-3
{(4\pi )}^{-4}
e^2
{1\over {\bar T}_0}
\int_{0}^{\infty}{dT\over T^2}e^{-m^2T}
{z\over \sinh(z)}
\label{tadpole}
\end{equation}
\no
Expanding the new integrand,
$K'(z,u_a)\equiv{z\over\sinh(z)}A'(z,u_a)$,
in $z$, 
we find a much simpler result than before,

\begin{equation}
K'(z,u_a)=
-6{1\over G_{Bab}}
+3
z^2
+O(z^4)
\label{scalintexpprime}
\end{equation}
\noindent
In particular, the absence of a subdivergence for 
the Maxwell term is now manifest. 

We delete the irrelevant 
constant term, and add
and subtract the Maxwell term. 
If we define

\begin{equation}
K_{02}(z,u_a)=-6{1\over G_{Bab}}
+3
z^2
\label{defK02}
\end{equation}
\no
the Lagrangian
becomes

\begin{eqnarray}
{\cal L}^{(2)}_{\rm scal}
[B]&=&
E({\bar T}_0)
-{\alpha\over
2{(4\pi )}^{3}}
\int_{0}^{\infty}{dT\over T^3}e^{-m^2T}
\,3z^2\nonumber\\
&&
-{\alpha\over
2{(4\pi )}^{3}}
\int_{0}^{\infty}{dT\over T^3}e^{-m^2T}
\int_0^1 du_a
\,
\Bigl[
K'(z,u_a)-K_{02}(z,u_a)
\Bigr]
\;\nonumber\\
\label{Gamma2scalBprime2}
\end{eqnarray}
\noindent
The second term, which we denote by $F$,
 is divergent when integrated over
the scalar proper-time
$T$. We regulate it
by introducing another
proper-time cutoff $T_0$
for the scalar proper-time
integral:

\bear
F(T_0)&:=&
-{\alpha\over
2{(4\pi )}^{3}}
\int_{2T_0}^{\infty}{dT\over T^3}e^{-m^2T}
\,3z^2
\non\\
\label{defF}
\ear
\noindent
(compare ~\cite{ditreu}).
The third term is convergent at $T=0$, but
still has a divergence at $u_a=u_b$, as it
contains negative powers of
$G_{Bab}$. 

Expanding the integrand in
a Laurent series in $G_{Bab}$, one finds

\bear
K'(z,u_a)-K_{02}(z,u_a)&=&
{f(z)\over G_{Bab}}+O\Bigl(G_{Bab}^0\Bigr)
\nonumber\\
f(z)&=&
3
\biggl[
2-{z\over\sinh(z)}-{z^2\cosh(z)\over\sinh(z)^2}
\biggr]
\nonumber\\
\label{laurentGab}
\ear
\noindent
Again the singular part of this expansion
is added and subtracted, yielding

\begin{eqnarray}
{\cal L}^{(2)}_{\rm scal}
[B]&=&
E({\bar T}_0)
+
F(T_0)
-{\alpha\over
2{(4\pi )}^{3}}
\int_{2T_0}^{\infty}{dT\over T^3}e^{-m^2T}
\int_{T_0\over T}^{1-{T_0\over T}} du_a
\,{f(z)\over G_{Bab}}
\nonumber\\
&&
-{\alpha\over
2{(4\pi )}^{3}}
\int_{0}^{\infty}{dT\over T^3}e^{-m^2T}
\int_0^1 du_a
\,
\Bigl[
K'(z,u_a)-K_{02}(z,u_a)-{f(z)\over G_{Bab}}
\Bigr]
\; \nonumber\\
\label{Gamma2scalBprime3}
\end{eqnarray}
\noindent
The last integral is now completely finite.
The third term,
which we call $G(T_0)$,
is finite at $T=0$, as $f(z)=O(z^4)$
by construction. 
Here we have introduced
$T_0$ for the purpose of
regulating the divergence at $u_a=u_b$.

Obviously this term cannot be made finite
by photon wave function renormalization,
so we must try to use the scalar mass
renormalization for the purpose. This will
be seen to work out in a quite nontrivial way.
The $u_a$ -- integral for this term is
readily computed and yields, in the limit
$T_0\rightarrow 0$, a contribution

\be
\int_{T_0\over T}^{1-{T_0\over T}}
du_a{1\over G_{Bab}}
=
-2{\rm ln}\Bigl({T_0\over T}\Bigr)
=-2{\rm ln}(m^2T_0)
+2{\rm ln}(m^2T)
\label{intua}
\ee
\noindent
We have rewritten this term for reasons
which will become apparent in a moment.
The upshot is that we can relate the function
$f(z)$ to the scalar one-loop Euler-Heisenberg
Lagrangian, eq.(\ref{eulheiscal}).
If we write this Lagrangian for the pure
magnetic field case, and subtract the
two divergent
terms lowest order in $z$, 
we obtain

\be
\bar{\cal L}^{(1)}_{\rm scal}[B]
=
{1\over {(4\pi)}^2}\int_0^{\infty}
{dT\over T^3}{\rm e}^{-m^2T}
\biggl[
{z\over\sinh(z)}
+{z^2\over 6}-1
\biggr]
\label{Gamma1scalmagren}
\ee
\noindent
On the other hand, we can write

\bear
f(z)&=&
3
\biggl[
2-{z\over\sinh(z)}-{z^2\cosh(z)\over\sinh(z)^2}
\biggr]\nonumber\\
&=&
3T^3{d\over dT}
\biggl\lbrace
{1\over T^2}
\biggl[
{z\over\sinh(z)}+{z^2\over 6}-1
\biggr]
\biggr\rbrace
\nonumber\\
\label{rewritef}
\ear
\no
By a partial integration over $T$, we can therefore
re-express

\bear
{1\over {(4\pi)}^2}
\int_0^{\infty}
{dT\over T^3}{\rm e}^{-m^2T}
f(z)
&=&
3{m^2\over {(4\pi)}^2}
\int_0^{\infty}
{dT\over T^2}{\rm e}^{-m^2T}
\biggl[
{z\over\sinh(z)}
+{z^2\over 6}-1
\biggr]\nonumber\\
&=&
-3m^2{{\partial}\over{\partial} m^2}
\bar{\cal L}^{(1)}_{\rm scal}[B]
\label{partintmqed}
\ear
\no
(there are no boundary contributions since
$f(z)=O(z^4)$).
Next note that,
at the two-loop level, the 
effect of mass renormalization
consists in  the following shift
produced by the one-loop 
mass displacement $\delta m^2$,

\be
\delta{\cal L}_{\rm scal}^{(2)}
[B]
=
\delta
m^2
{{\partial}\over{\partial} m^2}
\bar{\cal L}^{(1)}_{\rm scal}
[B]
\, 
\label{scaldeltaL}
\ee\no

To proceed, we need thus again the value of the
one-loop mass displacement in scalar QED,
but this time in proper-time regularization, and
including its finite part.
Again we give only the result,

\be
\delta m^2=
{3\alpha\over 4\pi}
m^2
\Bigl[
-{\rm ln}
\Bigl(m^2 T_0\Bigr)
-\gamma
+ c+
{1\over m^2{\bar T}_0}
\Bigr]
\label{deltamscal}
\ee
\no
where $\gamma$ denotes the Euler-Mascheroni
constant
\footnote{In comparing with 
\cite{ritusspin,ritusscal,ditreu,rescsc}
note that there this constant had been denoted by $\ln (\gamma)$.} 
, and $c$ is a 
renormalization scheme dependent constant.
Using this result, we may rewrite

\bear
G(T_0)&=&
\biggl[\delta m^2-3c{\alpha\over 4\pi}m^2
-3{\alpha\over 4\pi {\bar T}_0}
\biggr]
{{\partial}\over{\partial} m^2}
\bar{\cal L}^{(1)}_{\rm scal}[B]
\nonumber\\
&&
-{\alpha\over {(4\pi)}^3}
\int_0^{\infty}
{dT\over T^3}{\rm e}^{-m^2T}
\Bigl[
{\rm ln}(m^2 T)
+\gamma
\Bigr]
f(z)
\nonumber\\
\label{calcGchap8}
\ear
\no
As expected the ${1\over {\bar T}_0}$ -- term
introduced by the one-loop mass renormalization
cancels the tadpole term $E({\bar T}_0)$,
up to its constant and Maxwell parts.
Moreover, the complete divergence of
$G(T_0)$ for $T_0\rightarrow 0$
has been absorbed by $\delta m^2$.
This is precisely the mechanism which had been
found already
in Ritus' analysis ~\cite{ritusspin,ritusscal,ginzburg}.
Putting all pieces together, we can write
the complete two-loop
approximation to the
effective Lagrangian 
in the following way: 

\bear
{\cal L}_{\rm scal}^{(\le 2)}[B_0]
&=&
-\half B_0^2
-{1\over{(4\pi)}^2}
\int_{T_0}^{\infty}
{dT\over T^3}
{\rm e}^{-m^2_0T}
{z^2\over 6}
+
\bar{\cal L}^{(1)}_{\rm scal}[B_0]
+
\delta m_0^2
{{\partial}\over{\partial} m_0^2}
\bar{\cal L}^{(1)}_{\rm scal}
[B_0]\nonumber\\
&&-3c{\alpha_0\over 4\pi}m_0^2
{{\partial}\over{\partial} m_0^2}
\bar{\cal L}^{(1)}_{\rm scal}
[B_0]
-{\alpha_0\over {(4\pi)}^3}
\int_0^{\infty}
{dT\over T^3}{\rm e}^{-m_0^2T}
\Bigl[
{\rm ln}(m_0^2 T)
+\gamma
\Bigr]
f(z)
\nonumber\\
&&
-{\alpha_0\over
2{(4\pi )}^{3}}
\int_{0}^{\infty}{dT\over T^3}e^{-m_0^2T}
\int_0^1 du_a
\,
\Bigl[
K'(z,u_a)-K_{02}(z,u_a)-{f(z)\over G_{Bab}}
\Bigr]
\nonumber\\
&&
-{\alpha_0\over
2{(4\pi )}^{3}}
\int_{2T_0}^{\infty}{dT\over T^3}e^{-m_0^2T}
\,z^2
\Bigl(3-{T\over {\bar T}_0}\Bigr)
\nonumber\\
\label{Gammascalunrenorm}
\ear
We have rewritten this Lagrangian in
bare quantities, since up to now we have been
working in the bare regularized theory.
Only mass and photon
wave function renormalization are required
to render this effective Lagrangian finite:

\bear
m^2&=&m_0^2+\delta m_0^2\nonumber\\
e&=&e_0Z_3^{\half}\nonumber\\
B&=&B_0Z_3^{-\half}\nonumber\\
\label{renormalization}
\ear
\no
(note that this leaves $z=e_0B_0T$ unaffected).
Here $\delta m_0^2$ has already been introduced
in eq.(\ref{deltamscal}), while $Z_3$ is
chosen so as to absorb the diverging
one- and two-loop Maxwell terms in eq.
(\ref{Gammascalunrenorm}).
The final answer becomes
 
\bear
{\cal L}_{\rm scal}^{(\le 2)}[B]
&=&
-\half B^2
+
{1\over {(4\pi)}^2}\int_0^{\infty}
{dT\over T^3}{\rm e}^{-m^2T}
\biggl[
{z\over\sinh(z)}
+{z^2\over 6}-1
\biggr]
\nonumber\\
&&
+{3c}{\alpha\over 4\pi}m^2
{1\over {(4\pi)}^2}\int_0^{\infty}
{dT\over T^2}{\rm e}^{-m^2T}
\biggl[
{z\over\sinh(z)}
+{z^2\over 6}-1
\biggr]
\nonumber\\
&&
-{\alpha\over
2{(4\pi )}^{3}}
\int_{0}^{\infty}{dT\over T^3}e^{-m^2T}
\int_0^1 du_a
\,
\Bigl[
K'(z,u_a)-K_{02}(z,u_a)-{f(z)\over G_{Bab}}
\Bigr]
\nonumber\\
&&
-{\alpha\over {(4\pi)}^3}
\int_0^{\infty}
{dT\over T^3}{\rm e}^{-m^2T}
\Bigl[
{\rm ln}(m^2 T)+\gamma
\Bigr]
f(z)
\nonumber\\
\label{Gammascalrenorm}
\ear
$e$ now denotes the physical charge. However,
the result still contains the undetermined
constant $c$, which appeared in the
finite part of the one-loop
scalar mass renormalization
eq.(\ref{deltamscal}). 
What remains to be done is to determine the
value of $c$ for which the renormalized
mass $m$ becomes the physical mass.

The worldline formalism applies, at least
at the present stage of its development, only to the
calculation of bare regularized amplitudes. 
As we have seen already in our $\beta$ -- function
calculations, 
the
renormalization of those amplitudes has to rely
on auxiliary calculations in standard field theory.
Usually worldline calculations are done in dimensional
regularization, where there is no harm done in
using different formalisms for the calculation
of graphs and countergraphs, due to the
universality of the 
regulator and the minimal subtraction prescription
\footnote{
This may not hold in certain cases involving
$\gamma_5$ or spacetime supersymmetry.}.
The same is not true
for multiloop calculations
using a proper-time cutoff, where one must make
sure that the precise way of applying the cutoff
is chosen consistently between graphs and
countergraphs; otherwise one may have effectively
performed unwanted finite renormalizations.
This problem is well-known from multiloop
calculations
in scalar field theory performed with 
a naive momentum space cutoff (see \cite{vladimirov}
and refs. therein).

To ensure a correct identification of the physical
scalar mass, we find it easiest
to retrace the same calculation in dimensional
regularization.

If one keeps the integrands in
eqs.~(\ref{Gamma2scal}),
(\ref{Gamma2scalpI})
in $D$ dimensions, instead of eq.(\ref{simplifyremainder})
one obtains

\begin{eqnarray}
{\rm det}^{-{1\over 2}}
\biggl[{\sin({\cal Z})\over {{\cal Z}}}{\bigl(\bar T 
-{1\over 2}
{\cal C}_{ab}
\bigr )}
\biggr]
&=&
{z\over\sinh(z)}
\gamma^{{D\over 2} -1}\gamma^z\nonumber\\
{\rm tr}\Bigl[\ddot{\cal G}_{Bab}\Bigr]
&=&
2D\delta(\tau_a-\tau_b)
-2(D-2){1\over T}-{4\over T}
{z\cosh(z\dot G_{Bab})\over\sinh(z)}
\nonumber\\
{1\over 2}{\rm tr}\dot{\cal G}_{Bab}
{\rm tr}\biggl[
{\dot{\cal G}_{Bab}\over
{\bar T -{1\over 2}{\cal C}_{ab}}}
\biggr]
&=& \half
\biggl[
(D-2)\dot G_{Bab}+2\dot G_{Bab}^z
\biggr]
\biggl[
(D-2)\dot G_{Bab}\gamma
+2\dot G_{Bab}^z
\gamma^z
\biggr]
\nonumber\\
{1\over 2}{\rm tr}
\biggl[
{(\dot {\cal G}_{aa}-\dot {\cal G}_{ab})
(\dot {\cal G}_{ab}-\dot {\cal G}_{bb})
\over
{\bar T -{1\over 2}{\cal C}_{ab}}}
\biggr]
&=&
-\half (D-2)\dot G_{Bab}^2\gamma
-\biggl[
{\dot G^{z2}_{Bab}}
+{4\over T^2}z^2 G_{Bab}^{z2}
\biggr]
\gamma^z
\, 
\label{simplifyremainderdim}
\end{eqnarray}
\noindent
The term involving
$\delta(\tau_a-\tau_b)$
can now be omitted, since in
dimensional regularization 
it will not even contribute to the
unrenormalized
effective action (it leads to an
integral $\int_0^{\infty} d\bar T {\bar T}^{-{D\over 2}}$
which vanishes according to the rules of
dimensional regularization).

The linear combination eq.(\ref{Gamma2scaloptim})
generalizes to

\begin{equation}
{\cal L}^{(2)}_{\rm scal}[B]=
{{D-1}\over D}\times
{\rm eq.}
(\ref{Gamma2scal})
+{1\over D}\times
{\rm eq.}
(\ref{Gamma2scalpI})
\label{Gamma2scaloptimdim}
\end{equation}
\noindent
We rescale to the unit circle,
$\tau_{a,b}=Tu_{a,b}$,
set $\tau_b=0$ as usual, and also
rescale  $\bar T = T\hat T$.
This yields an integral

\begin{equation}
{\cal L}^{(2)}_{\rm scal}
[B]=-
{(4\pi )}^{-D}
{e^2\over 2}
\int_0^{\infty}{dT\over T}e^{-m^2T}T^{2-D} 
\int_0^{\infty}d\hat T 
\int_0^1 d u_a
\,
I(z,u_a,\hat T,D)\nonumber\\
\nonumber\\
\label{scaloptimint}
\end{equation}
\no
which is the $D$ -- dimensional version of 
eq.(\ref{Gamma2scalBprime}).
Note that
the rescaled integrand $I(z,u_a,\hat T,D)$ depends on
$T$ only through $z$.

In contrast to the calculation in proper-time regularization, 
the
$\hat T$ -- integration is nontrivial in dimensional regularization.
It will therefore now 
be easier to extract all subdivergences {\sl before}
performing this integral.
The analysis of the divergence structure shows that 
eqs.(\ref{scalintexpprime}),(\ref{laurentGab})
generalize to the dimensional case as follows,

\begin{equation}
K'(z,u_a,D)\equiv
\int_0^{\infty}
d\hat T\,
I(z,u_a,\hat T,D)
=
K_{02}(z,u_a,D)
+f(z,D)G_{Bab}^{1-{D\over 2}}
+
O\bigl(z^4,G_{Bab}^{2-{D\over 2}}\bigr)
\nonumber\\
\label{scaldivextr}
\end{equation}
\no
with
\bear
K_{02}(z,u_a,D)&=&
-4{{D-1}\over{D-2}}G_{Bab}^{1-{D\over 2}}
\non\\&&
+{2\over 3D(D-2)}
\biggl[
(D-1)(D-4)
G_{Bab}^{1-{D\over 2}}
+(
-2D^2+18D-4
)
G_{Bab}^{2-{D\over 2}}
\biggr]
z^2\nonumber\\
f(z,D)&=&
{D-1\over D(D-2)}
\biggl\lbrack
4D-{2\over 3}
(D-4)z^2
+(8-4D)
{z\over \sinh(z)}
-8
{z^2\cosh(z)\over \sinh^2(z)}
\biggr\rbrack
=O(z^4)\,
\non\\
\label{scaldivterms}
\end{eqnarray}
\noindent
After splitting off these two terms,
the
integral over the remainder is already finite,
so that one can set $D=4$ in its computation. 
The $\hat T$ - integral then becomes elementary,
and one is led back to
eq.(\ref{A1A2A3prime}),
since $K'(z,u_a,4) = K'(z,u_a)$.

Turning our attention to the second term
on the right hand side of eq.(\ref{scaldivextr}),
let us denote its contribution to the effective Lagrangian 
by $G_{\rm scal}(z,D)$,

\begin{equation}
G_{\rm scal}(z,D)
=
-
{(4\pi )}^{-D}
{e^2\over 2}
\int_0^{\infty}{dT\over T}e^{-m^2T}T^{2-D} 
\int_0^1 d u_a
\,
f(z,D)
G_{Bab}^{1-{D\over 2}}
\nonumber\\
\label{GzDint}
\end{equation}
\no
The equivalent of
eq.(\ref{intua}) now reads

\begin{equation}
\int_0^1\,du_a\,
G_{Bab}^{1-\Dhalf}
=
B\Bigl(2-\Dhalf,2-\Dhalf\Bigr)
=-{4\over\epsilon}+0+O(\epsilon )
\label{uaintdim}
\end{equation}
\no
where $B$ denotes the Euler Beta-function.

The identity eq.(\ref{rewritef})
generalizes to $D$ dimensions as
follows,

\bear
f(z,D)&=&
8{D-1\over D(D-2)}
T^{{D\over 2}+1}{d\over dT}
\biggl\lbrace
T^{-{D\over 2}}
\biggl[
{z\over\sinh(z)}+{z^2\over 6}-1
\biggr]
\biggr\rbrace
\, 
\label{rewritefdim}
\ear
\no
The partial integration over $T$ now produces
two terms,

\bear
\int_0^{\infty}
{dT\over T}{\rm e}^{-m^2T}
T^{2-D}
f(z,D)
&=&
8{D-1\over{D(D-2)}}
\biggl\lbrace
m^2
\int_0^{\infty}
{dT\over T}{\rm e}^{-m^2T}
T^{3-D}
\biggl[
{z\over\sinh(z)}
+{z^2\over 6}-1
\biggr]\nonumber\\
\hspace{30pt} &+&
{D-4\over 2}
\int_0^{\infty}
{dT\over T}{\rm e}^{-m^2T}
T^{2-D}
\biggl[
{z\over\sinh(z)}
+{z^2\over 6}-1
\biggr]
\biggr\rbrace
\non\\
\label{partintdim}
\ear
\no
We now need the complete one-loop mass displacement
calculated in dimensional regularization, which
is 
\footnote
{Note that this differs by a sign from $\delta m^2$
as used in eq.(\ref{masscount}) -
here $\delta m^2$ denotes the
mass displacement itself, while there 
it denoted the corresponding counterterm.}

\be
\delta m^2=
m^2
{\alpha_0\over 4\pi}
\Bigl[
-{6\over\epsilon}
+7
-3
\bigl[\gamma - \ln (4\pi)\bigr]
-3\ln (m^2)
\Bigr]+O(\epsilon)
\, 
\label{deltamscaldim}
\ee
\no
Expanding eqs.~(\ref{uaintdim}),
(\ref{partintdim}), and 
(\ref{scaldeltaL})
in $\epsilon$ one finds that,
up to terms of order $O(\epsilon)$,

\bear
G_{\rm scal}(z,D)&=&
\delta m^2 
{{\partial}\over{\partial} m^2}
\bar{\cal L}^{(1)}_{\rm scal}[B_0]
+m^2
{\alpha_0\over{(4\pi)}^3}
\int_0^{\infty}
{dT\over T^2}
\e^{-m^2T}
\biggl[
{z\over\sinh(z)}
+{z^2\over 6}-1
\biggr]
\nonumber\\
&&
\times
\biggl[
-3\gamma -3\ln (m^2T)+{3\over m^2T}
+{9\over 2}
\biggr]\,
\label{calcGzD}
\ear
\no
Note that again the whole divergence of
$G_{\rm scal}(z,D)$ for $D\rightarrow 4$
has been absorbed by $\delta m^2$.

Our final answer for the
two-loop contribution to the
finite renormalized scalar QED
Euler-Heisenberg thus becomes
\bear
{\cal L}_{\rm scal}^{(2)}[B]
&=&
-{\alpha\over
2{(4\pi )}^{3}}
\int_{0}^{\infty}{dT\over T^3}e^{-m^2T}
\int_0^1 du_a
\,
\Bigl[
K'(z,u_a)-K_{02}(z,u_a)-{f(z)\over G_{Bab}}
\Bigr]
\nonumber\\
&&
+{\alpha\over {(4\pi)}^3}m^2
\int_0^{\infty}
{dT\over T^2}{\rm e}^{-m^2T}
\biggl[
{z\over\sinh(z)}
+{z^2\over 6}-1
\biggr]
\biggl[
-3\gamma
-3\ln ( m^2T)+{3\over m^2T}
+{9\over 2}
\biggr]
\nonumber\\
\label{Gammascalrenormdim} 
\ear
This parameter integral representation is of a similar but
simpler structure than the one obtained in
~\cite{ritusscal,lebedev}, and we have not 
succeeded at a direct identification of both
formulas.
However, we have used MAPLE 
to
expand both formulas in a Taylor expansion in
$B$ up
to order $O(B^{20})$,
and found exact agreement for the coefficients.
Let us give the first few terms in this 
expansion,

\be
{\cal L}_{\rm scal}^{(2)}[B]
=
{\alpha m^4\over
{(4\pi )}^{3}}
{1\over 81}
\Biggl[
{\displaystyle \frac {275}{8}}
{\Bigl({B\over B_{\rm cr}}\Bigr)}^4
-
{\displaystyle \frac {5159}{200}}
{\Bigl({B\over B_{\rm cr}}\Bigr)}^6
+
{\displaystyle \frac {2255019}{39200}}
{\Bigl({B\over B_{\rm cr}}\Bigr)}^8
-
{\displaystyle \frac {931061}{3600}}
{\Bigl({B\over B_{\rm cr}}\Bigr)}^{10}
+\ldots \,
\Biggr]
\label{scalexpand}
\ee


\no
The expansion parameter has been rewritten 
in terms of
$B_{\rm cr}\equiv {m^2\over e}\approx 4.4 \cdot 10^{13}G$.

\subsubsection{Spinor QED}

The corresponding
calculation for fermion QED is completely
analogous, and we will present only
the version in dimensional regularization.

In the superfield formalism, the
formulas (\ref{Gamma2scal}), (\ref{wickscal})
immediately generalize to the following
integral representation
for the two-loop effective action
due to the spinor loop,

\begin{eqnarray}
{\cal L}^{(2)}_{\rm spin}
[F]&=&
(-2)
{(4\pi )}^{-D}
\Bigl(-{e^2\over 2}\Bigr)
\int_0^{\infty}{dT\over T}e^{-m^2T}T^{-{D\over 2}} 
\int_0^{\infty}d\bar T 
\int_0^T d\tau_a d\tau_b
\int d\theta_a d\theta_b
\nonumber\\
&\phantom{=}&\times
{\rm det}^{-{1\over 2}}
\biggl[{\tan({\cal Z})\over {{\cal Z}}}
\biggr]
{\rm det}^{-{1\over 2}}
\biggl[
\bar T 
-{1\over 2}
\hat{\cal C}_{ab}
\biggr]
\langle
-D_ay_a\cdot D_by_b\rangle
\quad \nonumber\\
\langle
-D_ay_a\cdot D_b y_b\rangle
&=&
{\rm tr}
\biggl[
D_aD_b\hat{\cal G}_{ab}+{1\over 2}
{D_a( \hat{\cal G}_{aa}- \hat{\cal G}_{ab})
D_b( \hat{\cal G}_{ab}- \hat{\cal G}_{bb})
\over
{\bar T -{1\over 2}\hat{\cal C}_{ab}}}
\biggr]\,
\non\\
\label{Gamma2spin}
\end{eqnarray}

Performing the Grassmann integrations, and
removing $\ddot {\cal G}_{Bab}$ by partial integration
over $\tau_a$, we obtain the equivalent of 
eq.~(\ref{Gamma2scalpI}),

\begin{eqnarray}
{\cal L}^{(2)}_{\rm spin}[F]&=&
{(4\pi )}^{-D}
e^2
\int_0^{\infty}{dT\over T}e^{-m^2T}T^{-{D\over 2}} 
\int_0^{\infty}d\bar T 
\int_0^T d\tau_a
\int_0^T d\tau_b
\nonumber\\
&\phantom{=}&\times
{\rm det}^{-{1\over 2}}
\biggl\lbrack{\tan({\cal Z})\over {{\cal Z}}}
\Bigl(
\bar T -{1\over 2}
{\cal C}_{ab}
\Bigr)
\biggr\rbrack
{1\over 2}
\Biggl\lbrace
{\rm tr}\dot{\cal G}_{Bab}
{\rm tr}
\biggl\lbrack
{\dot{\cal G}_{Bab}
\over
{\bar T -{1\over 2}{\cal C}_{ab}}}
\biggr\rbrack
-{\rm tr}{\cal G}_{Fab}
{\rm tr}\biggl[
{{\cal G}_{Fab}\over
{\bar T -{1\over 2}{\cal C}_{ab}}}
\biggr]
\nonumber\\
&\phantom{=}&
+{\rm tr}
\biggl[
{(\dot {\cal G}_{Baa}-\dot {\cal G}_{Bab})
(\dot {\cal G}_{Bab}-\dot {\cal G}_{Bbb}+2{\cal G}_{Faa})
+{\cal G}_{Fab}{\cal G}_{Fab}
-{\cal G}_{Faa}{\cal G}_{Fbb}
\over
{\bar T -{1\over 2}{\cal C}_{ab}}}
\biggr]
\Biggr\rbrace\; 
\non\\
\label{Gamma2spinpI}
\end{eqnarray}
\noindent
As we will discuss in more detail in the next section,
in the spinor loop case this partially integrated integral is
already a suitable starting point for renormalization.

Specializing to the magnetic field case,
it is again easy to
calculate the Lorentz determinants and traces.
In addition to the bosonic ones calculated in
eq.(\ref{simplifyremainderdim}) we now need
also

\bear
{\rm tr}{\cal G}_{Fab}
{\rm tr}\biggl[
{{\cal G}_{Fab}\over
{\bar T -{1\over 2}{\cal C}_{ab}}}
\biggr]&=&
G_{Fab}
\biggl[
(D-2)
+2 
{\cosh(z\dot G_{Bab})\over \cosh(z)}
\biggr]
G_{Fab}
\biggl[
(D-2)\gamma
+2 
{\cosh(z\dot G_{Bab})\over \cosh(z)}
\gamma^z
\biggr]
\nonumber\\
{\rm tr}
\biggl[
{(\dot {\cal G}_{Baa}-\dot {\cal G}_{Bab})
{\cal G}_{Faa}
\over
{\bar T -{1\over 2}{\cal C}_{ab}}}
\biggr]
&=&
2\gamma^z
\Bigl[
1-{\cosh(z\dot G_{Bab})\over \cosh(z)}
\Bigr]
\nonumber\\
\tr
\biggl[
{{\cal G}_{Fab}{\cal G}_{Fab}
\over
{\bar T -{1\over 2}{\cal C}_{ab}}}
\biggr]
&=&
(D-2)\gamma
+2\gamma^z
{\cosh^2(z\dot G_{Bab})+
\sinh^2(z\dot G_{Bab})
\over
\cosh^2(z)}
\nonumber\\
\tr
\biggl[
{{\cal G}_{Faa}{\cal G}_{Fbb}
\over
{\bar T -{1\over 2}{\cal C}_{ab}}}
\biggr]
&=&
2\tanh^2(z)\gamma^z
\nonumber\\
\label{remainingtraces}
\ear
\no
After rescaling to the unit circle, one
obtains a parameter integral

\begin{equation}
{\cal L}^{(2)}_{\rm spin}
[B]=
{(4\pi )}^{-D}
e^2
\int_0^{\infty}{dT\over T}e^{-m^2T}T^{2-D} 
\int_0^{\infty}d\hat T 
\int_0^1 d u_a
\,
J(z,u_a,\hat T,D)
\, 
\nonumber\\
\label{spinoptimint}
\end{equation}
\no
The extraction of the subdivergences
yields

\begin{equation}
L(z,u_a,D)\equiv
\int_0^{\infty}
d\hat T\,
J(z,u_a,\hat T,D)
=
L_{02}(z,u_a,D)
+g(z,D)G_{Bab}^{1-{D\over 2}}
+
O\bigl(z^4,G_{Bab}^{2-{D\over 2}}\bigr)
\nonumber\\
\label{spindivextr}
\end{equation}
\no
with
\be
L_{02}(z,u_a,D)=
-4(D-1)G_{Bab}^{1-{D\over 2}}
-{4\over 3D}
\biggl[
(D-1)(D-4)
G_{Bab}^{1-{D\over 2}}
+(D-2)(D-7)
G_{Bab}^{2-{D\over 2}}
\biggr]
z^2
\label{spinmaxdivterm}
\ee\no
(compare eq.(\ref{spinorintegrand})), and

\be 
g(z,D)=
-{4\over 3}
{D-1\over D}
\biggl\lbrack
6{z^2\over\sinh^2(z)}
+3(D-2)z\coth(z)-(D-4)z^2
-3D
\biggr\rbrack
=O(z^4)
\, 
\label{spinmassdivterm}
\ee
\noindent
$L_{02}$ is again removed by photon wave function renormalization.
Denoting the contribution of the second term by
$G_{\rm spin}(z,D)$, we note
that the $u_a$ - integral is the same as in the
scalar QED case, eq.~(\ref{uaintdim}).
Using the following identity 
analogous to eq.~(\ref{rewritef}),

\bear
g(z,D)&=&
8{D-1\over D}
T^{{D\over 2}+1}{d\over dT}
\biggl\lbrace
T^{-{D\over 2}}
\biggl[
{z\over\tanh(z)}-{z^2\over 3}-1
\biggr]
\biggr\rbrace
\label{rewriteg}
\ear
\no
we partially integrate the remaining integral over $T$. 
The $1\over\epsilon$ - part of $G_{\rm spin}$ is then
again found to be just right for absorbing the
shift induced by the one-loop mass displacement,

\be
\delta m_0=
m_0
{\alpha_0\over 4\pi}
\Bigl[
-{6\over\epsilon}
+4
-3
\bigl[\gamma - \ln (4\pi)\bigr]
-3\ln (m_0^2)
\Bigr]+O(\epsilon)
\, 
\label{deltamspin}
\ee
\no
Up to terms of order $\epsilon$ one obtains

\bear
G_{\rm spin}(z,D)&=&
\delta m_0 
{{\partial}\over{\partial} m_0}
\bar{\cal L}^{(1)}_{\rm spin}[B_0]
+m_0^2
{\alpha_0\over{(4\pi)}^3}
\int_0^{\infty}
{dT\over T^2}
\e^{-m_0^2T}
\biggl[
{z\over\tanh(z)}
-{z^2\over 3}-1
\biggr]
\nonumber\\
&&
\times
\biggl[
12\gamma +12\ln ( m_0^2T)-{12\over m_0^2T}
-18
\biggr]\,
\label{calcGspin}
\ear
\no
Our final result for the on-shell renormalized
two-loop spinor QED
Euler-Heisenberg Lagrangian is

\bear
{\cal L}_{\rm spin}^{(2)}[B]
&=&
{\alpha\over
{(4\pi )}^{3}}
\int_{0}^{\infty}{dT\over T^3}e^{-m^2T}
\int_0^1 du_a
\,
\Bigl[
L(z,u_a,4)-L_{02}(z,u_a,4)-{g(z,4)\over G_{Bab}}
\Bigr]\nonumber\\
&&-
{\alpha\over {(4\pi)}^3}
m^2\int_0^{\infty}
{dT\over T^2}{\rm e}^{-m^2T}
\biggl[
{z\over\tanh(z)}
-{z^2\over 3}-1
\biggr]
\biggl[18-12\gamma-
12\ln (m^2T)+{12\over m^2T}
\biggr]
\nonumber\\
\label{Gammaspinrenorm}
\ear\no
with

\bear
L(z,u_a,4)&=&{z\over\tanh(z)}
\Biggl\lbrace
B_1
{\ln ({G_{Bab}/G_{Bab}^z})
\over{(G_{Bab}-G_{Bab}^z)}^2}
+{B_2\over
G^z_{Bab}(G_{Bab}-G^z_{Bab})}
+{B_3\over
G_{Bab}(G_{Bab}-G^z_{Bab})}
\Biggr\rbrace\nonumber\\
B_1&=&4z
\Bigl(
\coth(z)-\tanh(z)
\Bigr)
G^z_{Bab}-4G_{Bab}
\quad  \nonumber\\
B_2&=&2\dot G_{Bab}\dot G^z_{Bab}+ 
z(8\tanh(z)-4\coth(z))
G^z_{Bab}-2
\quad \nonumber\\
B_3&=&4G_{Bab}
-2\dot G_{Bab}\dot G^z_{Bab}
-4z\tanh(z)G^z_{Bab}+2\nonumber\\
L_{02}(z,u_a,4)&=&-{12\over G_{Bab}}+2z^2\nonumber\\
g(z,4)&=&-6\biggl[
{z^2\over{\sinh(z)}^2}+z\coth(z)-2
\biggr]
\label{defLLg}
\ear\no
Comparing with the previous results by 
Ritus ~\cite{ritusspin,ginzburg}
and Dittrich-Reuter ~\cite{ditreu},
we have again not succeeded
at a direct identification with the 
more complicated parameter integral given
by Ritus.
However, as in the scalar QED case we have verified agreement
between both formulas up to the
order of $O(B^{20})$ in the weak-field
expansion in $B$. The first few coefficients are

\be
{\cal L}_{\rm spin}^{(2)}[B]
=
{\alpha m^4\over
{(4\pi )}^{3}}
{1\over 81}
\Biggl[
64 
{\Bigl({B\over B_{\rm cr}}\Bigr)}^4
-{1219\over 25}
{\Bigl({B\over B_{\rm cr}}\Bigr)}^6
+ {135308\over 1225}
{\Bigl({B\over B_{\rm cr}}\Bigr)}^8
-{791384\over 1575}
{\Bigl({B\over B_{\rm cr}}\Bigr)}^{10}
+\ldots \, 
\Biggr]
\label{spinexpand}
\ee


On the other hand, our formula {\sl almost}
allows for a term by term
identification with the result of Dittrich-Reuter
\cite{ditreu}, as given in eqs.~(7.21),(7.22),(7.37) there.
This requires 
a rotation to Minkowskian proper-time,
$T\rightarrow is$, a transformation of variables
from $u_a$ to $v:=\dot G_{Bab}$, 
the use of trigonometric identities,
and another partial
integration over $T$ for the second term in
eq.~(\ref{Gammaspinrenorm}). The only discrepancy arises in
the constant $18$, which reads $10$ in the Dittrich-Reuter
formula. 
One concludes that the results 
reached by
Ritus and Dittrich-Reuter
for the two-loop Euler-Heisenberg Lagrangian are
incompatible, and differ precisely by a finite electron
mass renormalization
\footnote{The two formulas had been compared in ~\cite{ditreu}
only in the strong field limit, which is not sensitive to this
discrepancy.}.
Moreover, it is clearly Ritus'
formula which correctly identifies the physical electron
mass.

The corresponding result for the case of a pure electric field is obtained
from this by the substitution $B^2\rightarrow -E^2$. 
This sign chance makes an important difference, since 
it creates an imaginary part for the effective
action, indicative of the possibility of pair
creation in an electric field \cite{schwinger51}.
In \cite{dunsch} an approximation for this imaginary
part was obtained from the weak
field expansion for the magnetic case, using 
Borel summation and a dispersion relation.

The generalization of this calculation to the case of a general
constant background field is straightforward \cite{korsch}.

Concerning the physical relevance of this type of calculation,
let us mention the experiment PVLAS in preparation at Legnaro,
Italy, which
is an optical experiment designed to yield the first experimental
measurement of the Euler-Heisenberg Lagrangian
~\cite{pvlas1,pvlas2}. 
It is conceivable that the
technology used there may even allow for the measurement of
the two-loop correction in the near future 
~\cite{bakalov,fermilab877}. 

\subsection{Some More Remarks on the 2-Loop QED $\beta$ -- Functions}

The calculation of the two-loop
Euler-Heisenberg Lagrangian has to teach us also something
about our previous $\beta$ -- function calculation.
Of course, the $\beta$ -- function
coefficients 
can be simply retrieved
from the Euler-Heisenberg Lagrangians.
Up to the contributions from mass renormalization,
they can be read
off from the expansions 
eq.(\ref{scalintexpprime}),
eq.(\ref{defLLg})

\bear
K'(z,u_a,4)
&=&
-{6\over G_{Bab}}+3z^2+O(z^4)\nonumber\\
L(z,u_a,4)
&=&
-{12\over G_{Bab}}+2z^2+O(z^4)
\nonumber\\
\label{spinintexpprime}
\ear\no
For example, the coefficient $2$ appearing in the
second line is nothing else but
the $-8$ which we found in 
eq.(\ref{threeint}) (up to the global factor of
$-2$, and another factor of $2$ which is due to
the different choice of field strength tensors).

Comparing with that calculation we see that the
use of the generalized Green's functions ${\cal G}_B,
{\cal G}_F$ has saved us two integrations: 
The same formulas
eq.(\ref{master}) which there had been employed
for executing the integrations over the 
points of interaction
$\tau_1,\tau_2$ with the external field,
have now entered already at the level of the
construction of those Green's functions.
Of course, for the $\beta$ -- function calculation
all terms of order higher than $O(F^2)$ are irrelevant,
so that one could then as well use the
truncations of those Green's functions given in
eq.(\ref{GB(F)expand}). Moreover, the choice of
an external field with the property $F^2\sim \Eins$
is then more convenient than a magnetic field
(this variant of the two-loop $\beta$ -- function
calculation was presented in ~\cite{zako}).

More interestingly, we
had noted before that, if the spinor QED
$\beta$ -- function is calculated
in a four-dimensional scheme,
a subdivergence-free integrand
is obtained proceeding directly from the
partially integrated version eq.(\ref{Gamma2spinpI}).
We are now in a position to see that this fact is not accidental,
but a consequence of renormalizability itself.

Let us retrace our two-loop Euler-Heisenberg calculations,
and analyze how the removal of all divergences worked
for the Maxwell term.
In the scalar QED case, there are
three possible sources of quadratic divergences for the
induced $z^2$ -- term:

\begin{enumerate}
\item
The contact term containing $\delta_{ab}$.

\item
The leading order term $\sim {1\over G_{Bab}^2}z^2$
in the ${1\over G_{Bab}}$ -- expansion
of the main term (see, e.g., eq.(\ref{scalintexp})).

\item
The explicit $1/{\bar T}_0$ appearing in the one--loop 
mass displacement eq.(\ref{deltamscal}).

\end{enumerate}
\no
The last one should cancel the other two in the
renormalization procedure, if those are regulated by the
same UV cutoff ${\bar T}_0$ for the photon proper-time,
and this is indeed the case, as we verified
in various versions of this calculation.
In the spinor QED case the fermion propagator has no quadratic 
divergence (this is, of course, manifest in the
standard first order formulation, while in the second
order formulation discussed in section \ref{comparefeynman}
there are several diagrams contributing to the electron
self energy, and the absence of a quadratic divergence is
due to a cancellation among them).
The third term is thus missing, and the other two have to
cancel among themselves. In particular, the completely partially
integrated version of the integrand has no $\delta_{ab}$ -- term
any more, and consequently the second term must also be absent.
But the structure of the integrals is such that,
if one does this calculation in $D=4$, 
the ${1\over G_{Bab}}$ -- expansion
of the main contribution to the Maxwell 
term is always of the form
shown in eq.(\ref{scalintexp}),

\be
\biggl[
{A\over G_{Bab}^2}+{B\over G_{Bab}}+C
\biggr]{\rm tr}(F^2)
\label{maxwellexp}
\ee
\no
with coefficients $A,B,C$.
In the partially integrated version first the absence of a 
quadratic subdivergence allows one to conclude that $A=0$, and then
the absence of a logarithmic subdivergence that $B=0$. 

Note that this argument does not apply
to the scalar QED case, nor does it to
spinor QED in dimensional
renormalization, due to the principal suppression of quadratic divergences by
that scheme. In both cases one would have only one constraint
equation for the two coefficients $A$ and $B$ appearing in the
partially integrated integrand, and indeed
they turn out to be nonzero in both cases.
In the present formalism, the fermion
QED two-loop $\beta$ -- function calculation 
thus becomes simpler when performed
not in dimensional regularization, but
in some four-dimensional scheme such as proper-time
or Pauli-Villars regularization.

\subsection{Beyond 2 Loops}

This cancellation mechanism
is interesting in
view of some facts known about the 
three-loop 
fermion QED $\beta$ -- function
\cite{rosner,brandt,brdekr}.
Apart from the well-known cancellation
of transcendental numbers occuring between diagrams
in the calculation of the
quenched (one fermion loop) contribution
to this $\beta$ -- function \cite{rosner,brdekr},
which takes place in any scheme and gauge, even more
spectacular cancellations were found in 
~\cite{brandt} where this
calculation was performed in 
four dimensions,
Pauli-Villars regularization, and
Feynman gauge. 
In that calculation all contributions 
from
non-planar diagrams happened to cancel out
exactly.

It appears that previously gauge invariance
was considered as the only source of
cancellations in this type of calculation.
The cancellation mechanism which we exhibited in the
previous section is clearly of a different type,
and specific to spinor QED.
Whether this mechanism
 has a generalization to the three-loop
level, or perhaps even relates to
the cancellation found by
Brandt, remains to be seen.
In a preliminary study ~\cite{fsswip}
we have computerized the
generation of the partially integrated integrand
for the three-loop spinor QED vacuum amplitude
in a constant field, using MAPLE and M ~\cite{M}.
In the three-loop
case the integrations over the proper-time
parameters for the two inserted photons are
still elementary. Therefore it turns
out to be relatively easy to generate the 
unrenormalized 3-loop Euler-Heisenberg Lagrangians,
which are now four-parameter integrals.
However the analysis of the divergence structure along
the above lines is a rather formidable problem,
and so far no conclusive results have been reached.
This is due not only to the large number of terms
generated at the three-loop level, but also
to the existence of several different 
subdivergences (in the notation of the 
example in section 
\ref{3loopexample} those are at $a\sim b$, $c\sim d$,
$(a,b)\sim (c,d)$, and $(a,b)\sim (d,c)$).

After discovering the rationality of the three - loop
quenched QED $\beta$ function many years ago,
Rosner ~\cite{rosner} conjectured that this
$\beta$ - function may perhaps provide a window
to high orders in perturbation theory.
Whether or not the worldline formalism will eventually
allow one to go beyond the four orders presently
accessible to other methods 
in this calculation,
is impossible
to say at present. Still we believe that this
is a question very much worth pursuing,
and that the
answer will be a good indicator for the
ultimate usefulness of this formalism.

\section{Conclusions and Outlook}
\renewcommand{\theequation}{10.\arabic{equation}}
\setcounter{equation}{0}

In this work we have reviewed the present status
of the ``string-inspired'' technique, and its
range of applications in perturbative quantum field theory.
Although we have given a sketch of the original 
derivation of the Bern-Kosower rules from string
theory, based on an analysis of the field theory limit of
the string path integral, 
our overall emphasis has been on Strassler's more
elementary ``worldline'' approach, using first-quantized
particle path integral representations for one-loop effective
actions. From our discussion of QCD amplitudes in chapter 4
it should have become clear that these two approaches 
complement each other nicely. The worldline approach
provides a simple and efficient method for computing either
the QCD effective action itself, or the one-particle irreducible
$N$ -- gluon vertex function. The original string -- based approach 
appears to be more powerful when it comes to the calculation of
the $N$ -- gluon scattering amplitude. Here the worldline approach
can still be used to correctly generate the input integrand for the
Bern -- Kosower rules, and also for a rederivation of the
``loop replacement'' part of those rules. However it
presently still falls short of yielding 
a complete rederivation of the ``pinch''
part of the Bern -- Kosower rules. On the other hand, the
string -- based approach is a priori restricted to the on-shell
case, due to the requirement of conformal invariance. The off-shell
continuation of string amplitudes in general leads to ambiguities,
although at the one -- loop level 
those seem now to be under control
\cite{dlmmr:plb388,dlmmr:npb469,limape}.
In particular, in \cite{dlmmr:npb469} a method of continuation
was given which in the field theory limit yields the
gluon amplitudes in background field Feynman gauge.
This line of work has recently led to the formulation
of a general algorithm for computing the
off-shell, one-loop multigluon amplitudes 
from the bosonic string \cite{frmaru00}. 

Some complementarity holds also concerning the present range 
of applications of both
methods. The string -- based method has so far, apart from scalar
amplitudes \cite{dlmmr:plb388,mape,frmaru}, 
essentially been applied only to one -- loop gluon \cite{bediko5glu} 
and graviton \cite{bedush,dunnor,djst} 
scattering amplitudes.
The worldline path integral approach has provided a simple
means to generate Bern -- Kosower type master formulas also for
field theories involving Yukawa \cite{mnss1,dhogag1} 
and axial \cite{mnss2,dhogag2,mcksch,dimcsc} couplings.
On the other hand, its application to gravitational backgrounds
requires the mastery of some technical subtleties
\cite{basvan,bpsv:npb446,bpsv:npb459,schvan,bascva,bascor,kleche1,kleche2,kleche3,baconi1,baconi2},
as we discussed in section \ref{otherbackgrounds}.
In this context most authors have concentrated on the
computation of anomalies 
\cite{alvarezgaume,alvwit,friwin,bastianelli,basvan,bpsv:npb446,bpsv:npb459} 
or of the effective action
\cite{dilmck,bascor,baconi1,baconi2}, 
although recently an interesting application was also
given to the calculation of scattering amplitudes in 
eleven-dimensional supergravity \cite{grgukw}.

Beyond the one -- loop level, our presentation has been
confined to the cases of scalar field theory and QED.
As we already mentioned in the introduction,
the construction of multiloop QCD amplitudes
in the string -- based formalism has turned out to be
a formidable technical challenge \cite{roland:unpubl}. 
Nevertheless, the recent progress in this line of work
\cite{dlmmr:plb388,dlmmr:npb469,magrus,limape,mape,frmaru,korsch:string} 
seems to indicate that at least a computation of the
two -- loop QCD $\beta$ -- function may now be in reach.
A parallel development using the worldline approach \cite{satsch1,satsch2}
has led to the derivation of a
two -- loop Euler-Heisenberg type action for pure
Yang-Mills theory \cite{sascza}, albeit not yet in a form which would allow
one to extract the two - loop Yang-Mills $\beta$ -- function.

As we have shown here, things are much easier in the abelian case.
Here the worldline path integral approach provides an
easy route to the construction of multiloop Bern-Kosower
type formulas, based on the concept of multi-loop worldline Green's
functions. Those Green's functions were explicitly constructed for 
Hamiltonian graphs, and
carry the full
information on the internal propagators
inserted into the Hamiltonian loop.
Concerning the significance of this concept,
from Roland and Sato's result eq.(\ref{Gasympt})
one can see that this treatment of scalar
propagator insertions is natural and
``stringy''.
Whatever the ultimate form of the multiloop
Bern-Kosower rules may turn out to be,
it seems likely that those functions will 
figure in them prominently.

The same cannot be said for our treatment of
photon insertions. Here it is rather clear
that a truly string -- based approach will
incorporate internal photons
in a more organic way.
Nevertheless, our experience with the formalism at the
two - and three - loop level clearly shows that
it has many of the properties which one expects 
from a string - derived approach.
It also seems to indicate that, at the multiloop
level, quantum electrodynamics 
is a field theory which should be particularly
suited to the application of
string-derived techniques. 
This impression is based on the following
properties:

\begin{itemize}
\item
Particularly extensive
cancellations are known to
occur in multiloop QED calculations, suggesting that
the standard field theory methods are far from optimized
for this task. 

\item
In contrast to scalar and non-abelian gauge theories,
for QED amplitudes
there exists a natural worldline parametrization 
exhausting the complete S-matrix.

\item
Sums of Feynman diagrams are always generated with the
correct statistical weights.

\item
No colour factors exist which may
prevent us from combining diagrams by ``letting legs slide
along lines''. 

\end{itemize}
\no
In the presentation of this formalism given here
we have tried to make maximal use of the
worldline supersymmetry (which comes as
a free gift from heaven in the worldline formalism).
There are several aspects to this. The existence
of this supersymmetry alone
already leads to functional 
relationships
between the parameter integrals for 
processes
with the same external states, but different types of 
virtual particles involved.
Moreover, the worldline superfield formalism
allows one to treat scalar, spinor and gluon loops
in a uniform manner, as well as
to avoid the introduction of two different types of vertex operators.

The introduction of multiloop worldline
Green's functions and of worldline superfields
have a principle in common, which is that
one should always try to absorb a maximum of 
information into the worldline propagators
themselves.
A third example of this principle 
was our introduction of worldline
Green's functions incorporating
constant external electromagnetic fields.
As we have seen in section \ref{qedmext} those
three instances of the principle can be freely combined
with each other.

We have discussed a large number of applications, often
in considerable detail.
Some of those calculations have been of
an illustrative nature, or mere consistency checks.
Others were state-of-the-art calculations, such as
our recalculations of the QED photon splitting amplitudes 
and of the two-loop Euler-Heisenberg
Lagrangians.
Those examples clearly display the technical advantages over standard
field theory techniques which one can hope to achieve
in this formalism, at least for certain types of calculations.
However, they also show that
the string -- inspired technique is presently still a rather specialized 
tool. All of our applications have been to processes involving a loop
{\sl with no change of the identity of the particle inside the loop}.
This class of amplitudes seems to be the most natural 
one to consider in this
context, although it is by no means the only one to which 
string -- inspired techniques can be applied.
Notably a generalization of the Bern-Kosower formalism to amplitudes
involving external quarks was constructed by L. Dixon
~\cite{dixon}, but in preliminary studies has not led
to as significant an improvement over field theory methods 
as was found in the calculation of the four and five gluon amplitudes
~\cite{dixonpc}
\footnote{This evaluation could possibly change due
to recent progress ~\cite{pasrol1,pasrol2} 
with regard to the computation of 
heterotic string amplitudes involving external fermions.}.
Similarly also the worldline formalism can be easily extended
to the computation of fermion self energies \cite{mckreb,karkto1,karkto2}.
However the resulting formalism seems somewhat less elegant than in the
photon self energy case, and has been too little explored yet to
be presented here. 

Another omission in the present review is the extension to
the finite temperature case, which has been considered by various
authors 
\cite{mckreb:therm1,mckreb:therm2,borkut,mckeon:therm,haasch,sato:therm}. 
The most general result which has been
reached in this line of work is a generalization of the
QED Bern -- Kosower master formula to the $N$ -- photon amplitude at
finite temperature and chemical potential \cite{sato:therm}. 
However, this result holds for the Euclidean amplitude, and for most
physical applications would still have to be analytically continued.
Here one encounters the same type of ambiguities
as in Feynman parameter calculations at finite temperature 
\cite{weldon,das}, which makes it presently difficult to judge
the practical usefulness of this generalization.

Finally, it is quite possible
that the use made of worldline path integrals
in the present work may appear
overly modest to the enterprising reader. 
Our whole aim here was to reproduce the
S-matrices of known renormalizable field theories
in a way which avoids some of the shortcomings of
conventional Feynman diagram calculations.
Clearly it is always possible to rewrite
a given field theory amplitude in terms of worldline
path integrals, although not in all cases this will
lead to calculational improvements. Much less obvious
is the converse question, which is whether
a ``sensible'' worldline Lagrangian must always be
induced by a spacetime Lagrangian, or whether
worldline path integrals can perhaps be used
to define physically relevant S-matrices 
which do not correspond to Lagrangian field theories.
The use of an additional worldline curvature term
\cite{pisarski,awazol,baclja} could be seen as
a step in this direction.

\vskip.25cm
{\bf Acknowledgements:}
I would like to thank M.G. Schmidt for a fruitful 
collaboration on many of the topics dealt with in
this work, as well as S.L. Adler, F.A. Dilkes, G. Dunne, 
D. Fliegner, P. Haberl, 
D.G.C. McKeon, M. Mondrag{\'o}n, L. Nellen and M. Reuter.  
Of the many other 
colleagues who supported this work in one
way or the other I wish to particularly thank: 
Z. Bern for kind
hospitality during a visit at UCLA, as well as for many 
explanations on the Bern-Kosower formalism;
D. Broadhurst for encouragement, and
for telling me many things about quantum electrodynamics
which I would have never learned otherwise;
L. Dixon for detailed informations on his unpublished extension of
the Bern-Kosower formalism to include external quarks,
as well as for explanations on \cite{dedima};
A. Laser for performing a 
recalculation of the two-loop spinor QED
$\beta$ -- function
in the second order formalism,
as a check on the worldline
calculation presented in section 9.4;
F. Bastianelli, T. Binoth, L. Dixon, R. Russo, H.-T. Sato,
and B. Zwiebach for useful comments on the
preprint version of this work.
Discussions and correspondence with
V.N. Baier,
F. Bastianelli,
J. Biebl, 
A. Davydychev, E. D'Hoker, 
W. Dittrich,
H. Dorn, 
D. Dunbar,
H. Gies,
H. Kleinert, D. Kreimer,
O. Lechtenfeld, 
N.E. Ligterink,
D. L{\"u}st, A. Morgan, U. M{\"u}ller, 
C. Preitschopf, 
M. Rausch, 
V.I. Ritus,
J.L. Rosner,
R. Stora,
M. J. Strassler, O. Tarasov, 
B. Tausk,
and 
P. van Nieuwenhuizen
are gratefully acknowledged
\footnote{
Special thanks to the referee for a large
number of useful suggestions and criticisms!}
.
I also thank the Institute for Advanced Study, Princeton,
and the Theory Group of Argonne National Laboratory 
for hospitality.


\begin{appendix}
\section{Summary of Conventions}
\label{conv}

\renewcommand{\theequation}{A.\arabic{equation}}
\setcounter{equation}{0}
\vskip10pt

At the path integral level, we work in the Euclidean throughout with
a positive definite metric
$(g_{\mu\nu})={\,\mathrm diag}(++\ldots +)$.
Our Euclidean Dirac matrix conventions are

\bear
\lbrace\gamma_{\mu},\gamma_{\nu}\rbrace &=& 2g_{\mu\nu}
\Eins,
\quad
\gamma_{\mu}^{\dag} = \gamma_{\mu},
\quad
\gamma_5 = \gamma_1\gamma_2\gamma_3\gamma_4,
\quad
\sigma_{\mu\nu} = \half [\gamma_{\mu},\gamma_{\nu}]
\label{convdirac}
\ear\no
The Euclidean field strength tensor is defined by
$F^{ij}= \varepsilon_{ijk}B_k, i,j = 1,2,3$,
$F^{4i}=-iE_i$, its dual by
$\tilde F^{\mu\nu} = \half 
\varepsilon^{\mu\nu\alpha\beta}F^{\alpha\beta}$
with $\varepsilon^{1234} = 1$.  

The corresponding Minkowski space amplitudes
are obtained by replacing 

\bear
g_{\mu\nu}&\rightarrow& \eta_{\mu\nu} \non\\
k^4&\rightarrow& -ik^0 \non\\
T&\rightarrow& is \non\\
\varepsilon^{1234}
&\rightarrow&
i\varepsilon^{1230} \non\\
F^{4i}&\rightarrow& F^{0i}=E_i \non\\
\tilde F^{\mu\nu}&\rightarrow& -i\tilde F^{\mu\nu} \non\\
\label{euctomink}
\ear\no
where
$(\eta_{\mu\nu}) = {\,\mathrm diag}(-+++),\quad \varepsilon^{0123}=1$.

\no
The non-abelian covariant derivative is
$D_\mu\equiv \partial_\mu+ig A^a_\mu  T^a$,
with $[T^a,T^b] = i f^{abc}T^c$. The adjoint
representation is given by $(T^a)^{bc}=-i f^{abc}$,
and the generators in the fundamental representation 
of $SU(N_c)$ are
normalized as 
${\rm tr}(T^aT^b) = \half \delta^{ab}$.

\no
Momenta appearing in vertex operators are {\sl ingoing}.

\section{Worldline Green's Functions}
\label{greendet}
\renewcommand{\theequation}{B.\arabic{equation}}
\setcounter{equation}{0}
\vskip10pt
\noindent

In this appendix we derive the generalized worldline
Green's functions $G^C_{P,A}$
needed for the evaluation of the
gluon -- path integral 
(section \ref{gluonloopgluonscatter})
and ${\cal G}_{B,F}$ for the scalar/spinor path integral in a constant
external field (chapter \ref{qed}). 

All those Green's functions
are kernels of certain integral operators, acting in 
the real Hilbert space of periodic or
antiperiodic functions defined on an interval
of length $T$. 
We denote by $\bar H_P$ the full space of periodic 
functions, by $H_P$ the same space with the constant
mode exempted, and by $H_A$ the space of antiperiodic
functions. The ordinary derivative acting on those
functions is correpondingly denoted by
$\partial_P$, $\bar\partial_P$ or $\partial_A$.
With those definitions, we can write our Green's functions
as

\begin{eqnarray}
{\cal G}_B(\tau_1,\tau_2)&=&2
\langle \tau_1\mid
{\Bigl(
{\partial_P}^2
-2iF\partial_P
\Bigr)}^{-1}
\mid
\tau_2\rangle\nonumber\\
{\cal G}_F(\tau_1,\tau_2)&=&2
\langle \tau_1\mid
{\Bigl(
{\partial_A}
-2iF\Bigr)}^{-1}\mid
\tau_2\rangle\nonumber\\
{G}_P^C(\tau_1,\tau_2)&=&
\langle \tau_1\mid
{\Bigl(
{\bar\partial_P}-C
\Bigr)}^{-1}
\mid \tau_2\rangle\nonumber\\
{G}_A^C(\tau_1,\tau_2)&=&
\langle \tau_1\mid
{\Bigl(
{\partial_A}-C\Bigr)}^{-1}
\mid \tau_2\rangle\nonumber\\
\label{collectGreen's}
\end{eqnarray}
\noindent
(in this appendix we 
absorb the coupling constant
$e$ into the external field $F$).
Note that ${G}^C_A$ is,
up to a conventional factor of 2,
formally identical with
${\cal G}_F$ under the replacement $C\rightarrow 2iF$.

${\cal G}_B$ and ${\cal G}_F$
are easy to construct
using the following representation of the 
integral kernels for inverse
derivatives \cite{ss3}

\begin{eqnarray}
\langle u
\mid {{\partial}_P}^{-n}\mid
u'\rangle
&=&
-{1\over n!}B_n(\mid u-u'\mid)
{\rm sign}^n(u-u')
\label{masterP}\\
\langle u
\mid {{\partial}_A}^{-n}\mid
u'\rangle&=&
{1\over{2(n-1)!}}
E_{n-1}(\mid u-u'\mid)
{\rm sign}^n(u-u' )
\label{masterA}
\end{eqnarray}
\noindent
Here $B_n$($E_n$) denotes the
$n^{th}$ Bernoulli (Euler) polynomial, and we have
set $T=1$.
Those formulas are valid for $\mid u-u'\mid \leq 1$.
Let us shortly prove the first identity; the proof
of the second one is completely analogous.

First observe that, 
by construction, ${1\over 2} \dot G_B$ is the
integral kernel inverting the first derivative 
${\partial}_P$ acting
on periodic functions. We may therefore write 

\begin{eqnarray}
K_n(u_1-u_{n+1})&:=&\int_0^1 du_2\ldots du_n
\dot G_{B12}\dot G_{B23}\ldots\dot G_{Bn(n+1)}
\nonumber\\
&=&
2^n<u_1\mid {{\partial}_P}^{-n}\mid u_{n+1}> 
\nonumber\\
\label{ninvert}
\end{eqnarray}

\noindent
This leads to the recursion relation

\begin{equation}
{\partial \over {\partial u}}
K_n(u-u') = 2^n<u\mid
{{\partial}_P}^{-(n-1)}\mid u'>
= 2 K_{n-1}(u-u')
\label{recurs}
\end{equation}

\noindent
We want to show that the same recursion relation
is fulfilled by the polynomial $\tilde K_n$,

\begin{equation}
\tilde K_n(u-u'):= 
-{2^n\over n!}B_n(\vert u-u'\vert)
{\rm sign}^n(u-u')
\label{recursprime}
\end{equation}

\noindent
Explicit differentiation yields

\begin{eqnarray}
{\partial \over {\partial u}}
\tilde K_n(u-u') &=&
-{2^n\over n!}
B_n'(\vert u-u'\vert)
{\rm sign}(u-u')
{\rm sign}^n(u-u')\nonumber\\
&=& -{2^n\over (n-1)!}
{B}_{n-1}(\vert u-u'\vert)
{\rm sign}^{n+1}(u-u')\nonumber\\
&=& 2\tilde K_{n-1}(u-u')\nonumber\\
\label{showrecurs}
\end{eqnarray}

\noindent
Here the recursion relation for the Bernoulli
polynomials was used, 
${d\over dx}{B_n}(x) = nB_{n-1}(x)$.
An additional term arising by differentiation of the signum
function for $n$ odd can be deleted due to the fact that

\begin{equation}
\delta(x)B_n(\mid x\mid)=\delta(x)B_n(0)=0
\label{deletedelta}
\end{equation}

\noindent
for $n$ odd, $n>1$.
The proof is completed by checking that the master identity works
for $n=1$ ($B_1(x)=x-{1\over 2}$), and on the diagonal
$u_1 = u_{n+1}$ for any $n$. The second statement is trivial
for odd $n$, since here  both sides vanish by antisymmetry.
For even $n$ it becomes

\begin{eqnarray}
Tr({\partial}_P^{-n})&=& -{B_n\over n!}
\label{tracechainP}
\end{eqnarray}\no
with $B_n=B_n(0)$ the $n^{th}$ Bernoulli number.
This identity is easily shown by writing the
trace in the eigenbasis
$\lbrace
\e^{2\pi iku},k\in {\bf Z}\backslash
\lbrace 0 \rbrace
\rbrace$, and using the well-known
relation between the Bernoulli numbers and the
values of the Riemann $\zeta$ - function
at positive even numbers,
$\zeta(2n)={(2\pi)^{2n}\over 2(2n)!}
|B_{2n}|$.

Using (\ref{masterP}), we can compute
${\cal G}_B$ for the unit circle as follows:
\vspace{-8pt}

\begin{eqnarray}
{\cal G}_B(u_1,u_2)&=&
2\langle u_1\mid {\Bigl(
{\partial_P}^2
-2iF\partial_P
\Bigr)}^{-1}\mid u_2\rangle
\nonumber\\
&=&2
\sum_{n=0}^{\infty}{(2iF)}^n
\langle u_1\vert
{\partial_P}^{-(n+2)}
\vert u_2\rangle\nonumber\\
&=&
-2\sum_{n=2}^{\infty}
{{(2iF)}^{n-2}{{\rm sign}^n(u_1-u_2)}
\over n!}
B_n(\mid u_1 -u_2\mid )\nonumber\\
&=&-{1\over iF}
{{\rm sign}(u_1-u_2)
{\rm e}^{2iF(u_1-u_2)}
\over
{{\rm e}^{2iF{\rm sign}(u_1-u_2)}
-1}}
+{{\rm sign}(u_1-u_2)\over iF}
B_1(\mid u_1-u_2\mid ) -{1\over 2F^2}\nonumber\\
&=& {1\over 2F^2}\biggl({F\over{{\rm sin}F}}
{\rm e}^{-iF\dot G_{B12}}+iF\dot G_{B12} -1\biggr)
\quad 
\label{calGB}
\end{eqnarray}
\noindent
In the next-to-last step we used the
generating identity for the
Bernoulli polynomials,

\begin{equation}
{t{\rm e}^{xt}\over{{\rm e}^t-1}}=
\sum_{n=0}^{\infty}  
B_n(x){t^n\over n!}
\label{defbernoullipol}
\end{equation}
\noindent
This is ${\cal G}_B$ as given in
eq.(\ref{calGBGF}) up to a simple
rescaling.
The computation of ${\cal G}_F$
proceeds in a completely analogous way.

This method does not work for the determination of
${G^C_P}$, since negative powers of
$\bar\partial_P$ are not even well-defined
in the presence of the zero mode.
In the following, we will calculate ${G^C_{P,A}}$
in a different way,
which corresponds to the usual construction of 
the Feynman propagator in
field theory.
In order to determine ${G}^C_A(\tau)$, say, we employ the
following set of basis functions over the circle with
circumference $T$:
\be\label{B.1}
f_n(\tau)=T^{-1/2}
\exp\Bigl[i\frac{2\pi}{T}(n+\frac{1}{2})\tau\Bigr],\ \
n\in Z\ee
They satisfy
\bear\label{B.2}
&&\int_0^Td\tau f_n^*(\tau)f_m(\tau)=\delta_{nm}\nonumber\\
&&\sum_{n=-\infty}^\infty f_n(\tau_2)f_n^*(\tau_1)=
\sum_{m=-\infty}^\infty\delta(\tau_2-\tau_1-mT)\ear
and $f_n(\tau+T)=-f_n(\tau)$. In this basis, the Green's function
(\ref{calcGCA}) becomes
\be\label{B.3}
G^C_A(\tau_1,\tau_2)=
{G}^C_A(\tau_1-\tau_2)=\frac{1}{T}\sum^\infty_{n=-\infty}
\frac{\exp\Bigl[i\frac{2\pi}{T}
(n+\frac{1}{2})(\tau_1-\tau_2)\Bigr]}
{i(2\pi/T)(n+\frac{1}{2})-C}\ee
By introducing an auxiliary integration in the form $(\tau\equiv\tau
_1-\tau_2)$
\be\label{B.4}
{G}_A^C(\tau)=\int^\infty_{-\infty}d\omega\ \frac{1}{T}
\sum^\infty_{n=-\infty}\delta
\biggl(\omega-\frac{2\pi}{T}(n
+\frac{1}{2})\biggr)\frac{e^{i\omega\tau}}{i\omega-C}\ee
and using Poisson's resummation formula, the Green's
function assumes
the suggestive form \cite{kleinert}
\be\label{B.5}
{G}_A^C(\tau)=\sum^\infty_{n=-\infty}(-1)^n{G}_\infty^C(\tau+nT)\ee
with
\be\label{B.6}
{G}^C_\infty(\tau)\equiv\int^\infty_{-\infty}
\frac{d\omega}{2\pi}
\frac{e^{i\omega\tau}}{i\omega-C}\ee
We verify that
\be\label{B.7}
\Bigl({d\over d\tau}-C\Bigr){G}_\infty^C(\tau)=\delta(\tau)\ee
\be\label{B.8}
\Bigl({d\over d\tau}-C\Bigr)
{G}^C_A(\tau)=\sum^\infty_{n=-\infty}
(-1)^n\delta(\tau+nT)\ee
which shows that ${G}^C_\infty$ is a Green's function on the
infinitely extended real line, while ${G}^C_A$ is defined
on the circle. The integral (\ref{B.6}) yields for $C>0$
\be\label{B.9}
{G}^C_\infty(\tau)=-\theta(-\tau)e^{C\tau}\ee
Hence, from (\ref{B.5})
\be\label{B.10}
{G}_A^C(\tau)=-e^{C\tau}\sum_{n=-\infty}^\infty(-1)^n
\theta(-\tau-nT)e^{nCT}\ee
For $\tau\in (0,T)$ only the terms $n=-\infty,...,-1$
contribute to the sum in (\ref{B.10}), while for $\tau
\in(-T,0)$ a nonzero contribution is obtained for $n=-\infty,
...,0$. Summing up the geometric series in either case
and combining the results we obtain the expression given
in eq. (\ref{calcGCA}). It is valid for
$-T<\tau<+T$. Using a basis of periodic functions the same
arguments lead to ${G}^C_P$ as stated in (\ref{calcGCP}).
Note that in the limit of a large period $T$
\be\label{B.11}
\lim_{T\to\infty}{G}^C_{A,P}(\tau)={G}^C_\infty(\tau)\ee
as it should be. For $C\to 0$, both ${G}^C_\infty$ and
${G}^C_A$ have a well-defined limit:
\bear\label{B.12}
&&{G}^0_\infty(\tau)=-\theta(-\tau)\nonumber\\
&&{G}^0_A(\tau)=\frac{1}{2}\ {\rm sign}\ (\tau)\ear
The periodic Green's function ${G}^C_P$ blows up in this limit
because $\bar\partial_P^{-1}$ does not exist in presence of the
constant mode. It is important to keep in mind that ${G}^C_P$
is defined in such a way that it includes the zero mode of
${d\over d\tau}$.

In the perturbative evaluation of the spin-1 path integral one has
to deal with traces over chains of propagators of the form
\be\label{B.13}
\sigma_{A,P}^n(C)\equiv
{{\Tr}}_{A,P}\Bigl[({d\over d\tau}-C)^{-n}\Bigr]\ee
Because
\be\label{B.14}
\sigma_{A,P}^n(C)=\frac{1}{(n-1)!}\left(\frac{d}{dC}\right)^{n-1}
\sigma_{A,P}^1(C)\ee
it is sufficient to know $\sigma_{A,P}^1(C)$. The subtle point which
we would like to mention here is that strictly speaking
the sum defining $\sigma_A^1$, say,
\be\label{B.15}
\sigma_A^1(C)=\sum^\infty_{n=-\infty}\frac{1}{i(2\pi/T)(n+\frac{1}
{2})-C}\ee
does not converge as it stands, and is meaningless without
a prescription of how to regularize it. The usual strategy
is to combine terms for positive and negative values of $n$,
and to replace (\ref{B.15}) by the convergent series
\bear\label{B.16}
\sigma_A^1(C)&=&-2C\sum^\infty_{n=0}\left[\left(\frac{2\pi}
{T}\right)^2(n+\frac{1}{2})^2+C^2\right]^{-1}\nonumber\\
&=&-\frac{T}{2}\tanh\left(\frac{CT}{2}\right)\ear
It is important to realize that this definition implies
a well-defined prescription for the treatment of
the $\theta$ functions in ${G}^C_{A,P}$ at $\tau=0$. In
fact,
\be\label{B.17}
\sigma_A^1(C)=\int^T_0d\tau\ {G}^C_A(\tau-\tau)=T\ {G}^C_A(0)
\ee
and by combining eqs. (\ref{B.16}) and (\ref{B.17}) we deduce
that we must set
\be\label{B.18}
\lim_{\tau\searrow0}\ \theta(\tau)=\lim_{\tau\searrow0}
\ \theta(-\tau)=\frac{1}{2}\ee
With (\ref{B.16}) we obtain
\be\label{B.19}
\sigma_A^n(C)=-\frac{1}{(n-1)!}\left(\frac{T}{2}\right)^n
\left(\frac{d}{dx}\right)^{n-1}\tanh(x)\Bigr|_{x=CT/2}\ee
The analogous relation in the periodic case is
\be\label{B.20}
\sigma^n_P(C)=-\frac{1}{(n-1)!}\left(\frac{T}{2}\right)^n
\left(\frac{d}{dx}\right)^{n-1}\coth(x)\Bigr|_{x=CT/2}\ee
if the zero mode of ${d\over d\tau}$ is included
in the trace (\ref{B.13}), and
\be\label{B.21}
{\sigma'}^n_P(C)=-\frac{1}{(n-1)!}\left(\frac{T}{2}\right)^n
\left(\frac{d}{dx}\right)^{n-1}\{\coth(x)-x^{-1}
\}\Bigr|_{x=CT/2}\ee
if the zero mode is omitted. For $C$ sufficiently small
one finds the power series expansions
\bear\label{B.22}
&&\sigma_A^n(C)=-\frac{1}{(n-1)!}\sum^\infty_{k=n/2}
\frac{(2^{2k}-1)B_{2k}}{2k(2k-n)!}T^{2k}C^{2k-n}\nonumber\\
&&{\sigma'}_P^n(C)=-\frac{1}{(n-1)!}\sum^\infty_{k=n/2}
\frac{B_{2k}}{2k(2k-n)!}T^{2k}C^{2k-n}\ear
$\sigma_A^n$ and ${\sigma'}_P^n$ have well defined limits
for $C\to0$:
\bear\label{B.23}
&&\sigma_A^n(0)=-\frac{(2^n-1)B_n}{n!}T^n
=\half{E_{n-1}(0)\over (n-1)!}T^n
\quad (n {\rm\; even})
\nonumber\\
&&{\sigma'}_P^n(0)=-\frac{B_n}{n!}T^n
\quad (n {\rm\; even})
\ear
\noindent
(those limits vanish for $n$ odd).
This brings us, of course, back to eqs.(\ref{masterP}),
(\ref{masterA}).

\section{Symmetric Partial Integration}
\label{q2to6}
\renewcommand{\theequation}{C.\arabic{equation}}
\setcounter{equation}{0}
\vskip10pt
\noindent

In this appendix we explain a partial integration
algorithm which allows one to remove all $\ddot G_{Bij}$'s
contained in the original numerator polynomial $P_N$
(\ref{defPN}) of the $N$ - photon amplitude, and which
preserves the full permutation symmetry in the $N$
photons. 

Such an ``impartial'' partial integration
algorithm can be defined in the following way:

\begin{enumerate}
\item
In every step, partially integrate away {\sl all}
$\ddot G_{Bij}$'s appearing in the
term under inspection {\sl simultaneously}.
This is possible since different $\ddot G_{Bij}$'s do
not share variables to being with, and this property
is preserved by all partial integrations. New
$\ddot G_{Bij}$'s may be created.

\item
In the first step, for every $\ddot G_{Bij}$ partially
integrate both over $\tau_i$ and $\tau_j$,
and take the mean of the results.

\item
At every following step, any $\ddot G_{Bij}$
appearing must have been created in the 
previous step. Therefore either both $i$ and $j$
were used in the previous step, or just
one of them. If both, the rule is to again use both
variables in the actual step for partial integration,
and take the mean of the results. If only one of them
was used
in the previous step, 
then the other one should be used in the actual
step.

\end{enumerate}
\no
For example, the term
$\ddot G_{B12}\ddot G_{B34}$
appearing in $P_4$ 
in the first step transforms as follows,

\bear
\ddot G_{B12}\ddot G_{B34}&\rightarrow&
\fourth
\dot G_{B12}\dot G_{B34}
\biggl\lbrace
\Bigl[
\dot G_{B1i}k_1\cdot k_i-\dot G_{B2i}k_2\cdot k_i
\Bigr]
\Bigl[
\dot G_{B3j}k_3\cdot k_j-\dot G_{B4j}k_4\cdot k_j
\Bigr]
\non\\
&&
-\ddot G_{B13}k_1\cdot k_3
 +\ddot G_{B14}k_1\cdot k_4 +\ddot G_{B23}k_2\cdot k_3
-\ddot G_{B24}k_2\cdot k_4
\biggr\rbrace
\non\\
\label{alg1step1}
\ear\no
The terms in the second line have to be further processed.
Considering just the first one of them, since both variables
appearing in $\ddot G_{B13}$ were active in the
first step, both must also be used in the second one.
This yields

\bear
-\fourth
\dot G_{B12}\dot G_{B34}
\ddot G_{B13}
&\rightarrow&
\eigth
\dot G_{B12}\dot G_{B34}
\dot G_{B13}
\Bigl[
\dot G_{B1i}k_1\cdot k_i
-\dot G_{B3i}k_3\cdot k_i
\Bigr]
\non\\
&&
+\eigth
\dot G_{B13}
\Bigl[
\ddot G_{B12}\dot G_{B34}
-\dot G_{B12}\ddot G_{B34}
\Bigl]
\non\\
\label{alg1step2}
\ear\no
Considering again the first term in the second
line, only $\tau_1$ was active in the previous step.
Therefore only $\tau_2$ must be used now, and the
third step is the final one,

\be
\eigth
\dot G_{B13}\ddot G_{B12}\dot G_{B34}
\rightarrow
\eigth
\dot G_{B13}\dot G_{B12}\dot G_{B34}
\dot G_{B2i}k_2\cdot k_i
\label{alg2step3}
\ee\no
This prescription treats all variables on the same
footing, and therefore must lead to a permutation
symmetric result.
The nontrivial fact is that the process terminates after a finite
number of steps, and does not become cyclic (as would be the case if,
for example, one would {\it always} treat the indices
in a $\ddot G_{Bij}$ symmetrically). 
This 
is not difficult to derive from the fact that, 
for any term in $P_N$,
the indices appearing in the $\ddot G_{Bij}$'s
and the
first indices of the $\dot G_{Bij}$'s
are associated to the polarization vectors, and thus
must all take different values.

\noindent
This algorithm transforms $P_4$ into
\bear
Q_4&=&
\dot G_{B1i}\varepsilon_1\cdot k_i
\dot G_{B2j}\varepsilon_2\cdot k_j
\dot G_{B3k}\varepsilon_3\cdot k_k
\dot G_{B4l}\varepsilon_4\cdot k_l
\non\\
&&
\!\!\!\!\!\!\!\!\!
+
\Biggl\lbrace
\half\dot G_{B12}
\varepsilon_1\cdot \varepsilon_2
\biggl\lbrace
\dot G_{B3i}\varepsilon_3\cdot k_i
\dot G_{B4j}\varepsilon_4\cdot k_j
\Bigl[
\dot G_{B1k}k_1\cdot k_k
-
\dot G_{B2k}k_2\cdot k_k
\Bigr]
\non\\
&&\!\!\!
+
\Bigl[
\dot G_{B3i}\varepsilon_3\cdot k_i
\bigl(
\dot G_{B41}\varepsilon_4\cdot k_1
-
\dot  G_{B42}\varepsilon_4\cdot k_2
\bigr)
\dot G_{B4k}k_4\cdot k_k
+ \bigl( 3 \leftrightarrow 4)
\Bigr]
\non\\
&&\!\!\!
+
\Bigl[
\bigl(
\dot G_{B31}\varepsilon_3\cdot k_1
-
\dot  G_{B32}\varepsilon_3\cdot k_2
\bigr)
\dot G_{B43}\varepsilon_4\cdot k_3
\dot G_{B4k}k_4\cdot k_k
+ \bigl( 3 \leftrightarrow 4)
\Bigr]
\biggr\rbrace
+ 5\,\, {\rm permutations} 
\Biggr\rbrace
\non\\
&&\!\!\!\!\!\!\!\!
+
\Biggl\lbrace
\fourth
\dot G_{B12}\dot G_{B34}
\varepsilon_1\cdot\varepsilon_2
\varepsilon_3\cdot\varepsilon_4
\biggl\lbrace
\Bigl[
\dot G_{B1i}k_1\cdot k_i
-\dot G_{B2i}k_2\cdot k_i
\Bigr]
\Bigl[
\dot G_{B3j}k_3\cdot k_j
-\dot G_{B4j}k_4\cdot k_j
\Bigr]
\non\\
&&\!\!\!
+\half
\Bigl[
\dot G_{B13}k_1\cdot k_3
-\dot G_{B23}k_2\cdot k_3
-\dot G_{B14}k_1\cdot k_4
+\dot G_{B24}k_2\cdot k_4
\Bigr]
\non\\
&&\!\!\!
\times
\Bigl[
\dot G_{B1i}k_1\cdot k_i
+\dot G_{B2i}k_2\cdot k_i
-\dot G_{B3i}k_3\cdot k_i
-\dot G_{B4i}k_4\cdot k_i
\Bigr]
\biggr\rbrace
+ 2\,\, {\rm perm.} 
\Biggr\rbrace
\label{Q4app}
\ear\no
This expression can be rewritten 
more compactly as follows,

\be
Q_4
=
q_4^4 + q_4^3 + q_4^2 - q_4^{22}
\non\\
\label{4photon}
\ee\no
where
\bear
q_4^4 &=& 
\dot G_{B12}
\dot G_{B23}
\dot G_{B34}
\dot G_{B41}
Z_4(1234)
+ 2 \,\, {\rm permutations}
\non\\
q_4^3 &=&
\dot G_{B12}
\dot G_{B23}
\dot G_{B31}
Z_3(123)
\dot G_{B4i}
\varepsilon_4\cdot k_i
+ 3 \,\, {\rm perm.}
\non\\
q_4^2 &=&
\dot G_{B12}\dot G_{B21}
Z_2(12)
\biggl\lbrace
\dot G_{B3i}
\varepsilon_3\cdot k_i
\dot G_{B4j}
\varepsilon_4\cdot k_j
+\half
\dot G_{B34}
\varepsilon_3\cdot\varepsilon_4
\Bigl[
\dot G_{B3i}
k_3\cdot k_i
-
\dot G_{B4i}
k_4\cdot k_i
\Bigr]
\biggr\rbrace
\non\\
&&
+ \, 5 \,\, {\rm perm.}
\non\\
q_4^{22} &=&
\dot G_{B12}\dot G_{B21}
Z_2(12)
\dot G_{B34}\dot G_{B43}
Z_2(34)
+ 2 \,\, {\rm perm.}
\non\\
\label{4photoncoeff}
\ear\no
and the ``Lorentz cycles'' $Z_n$,

\bear
Z_2(ij)&\equiv&
\varepsilon_i\cdot k_j
\varepsilon_j\cdot k_i
-\varepsilon_i\cdot\varepsilon_j
k_i\cdot k_j
\non\\
Z_n(i_1i_2\ldots i_n)&\equiv&
{\rm tr}
\prod_{j=1}^n 
\Bigl[
k_{i_j}\otimes \varepsilon_{i_j}
- \varepsilon_{i_j}\otimes k_{i_j}
\Bigr]
\quad (n\geq 3)
\nonumber
\ear\no
have already been introduced in (\ref{defZn}).
Note that the product of two-cycles $q_4^{22}$
appears with a minus sign in eq.(\ref{4photon}). 
The reason is that we corrected for an over-counting
here; $q_4^{22}$ is also contained twice
in $q_4^2$, and 
separating it out from there will change the
``-'' to a ``+''.

Before proceeding to higher point amplitudes,
it will be prudent to
further condense the notation.
We thus abbreviate

\bear
\dot G_{ij}&\equiv & \dot G_{Bij}
\varepsilon_i\cdot k_j\non\\
\Gund_{ij}
&\equiv &
\dot G_{Bij}\varepsilon_i\cdot\varepsilon_j
\non\\
{{\slG}_{ij}}
&\equiv &
\dot G_{Bij}
k_i\cdot k_j\non\\
{\dot G}(i_1i_2\ldots i_n)
&\equiv&
\dot G_{Bi_1i_2}
\dot G_{Bi_2i_3}
\cdots
\dot G_{Bi_ni_1}
Z_n(i_1i_2\ldots i_n)\non\\
\label{defabb}
\ear\no
As was already mentioned in the main text, it
is known from previous work
\cite{berkos:npb362,berkos:npb379,strassler2}
that a closed ``$\tau$ -- cycle''
$\dot G_{Bi_1i_2}
\dot G_{Bi_2i_3}
\cdots
\dot G_{Bi_ni_1}$
after the partial integration will
always appear multiplied by a 
factor of
$Z_n(i_1i_2\ldots i_n)$. 
This motivates the last one of the
abbreviations above, and also explains why the
formulation of the ``cycle substitution'' part
of the Bern-Kosower rules did not require the 
specification of a particular partial
integration
algorithm.

A given term in $Q_N$ thus will be a product
of ``bi-cycles'' $\dot G(\cdot )$,
multiplied by a remainder. Following
~\cite{strassler2} we call this remainder
``tail'', or ``$m$ - tail'', where $m$ denotes
the number of indices not appearing in 
any of the cycles. For example, $q_4^2$ is
the product of a $2$ -- bi-cycle and
a $2$ - tail. 
Only the tails depend on the
choice of the partial integration algorithm.
The tail generated by our specific 
symmetric algorithm
will be denoted by $T_m(i_1\ldots i_m)$.
The $1$ - tail is (unambiguously) given by
$T_1(i)=\dot G_{ij}$ ($i$ being fixed
and $j$ summed over).

With the above abbreviations, the result for $Q_5$ 
obtained by an application of the symmetric 
algorithm can be written as follows,

\be
Q_5
=
q_5^5 + q_5^4 + q_5^3 + q_5^2 
- q_5^{32} - q_5^{22}
\non\\
\label{5photon}
\ee\no
where
\bear
q_5^5 &=& 
\dot G(12345)
+ 11\,\, {\rm permutations}
\non\\
q_5^4 &=&
\dot G(1234)
\dot G_{5i}
+ 14\,\, {\rm perm.}
\non\\
q_5^3 &=&
\dot G(123)
\biggl\lbrace
\dot G_{4i}
\dot G_{5j}
+\half
\Gund_{45}
\Bigl[
\slG_{4i}
-
\slG_{5i}
\Bigr]
\biggr\rbrace
+ 9 \,\,{\rm perm.}
\non\\
q_5^2 &=&
\dot G(12)
\Biggl\lbrace
\dot G_{3i}\dot G_{4j}\dot G_{5k}
+\half \Gund_{34}
\biggl[
\Gd_{5k}
\Bigl[
\slG_{3i}-\slG_{4i}
\Bigr]
+\slG_{5i}
\Bigl[
\Gd_{53}-\Gd_{54}
\Bigr]
\biggr]
\non\\
&&
\phantom{\dot G(12)}
+\half \Gund_{35}
\biggl[
\Gd_{4k}
\Bigl[
\slG_{3i}-\slG_{5i}
\Bigr]
+\slG_{4i}
\Bigl[
\Gd_{43}-\Gd_{45}
\Bigr]
\biggr]\non\\
&&
\phantom{\dot G(12)}
+\half \Gund_{45}
\biggl[
\Gd_{3k}
\Bigl[
\slG_{4i}-\slG_{5i}
\Bigr]
+\slG_{3i}
\Bigl[
\Gd_{34}-\Gd_{35}
\Bigr]
\biggr]
\Biggr\rbrace
+ 9 \,\,{\rm perm.}
\non\\
q_5^{32}&=&
\Gd(123)\Gd(45) + 9 \,\,{\rm perm.}
\non\\
q_5^{22}&=&
\Gd(12)\Gd(34)\Gd_{5i}
+ 14 \,\,{\rm perm.}
\label{5photoncoeff}
\ear\no
Again we have an over-counting; $q_5^{32}$ is
contained once in both $q_5^3$ and $q_5^2$, and
$q_5^{22}$ is contained twice in $q_5^2$.

Comparing the 2 - and 3 - tails appearing in (\ref{5photoncoeff}) with
our earlier results (\ref{Q2}),(\ref{Q3}) for $Q_2$ and $Q_3$,
we note that there is a simple relation:
The tail $T_i$  can be obtained from $Q_i$, in
its un-decomposed form, by rewriting $Q_i$ in the tail variables, and
then extending the range of all dummy indices 
to run over the complete set of variables
$\tau_1,\ldots,\tau_5$. It is not difficult to see that this relation
generalizes to an arbitrary  $Q_m, T_m$.  Consider (the unpermuted term
of) $q_N^2$, which has a 2 - cycle $\dot G(12)$ and a tail
$T_{N-2}(3\ldots N)$. It suffices to consider those terms in $q_N^2$
having a $\varepsilon_1\cdot k_2\varepsilon_2\cdot k_1$ as their
$Z_2(12)$ -- component. From the master formula
eq.(\ref{scalarqedmaster}) one infers that for this part of $q_N^2$ the
partial integration procedure can have involved only  partial
integrations over the tail variables $\tau_3,\ldots,\tau_N$. Thus  the
calculation of $T_{N-2}$ and the lower order calculation of $Q_{N-2}$
are identical as far as the tail indices are concerned. The presence of
the cycle variables for the tail makes itself felt  only through an
extension of the momentum sums in the master formula, leading to the
stated extension rule for dummy indices. The same type of argument shows
that the structure of $T_m$ does not  depend on the number and lengths of
the cycles it multiplies.

At this point it should be noted that
every term in $Q_N$ must have 
at least one cycle factor (this is
a combinatorial consequence
of the fact that each such term contains
a total of $2N$ indices, of which only $N$
are different).
Thus the maximal tail occurring in $Q_N$
has length $N-2$. The above connection
between $T_N$ and $Q_N$ thus allows us to write down,
without going through the partial integration
procedure again, $Q_6$ as follows,

\be
Q_6
=
q_6^6 + q_6^5 + q_6^4 + q_6^3
+ q_6^2
-q_6^{42} 
- q_6^{33} - q_6^{32}
- q_6^{22}
+ q_6^{222}
\non\\
\label{6photon}
\ee\no
where
\bear
q_6^6
 &=& 
\dot G(123456)
+  {\rm permutations} 
\quad\Bigl(
{5!\over 2} = 60 {\rm \,\,in\,\, total}
\Bigr)
\non\\
q_6^5 &=&
\dot G(12345)
T_1(6)
+  {\rm perm.}
\quad\Bigl(
{4!\over2}{6\choose 1}
=72
 {\rm \,\,in\,\, total}
\Bigr)
\non\\
q_6^4 &=&
\dot G(1234)
T_2(56)
+  {\rm perm.}
\quad\Bigl(
45
 {\rm \,\,in\,\, total}
\Bigr)
\non\\
q_6^3 &=&
\dot G(123)
T_3(456)
+  {\rm perm.}
\quad\Bigl(
20
 {\rm \,\,in\,\, total}
\Bigr)
\non\\
q_6^2 &=&
\dot G(12)
T_4(3456)
+  {\rm perm.}
\quad\Bigl(
15
 {\rm \,\,in\,\, total}
\Bigr)
\non\\
q_6^{42}&=&
\Gd(1234)\Gd(56) 
+  {\rm perm.}
\quad\Bigl(
45
 {\rm \,\,in\,\, total}
\Bigr)
\non\\
q_6^{33}&=&
\Gd(123)\Gd(456)
+  {\rm perm.}
\quad\Bigl(
10
 {\rm \,\,in\,\, total}
\Bigr)\non\\
q_6^{32}&=&
\Gd(123)\Gd(45)T_1(6)
+  {\rm perm.}
\quad\Bigl(
60
 {\rm \,\,in\,\, total}
\Bigr)\non\\
q_6^{22}&=&
\Gd(12)\Gd(34)T_2(56)
+  {\rm perm.}
\quad\Bigl(
45
 {\rm \,\,in\,\, total}
\Bigr)
\non\\
q_6^{222}&=&
\Gd(12)\Gd(34)\Gd(56)
+  {\rm perm.}
\quad\Bigl(
15
 {\rm \,\,in\,\, total}
\Bigr)
\label{6photoncoeff}
\ear\no
Here the only new ingredient, $T_4$, according to the
above is related to the un-decomposed $Q_4$ of
eq.(\ref{Q4}) simply by a relabelling, and an
extension of the range of
all dummy indices to run from $1$ to $6$:

\bear
T_4(abcd)&=&
\dot G_{ai}\dot G_{bj}\dot G_{ck}\dot G_{dl}
+\biggl\lbrace
\half\Gund_{ab}
\Bigl\lbrace
\dot G_{ci}\dot G_{dj}
(
\slG_{ak}-\slG_{bk}
)
+\Bigl[
\dot G_{ci}
(\dot G_{da}-\dot G_{db})
\slG_{dk}+(c\leftrightarrow d)
\Bigr]
\non\\
&&
+\Bigl[
(\dot G_{ca}-\dot G_{cb})
\dot G_{dc}
\slG_{dk}
+(c\leftrightarrow d)
\Bigr]
\Bigr\rbrace
+ 5\quad{\rm perm.}
\biggr\rbrace
+\biggl\lbrace
\fourth\Gund_{ab}\Gund_{cd}
\biggl\lbrace
\Bigl[
\slG_{ai}-\slG_{bi}
\Bigr]
\Bigl[
\slG_{cj}-\slG_{dk}
\Bigr]
\non\\
&&
+
\half
\Bigl[
\slG_{ac}-\slG_{bc}-\slG_{ad}+\slG_{bd}
\Bigr]
\Bigl[
\slG_{ai}+\slG_{bi}-\slG_{ci}-\slG_{di}
\Bigr]
\biggr\rbrace
+ 2\;{\rm perm.}
\biggr\rbrace\non\\
\label{T4}
\ear

Note that the integrand is
not yet quite suitable for the
application of the cycle substitution
rules, since the tails still contain cycles. 
For this purpose, one should further rewrite 
$Q_6$ as
\footnote{When comparing with \cite{menphoton} note
that our present definition of  $q_N^{(\cdot)}$ ($Q_N^{(\cdot)}$) 
there corresponds to  $Q_N^{(\cdot)}$ ($\hat Q_N^{(\cdot)}$).}

\be
Q_6=
 Q_6^6 +  Q_6^5 +  Q_6^4 +  Q_6^3
+  Q_6^2
+ Q_6^{42} 
+  Q_6^{33} +  Q_6^{32}
+  Q_6^{22}
+ Q_6^{222}
\label{Q6altern}
\ee\no
where, by definition, $Q_6^{(\cdots)}$ is obtained from
the corresponding $q_6^{(\cdots)}$ by restricting the
range of the dummy indices appearing in its tail
so as to precisely eliminate all
additional cycles. This also removes the over-counting,
so that now all coefficients are unity.

It is now obvious that in this way one arrives
at a canonical permutation symmetric
version of the Bern-Kosower integrand for
the one-loop $N$ - photon/gluon amplitude. 
Moreover, in \cite{menphoton} it was 
shown that the cycle decomposition has the additional
advantage of 
constituting a maximal gauge invariant decomposition of 
this amplitude.

\section{Proof of the Cycle Replacement Rule}
\label{proof}
\renewcommand{\theequation}{D.\arabic{equation}}
\setcounter{equation}{0}
\vskip10pt
\noindent

Combining the results of the previous appendix 
with the worldline superfield formalism
we are now in a position to give
a direct and simple proof of the basic 
cycle replacement rule
(\ref{fermion}) connecting the
scalar and fermion loop integrands for the $N$ -- photon/gluon
amplitudes.

Expanding out the superfield master formula (\ref{supermaster}) for
the fermion loop we obtain a formula isomorphic to the scalar
case,

\bear
\exp\biggl\lbrace
\sum_{i,j=1}^N
\Biggl\lbrack
\half\hat G_{ij} k_i\cdot k_j
+iD_i\hat G_{ij}\varepsilon_i\cdot k_j
+\half D_iD_j\hat G_{ij}\varepsilon_i\cdot\varepsilon_j\Biggr]
\biggr\rbrace 
\mid_{\varepsilon_1\ldots\varepsilon_N}
&=&
\non\\
\hspace{-80pt}
(-i)^n
P_N(-D_i\hat G_{ij},D_iD_j\hat G_{ij})
\exp\biggl\lbrace
\sum_{i,j=1}^N
\half\hat G_{ij} k_i\cdot k_j
\biggr\rbrace
\label{superPN}
\ear
Here $P_N$ is the same polynomial which appears in 
the expansion (\ref{defPN}) of the ordinary master formula.
We now apply to this integrand the same partial algorithm
as in the previous appendix, just with ordinary derivatives
replaced by super derivatives. 
In this way we can remove all 
second derivatives 
$D_iD_j\hat G_{ij}$. 
A recursive analysis analogous to the one performed 
in appendix \ref{q2to6} shows, that
the final result of the partial
integrations is almost isomophic to the ordinary
$Q_N$, except that we have to take into account that
$D_i\hat G_{ij}\ne -D_j\hat G_{ij}$. This means that, for example,
the formula for the ordinary $2$ - tail

\bear
T_2(ab)&=&\dot G_{ai}
\dot G_{bj}
+\half
\Gund_{ab}
(\slG_{ai}-\slG_{bi})
\label{2tail}
\ear
in the super case has to be written differently,

\bear
\hat T_2(ab) &=&
D_a\hat G_{ai}\varepsilon_a\cdot k_i
D_b\hat G_{bj}\varepsilon_b\cdot k_j
+\half\Bigl[
D_a\hat G_{ab}\varepsilon_a\cdot\varepsilon_b
D_b\hat G_{bi}k_b\cdot k_i
+ (a\leftrightarrow b)\Bigr]
\label{super2tail}
\ear
The general structure found above remains the same, i.e. the
final integrand $Q_N$ can be decomposed into a sum of terms 
$Q_N^{(\cdots)}$ which individually are products of bi-cycles and
tails, where the tails do not contain closed cycles of indices.
Each term contains precisely $N$ factors of $D_i\hat G_{ij}$'s.
Now, observe that, since in the master formula
(\ref{supermaster}) $D_i$ appears only together with 
$\varepsilon_i$, all terms in the original integrand $P_N$  
contain every $D_i$ precisely once. Since this property
is preserved by the partial integrations, and the cycles 
contain only $D_i$'s in the cycle variables, it follows
that the tails can contain only $D_i$'s in the tail variables.
From

\bear
D_i\hat G_{ij} &=& \theta_i\dot G_{Bij}-\theta_j G_{Fij}
\label{DGhat}
\ear
it is then clear that a $G_{Fij}$ produced by a
$D_i\hat G_{ij}$ in a tail cannot survive the
Grassmann integrations if one of the indices
$i,j$ is a cycle index; terms of this kind have too
many cycle - $\theta$'s and too few tail - $\theta$'s.
Therefore after the $\theta$ - integrations 
for all surviving $G_{Fij}$'s the indices $i,j$
are either both cycle variables, or both tail variables. 
Since from the structure of the component field integral
${\cal D}\psi$ it is clear that $G_{Fij}$'s can
generally only appear in cycles, $G_{Fij}$'s from tails
would therefore have to form cycles among themselves; but
this is not possible, since the tails do not contain closed chains
of indices. We conclude that, in fact, all $G_{Fij}$'s
coming from tails must drop out in the $\theta$ - 
integrations.

This leaves us with the basic super cycle integrals, for which
we can directly verify that

\bear
\int d\theta_{i_1}\cdots\theta_{i_n} 
D_{i_1}\hat G_{i_1i_2}
D_{i_2}\hat G_{i_2i_3}
\cdots
D_{i_n}\hat G_{i_ni_1}
&=&
(-1)^{n(n+1)\over 2}
\Bigl(
\dot G_{Bi_1i_2} 
\dot G_{Bi_2i_3} 
\cdots
\dot G_{Bi_ni_1}
\non\\&&\hspace{15pt}
-
G_{Fi_1i_2}
G_{Fi_2i_3}
\cdots
G_{Fi_ni_1}
\Bigr)
\label{intsupercycle}
\ear

\section{Massless 1 -- Loop 4 -- Point Tensor Integrals}
\label{boxint}

\renewcommand{\theequation}{E.\arabic{equation}}
\setcounter{equation}{0}
\vskip10pt
\noindent

In the worldline parametrization the
massless scalar box becomes 

\begin{eqnarray}
B[k_1,k_2,k_3,k_4]
&=&
\Tint
T^{4-{D\over 2}}
\int_0^1du_1\int_0^{u_1}du_2\int_0^{u_2}du_3
\,\,{\rm exp}
\biggl\lbrace
T\sumij G_{Bij}k_i\cdot k_j
\biggr\rbrace
\nonumber\\
&=&
\Gamma(2-{\epsilon\over 2})
\int_0^1du_1\int_0^{u_1}du_2\int_0^{u_2}du_3
{1\over
{\Bigl[-\sumij G_{Bij}k_i\cdot k_j\Bigr]}^{2-{\epsilon\over 2}}
}
\nonumber\\
\label{box}
\end{eqnarray}
\noindent
This is essentially eq.(\ref{scalarmaster}) specialized to
$N=4$. We have already rescaled to the unit circle and
put $u_4=0$. 
We return to Feynman parameters $a_i$ via eq.(\ref{trafotaualpha}),

\be
a_1 = 1-u_1,\quad a_2 = u_1-u_2, \quad
a_3 = u_2 - u_3, \quad a_4 = u_3
\non\\
\label{trafoutoa}
\ee
\no
($\alpha_i = Ta_i$)
so that

\be
\int_0^1du_1\int_0^{u_1}du_2\int_0^{u_2}du_3
=
\int_0^1 da_1da_2da_3da_4
\delta (1-\sum_{i=1}^4 a_i)
\non\\
\label{trafoutoaint}
\ee
\noindent
Also we introduce the Mandelstam variables

\be
s={(k_1+k_2)}^2,\quad
t={(k_2+k_3)}^2
\label{mandelstam}
\ee
\non
With all external legs massless and on-shell, $k_i^2=0$,
the Feynman denominator simplifies to

\be
\sum_{i<j}G_{Bij}k_i\cdot k_j = -a_1a_3s - a_2a_4t \non\\
\label{onshelldenominator}
\ee
\noindent
Generally we will also have a numerator polynomial
$P(\lbrace a_i\rbrace)$. Let us thus define

\be
I[P(\lbrace a_i\rbrace)]\equiv\Gamma (2-\epshalf)
\int_0^1\prod_{i=1}^4 da_i
\,\delta\Bigl(1-\sum_{i=1}^4a_i\Bigr)
{P(\lbrace a_i\rbrace)\over
{\Bigl[
a_1a_3s+a_2a_4t
\Bigr]
}^{2-\epshalf}
}
\non\\
\label{defI[P]}
\ee
\no
In the rest of this subsection 
we will perform some manipulations which allow us to
generate the numerator polynomial by differentiations
performed on the denominator polynomial. 

First it is useful to homogeneize the denominator by the following
well-known transformation due to 't Hooft and Veltman
\cite{thovel2},

\bear
a_i &=& {\alpha_ix_i\over \sum_{j=1}^4\alpha_jx_j}
\qquad\qquad\qquad\qquad i=1,2,3\non\\
a_4 &=& {\alpha_4(1-x_1-x_2-x_3)\over \sum_{j=1}^4\alpha_jx_j}
\non\\
\label{trafoatoalpha}
\ear
\noindent
with the constraints

\be
\alpha_1\alpha_3 = {1\over s},
\quad
\alpha_2\alpha_4 = {1\over t}
\non\\
\label{alphaconstraint}
\ee
\no
Note that this implies
\be
\sum_{j=1}^4
{a_j\over\alpha_j}
=
{1\over \sum_{j=1}^4\alpha_jx_j}
\non\\
\label{sumalphax}
\ee
\no
This transformation leads to

\be
I[P(\lbrace a_i\rbrace)] = \Gamma(2-\epshalf)
\int_0^1\prod_{i=1}^4dx_i
\,\delta\Bigl(1-\sum_{j=1}^4x_j\Bigr)
{
\Bigl(\prod_{i=1}^4\alpha_i\Bigr)
{(\sum_{j=1}^4\alpha_jx_j)}^{-\eps}
P(\lbrace a_i\rbrace)
\over
{\Bigl[
x_1x_3+x_2x_4\Bigr]}^{2-\epshalf}
}
\non\\
\label{I[P]xvar}
\ee\no
If we now specialize $P$ to be a polynomial
$P_m$ of definite degree $m$, this can be
further rewritten as

\bear
I[P] &=& \Bigl(\prod_{i=1}^4 \alpha_i\Bigr)
\Gamma(2-\epshalf)
\int_0^1\prod_{i=1}^4dx_i
\,\delta\Bigl(1-\sum x_j\Bigr)
{P_m(\lbrace\alpha_ix_i\rbrace)
{\Bigl(
\sum_{j=1}^4\alpha_jx_j\Bigr)}^{-m-\eps}
\over
{\Bigl[
x_1x_3+x_2x_4\Bigr]
}^{2-\epshalf}
}
\non\\
&=&
\Bigl(\prod_{i=1}^4 \alpha_i\Bigr)
\Gamma(2-\epshalf)
{\Gamma(1-\eps-m)\over\Gamma(1-\eps)}
\int_0^1\prod_{i=1}^4dx_i
\,\delta(1-\sum x_j)
{
P_m(\lbrace\alpha_i{\partial\over\partial\alpha_i}\rbrace)
{\Bigl(\sum_{j=1}^4\alpha_jx_j\Bigr)}^{-\eps}
\over
{\Bigl[
x_1x_3+x_2x_4\Bigr]
}^{2-\epshalf}
}
\non\\
&=&
{\Gamma(1-\eps-m)\over\Gamma(1-\eps)}
\Bigl(\prod_{i=1}^4\alpha_i\Bigr)
P_m(\lbrace\alpha_i{\partial\over\partial\alpha_i}\rbrace)
\biggl({I[1]\over \prod_{i=1}^4\alpha_i}\biggr)
\non\\
\label{I[P]final}
\ear
\no
Here it is understood that 

\be
{\Bigl(\alpha_i{\partial\over\partial\alpha_i}\Bigr)}^n
\equiv
\alpha_i^n
{\Bigl({\partial\over\partial\alpha_i}\Bigr)}^n
\label{partialalphaconvention}
\ee
\no 
It remains to calculate $I[1]$,

\bear
I[1] &=& 
 \Bigl(\prod_{i=1}^4 \alpha_i\Bigr)
\Gamma(2-\epshalf)
\int_0^1d^4x
\,\delta(1-\sum x_j)
{{\Bigl(
\sum_{j=1}^4\alpha_jx_j\Bigr)}^{-m-\eps}
\over
{\Bigl[
x_1x_3+x_2x_4\Bigr]
}^{2-\epshalf}
}
\non\\
&=&
\Gamma(2-\epshalf)
\int_0^1
d^4a
\,\delta(1-\sum a_i)
{1\over 
{\Bigl[
a_1a_3s+a_2a_4t
\Bigr]
}^{2-\epshalf}
}
\non\\
\label{calcI[1]}
\ear\no
This can be easily done using the following transformation
of variables going back to Karplus and Neuman
\cite{karneu},

\be
a_1 = (1-y)(1-z),\quad
a_2 = z(1-y),\quad
a_3 = y(1-x), \quad
a_4 = xy
\non\\
\label{trafoatoxyz}
\ee\no
which has a Jacobian
\be
{\Large{\mid}}
{\partial(a_1,a_2,a_3)
\over
\partial(x,y,z)
}
{\Large{\mid}}
= y(1-y)
\label{detatoxyz}
\ee\no
This transformation leads to

\be
I[1]=\Gamma(2-\epshalf)
\int_0^1dx
\int_0^1dy
\int_0^1dz
{1\over
{[y(1-y)]}^{1-\epshalf}
{\Bigl[
(1-x)(1-z)s+xzt\Bigr]}
^{2-\epshalf}
}
\non\\
\label{I[1]xyz}
\ee\no
The $y$ - integral factors out and gives an Euler
Beta function,

\be
\int_0^1 dy {\Bigl[
y(1-y)\Bigr]}
^{-1+\epshalf}
=
B(\epshalf,\epshalf)
\non\\
\label{eulerbeta}
\ee\no
For the remaining double integral the $x$ - integration is
elementary, and the resulting
z - integral can be expressed in terms of
hypergeometric functions using the formula

\be
\int_0^1dt\,
t^{a-1}{(1-t)}^{b-a-1}{(1-\lambda t)}^{-c}
=
{
\Gamma(a)\Gamma(b-a)
\over
\Gamma(b)
}
\,
\phantom{}_2
F_1
(a,c;b;\lambda)
\non\\
\label{getF21}
\ee\no
In this way one arrives at

\be
I[1] = {8\over\eps^2}r_{\Gamma}s^{-1}
\biggl\lbrace
t^{\sy{\epshalf -1}}
\phantom{}_2F_1
(\epshalf,1;1+\epshalf;1+{t\over s})
-
s^{\sy{\epshalf -1}}
\phantom{}_2F_1
(1,1;1+\epshalf;1+{t\over s})
\biggr\rbrace\non\\
\label{I[1]final}
\ee
where

\be
r_{\Gamma} \equiv
{\Gamma(1-\epshalf)\Gamma^2(1+\epshalf)
\over
\Gamma(1+\eps)
}
\non\\
\label{defrGamma}
\ee\no
The $\epsilon$ - expansion of this expression is 

\be
I[1] = r_{\Gamma}\alpha_1\alpha_2\alpha_3\alpha_4
\biggl[
{8\over\eps^2}
\Bigl(
{(\alpha_1\alpha_3)}^{-\epshalf}
+
{(\alpha_2\alpha_4)}^{-\epshalf}
\Bigr)
-
\ln^2
\Bigl(
{\alpha_1\alpha_3\over \alpha_2\alpha_4}
\Bigr)
-\pi^2
\biggr]
+
{\rm O}
(\eps)
\non\\
\label{I[1]epsexpand}
\ee\no
This is sufficient as far as the naked box integral is concerned,
but not if used in formula (\ref{I[P]final}), since for nontrivial
$P$ the factor $\Gamma (1-\epsilon-m)$ in front has a pole,
so that $I[1]$ here would be needed to order ${\rm O}(\eps)$.
Rather than using this formula as it stands, it is simpler to
start the recursion with the four polynomials of degree $m=1$.
In this case there is no singular prefactor, and it suffices
to give those polynomials to order $\epsilon^0$: 

\bear
I[a_1] &=& I[a_3] = r_{\Gamma}\alpha_1\alpha_2\alpha_3\alpha_4
\biggl\lbrace{4\over\eps^2}{(\alpha_2\alpha_4)}^{-\epshalf}
-{\alpha_1\alpha_3\over 2(\alpha_1\alpha_3+\alpha_2\alpha_4)}
\Bigl[
\ln^2
\Bigl(
{\alpha_1\alpha_3\over \alpha_2\alpha_4}
\Bigr)+\pi^2 \Bigr]
\biggr\rbrace
+
{\rm O}
(\eps)
\non\\
I[a_2] &=& I[a_4] = r_{\Gamma}\alpha_1\alpha_2\alpha_3\alpha_4
\biggl\lbrace{4\over\eps^2}{(\alpha_1\alpha_3)}^{-\epshalf}
-{\alpha_2\alpha_4\over 2(\alpha_1\alpha_3+\alpha_2\alpha_4)}
\Bigl[
\ln^2
\Bigl(
{\alpha_1\alpha_3\over \alpha_2\alpha_4}
\Bigr)+\pi^2 \Bigr]
\biggr\rbrace
+
{\rm O}
(\eps)
\non\\
\label{I[a]epsexpand}
\ear
\section{Some Worldloop Formulas}
\label{formulas}
\renewcommand{\theequation}{F.\arabic{equation}}
\setcounter{equation}{0}
\no
In this appendix we abbreviate $G_B=G$,
${\cal G}_B={\cal G}$. All formulas
are written for the unit circle, $T=1$.

\vspace{15pt}

{\bf Chain Integrals involving $\dot G, G_F$}
\vspace{10pt}
\begin{eqnarray}
\int_0^1 du_2\ldots du_n
\dot G_{12}\dot G_{23}\ldots\dot G_{n(n+1)} &=&
2^n <u_1\mid {\partial}^{-n}_P \mid u_{n+1}>\nonumber\\
&=& -{2^n\over n!}B_n(\vert u_1-u_{n+1}\vert)
{\rm sign}^n(u_1-u_{n+1})
\label{chainmasterP}\\
\int_0^1 du_2\ldots du_n
{G_F}_{12}{G_F}_{23}\ldots{G_F}_{n(n+1)} &=&
2^n <u_1\mid {\partial}^{-n}_A \mid u_{n+1}>\nonumber\\
&=& {2^{n-1}\over{(n-1)!}}
E_{n-1}(\vert u_1-u_{n+1}\vert)
{\rm sign}^n(u_1-u_{n+1})
\non\\
\label{chainmasterA}
\end{eqnarray}

\no
Here $B_n(x)$ ($E_n(x)$) denotes the $n^{th}$ Bernoulli
(Euler) polynomial. The right hand sides can be rewritten
in term of $G, \dot G, G_F$:

\begin{eqnarray}
<1\mid {\partial}_P^{-1}\mid 2> &=& \half \dot G_{12}
\\
<1\mid {\partial}_P^{-2}\mid 2> &=& \half (G_{12}-\sixth )
\\
<1\mid {\partial}_P^{-3}\mid 2> &=& -{1\over 12} \dot G_{12}G_{12}
\\
<1\mid {\partial}_P^{-4}\mid 2> &=& -{1\over 24} (G_{12}^2 - {1\over 30})\\
<1\mid {\partial}_P^{-5}\mid 2> &=& 
{1\over 5!} \dot G_{12}(\half G_{12}^2 + \sixth G_{12})\\
<1\mid {\partial}_P^{-6}\mid 2> &=& {1\over 6!}
(G_{12}^3+\half G_{12}^2 - {1\over 42} )
\\
\label{firstboson}\nonumber
\end{eqnarray}

\begin{eqnarray}
<1\mid {\partial}_A^{-1}\mid 2> &=& \half {G_F}_{12}
\\
<1\mid {\partial}_A^{-2}\mid 2> &=& -\fourth {G_F}_{12}\dot G_{12}
\\
<1\mid {\partial}_A^{-3}\mid 2> &=& -\fourth {G_F}_{12}G_{12}
\\
<1\mid {\partial}_A^{-4}\mid 2> &=& {1\over 24}{G_F}_{12}
\dot G_{12} (G_{12} + \half )
\\
\label{firstfermion}\nonumber
\end{eqnarray}
etcetera.

\vfill\eject

{\bf Chain Integrals involving $G$}

\begin{eqnarray}
\int_0^1 du_2 \, G_{12}G_{23} &=& -\sixth G_{13}^2 + {1\over 30}\\
\int_0^1 du_2\int_0^1 du_3 \, G_{12}G_{23}G_{34} &=&
{1\over 90}(G_{14}^3 + \half G_{14}^2) + {1\over 180} - {6\over 7!}\\
\label{G-chains}\nonumber
\end{eqnarray}

{\bf Chain Integrals involving $\dot {\cal G}$}

\vspace{6pt}

\bear
\int_0^1 du\,\dot {\cal G}_{1u}\dot {\cal G}_{u2}
&=& -i{\partial\over \partial {\cal Z}}
\dot {\cal G}_{12}
=
-{\e^{-i{\cal Z}\dot G_{12}}\over \sin^2({\cal Z})}
\Bigl[
\cos({\cal Z})+i\dot G_{12}\sin({\cal Z})
\Bigr]
+
{1\over {\cal Z}^2}
\nonumber\\
\label{intcalGpcalGp}
\ear\no

\vspace{14pt}

{\bf Integrals appearing in the Calculation of the 2-Loop
Spinor QED $\beta$ - Function}

\begin{eqnarray}
\int_0^1 du_1\int_0^1 du_2 
(\dot G^2_{12} - G_{F12}^2 )
  &=& -{2\over 3}\\
\int_0^1 du_1 (G_{13}-G_{14})^2 &=& \third G_{34}^2
  \\
\int_0^1 du_1\int_0^1 du_2 (\dot G_{12}\dot G_{23}
\dot G_{34}\dot G_{41} -&&\nonumber\\
-G_{F12}G_{F23}G_{F34}G_{F41} )
&=& -{8\over 3} G_{34}^2 - {4\over 3}G_{34}\\
\int_0^1 du_1\int_0^1 du_2 
(\dot G_{13}\dot G_{32}\dot G_{24}\dot G_{41} 
-&&\nonumber\\
-G_{F13}G_{F32}G_{F24}G_{F41})
&=& 4G_{34}^2 + {8\over 3}G_{34} - {8\over 9}
\\
\label{somesuper}\nonumber
\end{eqnarray}

{\bf 2-Point Integrals}

\begin{eqnarray}
\int_0^1 du{[G(u,u_1)-G(u,u_2)]}^{2n}
&=&{G(u_1,u_2)^{2n}\over 2n +1}
\qquad (n\in{\bf N}) 
\\
\int_0^1 du \, {\rm exp}
\lbrace {c[G(u,u_1)-G(u,u_2)]}\rbrace
&=& {{\rm sinh}[cG(u_1,u_2)]\over
{cG(u_1,u_2)}}
\label{equmir}\\
\int_0^1 du\, \dot G^k(u_1,u)\dot G^l(u,u_2)
&=&
{k!l!\over 2}\sum_{i=0}^l
{\Bigl(1-(-1)^{k+l-i+1}\Bigr)\dot G_{12}^i
- \Bigl(1-(-1)^i\Bigr)\dot G_{12}^{k+l-i+1}
\over i!(k+l-i+1)!}\non\\
\label{GdottokGdottol}
\ear

\vfill\eject

{\bf 3-Point Integrals}

\begin{eqnarray}
\int_0^1 du\, 
\dot G(u,u_1)\dot G(u,u_2)\dot G(u,u_3) 
&=& -{2\over 3}
\lbrace
\dot G_{23}[G_{12}-G_{13}]
+\dot G_{31}[G_{23}-G_{21}]
+\dot G_{12}[G_{31}-G_{32}]
\rbrace\non\\
&=&
-\sixth
(\Gd_{12}-\Gd_{23})(\Gd_{23}-\Gd_{31})(\Gd_{31}-\Gd_{12})
\nonumber\\
\label{GdotGdotGdot}
\end{eqnarray}

{\bf $n$-Point Integrals}

\bear
\int_0^1 du\,
\e^{\sum_{i=1}^n c_i \dot G(u,u_i)}
&=&
{\sum_{i=1}^n \sinh (c_i)\,
\e^{\sum_{j=1}^n c_j \dot G_{ij}}
\over\sum_{i=1}^n c_i}
\non\\
\label{intexpdotGnpoint}
\ear

{\bf Miscellaneous Identities}

\begin{eqnarray}
{\dot G}^2_{ij} &=& 1-4G_{ij}\label{Gdotsquare}\\
\dot G_{12}+\dot G_{23}+\dot G_{31} &=&
-{G_F}_{12}{G_F}_{23}{G_F}_{31} 
\label{Gdotsum}\\
{d^n\over du_i^n}(G^n_{ij})&=&n! P_n(\dot G_{ij})
\label{legendre}
\end{eqnarray}
\noindent
where $P_n$ denotes the $n^{th}$ Legendre polynomial.

\end{appendix}
\end{document}